\documentclass{SVProc}
\usepackage{makeidx}  
\makeindex
\usepackage{url}

\usepackage{graphicx}
\usepackage{amsfonts,amsmath,amssymb}
\usepackage[english]{babel}
\usepackage[a4paper,margin=1.5in]{geometry}
\usepackage[some]{background}
\usepackage{framed}

\usepackage{amsmath,mathrsfs,wasysym}
\usepackage{textcomp, gensymb}
\usepackage[sectionbib]{chapterbib}
\usepackage{hyperref}
\usepackage{xcolor}
\definecolor{p-r}{RGB}{171, 40, 52}
\hypersetup{colorlinks=true, citecolor=p-r, linkcolor=p-r,urlcolor=p-r}
\newcommand{\calR}{{\cal R}}

\def\TDH {\emph{T-DH}\, }
\def\ATDH {\emph{AT-DH}\, }

\def\T{\mathcal{T}}
\newcommand{\f}{\frac}
\def\rmd{\rm d}
\def\lp{\ell_{\rm Pl}}

\def\d{{\rm d}}

\def\t{\tilde}
\def\h{\hat}
\def\uDelta{\underline{\Delta}}
\def\inot{i^\circ}
\newcommand*{\scri}{\ensuremath{\mathscr{I}}} 
\newcommand*{\scrip}{\ensuremath{\mathscr{I}^{+}}} 
\newcommand*{\scrim}{\ensuremath{\mathscr{I}^{-}}}

\definecolor{titlepagecolor}{RGB}{171, 40, 52}

\backgroundsetup{
scale=1,
angle=0,
opacity=1,
contents={\begin{tikzpicture}[remember picture,overlay]
 \path [fill=titlepagecolor] (current page.west)rectangle (current page.north east); 
 \draw [color=white, line width=2pt] (5,0)--(5,0.5\paperheight);
\end{tikzpicture}}
}

\makeatletter                   
\def\printauthor{%
    {\large \@author}}          
\makeatother

\author{%
    \begin{flushright}
    Editors:
    \end{flushright}
    Luca Buoninfante$^{1}$\\ Ra\'ul Carballo-Rubio$^{2}$\\ Vitor Cardoso$^{3,4}$\\ Francesco~Di~Filippo$^{5}$\\Astrid Eichhorn$^{6}$\\
    }

\begin{document}

\begin{titlepage}
\BgThispage
\newgeometry{left=1cm,right=6cm,bottom=2cm}
\vspace*{0.0\textheight}
\begin{figure}
\begin{center}
\hspace{-0.5cm}\includegraphics[width=0.42\textwidth]{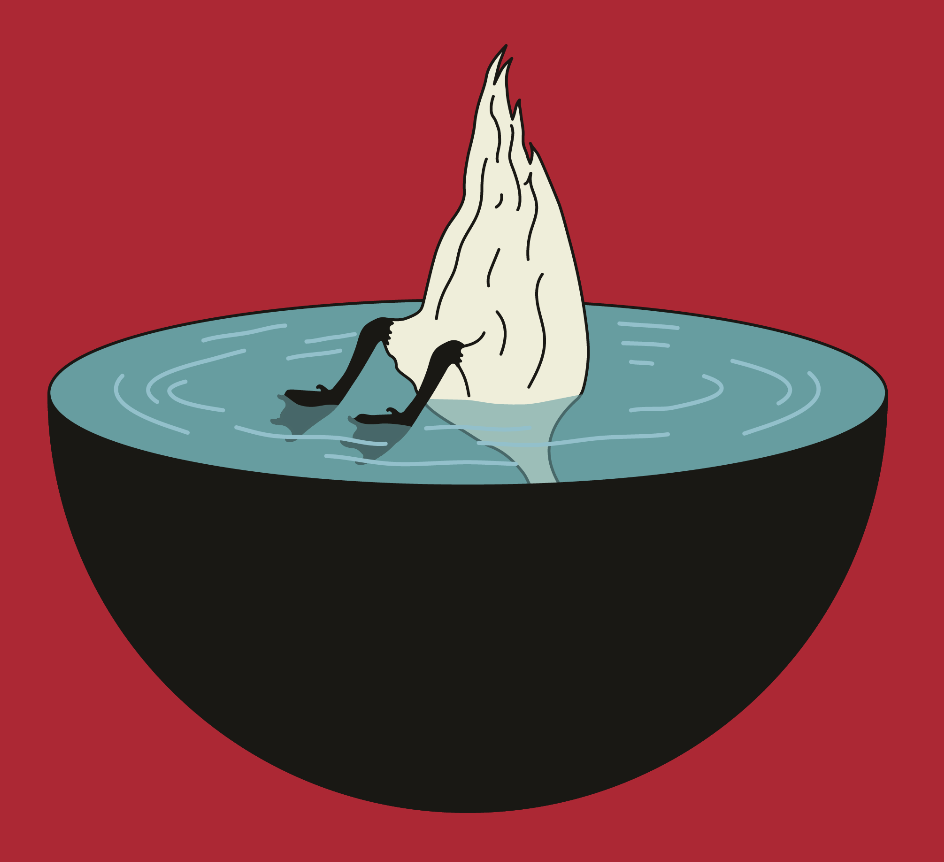}
\end{center}
\end{figure}
\vspace{1.2cm}
\begin{flushright}
\noindent
\textcolor{white}{\Huge\textbf{\textsf{Black Holes Inside and Out 2024}}}

\vspace{0.4cm}

\textcolor{white}{\LARGE\textbf{\textsf{Visions for the future of black hole physics}}}
\end{flushright}
\vspace*{7cm}\par
\noindent
\hspace{1.2cm}
\begin{minipage}{0.35\linewidth}
    \begin{flushright}
        \printauthor
    \end{flushright}
\end{minipage} \hspace{15pt}
\begin{minipage}{0.02\linewidth}
    \rule{1.5pt}{125pt}
\end{minipage} \hspace{20pt}
\begin{minipage}{0.63\linewidth}
\vspace{5pt}
\textbf{Niayesh Afshordi\inst{7,8,9}, Abhay Ashtekar\inst{10,9}, Enrico\linebreak Barausse\inst{11,12}, Emanuele Berti\inst{13}, Richard Brito\inst{4}, Gregorio Carullo\inst{3,14}, Mihalis Dafermos\inst{15,16}, \linebreak Mariafelicia De Laurentis\inst{17}, Adrian del Rio\inst{18}, Roberto Emparan\inst{19,20}, Ruth Gregory\inst{21,9}, Carlos A. R. Herdeiro\inst{22}, Jutta Kunz\inst{23}, Luis Lehner\inst{9}, Stefano Liberati\inst{11,12}, Samir D. Mathur\inst{24}, Samaya Nissanke\inst{25,26}, Paolo Pani\inst{27}, Alessia Platania\inst{3}, Frans Pretorius\inst{28}, Misao Sasaki\inst{29,30,31}, Paul Tiede\inst{32,33}, William G. Unruh\inst{34}, Matt Visser\inst{35}, Robert M. Wald\inst{36}}
\end{minipage}
\end{titlepage}
\restoregeometry

\frontmatter          
\pagestyle{headings}  
%
\chapter*{Preface}
The gravitational physics landscape has been evolving rapidly, driven in good part by our ability to study strong-field regions, in particular black holes. Black holes are the most amazing macroscopic objects in the universe. According to Einstein’s General Relativity, their exterior is extremely simple, while their interior holds the failure of the underlying theory. There are thus strong reasons to test black hole spacetimes, an endeavor made possible by gravitational-wave astronomy, infrared interferometry
and radio very large baseline interferometry. Black hole physics is an active field of research, which witnessed a number of important results in the last few years.

Black Holes Inside and Out (BHIO) 2024 gathered some of the world experts in black hole physics, numerical relativity, black hole perturbation theory, data analysis and astronomy to discuss the status of the field and prospects ahead. The meeting took place at the iconic Black Diamond building from August 26 to August 30, 2024. With about 250 on-site participants and 200 online participants, it was the largest conference devoted to the study of black holes up to date, and the second in a series that started with a fully online edition in 2021.

All 32 invited speakers and round table moderators were asked for a “vision” document. No style constraints were imposed on these documents, with the aim of offering the authors the possibility of freely expressing their thoughts about the state of the art and its possible future evolution, and maybe adding a personal touch.

We are extremely grateful to all invited speakers and round table moderators for their commitment towards making BHIO 2024 a memorable event and contributing to ensuring its legacy through their participation in this collection. We hope that the ideas, arguments and perspectives presented here will be a source of inspiration for new and experienced researchers alike in the years to come.

\vspace{1cm}
\begin{flushright}\noindent
October 2024\hfill Luca Buoninfante\\
Ra\'ul Carballo-Rubio\\
Vitor Cardoso\\
Francesco Di Filippo\\
Astrid Eichhorn\\
\textit{SOC, BHIO 2024}
\end{flushright}
\chapter*{Organization}
BHIO 2024 was locally organized by the Niels Bohr Institute, University of Copenhagen.\\ Website hosted online at \url{https://strong-gr.com/black-holes-inside-and-out/}.
\section*{Scientific Organizing Commitee}
Luca Buoninfante {\small (Radboud University, Netherlands)}\\
Ra\'ul Carballo-Rubio {\small (Southern Denmark University, Denmark)}\\
Vitor Cardoso {\small (Niels Bohr Institute, Denmark \& Instituto Superior T\'ecnico, Portugal)}\\
Francesco Di Filippo {\small (Charles University, Czech Republic)}\\
Astrid Eichhorn {\small (Heidelberg University, Germany)}
\begin{multicols}{2}[\section*{Local Organizing Commitee}]
Vitor Cardoso\\
Gregorio Carullo\\
Yifan Chen\\
Julie de Molade\\
Jose Ezquiaga\\
Takuya Katagiri\\
David Pereñiguez\\
Luka Vujeva\end{multicols}
\begin{multicols}{2}[\section*{Invited Speakers}]
Abhay Ashtekar\\
Emanuele Berti\\
Richard Brito\\
Alessandra Buonanno\\
Gregorio Carullo\\
Mihalis Dafermos\\
Adrian del Rio\\
Roberto Emparan\\
Netta Engelhardt\\
Stefan Gillessen\\
Carlos Herdeiro\\
Samir Mathur\\
Ramesh Narayan\\
Alessia Platania\\
Eric Poisson\\
Frans Pretorius\\
Misao Sasaki\\
Paul Tiede\\
Will Unruh\\
Matt Visser\\
Bob Wald\\
Helvi Witek\end{multicols}
\begin{multicols}{2}[\section*{Invited Round Table Moderators}]
Niayesh Afshordi\\
Enrico Barausse\\
Mariafelicia de Laurentis\\
Ruth Gregory\\
Tanja Hinderer\\
Jutta Kunz\\
Luis Lehner\\
Stefano Liberati\\
Samaya Nissanke\\
Paolo Pani\end{multicols}
\section*{Sponsoring Institutions}
Niels Bohr Institute\\
Villum Fonden\\
European Research Council\\
Danish National Research Foundation

\chapter*{}
\definecolor{shadecolor}{RGB}{244,244,244}
\begin{figure}
\begin{shaded}
\begin{center}
\includegraphics[width=1.0\textwidth]{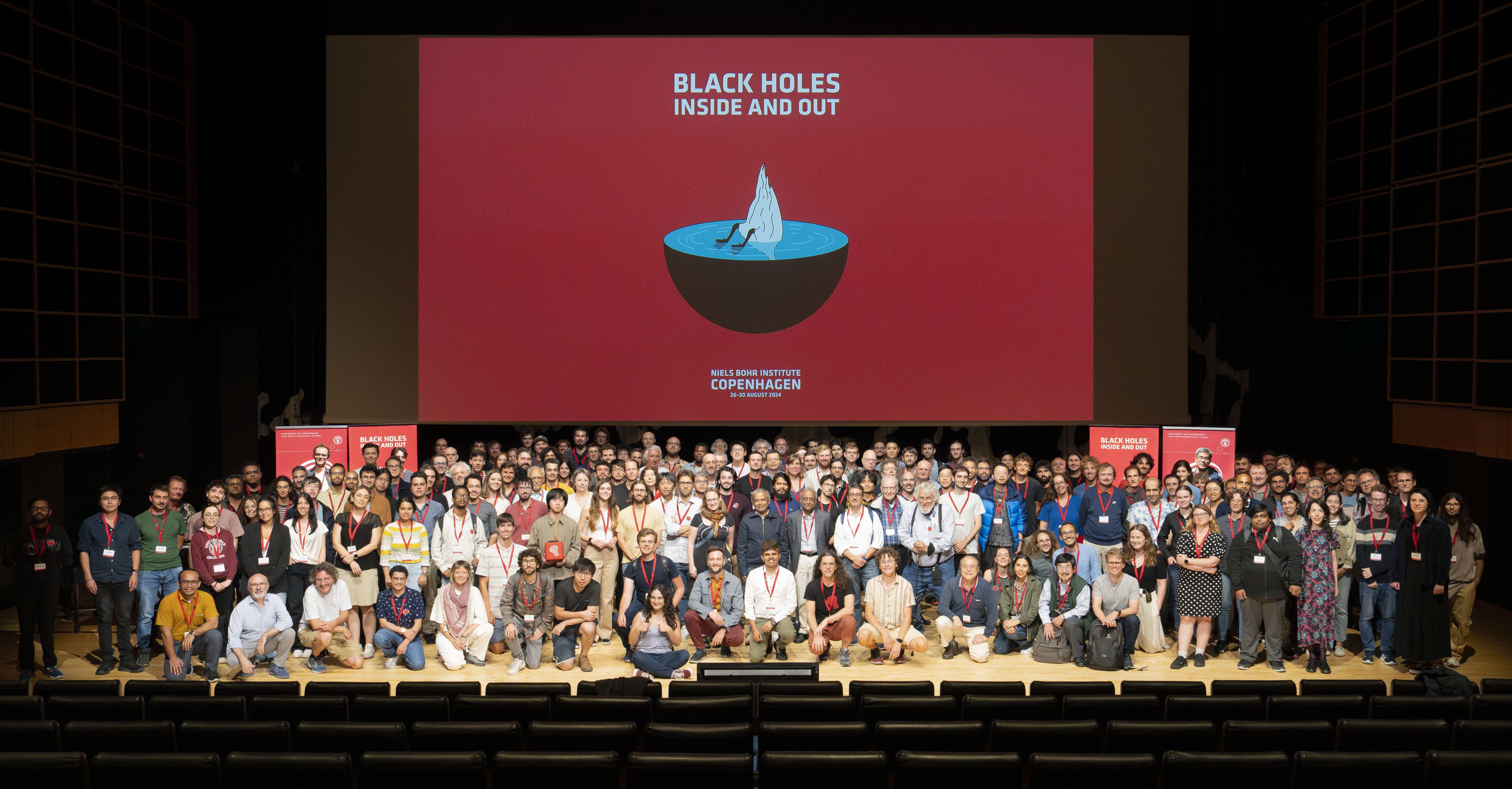}
\end{center}
\vspace{-0.5cm}
\begin{flushright}
{\textit{Group photo taken by artist Sjoerd Knibbeler in Queen’s Hall.}}
\end{flushright}
\vspace{-0.3cm}
\end{shaded}

\vspace{1cm}

\begin{shaded}
\vspace{-0.5cm}
\begin{center}
\includegraphics[width=0.4\textwidth]{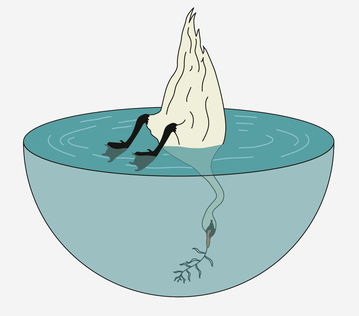}
\end{center}
\vspace{-0.6cm}
\begin{flushleft}
\textit{\textbf{About our mascot:}}
\end{flushleft}
\vspace{-0.6cm}
\begin{flushright}
{The Cygnus olor, commonly known as the Mute Swan is the national bird of Denmark. Despite their common name, they are not mute, but neither can they sing as may be implied by the ancient belief in the “Swan song”, according to which a swan will sing a beautiful song as it dies. To this day, no one has offered proof of the existence of a singing mute swan. Already in 77 A.D., Pliny the Elder set about investigating the matter and later wrote in his “Natural History” (Book 10, Chapter xxxii): “Observation shows that the story that the dying swan sings is false.”}

\vspace{0.2cm}
{Singing or not, its beauty has inspired artists and writers throughout history. Hans Christian Andersen’s “The Ugly Duckling”, which tells the story of an young ungainly cygnet who grows up to be a beautiful swan, is one of the most popular fairy tales of our time.}

\vspace{0.2cm}
{They are a constant and ubiquitous presence in Copenhagen. Voracious eaters, they can eat up to 3 kg of food per day. In the Copenhagen Lakes, they can often be seen upside down, as they submerge their neck and head to feast on the aquatic vegetation.}

\vspace{0.1cm}
\textit{Illustration and design: Ana Carvalho.}
\end{flushright}
\vspace{-0.4cm}
\end{shaded}
\end{figure}
\tableofcontents
\mainmatter              

\part{Classical black holes}
%
\title{Black Hole Evaporation -- 50 Years}
\author{William G. Unruh}
\institute{\textit{Department of Physics and Astronomy, University of British Columbia, Vancouver, BC, Canada V6T 1Z1}}

\maketitle 

\begin{abstract}
Personal reflections (this is not a scholarly history but my own memories and work on this topic). I apologize beforehand to everyone whose work I do not mention. Note that there is lots of such work, much brilliant.

	[This document is a transcription of the slides used at the
	conference, and as a result is rather rough as a document.]
\end{abstract}

1967 -- I began my work on my PhD at Princeton. After my Generals I asked
Wheeler if I could work with him. This was a hayday for his group, about
9 graduate students [Robert Wald, Bei-Lok Hu, Claudio Teitelboim (now Bunster),
Jacob Bekenstein, Steve Fulling (Wightman's student), Demetrios Christodoulou
(Ruffini), Nial O'Murachadha (York), Cliff Rhoades, Brendan Godfry]
1 postdoc (Remo Ruffini), 2 assistant professors
(Karel Kuchar, Jimmy York) and a sabbatical visitor (Charlie Misner).

The key topic of all of almost all of their research was quantum mechanics and gravity.

Key people whose research really influenced me:\\
Steve Fulling -- gave infomal lectures on quantum field theory (he had translated Bugoliubov’s text
from Russian to English), worked on quantum field theory in curved spacetime (strongly influenced by
L. Parker’s work on quantum fields in cosmological spacetimes~\cite{Parker:1968mv}) and worked on
quantization of quantum fields in Minkowski spacetime and in ``Rindler"
spacetime (he named the latter)~\cite{Fulling:1972md}.

Demetrios Christodoulou -- showed that there was for Schwarzschild and Kerr, a quantity called the ``irreducible mass'', which had to increase if positive energy flowed through the horizon~\cite{Christodoulou:1970wf,Christodoulou:1971pcn}.
At the same time, and independently, Hawking showed that
the surface area of a sequence of cross-sections of the horizon
of a black hole had to increase if the energy-momentum flux
through the horizon was positive~\cite{Hawking:1972hy}.
Both acted like entropy (i.e. increased with time).

Meanwhile Wheeler had given Bekenstein the problem of what
happened to the second law of thermodynamics if entropy
disappeared down the black hole.

Bekenstein chose Hawking’s more general formulation as a
better analogy of entropy.
This led him and then Bardeen, Carter and Hawking to formulate
the ``Thermodynamics of Black holes”~\cite{Bardeen:1973gs}. It was silly because
there was no temperature. Black Holes absorb. They do not
emit anything.
Bob Geroch threw another spanner in the works. Bekenstein
had postulated a heat engine with black holes.
Fill a box with high entropy radiation, lower it slowly on a rope
toward the black hole horizon extracting its energy through
the tension on the rope.
However, by lowering to the horizon one could extract all of
its energy violating the 2nd law~\cite{Bekenstein:1972tm}.

Disaster!
Bekenstein’s solution~\cite{Bekenstein:1973ur}: You cannot lower the box all the way
to the horizon. There is another new law of physics which says that
the entropy to energy ratio of the contents of the box is limited
by the radial dimension of the box (he cited many many examples
of quantum fields in boxes for which this is true).
Hawking decided to get involved in the game of the behaviour of
quantum fields (scalar fields) in the vicinity of a Schwarzschild black hole formed by collapse~\cite{Hawking:1974rv,Hawking:1975vcx}.
I recall in early 1973 visiting a conference at Cambridge in the UK
(my girlfriend, had been left behind when I moved from Birkbeck
college as Penrose’s postdoc, to a Miller Fellowship at Berkeley, so I found every
excuse I could to go to the UK) and overhearing Don Page 
talking to Hawking about some calculation
about quantum fields near black holes that Hawking was doing.

In the meantime I was getting interested in Fulling’s thesis,
where he asked, but did not answer, the question as to whether
or not field quantization was the same in Minkowski and Rindler
spacetimes. I showed (to myself) that they were different, by
showing that by using the horizon in Rindler spacetime as an
initial value surface for the quantum field, that one corresponded
to using the parameter along the horizon to define
``positive frequency” for field quantization gave the usual
Minkowski quantization, while using the Killing parameter
gave Rindler quantization. I also showed the relation between
them. I did not notice that the relationship was thermal.

I wrote Fulling a letter outlining this result, but did not publish it
till 1976~\cite{Unruh:1976db}, much to Stephen’s disgust (and my future chagrin).
In early 1974, I went to the conference at the Rutherford labs
where Hawking first presented his results (he had sent out
preprints in late 1973, and I got one just before Christmas via
Abe Taub and Vince Moncrief, Taub’s postdoc). Just before the conference, I had showed
my paper on quantization in the Kerr metric to Sciama, and he told
me I could have 10 minutes in the conference to present it. Instead
I tried to present the Minkowski-Rindler results, and made a
complete hash of the presentation.

So, in early 1974 Hawking’s result that the quantization of a
scalar field around a star which collapses to a black hole would
produce a flux of radiation with temperature~\cite{Hawking:1974rv,Hawking:1975vcx}
\begin{equation}
   T= \frac{1}{8\pi M}\frac{\hbar c^3}{G k_B}.
\end{equation}
Well not quite -- because the black hole has an albedo to absorption and emission due to
the angular momentum of the field and the curvature of the
spacetime which filters out the low frequencies of the thermal spectrum,
but Hawking’s result is the temperature of the (redshift corrected) black hole boundary before that filtering.

In 1976 I published my paper, ``Notes on black hole evaporation”~\cite{Unruh:1976db}
in which I tried to clarify what was happening in the emission
process by the black hole, also using the affinea and Killing parameterisations of the horizon
to show that in the affine vacuum (which would be the
Minkowski vacuum in the flat spacetime, and is roughly the
state one would expect after the collapse of a star to a
black hole). An accelerated observer (i.e. someone held at
constant radius near a black hole, or at constant acceleration in flat
spacetime) would see itself surrounded by a fluid of particles
with a temperature proportional to the acceleration
\begin{equation}
   T= \frac{a}{2\pi}\frac{\hbar}{ck_B}.
\end{equation}

Near the black hole, the Hawking radiation looks just like this
acceleration radiation. Further away (greater than about 1.5
times the radius of the black hole) the Hawking radiation looks
like the radiation from a finite sized hot body with a low frequency non-zero albedo.
This led to my favourite paper~\cite{Unruh:1982ic} (written with Wald), because of one
of the references. I had always found
Bekenstein’s entropy-to-energy ratio argument problematic.
However, the acceleration radiation gave an
alternative argument for why black holes do not violate the second law. Near the black hole, as one lowers the
box, it finds itself in a fluid which exerts a buoyant force according
to a pre-print written by a Greek by name of Archimedes.
That buoyant force reduces the energy extracted in lowering
and saves the second law, without the need for new physics.

One of the questions which bothered me from the beginning~\cite{Unruh:1977ga}:
\textit{``Where are the particles in black hole evaporation created?”}\\ 

\textit{What is a particle?}\\
A quantum field is a field. It is defined everywhere and at all times.
A particle is something with a well defined location at any time
carrying a well defined energy and momentum.
Traditional approach is to quantize the amplitude of the modes of the field (like the plane wave modes) as harmonic oscillators and call the resultant harmonic discrete energy levels as particles.
\begin{equation}
    \phi(t,x)=\int d^3k\frac{a_k}{\sqrt{\left(2\pi\right)^32\omega}}e^{-i\omega t }e^{ik\cdot x}+\mbox{H.C.}
\end{equation}
But this is silly -- This is totally delocalised definition and does not represent a localised particle.\\
My attitude was: \emph{``A particle is what a particle detector detects”}\\
This is a localised definition since detectors (Geiger detector, cloud chamber, bubble chamber spark chamber...) are localised. In particular one can look near a black hole to see what is happening there and perhaps see the particles being created.

Wald and I~\cite{Unruh:1983ms} wrote -- \textit{``What happens when an accelerated
particle detector detects a particle?”}
One answer would be that in the Minkowski vacuum,
the detector does never  clicks--vacuum has no particles.
\emph{False:} In $1+1$ dimensions,
\begin{equation}
    \psi_{M\omega,k}=\frac{1}{\sqrt{(2\pi)2\omega}}e^{-i\omega t}\,;\qquad \omega>0\,,
\end{equation}
\begin{equation}
    \psi_{R\nu,k}=\frac{1}{\sqrt{(2\pi)2\nu\sinh{\nu\mathcal{M}}}}\left[e^{\nu\mathcal{M}/2}\Theta(z)e^{-i\nu(\hat{\tau}-\rho_+)}+e^{-\nu\mathcal{M}/2}\Theta(\hat{\tau}+\rho)e^{i\nu(\hat{\tau}-\rho_-)}\right]\,;\quad \forall \nu\,.
\end{equation}
Identical set of positive norm modes:
\begin{equation*}
    e^{i\Omega t};\,\qquad\text{Minkowski detector creation operator mode,}
\end{equation*}
\begin{equation*}
    e^{iN \tau};\,\qquad\text{Rindler detector creation operator.}
\end{equation*}
If an accelerated detector detects a particle the probability of the emitted particle being in the acausal region of the detection is much higher than in the region containing the detector.

This has led people to say that this represents a tunneling
process (like the Schwinger EM particle creation). The evidence is against this.

\begin{figure}[ht]
\centering
\includegraphics[width=0.7\linewidth]{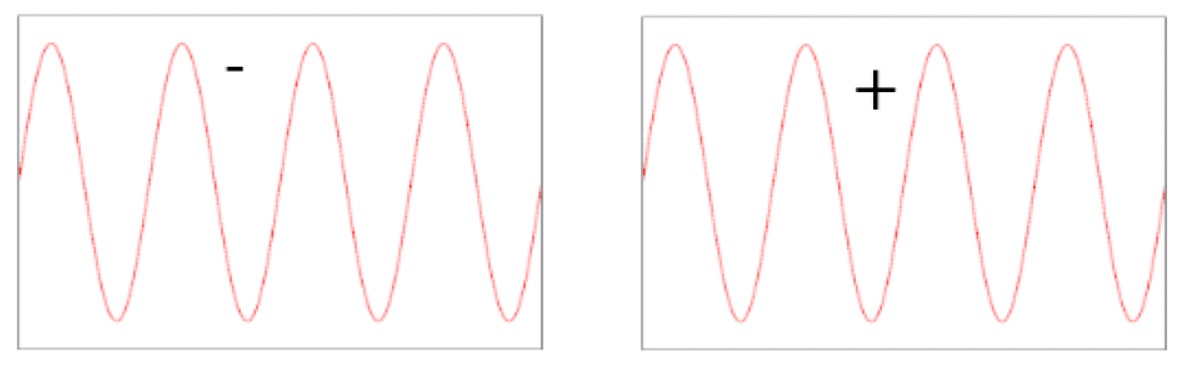}
\caption{Two impenetrable boxes with different vacua.}
\end{figure}

\emph{Inside affects outside?}\\
\emph{False:} Consider two impenetrable boxes, length $L$, inside coordinates $x_\pm $ with fields inside the boxes. Assume that the walls are completely impermeable to the fields. One can have the usual plane wave modes to define two separatee vacuum state, one in each box.
A static detector will see nothing.

However one can put the field into another gaussian pure state in which, using a Bogoliubov two mode set of squeezed states for each frequency in the two boxes, one gets a very different result: 
\begin{equation}
    \phi=a_n\frac{e^{-i\omega_nt}\sin{n\pi x_\pm/L}}{\sqrt{2\pi L2\omega_{in}}}+\mbox{H.C.}\,
\end{equation}
\begin{equation}
    a_{n\pm}\left|0_M\right>=0\,,\qquad \omega_n>0\,.
\end{equation}
This is a squeezed state Rindler analog. Now a detector at rest placed into either box will be thermally excited, exactly like an accelerated detector at rest in Rindler coordinates is thermally excited:
\begin{equation}
    b_{n+}=\cosh{\theta_n}a_{n+}+\sinh{\theta_n}a_{n-}^\dagger\,,\qquad \sinh{\theta_n}=\frac{e^{-\omega_n/4T}}{\sqrt{\sinh{\omega_n/2T}}}
\end{equation}
\begin{equation}
    b_{n\pm}\left|0_R\right>=0\,.
\end{equation}

If the detector is placed into the usual vacuum will not be excited. If the detector is placed into squeezed vacuum, it can be excited while emitting a particle which is dominantly in the
opposite box. E.g., if the detector is in the $+$ box and the detector frequency is much larger than the temperature $T$, then the probability is exponentially larger that the particle is in the $-$ box than in the $+$ box. This is precisely the situation (except for boundaries) for and accelerated detector in the Minkowski vacuum.

\begin{figure}[ht]
\centering
\includegraphics[width=0.35\linewidth]{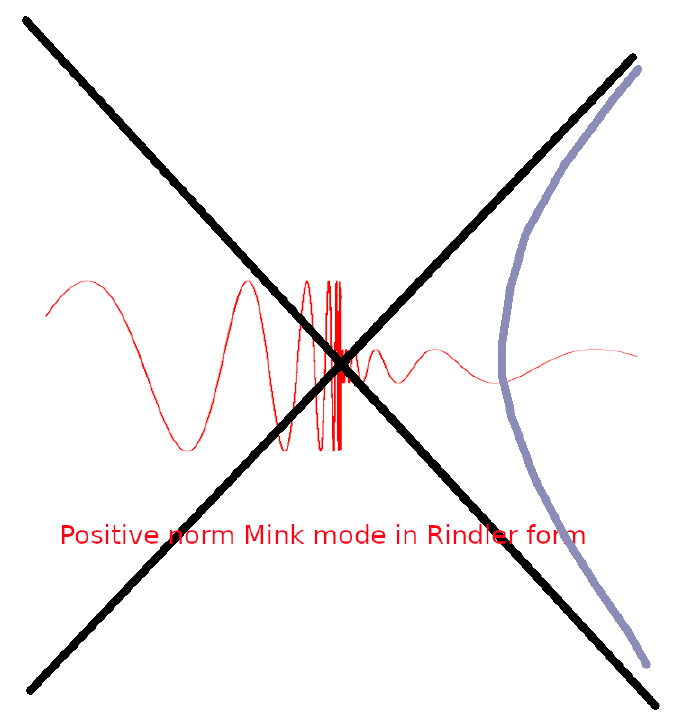}
\caption{Representation of a Minkowski mode on the Rindler wedge.}
\end{figure}

This also applies to a black hole spacetime, where the
state after collapse is close (except at exponentially low
final frequencies) to the Minkowski vacuum.
If your detector clicks and thus detects a particle, the
probability is much higher that the particle is inside the
horizon if the detector is near the horizon.\\

\textit{Boundary condition on singularity?}\\
To solve what is called the information paradox one needs to solve a problem. The information that went into the black hole, flows out of the black hole eventually (somehow). But it also flows into the singularity. This clones the inflowing information into to both outside the black hole and into the singularity. However one cannot clone a quantum state. Proposed solution: place boundary condition on the singularity.

However quantum mechanics is not classical mechanics. Future boundary conditions are not equivalent to past conditions. The predictions of quantum theory become very strange. 

The simplest example is the following. Consider a two level system (e.g. a 
spin $1/2$ particle). At 9AM, the spin is measured in the x direction and found to be $+1/2$. At 11 it is measured in the y direction and found to be $+1/2$. At 10 a student snuck into the lab and measured it in a direction between $x$ and $y$. What is the probability that she measured $+1/2$? Quantum mechanics answers this simply. It is a probability distribution such that if she measured it in the $x$ direction, the probability must be unity. If she measured it in the y direction, the probability is also $1$. Other angles give probabilities given by the following graph. Note that is no state, or density matrix which gives this resultant distribution, and yet it is almost trivially calculated by quantum mechanics.

\begin{figure}[ht]
\centering
\includegraphics[width=0.35\linewidth]{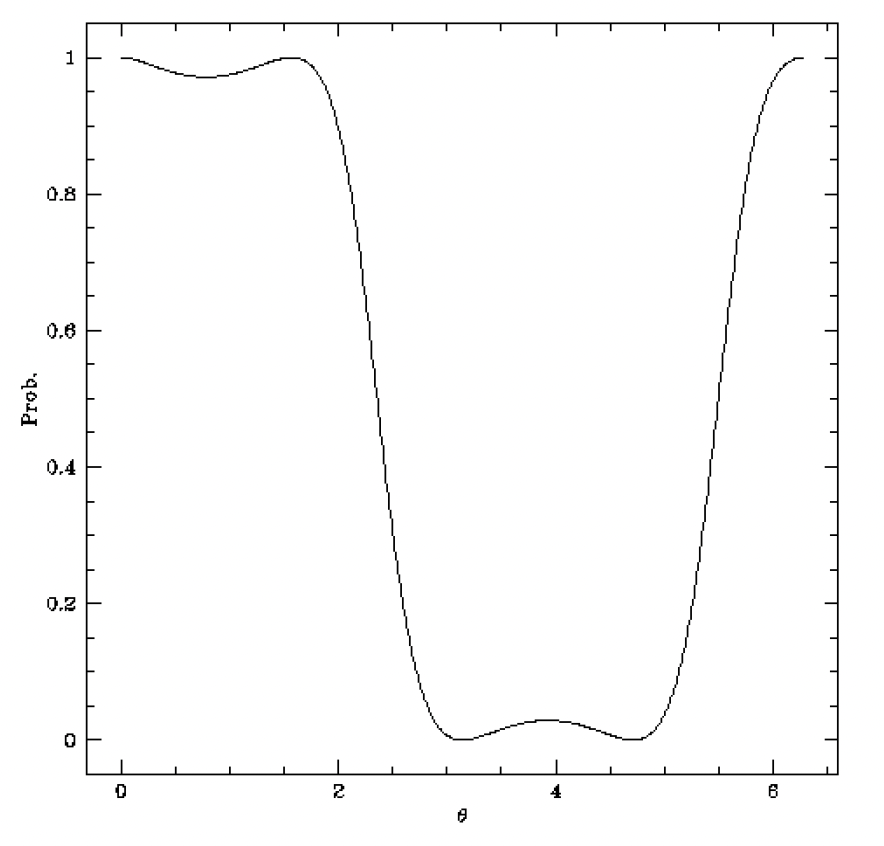}
\caption{Probability distribution of finding the spin in direction $\theta$ to be $1/2$ given initial and final knowledge of spin value in $x$ and $y$ directions.}
\end{figure}

Yakir Aharonov and his collaborators have looked at the strange results standard quantum mechanics gives for such initial and final state. As an example, let us take a spin 12.5 particle and sets the condition that the initial $x$ and final $y$ components are 12.5. At the intermediate time, measure the spin in the $(x+y)/\sqrt{2}$ direction. If one uses a measuring device which is inaccurate --for example which has a gaussian initial distribution of width $\sigma$-- then the distribution for $\sigma$ is a bunch of delta functions at 12.5,11.5,10.5,...-12.5.
As $\sigma$ becomes large however, the peaks in the probability distribution become fewer, and more to higher values. By the time $\sigma$ becomes 4, that is only one single peak, of width 4, and centered on a value of about 18 ($\sqrt{2}\cdot 12.5$). This is a value substantially larger than the maximum value of the spin in any direction (12.5)

\begin{figure}[ht]
\centering
\includegraphics[width=.85\linewidth]{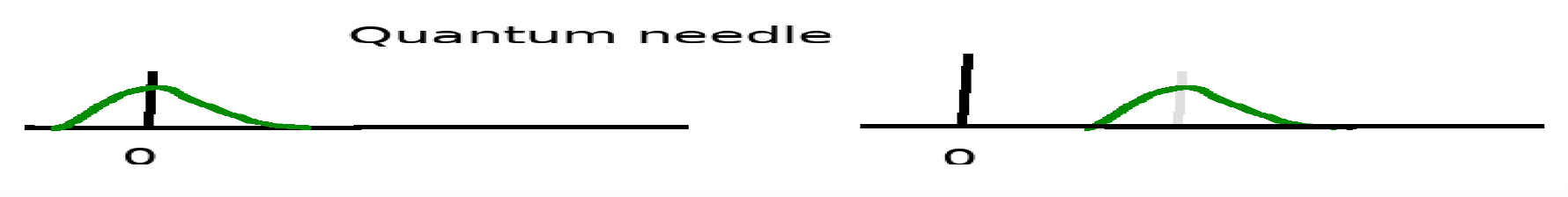}
\caption{Probability of measuring values of spin 12.5 particle if initial  x and final y values were  12.5 and intermediate measurements had uncertainty $\sigma$.}
\end{figure}

Hence, the behaviour of the quantum universe outside the horizon would be substantially different than its behaviour under quantum mechanics with initial conditions only.\\

\textit{Where are particles created?}\\
The notion of particle depends on state of motion of particle detector.
Particles are carriers of energy.
In 1974, Davies, Fulling and I~\cite{Davies:1976ei} calculated the massless field energy-momentum tensor in $(1+1)$ dimensions.
Conformal anomaly --the non-zerotrace of the tensor is proportional to curvature
$R$-- acts as a source for the flux of energy in 1+1 dimensions. In writing the paper I pushed for interpretation of the curvature  $R$ as the source of energy flux. It was an argument which did not catch on. So let me try again.\\

\begin{figure} 
  \begin{center}
     \begin{minipage}{1.5in}
      \begin{center}
       \includegraphics[width=1.7in,height=1.9in]{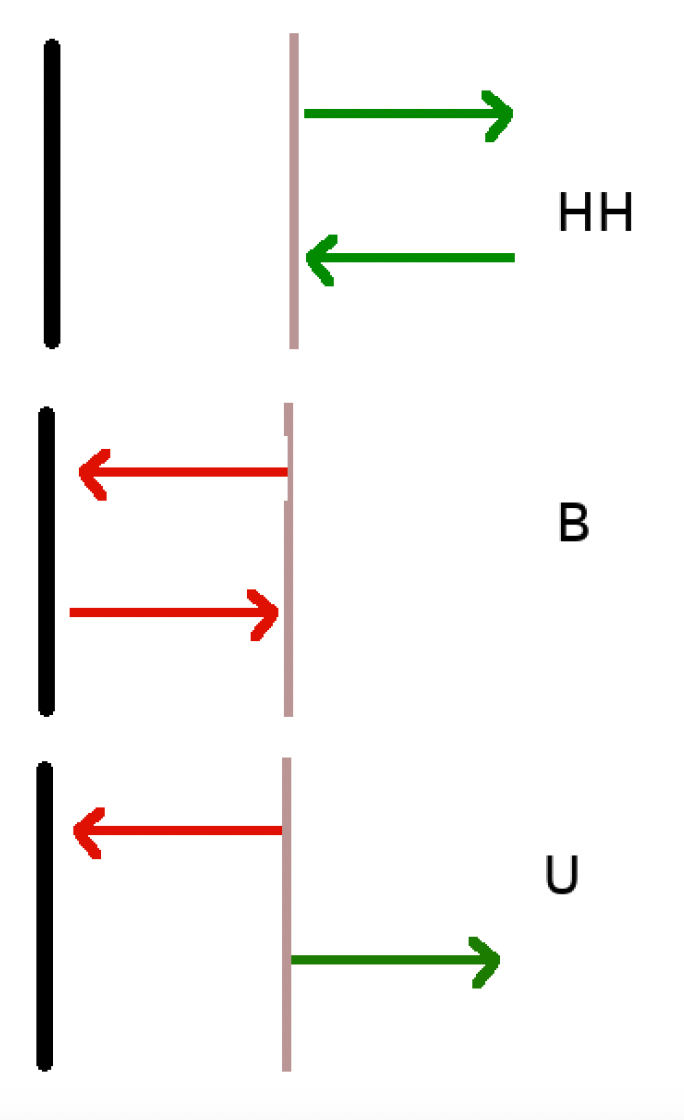}\\ (a)
         \end{center}
    \end{minipage}
   \hspace{.5in}
    \begin{minipage}{1.7in}
      \begin{center}
       \includegraphics[width=1.7in,height=1.7in]{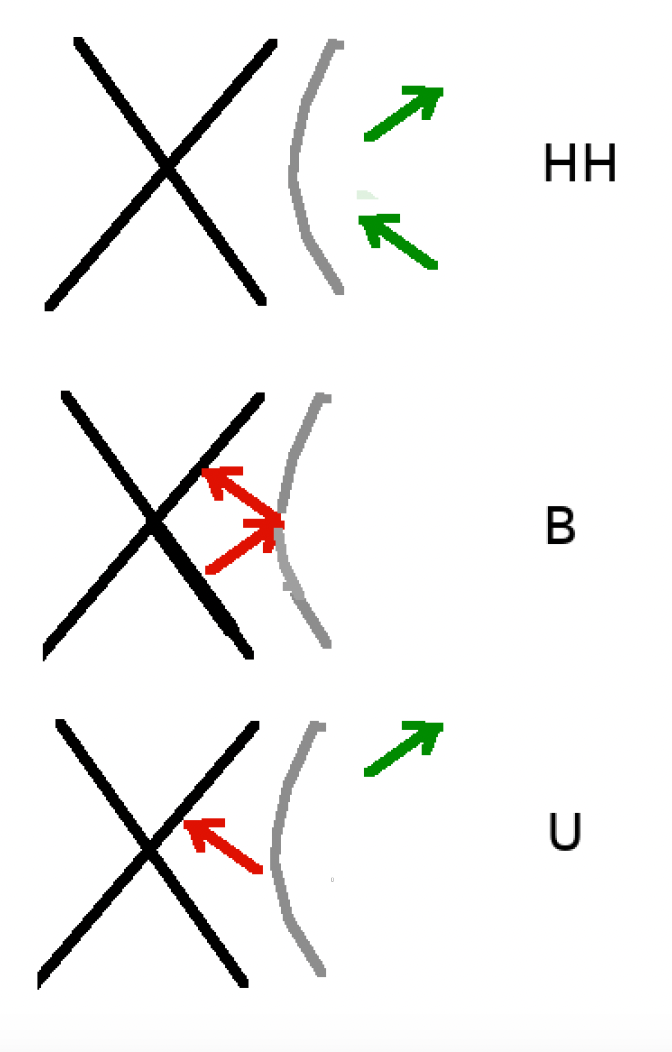} \\ {(b)}
        \end{center} 
   \end{minipage}
\caption{Non-zero average energy fluxes. Thermal fluxes are depicted in green if positive and in red if negative. Left figure shows a spatial representation while right figure shows a null representation. There are different states. \textit{Hartle Hawking (HH):} No fluxes across horizons, near horizon like Minkowski vacuum. \textit{Boulware (B):} No fluxes across
infinity, near infinity like Minkowski vacuum. \textit{Unruh (U):} No fluxes across past horizon or past infinity, like state from collapsing matter forming black hole.}\label{fig:bhemission}
\end{center}
\end{figure}

\textit{A Model of black hole emission:}\\
The model will be that of a two flat spacetimes sewn together along a timelike line which is an accelerated curve in one spacetime, and is a timelike geodesic as seen from the other spacetime. This junction lies along the coordinate  $r=1$, where there is a $\delta$-function curvature (see Fig.~\ref{fig:bhemission}):
\begin{equation}
    ds^2=\begin{cases}
        r^2dt^2-dr^2\,\qquad r<1\\
        dt^2-dr^2\,\qquad r>1
    \end{cases}
\end{equation}
\begin{equation}
    \begin{cases}
        \partial_t^2\phi-r\partial\left(r\partial_r\phi\right)=0\,\qquad r<1\\
         \partial_t^2\phi- \partial_r^2\phi=0\,\qquad r>1
    \end{cases}
\end{equation}

In one section, we take $(t,r)$ as rindler coordinates, while in the other they are Minkowski coordinates. The Rindler coordinates have a horizon at $r=0$, and are flat for $r>1$. One can now choose a number of quantizations of the scalar field in these coordinates, corresponding to ``vacuums” in eternal Schwartzschild -- The Hartle Hawking (HH) state, the Boulware state (B), and the Unruh (U) state
The energy fluxes in these states in this model take the form as in Fig.~\ref{fig:bhemission}.

This is also true of Dirac field.
Is this also true of massive field in 1+1, or of fields in 3+1
in which the inner metric would be a cylindrical Rindler spacetime, and the outer would be flat Minkowski spacetime in polar coordinates?

My suspicion is yes. But there are many ways I could be wrong.\\

\textit{Planck scale problem:}\\
Thermal radiation comes from exponentially super-Planckian
scale vacuum fluctuations (1 second after solar mass black hole
emission comes from amplification of $10^{10^5}\mbox{Hz}$). I could argue (and I have) that adiabaticity explains why the super-Planckian origin of the radiation has no effect. 
Analog gravity calculations and experiments support this
argument~\cite{Weinfurtner:2010nu,Steinhauer:2015saa}. But can we find a simple derivation of Hawking
radiation which shows this?
In black hole thermodynamics the only ``reliable” calculation is that which associates a temperature with black holes. But most people worry about entropy, not temperature. There is no evidence that temperature has anything
to do with Planck scale physics, so why would entropy? It would be wonderful to get a derivation of the back-hole emission which did not rely on super Planckian modes.\\

\textit{Experimental black-hole temperature:}\\
In 1980 I suggested that it might be possible to carry out experiments in the lab to see the analog of the quantum instability of black holes. Although slow, this has become a sizeable area of research. The history of these analogs runs something like the following:
\vspace{-0.7cm}
\begin{itemize}
\item[1972:] Oxford  -- Sonic analog of black holes to give physical feel for black holes.
\item[1980:]PRL ``Experimental Black Hole Evaporation?”~\cite{Unruh:1980cg}.
I tried to call them ``dumb holes"~\cite{Unruh:2008zz} – but that name  failed.\\
This field is now called ``Analog Gravity"~\cite{Barcelo:2005fc}.
\item[1990:] Jacobson – Quantization of sound wave solutions of the inviscid fluid equations ~\cite{Jacobson:1996zs}.
\item[1995:]Numerical solution shows that high-frequency dispersion
does not alter the low-frequency results~\cite{Unruh:1994je}.
\item[Late 1990:] With Sch{\"u}tzhold and others -- many examples of possible experimental realizations~\cite{Schutzhold:2002rf}.
\item[2010:]Weinfurtner \emph{et al.}, stimulated emission of Hawking radiation~\cite{Weinfurtner:2010nu}.
\item[2015-2020:]Quantized soundwaves in BEC -- demonstration of
trans horizon entanglement and temperature by Steinhauer~\cite{Steinhauer:2014dra,Steinhauer:2015saa,MunozdeNova:2018fxv,Kolobov:2019qfs}.
\item[2020:] Proposed measurement of  acceleration temperature with bifrequency laser microphone, Weinfurtner \emph{et al.}~\cite{Gooding:2020scc}.
\end{itemize}

\textit{String Theory?}\\
Of course another of the developments has been the attempt to unify string theory with black holes, e.g., AdS/CFT, claiming
black hole evaporation is reversible (a black hole remembers its
origins). This is often called the resolution to the ``Information Paradox”.

I am still far from convinced that there is any such paradox, or that a solution is needed or is solved by quantum gravity. It is possible, but more probably is that black hole entropy is fundamental (emitted uncoordinated
with original). But I do not have the time to go into this.

What IS clear is that black hole quantum mechanics is still a far from finished topic.
50 years after that initial discovery by Hawking, which we are celebrating here,   there is still a massive amount that we do not know or understand about black hole quantum mechanics.

\bibliographystyle{utphys}
\bibliography{references}

\title{The Stability Problem for Extremal Black Holes}
\author{Mihalis Dafermos}

\institute{\textit{Princeton University, Department of Mathematics, Fine Hall, Washington Road, Princeton~NJ 08544, United States of America} \and \textit{University of Cambridge, Department of Pure Mathematics and Mathematical
Statistics, Wilberforce Road, Cambridge CB3 0WA, United Kingdom}}

\maketitle

\begin{abstract}
I present a  series of conjectures aiming to describe the general dynamics of the Einstein equations of classical general relativity in the vicinity of extremal black holes. I will reflect upon how
these conjectures transcend older paradigms concerning extremality and near-extremality, in particular,
the so-called ``third law of black hole thermodynamics'', which viewed  extremality as an unattainable limit, and the ``overspinning/overcharging'' scenarios, which viewed  extremality as a harbinger of naked singularities. 
Finally, I will outline some of
 the difficulties in proving these conjectures and speculate on what it could mean if
the conjectures turn out not  to be true.
\end{abstract}

\section{Introduction}
General relativists have long made their peace with ``normal'' black holes, even highly spinning ones.  
\emph{Extremal black holes}, on  the other hand,
remain  a source of  extreme uneasyness. Though their name arises from the fact that they are extreme in the literal sense of the word, occupying the extreme end of the allowed  parameter space, they turn out to be so also in the figurative sense, exhibiting  properties ``extreme'' even by the standards of black hole physics, in particular \emph{instabilities}. Indeed, both these two facets of the word ``extreme'' will be essential to the story I want to tell, and in fact it is precisely the coexistence of the two which is responsible for much of the story's rich complication---and the surrounding confusion. For, in this author's view at least, {\bf \emph{there is no object more misunderstood in classical general relativity than extremal black holes!}}

In this short essay, accompanying a talk given at the ``Black holes:~inside and out'' conference held in August 2024 in Copenhagen, I will describe some recent progress which has conditioned my 
own---still provisional!---expectations for the general dynamics of the Einstein equations in a neighbourhood of extremal black holes, and I will try to organise these in a set of precisely stated conjectures. I emphasise that the conjectures may or may not be true (hence the word conjecture!), but they represent a mix of reasonable extrapolation from recently proven theorems with a dose of sheer wishful thinking. In particular, I will also discuss the various ways in which the conjectures may turn out to be false, a scenario which would be even more interesting but also more complex than anything previously entertained.  Nonetheless, I will  stick my head out and commit the conjectures to paper, because having a definite goal to prove or disprove provides a useful starting point for further rational study.  If they do in fact turn out to  be false, so much the better!

To set the ``ground rules'' of my discussion, let me say at the outset that I will remain firmly within the confines of \emph{classical} general relativity,
indeed, much of what I say will  concern the Einstein \emph{vacuum} equations alone, 
though sometimes it will be useful to also invoke well-established classical matter models, 
like the Einstein--Maxwell equations governing electrovacuum. I will assume moreover that
my reader
is familiar with the basic black hole solutions of these equations---those of
Schwarzschild, Reissner--Nordstr\"om,
Kerr and Kerr--Newman, including their extremal cases---and will refer to their standard properties without comment and without ever explicitly writing down the metric. 
I will also assume the reader has at least nodding familiarity with the fundamental principle that
classical general relativity can be understood \emph{dynamically}, i.e.~it has a well-posed initial value 
problem, where an appropriate notion of Cauchy (or alternatively, characteristic) initial data gives rise to a unique solution of the equations of motion.
For all this, the reader may refer to standard textbooks, for instance~\cite{wald2010general}.

Of course, extremal black holes are important in considerations connected to \emph{quantum} 
gravity.  
Indeed, it is common in interactions with my high energy physics colleagues that even before 
you finish your sentence they are conjuring up (in real time!)~quantum effects that will change  whatever classical story you were about to tell them.  Let me appeal only for a little bit of the reader's patience!
I have no doubt that other essays in this collection, reflecting the wide spectrum of talks at the Copenhagen conference, will dedicate ample space for quantum speculations of various sorts. My purpose here is simply to get the classical story right. 

In a similar vein, I will also stay clear from any discussion of extremal or near extremal black holes  in real astrophysical environments. Though these of course lie well within the domain of classical general relativity, astrophysics is messy and complicated, whereas the Einstein vacuum equations are clean and simple(r), and much more amenable to mathematical analysis.  When it comes to issues of principle, the point of view of the present essay is: 
the cleaner, the better! I will thus for now leave aside speculation as to what implications the conjectural picture for the vacuum   to be described here would have for more realistic astrophysical settings.

Though I hope to eventually convince the reader that the conjectures stated here are reasonable, indeed in some sense inevitable as the ``minimalist'' statements to hope for, I want to emphasise at the outset  that they in fact would completely contradict
two paradigms of  how to think about extremality which have dominated the literature, what I will call the   \emph{third law paradigm} and the \emph{overspinning/overcharging paradigm}. Indeed, I believe  that these two---false, in my view!---paradigms have been the biggest hindrance to understanding the problem of extremality, and---\emph{independently of the truth of the conjectures I will propose in their place}---in order to make further progress we must transcend both these paradigms.  Thus, I  will in fact begin my discussion already in Section~\ref{received} below from a description of these, 
followed in Section~\ref{critical} by a critical analysis.

Finally, though the present reflections are in my own words (and thus any errors or unfortunate formulations here are mine alone!), the conjectured picture which I will attempt to sketch  is  very much influenced by the work of others (especially some surprising recent developments to be described in more detail below), and my own understanding  arose in the context of many years of  discussion, indeed sometimes vigorous debate, and collaboration. Let me mention, in addition to my own teacher Demetrios Christodoulou, especially the influence of Yannis Angelopoulos, Stefanos Aretakis, Dejan Gajic,  Christoph Kehle, Jonathan Luk, Frans Pretorius, Harvey Reall,  Rita Teixeira da Costa, Ryan Unger, Bob Wald and Claude Warnick, all of who have worked directly on aspects of this problem.
I am of course greatly  indebted to my collaborators Gustav Holzegel, Igor Rodnianski and Martin Taylor and have adapted Conjectures~\ref{lcodimstabext} and~\ref{phasespace} from the discussion in Section~IV.2 of our joint 
paper~\cite{dhrtplus}.

\section{The two received paradigms}
\label{received}

The study of extremal black holes over the last fifty years has been largely shaped by two distinct paradigms which---implicitly and explicitly---have dominated the way these objects are discussed in the literature.
\emph{I believe that both these paradigms are wrong.}
In order to explain my expectations for the actual dynamics of the Einstein equations in a neighbourhood of extremal black holes, the first order of business is to describe these paradigms.

\subsection{The third law paradigm: \emph{extremal black holes as the unattainable}}
\label{TLP}

The first paradigm, which I will dub the ``third law paradigm'', goes back to the seminal paper~\cite{bardeen1973four} 
of Bardeen, Carter and Hawking which explicitly initiated the thermodynamic analogy in black hole physics. I will not discuss here the history of this analogy (the first hints of which go back to Christodoulou's work~\cite{christodoulou1970reversible}) or its incredible successes. I will focus entirely on the final part of the analogy they propose, what they call the third law of black hole mechanics, also known as the third law of black hole \emph{thermodynamics}.

I recall of course that in the analogy with classical thermodynamics, \emph{temperature} corresponds to \emph{surface gravity}, which vanishes in the extremal case. Motivated by the ``unattainability'' formulation of
the third law of thermodynamics, the authors of~\cite{bardeen1973four}  put forth a  conjectured ``third law'' for black holes, which 
in their own words read:
\begin{quotation}
\emph{ ``It is impossible, by any procedure, 
no matter how idealized, to reduce [the surface gravity] $\kappa$  to zero in a finite sequence of operations.''}
\end{quotation}
The law was given a more precise formulation by Israel~\cite{israel1986third}, 
who interpreted ``finite sequence of operations'' to mean ``finite affine time along the event horizon'', added the stipulation that the matter model be reasonable, and, importantly, the condition of regularity, because without the latter, there were already counterexamples, in fact, counterexamples he himself had constructed! Indeed, Israel seems to have explicitly related the existence of his ``counterexamples'' to failure of regularity---a point to which I will return in Section~\ref{deathsection}.

The interesting thermodynamic analogy aside, at first glance, the question of the ``third law'' per se, namely, whether one can produce an exactly extremal black hole in finite time, may seem like an academic one.  
After all, even if one could do this, it 
is clearly something very exceptional (hence the explicit emphasis ``no matter how idealized'' in the formulation!), not only because one must obtain exact extremality, but one then must keep the black hole at extremality for infinite time thereafter.
The point, however, is that the analogous procedure---exceptional  and idealised though it may be---is indeed theoretically realisable in the subextremal case, as one can see from elementary examples.   
The significance of the unattainability of extremality in finite time is thus best understood \emph{\underline{relative} to the subextremal (fixed spin-to-mass or charge-to-mass ratio) case.} I
shall return to this point in Section~\ref{generalisedparadigm} where~I introduce the ``generalised
third law paradigm''.

\subsection{The overspinning/overcharging paradigm: \emph{extremal black holes as the harbinger of super-extremal naked singularities}}

The second paradigm I will discuss can also be said to originate from a quote from Bardeen, Carter and Hawking's paper~\cite{bardeen1973four}:
\begin{quotation}
\emph{``Another reason for believing the third law is that if one could reduce $\kappa$ to zero by a finite sequence of operations, then presumably one could carry the process further, thereby creating a naked singularity.''}
\end{quotation}
Wald was the first to entertain this possibility in his paper~\cite{wald1974gedanken}. What Wald showed, however,
was that  neither overspinning nor overcharging were possible in the ``test particle'' approximation.

\begin{figure}[ht]
\centering
\includegraphics[width=0.4\linewidth]{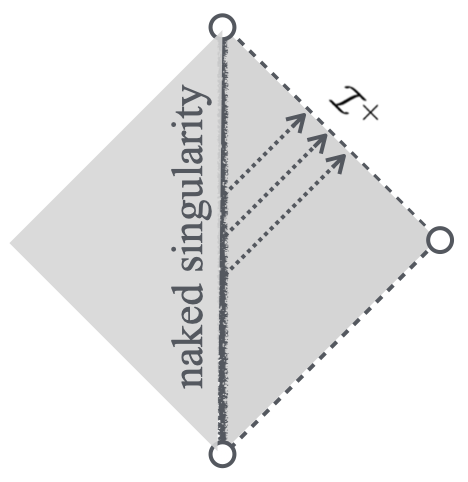}
\caption{The Penrose diagram of superextremal Kerr ($|a|>M$).
 \label{fig:naked}}
\end{figure}

Let us note that the idea that overspinning or overcharging a black hole has anything in principle to do with 
 naked singularities in the first place arises from the fact that in the global Penrose diagram of
\emph{super-extremal} Kerr or Reissner--Nordstr\"om, there is indeed a naked singularity as
in Fig.~\ref{fig:naked}.
We will revisit this implicit connection later on in Section~\ref{nearextrem}.

The story could have ended with Wald's work, but
the idea of overspinning/overcharging proved too tempting to be given up so easily!
The idea was revived in~\cite{PhysRevD.59.064013, jacobson2009overspinning}, and 
more recently has
 been taken up again and again by many authors, the scenarios becoming more and more elaborate and the approximations used harder and harder to make sense of.  I will not try to give a survey
 of these works, but distil from these what I will dub 
 the ``overspinning/overcharging paradigm'',  namely:
\begin{quotation}
\emph{Near-extremal dynamics can indeed lead to the formation of a naked singularity
as in super-extremal Kerr or Reissner--Nordstr\"om via an overspinning or overcharging mechanism}.
\end{quotation}

 Of course,  the formation of naked singularities---at least if their ``nakedness'' is moreover  stable to perturbation--- would contradict another well-known conjecture, Penrose's \emph{weak cosmic censorship} conjecture~\cite{penrose1968battelle}.  
Some overspinning/overcharging papers embrace the prospect of falsifying weak cosmic censorship, whereas others present a failed attempt at overspinning/overcharging as more evidence for the validity of the conjecture.

\section{Questioning the paradigms?}
\label{critical}

In this section, I will attempt a first critical analysis of the two paradigms. 
Let me first dispose of  the third law, for which  a very  definitive  theorem can now be stated:

\subsection{The death of the third law}
\label{deathsection}

There is  in fact not much to say here, other than quote directly a remarkable recent  theorem of Kehle and Unger:
\begin{theorem}[Kehle--Unger~\cite{kehle2024gravitationalcollapseextremalblack}]
\label{kehleungertheorem}
There exist regular one-ended Cauchy data for the Einstein–Maxwell-charged scalar field system which undergo gravitational collapse and form an exactly Schwarzschild apparent horizon, only for the spacetime to form an exactly extremal Reissner–Nordstr\"om event horizon at a later advanced time.
\end{theorem}

Thus, initially subextremal black holes can become extremal in finite time after all, 
evolving from regular initial data. The matter model is reasonable by all measures
(there is now in fact a similar construction for the Einstein--Maxwell--charged Vlasov system~\cite{KU23}),
and the process is completely regular, meeting all of the requirements of~\cite{israel1986third}.
{\bf The “third law of black hole thermodynamics”, as formulated in}~\cite{bardeen1973four, israel1986third}{\bf,  is simply false!}

\begin{figure}[ht]
\centering
\vspace*{-.5cm}
\includegraphics[width=0.7\linewidth]{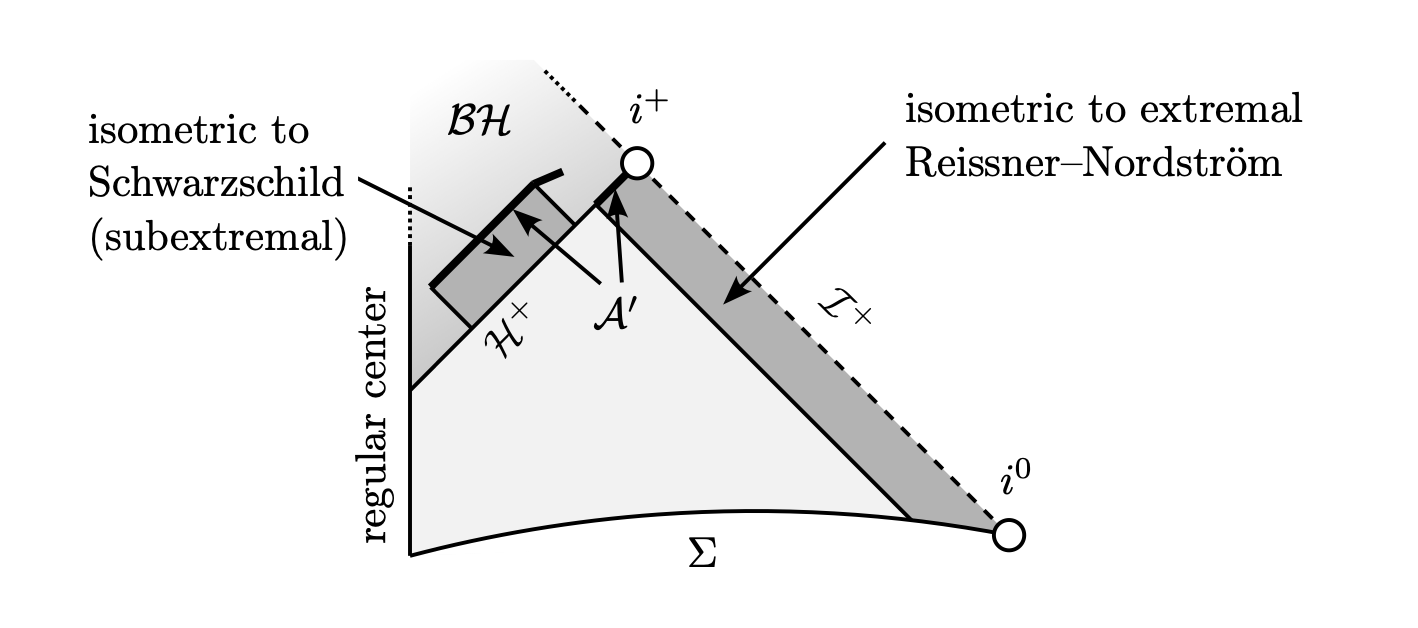}
\caption{The third-law violating spacetimes of Kehle--Unger, taken from~\cite{kehle2024gravitationalcollapseextremalblack} with permission. \label{fig:logo}}
\end{figure}

The Penrose diagram of the spacetimes constructed is given by  Fig.~\ref{fig:logo}.

I can't do justice to the history of the third law here, but I do want to point out one important point: In the Penrose diagram of Fig.~\ref{fig:logo}, the thicker curve $\mathcal{A}'$ represents an outermost apparent horizon, and as noted, this is disconnected, the event horizon $\mathcal{H}^+$ containing an isolated component. Israel's attempted proof in~\cite{israel1986third} had implicitly used the global connectivity of an  apparent horizon (here understood as a tube of marginally trapped surfaces).

(By unfortunate coincidence, in previous singular ``counterexamples'' discussed earlier, the existence of which indeed motivated Israel's explicit added assumption of regularity in~\cite{israel1986third},  there were again discontinuities of the outermost apparent horizon which occurred exactly when the singular matter shells crossed it. This may have suggested that for sufficiently regular solutions,
apparent horizons would necessarily be connected, something which is not in fact true!)

\subsection{Status of the overspinning/overcharging paradigm}

At first glance, the fact that extremal black holes can indeed be created in finite time may suggest that the overspinning/overcharging paradigm is all the more relevant. After all, it was 
precisely the menace of overspinning which provided the authors of~\cite{bardeen1973four} one of their motivations for proposing the third law in the
first place. 

Before discussing the status of the  overspinning/overcharging  paradigm, it may be useful to emphasise the following  point of logic in comparison to the status of the third law:
The third law had claimed the  \emph{impossibility} of something.
Thus, it would have been potentially hard to prove (even had it been true, that is!), but was ``easy'' to disprove (given that it was false), as it was  falsifiable by providing a \underline{single} example.
In contrast, overspinning/overcharging claims the \emph{possibility} of something.
Thus,  the situation is reversed. It would be  proveable by providing a single example, but  to falsify one would require a very general kind of mathematical theorem,
applying to an infinite dimensional set of solutions of the full non-linear Einstein equations.

In the \underline{restricted context} of spherical symmetry, one does actually have such a result:
\begin{theorem}[\cite{dafermos2005spherically}]
\label{noovercharg}
For any reasonable matter model in spherical symmetry (e.g.~charged scalar field~\cite{kommemi2013global}, charged Vlasov~\cite{KU23}), then if spacetime contains
\underline{at least one}
trapped or marginally trapped surface, then there are no naked singularites. 
\end{theorem}

It follows in particular from the above that one cannot create a naked singularity by overcharging
either an exactly extremal or near extremal black hole in an entirely spherically symmetric process.
(In particular, Theorem~\ref{noovercharg} can be thought to already give a definitive answer, in the negative, to the attempts
at overcharging described in~\cite{PhysRevD.59.064013}.)

%
%
As we shall see much later, the definitive disproof of overspinning in a very general, fully nonlinear setting \emph{without symmetry} assumptions would follow as  a corollary (Corollary~\ref{nooverspincor}) to  a positive resolution of Conjecture~\ref{phasespace}, which I will propose in Section~\ref{conjecturesection}.  
But even if that conjecture turns out to be false, I claim (see already Section~\ref{whatif}) 
that there is no sense that this could be reasonably  interpreted as being  due to ``overspinning''! 
In any case, to arrive at these conjectures we essentially have to revisit the problem of the near extremal dynamics of the Einstein equations with a fresh lens.

\subsection{A generalised third law paradigm?}
\label{generalisedparadigm}

Before proceeding, however, let us return to the issue of the third law.  
Though Theorem~\ref{kehleungertheorem} definitively disproves the third law, one could wonder
whether its ``spirit'', as captured  in the last paragraph of
Section~\ref{TLP}
by the idea of extremality as representing ``the (relative) unattainable'', might somehow live on. 
To humour this idea, 
let me formulate what I will dub the ``generalised third law paradigm'':
\begin{quotation}
\emph{Forming extremal black holes (whether in finite or infinite time, and whether starting from a subextremal black hole or directly in collapse) should somehow be more difficult than forming a black hole of any other fixed subextremal spin-to-mass or charge-to-mass ratio.}
\end{quotation}

Though 
the remarkable paper~\cite{kehle2024gravitationalcollapseextremalblack} put an end
to the third law itself, it does not of course address  the ``generalised third law paradigm'' above.
Like with the overspinning/overcharging paradigm, such a statement can only be falsified by a general mathematical theorem pertaining to general dynamics of the Einstein equations in a neighbourhood of extremal Kerr, and no such 
statement is yet available.
Indeed, I have conjured up this generalised third law paradigm precisely 
in order to give the ``spirit'' of the third law, as exemplified by the idea of ``unattainability'', a second fighting chance!

I must warn the reader already, however, that  one who is still betting for
this generalised third law paradigm does so at their own risk!
In analogy with Theorem~\ref{noovercharg}, 
there is in fact already some evidence in plain view---coming from a pioneering numerical study of Murata--Reall--Tanahashi~\cite{murata2013happens}---that even this generalised third law paradigm is again false, at least when restricted to spherical symmetry. 
I will discuss this work a little bit later. 
Indeed,  as with  the
overspinning/overcharging paradigm,
a \emph{definitive disproof} of the generalised third law paradigm would   in fact follow from Conjecture~\ref{lcodimstabext} to be discussed later.

\section{The Schwarzschild case: a blueprint for extremal Kerr?}

To get a first glimpse of the conjectured picture of near extremal dynamics which I will describe in Section~\ref{conjecturesection}, it is useful to first examine  the other ``extremal'' member of the Kerr family, namely Schwarzschild (extremal now only in the literal ``boundary'' sense as described in the opening paragraph of this essay, characterised by the rotational parameter taking its \emph{minimum} possible modulus, i.e.~vanishing!).

\begin{figure}[ht]
\centering
\vspace*{-.6cm}
\includegraphics[width=0.5\linewidth]{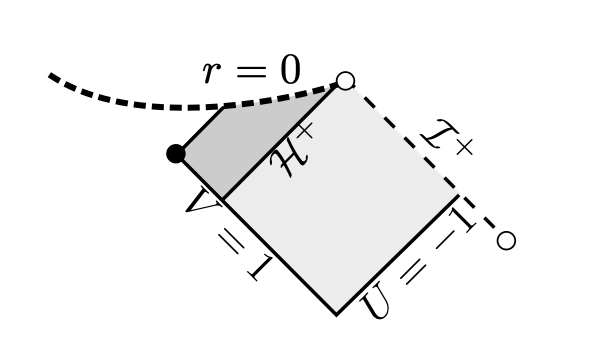}
\vspace*{-.6cm}
\caption{Schwarzschild as a Cauchy evolution of characteristic initial data, taken from~\cite{dhrtplus}.  \label{fig:dhrt}}
\end{figure}

Concerning thus near-Schwarzschild dynamics, it is of course a truism that one can perturb Schwarzschild into the Kerr family by adding a little bit of angular momentum. Thus, Schwarzschild is not asymptotically stable in the strict sense, i.e.~if we view
Schwarzschild as the Cauchy evolution of ``Schwarzschild initial data'' prescribed say on 
two null cones $V=1$ and $U=-1$ (here $U,V$ are global double null coordinates
covering Schwarzschild) as in Fig.~\ref{fig:dhrt}, 
then the generic small perturbation of the data  will  evolve under the Einstein vacuum equations
\begin{equation}
\label{vac}
{\rm Ric}(g)=0
\end{equation}
 to a spacetime which will not settle back down to a Schwarzschild metric.
 \emph{What is then the best asymptotic stability statement that can be true about the Schwarzschild family?} 
 
 It turns out that at the linear level, one can explicitly identify ``Kerr perturbations'' as a $3$-dimensional subfamily~\cite{10.4310/ACTA.2019.v222.n1.a1}, spanned by a set of three quantities with fixed spherical harmonic frequency $\ell=1$.   (Note that the dimension is $3$ and not $1$ because to parametrise smoothly the moduli space near Schwarzschild, one must also encode the unique symmetry axis of $a\ne0$ Kerr solutions.)
Given a linear perturbation for which  these three quantities vanish, then it was proven in~\cite{10.4310/ACTA.2019.v222.n1.a1} that the
perturbation decays as time goes to infinity to a pure gauge solution, and moreover, gauge normalisations may be chosen (``teleologically'', i.e.~from the future!)~so that this pure gauge solution in fact vanishes identically.  This can be recast as the following statement:
\begin{quotation}
\emph{In the infinite dimensional space of all linear vacuum perturbations around Schwarzschild, after suitable teleological double null gauge normalisations, there is a unique \underline{co}dimension-3 subspace such that all linear perturbations lying in this space in fact decay to zero as time goes to infinity.}
\end{quotation}

The best fully nonlinear asymptotic stability statement concerning the Schwarzschild family that could be expected under evolution by~\eqref{vac} would then be an analogous nonlinear codimension-3 asymptotic stability statement. This is precisely the main result  proven in~\cite{dhrtplus}:

\begin{theorem}[\cite{dhrtplus}]
\label{DHRTtheorem}
For all vacuum characteristic initial data prescribed on cones as in Fig.~\ref{fig:dhrt}, assumed sufficiently close to Schwarzschild data with mass $M_{\rm init}$
and lying on a codimension-$3$ “submanifold” $\mathfrak{M}_{\rm stable}$ of 
the moduli space $\mathfrak{M}$ of initial data, the arising vacuum solution $(\mathcal{M}, g)$ 
satisfies the following properties:
\begin{enumerate}
\item[(i)]
 $(\mathcal{M}, g)$
possesses a complete future null infinity $\mathcal{I}^+$,
and in fact the future boundary of $J^-(\mathcal{I}^+)$ in $\mathcal{M}$ is a regular, future affine complete event horizon $\mathcal{H}^+$. 
\item[(ii)]
The metric $g$ remains close to the Schwarzschild metric with mass $M_{\rm init}$ in 
$J^-(\mathcal{I}^+)$.
\item[(iii)]
The metric $g$ asymptotes, inverse polynomially, to a Schwarzschild metric with mass $M_{\rm final} \approx M_{\rm init}$ as $u \to \infty $ and $v \to  \infty$, 
in particular along  $\mathcal{I}^+$ and $\mathcal{H}^+$, where $u$ and $v$ are suitably normalised null coordinates.
\end{enumerate}
\end{theorem}

Statement (i) asserts  the absence of naked singularities,  statement (ii) asserts ``orbital stability'', while statement (iii) is the statement of
asymptotic stability, all for data on~$\mathfrak{M}_{\rm stable}$.

We note that, unlike linear theory, where the codimension-$3$ subspace is explicit (characterised by vanishing Kerr perturbations), the codimension-$3$ ``submanifold'' $\mathfrak{M}_{\rm stable}$
above \emph{is itself only teleologically determined} (as is the exact value of the final Schwarzschild mass $M_{\rm final}$ and the location of the event horizon $\mathcal{H}^+$). This is reminiscent of the
fact mentioned earlier that already in linear theory,  the gauge  normalisations necessary to obtain decay were themselves determined only teleologically.

For discussion of stability results for Reissner--Nordstr\"om as a solution of Einstein--Maxwell, see~\cite{giorgi2020linear}.
For  the fate of data as 
in Theorem~\ref{DHRTtheorem} but  \emph{not} lying on $\mathfrak{M}_{\rm stable}$ above---as expected, they settle down to a very slowly rotating  (i.e.~$|a|\ll M$) Kerr exterior---see~\cite{giorgi2022wave}.
See also~\cite{hintz2018global} for nonlinear stability  for black holes in the $\Lambda>0$
case.

\section{Instabilities}
\label{Instabsec}

Could the above picture of Theorem~\ref{DHRTtheorem} carry over directly to the  extremal Kerr case
$a=M$, i.e.~can we simply replace \emph{Schwarzschild} above by \emph{extremal Kerr}? Before entertaining this issue, we have to deal with the elephant in the room, related to the 
other---figurative---meaning of the word ``extreme'', mentioned in the opening paragraphs of this essay: 
\emph{Extremal black holes are characterised by extreme behaviour, in particular, by the
presence of several well-known \underline{instabilities}.}

\subsection{The Aretakis instability}
\label{Aretakis}

The Aretakis instability~\cite{Aretakis2} is a remarkable very general property which can be associated to any extremal Killing horizon. It already applies to the linear massless wave equation
\begin{equation}
\label{linwaveeq}
\Box_g\psi =0,
\end{equation}
but more generally applies to the linearised Einstein vacuum equations and Einstein--Maxwell equations, around extremal Kerr and extremal Reissner--Nordstr\"om respectively~\cite{PhysRevD.86.104030, luciettietal}.
According to this instability, translation-invariant transversal first derivatives of $\psi$ (say $\partial_r\psi$) along the
 horizon $\mathcal{H}^+$ \emph{fail to decay}, while second derivatives {\bf blow up polynomially}
\begin{equation}
\label{Aretakisblowup}
|\partial^2_r\psi|(r, v)\to \infty 
\end{equation}
as advanced time $v$ goes to infinity along $\mathcal{H}^+$. (Here $(r,v)$ denote Eddington--Finkelstein type coordinates regular through $\mathcal{H}^+$.) The seed of the instability is an exact conservation law along the horizon $\mathcal{H}^+$, which turns out to have a similar origin to the Newman--Penrose constants at null infinity $\mathcal{I}^+$ (see~\cite{aretakis2017characteristic}).

\subsection{Weak stability for Reissner--Nordstr\"om}
\label{weakstabRN}
An important aspect of the Aretakis instability is that it is \emph{weak}. The blow-up~\eqref{Aretakisblowup} along~$\mathcal{H}^+$ is still compatible with
good decay properties \emph{away from the horizon} $\mathcal{H}^+$, and moreover,  the amplitude of $\psi$ itself (and its \emph{tangential} derivatives, say $\partial_v\psi$) may still decay along $\mathcal{H}^+$ itself. Indeed, in the case of~\eqref{linwaveeq}
on Reissner--Nordstr\"om, Aretakis~\cite{aretakis2011stability, aretakis2011stability2} already showed precisely such stability statements, complementing~\eqref{Aretakisblowup}.  For the full linearised
Einstein--Maxwell system around extremal Reissner--Nordstr\"om, such weak stability statements, as well as upper bounds for the growth of the unstable, transversal quantities, were shown recently
by Apetroaie~\cite{apetroaie2022instability}.

\subsection{Nonlinear model problems on fixed extremal Reissner--Nordstr\"om backgrounds}
\label{nonlinearmodels}

Of course, it is not clear at all whether weak stability results at the linear level are sufficient to ensure \emph{nonlinear stability}. For instance, Aretakis has shown~\cite{aretakis2013nonlinear} 
that for nonlinear (``semilinear'') wave equations of the form (for $n\ge1$)
\begin{equation}
\label{naivemodel}
\Box_g\psi = \psi^{2n}+(\partial_{v}\psi)^{2n} +(\partial_r\psi)^{2n}
\end{equation}
on fixed extremal Reissner--Nordstr\"om backgrounds,  
arbitrarily small spherically symmetric data lead to solutions which blow up in \underline{finite} advanced time
on the horizon $\mathcal{H}^+$, whereas it follows from~\cite{dafermos2005small} that  in the subextremal case, for sufficiently high $n$, such solutions  exist for all time on the black hole exterior, up to and
including the horizon. (In~\eqref{naivemodel}, coordinate derivatives are again with respect to Eddington--Finkelstein coordinates $(v,r)$ regular across~$\mathcal{H}^+$.)

The equation~\eqref{naivemodel} is not so good a model, however, for the nonlinearities occurring in the Einstein vacuum equations~\eqref{vac}.
A better model would perhaps be a nonlinear (again ``semilinear'') equation of the form
\begin{equation}
\label{bettermodel}
\Box_g\psi= A(\psi, x)\,g^{\alpha\beta}\partial_\alpha\psi\,\partial_\beta\psi ,
\end{equation}
where $A(0,x)$ is potentially nonzero and the nonlinearity is quadratic in first derivatives of~$\psi$.
In a remarkable series of works~\cite{angelopoulos2016global, angelopoulos2020nonlinear}, it was shown that solutions of~\eqref{bettermodel} arising from sufficiently small initial data exist globally in the black hole exterior of extremal Reissner--Nordstr\"om, up to and including the horizon $\mathcal{H}^+$, despite the Aretakis instability~\eqref{Aretakisblowup}.

We note that, even in the case where $A(\psi,x)$ vanishes for large $x$, the results of~\cite{angelopoulos2020nonlinear} depend on the fact that it is precisely the expression  $g^{\alpha\beta}\partial_\alpha\psi\,\partial_\beta\psi$ which appears in~\eqref{bettermodel}, and not some other quadratic combination of
second derivatives of $\psi$.  (This is again in contrast to the subextremal Reissner--Nordstr\"om or Kerr case  where, for such  
$A$, 
global existence  holds replacing $g^{\alpha\beta}\partial_\alpha\psi\,\partial_\beta\psi$  with a general quadratic expression; 
see for instance~\cite{dafermos2022quasilinear}.)
The importance of structure for the quadratic terms is analogous to the well known fact that, if $A$ does \underline{not} vanish for large $x$, say if $A(\psi,x)=1$ identically, then even if $g$ denotes the  Minkowski metric, while  small data solutions
of~\eqref{bettermodel} indeed exist globally (as first observed by Nirenberg), replacing 
  $g^{\alpha\beta}\partial_\alpha\psi\,\partial_\beta\psi $
with say $(\partial_t\psi)^2$ leads to blow up~\cite{john1979blow}.
We will return to this point in Section~\ref{whatif}.

\subsection{Extremal Kerr and higher azimuthal instabilities}
\label{higherazi}

Turning to extremal Kerr, however, the situation is much less clear. Whereas restricting
to axisymmetric solutions of~\eqref{linwaveeq},
one has results analogous to that of extremal Reissner--Nordstr\"om 
(see~\cite{aretakis2012decay}),
a recent theorem of Gajic~\cite{gajic2023azimuthal} proves that for fixed  higher azimuthal $m$-mode solutions, \emph{worse} instabilities arise, confirming  a previous heuristic study by Casals, Gralla and Zimmerman~\cite{casals2016horizon}.
Even for such fixed higher $m$-modes, however, it is not known whether  weak stability statements analogous to 
those of~\cite{aretakis2012decay} hold.
Indeed, the only
general statement presently known is so-called ``mode stability'', recently shown in the extremal case by Teixeira da 
Costa~\cite{teixeira2020mode}.
The situation for general data, which would concern the sum over \emph{infinitely} many $m$-modes,
is even more unclear. Thus,  even the question of whether all solutions of the linear homogeneous scalar wave equation~\eqref{linwaveeq} on extremal Kerr, arising  from regular localised initial data, remain bounded for all time (cf.~\cite{dafermos2016decay}, \cite{shlapentokh2020boundedness, shlapentokh2023boundedness} where this is shown in the general subextremal case  $|a|<M$ for the wave and Teukolsky equations, respectively), even when one restricts considerations to a region well  outside the horizon, 
remains very much open!  

\section{The stability conjecture for extremal black holes and the phase portrait of near extremal dynamics}
\label{conjecturesection}

The current uncertainty regarding linear theory on extremal Kerr described above might make it feel a bit premature to conjecture anything at the nonlinear level. Indeed,  precisely for this reason, in our~\cite{dhrtplus}, we shied away from any such conjecture, preferring to volunteer a conjecture only for the Einstein--Maxwell theory around extremal Reissner--Nordstr\"om (see Conjecture IV.2 of~\cite{dhrtplus} and the subsequent paragraph), for which, as described
in Section~\ref{weakstabRN}, the linear theory is better understood.
 Nonetheless, I will here state 
the analogous conjectures in the extremal Kerr case.

\subsection{Codimension-$1$ stability of extremal Kerr with horizon hair}
\begin{figure}[ht]
\centering
\includegraphics[width=0.5\linewidth]{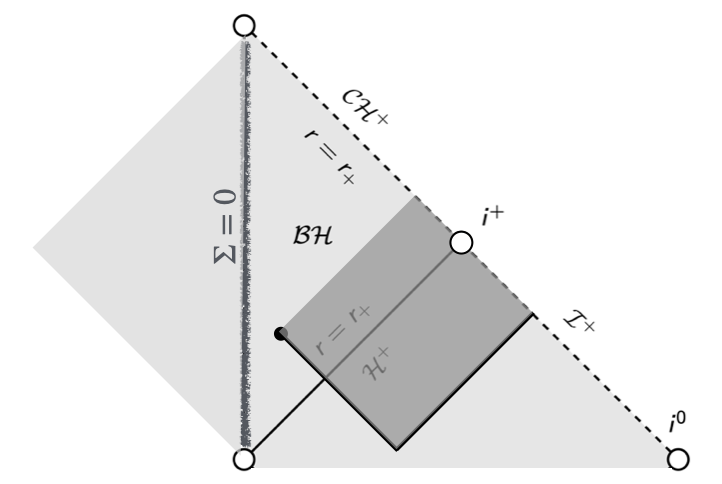}
\caption{Extremal Kerr as a Cauchy evolution of characteristic initial data. \label{fig:extkerrivp}}
\end{figure}

As with Schwarzschild, we may view extremal Kerr as the
Cauchy evolution of characteristic initial data posed on two null cones.
 Refer to Fig.~\ref{fig:extkerrivp}.
Unlike in the Schwarzschild case, however, when  parametrising linear perturbations around
extremal Kerr, non-extremal Kerr perturbations are  spanned by a single quantity, not 3. Thus,
if one wishes to show asymptotic stability of extremal Kerr, the natural conjecture would now be   
a codimension-$1$  nonlinear stability statement.  At the same time, the instabilities described in 
Sections~\ref{Aretakis} and~\ref{higherazi} 
must at the very least also somehow appear. The most optimistic scenario is then:
\begin{conjecture}
\label{lcodimstabext}
For all vacuum characteristic initial data prescribed on cones as in Fig.~\ref{fig:extkerrivp}, assumed sufficiently close to extremal Kerr data with mass $M_{\rm init}=a_{\rm init}$
and lying on a codimension-$1$ “submanifold” $\mathfrak{M}_{\rm stable}$ of 
the moduli space $\mathfrak{M}$ of initial data, the arising vacuum solution $(\mathcal{M}, g)$ 
satisfies the following properties:
\begin{enumerate}
\item[(i)]
 $(\mathcal{M}, g)$
possesses a complete future null infinity $\mathcal{I}^+$,
and in fact the future boundary of $J^-(\mathcal{I}^+)$ in $\mathcal{M}$ is a regular, future affine complete event horizon $\mathcal{H}^+$. 
\item[(ii)]
The metric $g$ remains close to the extremal Kerr metric with mass $M_{\rm init}$ in 
$J^-(\mathcal{I}^+)$.
\item[(iii)]
The metric $g$ asymptotes, inverse polynomially, to an extremal Kerr metric with mass $M_{\rm final}=a_{\rm final} \approx M_{\rm init}$ as $u \to \infty $ and $v \to  \infty$, 
in particular along  $\mathcal{I}^+$ and $\mathcal{H}^+$, where $u$ and $v$ are suitably normalised null coordinates.
\item[(iv)] For generic initial data \underline{conditioned to lie on $\mathfrak{M}_{\rm stable}$}, then suitable
quantities associated to derivatives of the metric grow without bound (``horizon hair'')  along $\mathcal{H}^+$ as $v\to \infty$.
\end{enumerate}
\end{conjecture}

The statement is thus in complete analogy with Theorem~\ref{DHRTtheorem}, \emph{except for the
additional weak instability statement (iv).}  For  near extremal  Kerr black holes with fixed spin-to-mass ratio, one would again expect a similar codimension-$1$ stability statement (without of course the instability part (iv)). In particular, the codimension in moduli space $\mathfrak{M}$ of the set $\mathfrak{M}_{\rm stable}$ of spacetimes evolving to extremal Kerr would be \emph{exactly the same} as those evolving to nearby fixed subextremal spin-to-mass ratio Kerrs. 
In this sense, {\bf it would be no more difficult  to form an extremal black hole than it would be a subextremal black hole of some other fixed spin-to-mass ratio.}
Thus, according to the above, there would be no saving even the ``generalised'' third law paradigm from  the fate of the original  third~law.

In addition to the semilinear problem~\eqref{naivemodel} discussed in Section~\ref{nonlinearmodels}, 
which concerned a fixed extremal Reissner--Nordstr\"om background, there  is in fact a nonlinear self-gravitating model problem (i.e.~where the spacetime is itself dynamical) where a much simplified analogue of the above conjecture can already be studied, namely the Einstein--Maxwell-scalar field system under spherical symmetry. Indeed, a numerical study of this system was conducted by 
Murata--Reall--Tanahashi~\cite{murata2013happens},  and the results reported are entirely consistent with a statement analogous  to 
Conjecture~\ref{lcodimstabext}.
Proving these numerical results for the spherically-symmetric system would be an important first step towards a proof
of the far more ambitiuous Conjecture~\ref{lcodimstabext}. 
See also~\cite{gajic2019interior}. \emph{(Note added: A proof of the analogue of 
Conjecture~\ref{lcodimstabext} for the above spherically symmetric system has in fact just been announced (see the upcoming~\cite{upcoming}).)}

\subsection{The phase portrait of near extremal dynamics}
\label{nearextrem}

\begin{figure}[ht]
\centering
\includegraphics[width=1.0\linewidth]{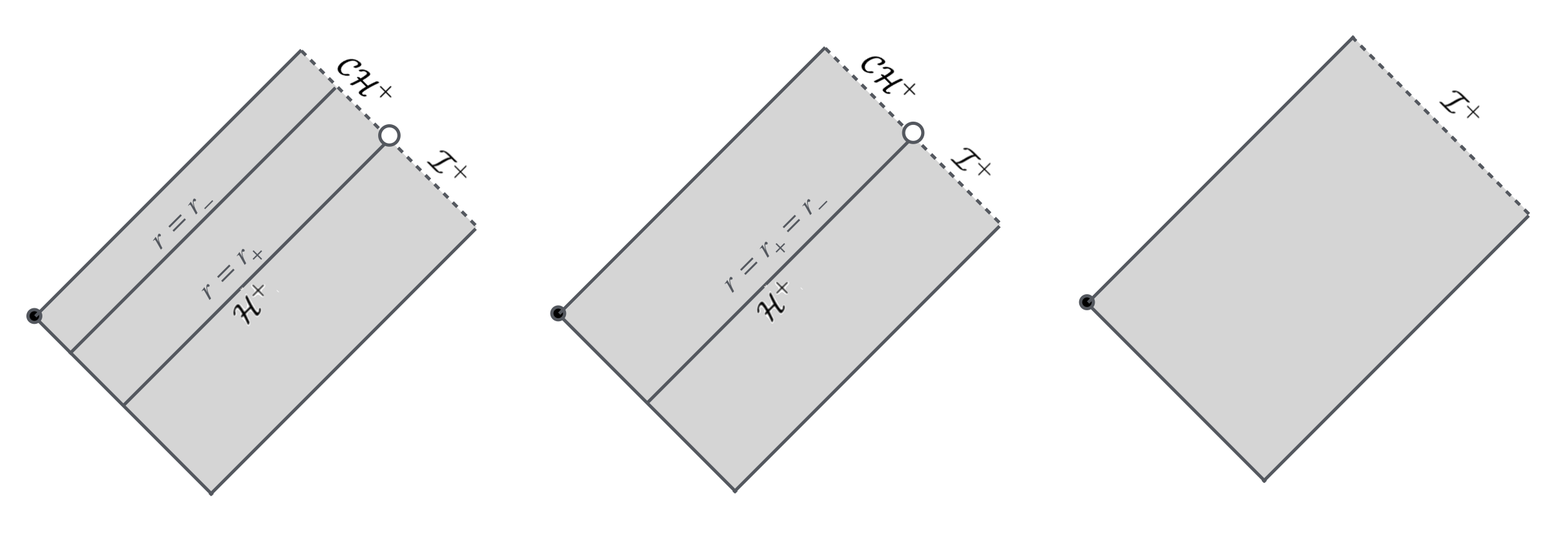}
\caption{The Kerr family  as Cauchy evolutions of a smooth family characteristic initial data: the subextremal, extremal and superextremal cases, respectively. \label{fig:kerrfam}}
\end{figure}

If Conjecture~\ref{lcodimstabext} is indeed true, then it makes sense to ask what nearby solutions \emph{not lying on this codimension-$1$ hypersurface $\mathfrak{M}_{\rm stable}$} evolve to (cf.~the comments after Theorem~\ref{DHRTtheorem}). 
It is instructive to first understand the Kerr family itself 
from the point of view of this initial value problem. Indeed, when we realise the Kerr family (through extremality!) as
a smooth family  of solutions evolving from initial data as in 
Conjecture~\ref{lcodimstabext}, 
 then we see that as we approach extremality, the event horizon and the ``outgoing'' component of the inner horizon coalesce into  the extremal event horizon, which then disappears once we pass
 into the superextremal range. See
 Fig.~\ref{fig:kerrfam}.
Note that the superextremal case has no naked singularity in the domain of development of the characterstic initial data depicted (see already Fig.~\ref{fig:notsonaked} for the relation of this region
to the global Penrose diagram). In this case, the solution lies entirely in $J^-(\mathcal{I}^+)$, but $\mathcal{I}^+$ is now incomplete, since after finite Bondi time observers see the final sphere
of the initial ingoing cone.

I claim that the ``minimalist'' expectation consistent with what we know would be the following:
\begin{conjecture}
\label{phasespace}
Under the assumptions of Conjecture~\ref{lcodimstabext},
the codimension-$1$ ``submanifold'' $\mathfrak{M}_{\rm stable}$ is in fact a regular hypersurface  which separates the moduli space $\mathfrak{M}$ into two open regions, each with boundary $\mathfrak{M}_{\rm stable}$: the set of initial data $\mathfrak{M}_{\rm subextremal}$ evolving
to subextremal black holes and the set of initial data $\mathfrak{M}_{\rm noncollapse}$ not
collapsing in the domain of dependence of the data, i.e.~such that the domain of dependence of the data is entirely contained in $J^-(\mathcal{I}^+)$  (but with incomplete $\mathcal{I}^+$).
\end{conjecture}

\begin{figure}[ht]
\centering
\includegraphics[width=0.5\linewidth]{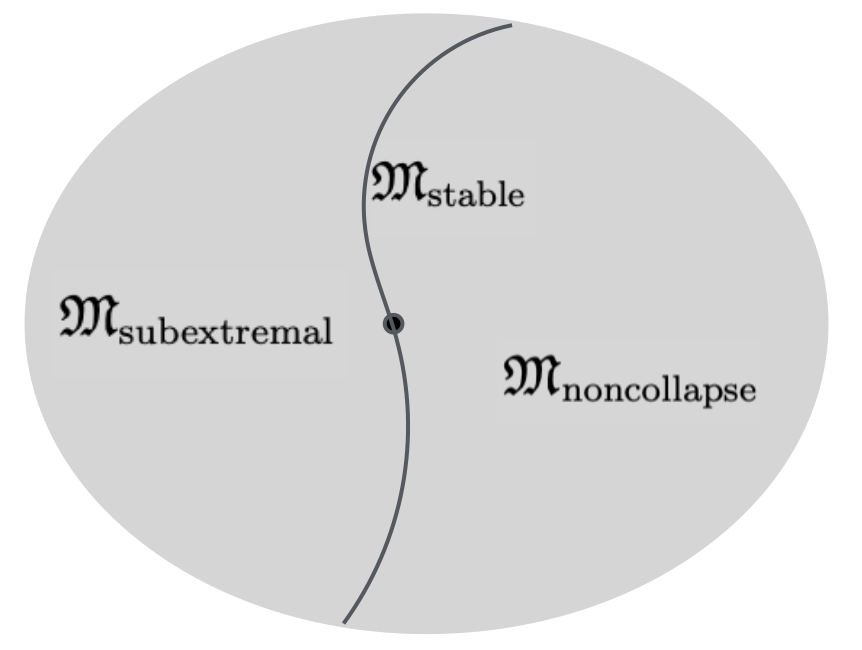}
\caption{The moduli space $\mathfrak{M}$ of characteristic initial data near extremal Kerr
partitioned as in the statement of Conjecture~\ref{phasespace}. \label{fig:modulispace}}
\end{figure}

The structure of the moduli space $\mathfrak{M}$ according to the  above conjecture is depicted schematically in
 Fig.~\ref{fig:modulispace}.

Let us note that a \emph{necessary} condition for the above to hold is that for all data lying on
$\mathfrak{M}_{\rm stable}$, there is no (strictly) trapped surface in the domain of
dependence of the data. For otherwise, by Cauchy stability any nearby solution
would necessarily also have such a trapped surface, and thus, by
Penrose type monotonicity arguments involving the Raychaudhuri equation (see for instance~\cite{wald2010general}),
could not lie entirely in $J^-(\mathcal{I}^+)$.

Note how the three Kerr spacetimes in Fig.~\ref{fig:kerrfam} provide explicit examples
of data lying
on $\mathfrak{M}_{\rm subextremal}$, $\mathfrak{M}_{\rm stable}$ and $\mathfrak{M}_{\rm noncollapse}$, respectively. According to Conjecture~\ref{phasespace}, 
whereas for data lying in~$\mathfrak{M}_{\rm subextremal}$, the above development includes a complete null infinity $\mathcal{I}^+$, and in this sense describes the complete future of far away observers, in the case of  data lying in~$\mathfrak{M}_{\rm noncollapse}$,  null infinity $\mathcal{I}^+$ is incomplete.  This is not because there is anything singular in the evolution, but  simply because, just as for superextremal Kerr itself, the ingoing cone of the data themselves is incomplete,  yet no horizon formed to shield the end of this cone from $J^-(\mathcal{I}^+)$.  These solutions would in fact  be smoothly
extendible beyond a complete outgoing null cone emanting from the final sphere of the initial ingoing
null cone.

In order to determine what happens ``next'' to far away observers in spacetimes evolving from data lying in~$\mathfrak{M}_{\rm noncollapse}$, one must ``complete'' the initial data.
An even more amibitious conjecture, due to Kehle and Unger, 
considers the ``larger'' moduli space \scalebox{1.3}{${{\mathfrak{M}}}$} consisting now of the set of one-ended (complete) asymptotically flat initial data.  Their conjecture asserts the existence of a codimension-$1$ submanifold
\scalebox{1.3}{$\mathfrak{M}_{\rm critical}\subset \mathfrak{M}$} consisting of data evolving to spacetimes settling down to extremal black holes such that \scalebox{1.4}{$\mathfrak{M}_{\rm critical}$}  \emph{locally} separates 
\scalebox{1.3}{$\mathfrak{M}$}
 into data evolving to spacetimes settling down to subextremal black holes and data evolving to spacetimes $(\mathcal{M},g)$ with a \underline{complete} $\mathcal{I}^+$ and no horizon, i.e.~with $\mathcal{M}=J^-(\mathcal{I}^+)$, such that $g$ moreover asymptotically settles down to the Minkowski metric. Thus, according to Kehle and Unger's
 conjecture, {\bf extremal black holes can arise at the threshold between black hole formation and
 dispersion}. The authors dub this ``extremal critical collapse'', in analogy with the more familiar threshold naked singularity solutions which have been studied numerically in the ``critical collapse'' 
 literature~\cite{choptuik1993universality, gundlach2007critical}, solutions which presumably
 lie on  a separate codimension-$1$ submanifold of  \scalebox{1.3}{${{\mathfrak{M}}}$} with a similar local separation property.  We emphasise, however, that not 
 all extremal black holes are threshold solutions!  
 The data $\mathfrak{M}_{\rm noncollapse}$ in Conjecture~\ref{phasespace} can also be suitably completed as one-ended asymptotically flat data in \scalebox{1.3}{${{\mathfrak{M}}}$} so as to themselves evolve to black hole spacetimes, where the horizon $\mathcal{H}^+$ however does not lie in the domain of
 dependence of $\mathfrak{M}_{\rm noncollapse}$.
In this case, 
 crossing the
submanifold $\mathfrak{M}_{\rm stable}$ from $\mathfrak{M}_{\rm subextremal}$ to
$\mathfrak{M}_{\rm noncollapse}$  would correspond not to the horizon $\mathcal{H}^+$ disappearing
but to the location of the horizon 
 ``jumping''.  For more details, see~\cite{KU23}.

\subsection{How  would Conjecture~\ref{phasespace} disprove ``overspinning''?}
As we have pointed out already, just as in Theorem~\ref{DHRTtheorem}, 
the set $\mathfrak{M}_{\rm stable}$
of Conjecture~\ref{phasespace} depicted in Fig.~\ref{fig:modulispace}
would only be teleologically defined. Thus, in general one would not know which ``side'' of  the $\mathfrak{M}_{\rm stable}$ hypersurface an initial data set lies on 
without evolving the data to the future. There would however be an obvious \emph{sufficient} condition on initial data that ensures that the data indeed lie in $\mathfrak{M}_{\rm subextremal}\cup \mathfrak{M}_{\rm stable}$, namely the existence of at least one trapped or marginally trapped surface. This is because by
Penrose type monotonicity 
arguments
mentioned already in Section~\ref{nearextrem},
 such a surface could not lie in the past of $\mathcal{I}^+$. 
Thus, we may infer from a positive resolution of Conjecture~\ref{phasespace} the following Corollary:
\begin{corollary}[Given Conjecture~\ref{phasespace}]
\label{nooverspincor}
Consider a vacuum solution that contains a Cauchy hypersurface with a
marginally trapped surface such that outside that surface, the spacetime is close
to extremal Kerr in a suitable sense. Then the spacetime has a black hole bounded
by a regular event horizon $\mathcal{H}^+$ and with complete null infinity $\mathcal{I}^+$. In particular, \underline{no naked singularity forms}.
\end{corollary}

To say it more colloquially, no matter what small incoming gravitational radiation one tries to ``throw'' into an initially exactly extremal or slightly subextremal black hole, one can in particular never produce a naked singularity. The spacetime will settle back down to a subextremal or extremal black hole. 
{\bf Thus, Corollary~\ref{nooverspincor}, if indeed true, can be viewed as definitively disproving
the ``overspinning'' paradigm.}

An analogous statement to Corollary~\ref{phasespace} 
replacing Kerr with Kerr--Newman, and the vacuum equations with those of a suitable self-gravitating charged matter model, would thus similarly definitively disprove ``overcharging'' scenarios.
(One can also of course consider overspinning more generally for Einstein-matter systems, where
the analogous statement can again be conjectured, but we emphasise that overspinning, unlike overcharging, is an issue  which may be studied for the vacuum equations~\eqref{vac}.)

One might ask what about ``large perturbations'' of  extremal Kerr? Remember, however, that in nonlinear theories, \emph{large perturbations are no longer ``perturbations''}!
If there is some self-gravitating system which collapses to  form a  naked singularity, then  it will 
presumably still form a naked singularity
if the whole system is ``thrown'' into a large extremal black hole.  This would have nothing to do with overspinning/overcharging.

\subsection{What if Conjectures~\ref{lcodimstabext} and~\ref{phasespace} are not true?}
\label{whatif}

Already Conjecture~\ref{lcodimstabext} is based on two  pieces of wishful thinking: Firstly, that the linear instabilities of extremal Kerr described in Section~\ref{higherazi}, though known to be \emph{stronger} than those of extremal Reissner--Nordstr\"om, are still in some sense weak, and in particular are still complemented by \emph{stability} statements away from the horizon (and, for tangential quantities, along the horizon). And secondly, that the nonlinearities in the Einstein vacuum equations~\eqref{vac}  near the horizon are indeed well modelled by~\eqref{bettermodel}. The worse the linear behaviour around extremal Kerr turns out to be, the more compensating the nonlinear structure of~\eqref{vac}
near the horizon
would have to be  for there to be any hope to prove Conjecture~\ref{lcodimstabext}. 

In view of the fact that there is a lot of uncertainty as to all of the above, we should already also entertain the possibility that Conjectures~\ref{lcodimstabext} and~\ref{phasespace} may turn out to 
be {\bf false}. \emph{What could the situation look like in that case?}

Perhaps the most exciting possibility to entertain is that the linear instabilities described in Section~\ref{higherazi} indeed lead  in the full nonlinear theory to the formation of naked singularities.  

In this case, we may further distinguish  two  scenarios: (i) the case where such naked singularities occur only for an exceptional non-generic set of data, in which case ``weak cosmic censorship'' would be saved and in fact a modified  version of~Conjecture~\ref{phasespace} could still hold, with a codimension-$1$ submanifold $\mathfrak{M}_{\rm threshold}$ in place of $\mathfrak{M}_{\rm stable}$, now containing 
both all data $\mathfrak{M}_{\rm extremal}$ evolving to extremal Kerr but 
also all data  $\mathfrak{M}_{\rm naked}$ leading to naked singularities, and (ii) the case where these naked singularities are in fact stable and thus the set
$\mathfrak{M}_{\rm naked}$
 of data leading to naked singularities would have non-empty interior in $\mathfrak{M}$.

\begin{figure}[ht]
\centering
\includegraphics[width=0.5\linewidth]{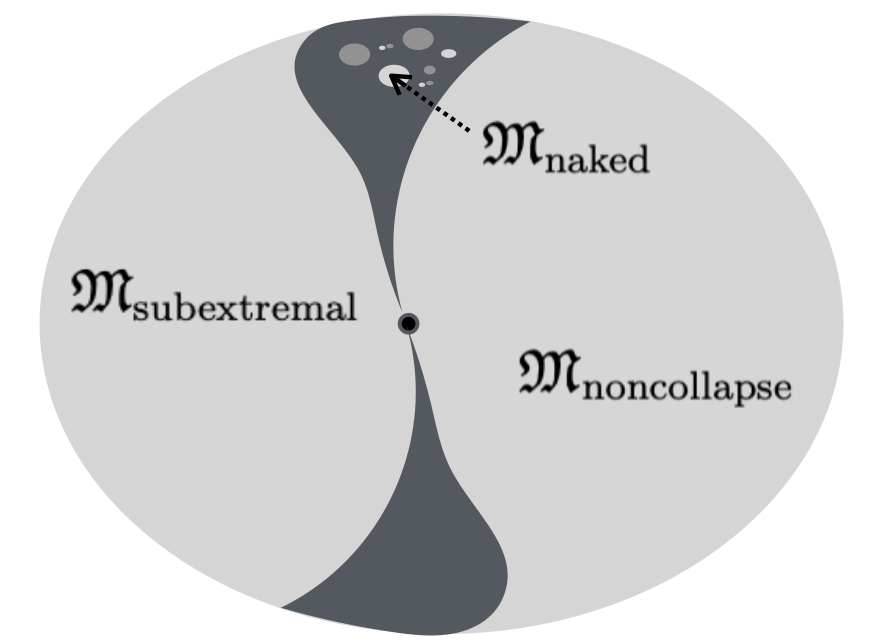}
\caption{A scenario where a chaotic region of the moduli space $\mathfrak{M}$ opens up from extremal Kerr,
with alternating regions including a set $\mathfrak{M}_{\rm naked}$ of data leading to naked singularities.}
 \label{fig:morecomplicated}
\end{figure}

Indeed, scenario (ii) would be  the even more exciting one, as it would in particular falsify \emph{weak cosmic censorship}.  One can for instance
imagine a picture of the moduli space~$\mathfrak{M}$ as depicted in Fig.~\ref{fig:morecomplicated}.
Here, a chaotic region  
``opens up'' from the extremal Kerr solution, containing infinitely many disconnected components of $\mathfrak{M}_{\rm subextremal}$ and $\mathfrak{M}_{\rm noncollapse}$ alternating with a (possibly disconnected) set
$\mathfrak{M}_{\rm naked}$ of open interior consisting of data evolving to  naked singularities (depicted say as the lightest  shaded regions), all  surrounded by a complicated residual set $\mathfrak{M}_{\rm extremal}$  of
data evolving to extremal Kerr.

\emph{Would any of the above scenarios vindicate the original generalised third law paradigm or the  overspinning/overcharging paradigms?}

Concerning the generalised third law paradigm, if the scenario was indeed something as depicted in Fig.~\ref{fig:morecomplicated}, then, clearly the ``residual'' set $\mathfrak{M}_{\rm extremal}$ of data evolving to extremal black holes would have a very different geometry from the analogous set for fixed subextremal spin-to-mass ratio. It isn't clear, however, even in this scenario, whether $\mathfrak{M}_{\rm extremal}$ would be ``smaller'' (in the sense of codimension) than the set of analogous subextremal fixed spin-to-mass ratio ones. It could of course indeed be smaller, but, in principle, it could even be that $\mathfrak{M}_{\rm extremal}$ is in fact larger, filling part of the darker shaded region, say  as a  fractional codimension subset.

Concerning the overspinning paradigm, whereas according to Fig.~\ref{fig:morecomplicated}, naked singularities would indeed occur in an arbitrary small neighbourhood of extremal Kerr, these naked singularities would have nothing to do with the naked singularities of superextremal Kerr,  
and in fact, they would not arise from ``overspinning'' \emph{but precisely from trying to preserve extremality as closely as possible.} 

\begin{figure}[ht]
\centering
\includegraphics[width=0.4\linewidth]{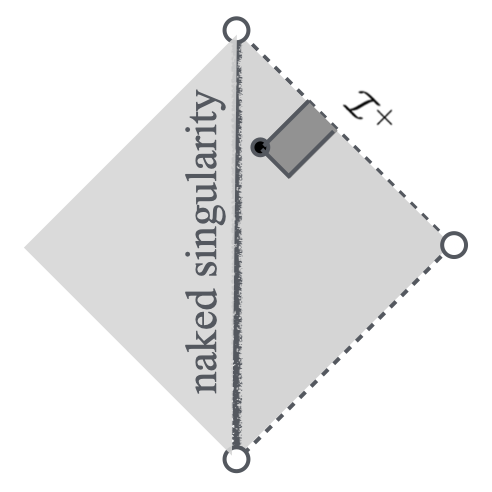}
\caption{The evolution of the initial data of Fig.~\ref{fig:kerrfam} in the superextremal case, superimposed on the spacetime of Fig.~\ref{fig:naked}.}
 \label{fig:notsonaked}
\end{figure}

Indeed, the connection of ``overspinning'' with the ``naked singularity'' of Fig.~\ref{fig:naked} in the first place was always based on what I believe is simply a misreading of the Penrose diagram, and it is perhaps useful to elaborate some more on this point.  If the ``overspinning'' scenario really envisions a ``quasistationary'' transition from subextremal to superextremal, then 
the only relevant regions would be the regions of Fig.~\ref{fig:kerrfam}, as these are the domain of dependence of the relevant initial data. (See also Fig.~\ref{fig:notsonaked} where
this region is superimposed on Fig.~\ref{fig:naked} in the superextremal case.)
 For as long as the solution indeed remains near the Kerr 
family in this domain, the monotonicity of Raychaudhuri's equation would clearly exclude such 
a transition, just as in the proof of Corollary~\ref{nooverspincor}. 
One often reads that, if a naked singularity were to form, one
cannot apply this monotonicity, hence evading the argument. For this, however, \emph{long before this purported naked singularity}
has formed, one would have to have
evolved \emph{far from the Kerr family}, 
thus thwarting the premise of this having anything to do with ``overspinning'' in the first place.

But to humour ``overspinning'' even more, let us  even suppose that the Einstein equations didn't happen to enjoy the monotonicity of the Raychaudhuri equation, and something like a quasistationary transition were indeed possible from
subextremal to superextremal. For instance, redefine the energy momentum tensor of a usual matter model to be its negative and
consider the resulting Einstein-matter system, and assume the validity of Conjecture~\ref{phasespace}. 
Corollary~\ref{nooverspincor} would \underline{not} now follow, and indeed, initial data with a marginally trapped surface  could in principle now be contained
in the  set $\mathfrak{M}_{\rm noncollapse}$ of Conjecture~\ref{phasespace}. 
 This in turn could mean that at the final outgoing cone emanating from the data,  the solution would be 
close to an outgoing cone of superextremal Kerr.  (Refer again to Fig.~\ref{fig:kerrfam} or Fig.~\ref{fig:notsonaked}.)
This would indeed represent in some sense successful ``overspinning''!
But what of it?
This would be  in no sense inconsistent with  the solution subsequently dispersing in the future, or recollapsing
later to a sub-extremal (or even exactly extremal!) black hole, as angular momentum can happily 
radiate to infinity (cf.~the fate of initially superextremal data in~\cite{KU23}).  Thus, even in the absence of the Einstein equations' monotonicity properties,  in no scenario is there a ``mechanism'' by which the naked singularity of Fig.~\ref{fig:notsonaked} becomes relevant for the evolution of data as in Conjectures~\ref{lcodimstabext} and~\ref{phasespace}.

In summary, even if it turns out that, in the complexity of   Fig.~\ref{fig:morecomplicated}, extremal black holes
are indeed ``more exceptional'' than fixed spin-to-mass subextremal ones, or that naked singularities do arise in a neighbourhood of extremality,
the considerations leading to this would be entirely different from those underlying  the generalised third law and the overspinning/overcharging paradigms. 

So as not to end on such a note of criticism, however, and 
lest I give the impression that I believe the legacy of
these two received paradigms to be entirely negative,
let me say one important thing in their favour:
 Both these paradigms, in their slightly contradictory ways, did succeed nonetheless in  drawing attention early on to the importance---and the potential complications---of near extremal dynamics, in a period where the stability considerations of black holes
 in general, even subextremal ones, 
 were not yet well understood. This is unquestionably a positive historical legacy, despite
 whatever confusion accompanied it.
As should be clear, however, going forward, I don't find the paradigms particularly useful any more,
and 
the considerations that led to their formulation are in my view completely  transcended by the considerations discussed
in Section~\ref{Instabsec}. 
Indeed, irrespectively of whether one is ``rooting'' for Conjectures~\ref{lcodimstabext}  and~\ref{phasespace}, or one is attracted to the more complicated alternative scenarios described in this section, I hope that I have convinced the reader that the fundamental issue which will  determine what is true  can be nothing other than the \emph{precise analysis of the interaction between the instabilities of Sections~\ref{weakstabRN} and~\ref{higherazi} and the nonlinearities of the Einstein equations}. 

\section{Epilogue}
I opened this article remarking how the two natures of the word ``extreme'', its literal 
and its figurative sense, are both central to the story of extremal black holes. I hope the conjectural picture 
of Section~\ref{conjecturesection} succeeded in describing a scenario where these two facets of extremality may ``peacefully'' coexist, Conjecture~\ref{phasespace} capturing its literal  ``boundary'' aspect, while the instability statement (iv)  of Conjecture~\ref{lcodimstabext} (the ``horizon hair'') capturing one figurative aspect of these black holes' extreme behaviour.  
We must not forget, however, that this is the optimistic scenario.
The ``coexistence'' might turn out to be a lot  more chaotic, and Fig.~\ref{fig:morecomplicated} gives just one idea of what we might have to come to terms with if these conjectures turn out to be false.
While I hope the conjectures described here provide a fruitful framework for further study, given the surprising twists and turns of the story of extremal black holes so far, it is  extremely unlikely that anything written here will end up being the last word!

\bibliographystyle{utphys}
\bibliography{references}

%





\title{The Entropy of Black Holes}
\author{Robert M. Wald}
\institute{\textit{Enrico Fermi Institute and Department of Physics, University of Chicago 933 E. 56th St. Chicago, IL 60637}}

\maketitle 

\begin{abstract}
The remarkable connection between black holes and thermodynamics provides the most significant clues that we currently possess to the nature of black holes in a quantum theory of gravity. The key clue is the formula for the entropy of a black hole. I briefly review some recent work that provides an expression for a dynamical correction to the entropy of a black hole and briefly discuss some of the implications of this new formula for entropy.
\end{abstract}

\label{sec:wald}


More than fifty years ago, Bekenstein \cite{bek} proposed that a black hole in general relativity should be assigned an entropy proportional to its area, $A$. Shortly thereafter, Bardeen, Carter, and Hawking \cite{bch} showed that black holes satisfy laws of ``black hole mechanics'' that are precise mathematical analogs of the laws of thermodynamics, with $A$ playing the role of entropy, $S$, and the surface gravity, $\kappa$, of the black hole playing the role of temperature, $T$. However, at that time it was not consistent to identify the laws of black hole mechanics as physically corresponding to the laws of thermodynamics because the physical temperature of a black hole in classical general relativity is absolute zero. Soon thereafter, this situation changed dramatically: Hawking \cite{haw} discovered that when quantum field theory effects are considered, black holes radiate as black bodies at temperature $T = \kappa/2\pi$. From this point on, it was clear that the correspondence of the laws of black hole mechanics and the laws of thermodynamics is truly a physical correspondence, and that the entropy assigned to a black hole must represent its true thermodynamic entropy. The formula for the entropy of a black hole thus gives us an important clue to how black holes will be described in a quantum theory of gravity.

Thirty years ago  Iyer and I \cite{iw} derived a first law of black hole mechanics for perturbations of a stationary black hole in an arbitrary theory of gravity obtained from a diffeomorphism covariant Lagrangian, thereby enabling us to obtain a formula for black hole entropy in these much more general theories. However, our derivation required evaluation of the black hole entropy on the bifurcation surface, $\mathcal B$, of the black hole, thus restricting the validity of our formula for entropy to stationary black holes and their linear perturbations, evaluated at the ``time'' represented by $\mathcal B$. It is of considerable interest to obtain an expression for the entropy of a non-stationary black hole in a general theory of gravity at a ``time'' represented by an arbitrary cross-section, $\mathcal C$, since, in particular, this would allow one to investigate whether, classically, black hole entropy satisfies a second law (i.e., whether it is non-decreasing with time) and whether, semiclassically, black hole entropy satisfies a generalized second law (i.e., whether the sum of black hole entropy and a matter contribution to entropy is non-decreasing). Iyer and I had proposed a formula for dynamical black hole entropy in a general theory of gravity, but our formula was not field redefinition invariant, and we retracted our proposal in a ``note in proof'' in the published version of our paper \cite{iw}. About ten years ago, Dong \cite{dong} proposed a formula for dynamical black hole entropy in theories whose Lagrangian is an arbitrary function of curvature (but not derivatives of curvature) based on holographic entanglement arguments. Shortly thereafter, Wall \cite{wall} proposed a general prescription for dynamical black hole entropy for perturbations of a stationary black hole. When evaluated in the case of a Lagrangian that is a function of curvature, it was found to agree with Dong's formula \cite{wall}.

Very recently, Hollands, Zhang, and I \cite{hwz} have proposed a new definition of dynamical black hole entropy. As the derivation of \cite{iw} makes clear, the formula for the entropy of a black hole evaluated at its bifurcation surface $\mathcal B$ is closely analogous to the formula for ADM mass at spatial infinity. On the other hand, the desired properties of black hole entropy on an arbitrary cross section $\mathcal C$ are closely analogous to the desired properties of Bondi mass at null infinity. Zoupas and I had previously proposed a strategy for defining Bondi mass and other BMS charges at null infinity \cite{wz}. In \cite{hwz}, Hollands, Zhang, and I applied an analog of this strategy to the definition of dynamical black hole entropy. Our resulting formula is intended to be applied only at leading nontrivial order in perturbation theory about a stationary black hole. By its construction, it is guaranteed to satisfy a ``local physical process version'' of the first law of black hole mechanics \cite{hwz}.

We had no intention of modifying the Bekenstein-Hawking formula, $S_{\rm BH} = A/4$, for the entropy of a black hole in general relativity. However, in general relativity, our prescription yields a nontrivial dynamical correction to the Bekenstein-Hawking entropy formula on an arbitrary cross-section, $\mathcal C$, namely \cite{hwz}
\begin{equation}
S_{\mathcal C} = \frac{A[\mathcal C]}{4} - \frac{1}{4} \int_{\mathcal C} V \vartheta 
\label{grent}
\end{equation}
where $V$ is an affine parameter of the null generators of the horizon (with $V=0$ corresponding to the bifurcation surface $\mathcal B$), and $\vartheta$ is the expansion of these generators. For more general theories of gravity, our formula differs from the formula proposed by Dong and Wall by a similar dynamical correction term.

An important consequence of the fact that our notion of entropy satisfies a local physical process version of the first law of black hole mechanics is that our entropy will change only when matter or radiation enters the black hole. This contrasts with the Bekenstein-Hawking entropy, which increases before matter enters the black, since the teleological nature of the event horizon requires that it must move outward in anticipation of matter being thrown in at a later time. However, for the same reason, the horizon will also start expanding in anticipation of the matter that will later be thrown in. Our dynamical correction term $ - \frac{1}{4} \int_{\mathcal C} V \vartheta$  exactly compensates for the area increase (to leading order in perturbations about a stationary black hole), so that $S_{\mathcal C}$ does not change until matter or radiation actually enters the black hole. Furthermore, although eq.~(\ref{grent}) is evaluated on a cross-section $\mathcal C$ of the true event horizon, we have shown that the entropy $S_{\mathcal C}$ is, in fact, the area of the apparent horizon corresponding to ``time'' $\mathcal C$ \cite{hwz} and thus can be determined locally, without knowledge of the future behavior of the spacetime.

Another important consequence of the fact that our notion of entropy satisfies a local physical process version of the first law in an arbitrary theory of gravity is that our entropy satisfies the second law---i.e., it is nondecreasing---when matter satisfying the null energy condition is thrown into the black hole \cite{hwz}. However, it will satisfy the second law when gravitational radiation enters the black hole if and only if the ``modified canonical energy flux'' of this radiation is positive \cite{hwz}. This is the case for general relativity but would not be expected to hold in more general theories of gravity.

The quantum null energy condition (QNEC) \cite{bou} asserts that if ${\mathcal C}$ is a cross section of a null hypersurface with vanishing expansion and shear at ${\mathcal C}$, then the expectation value of the stress-energy flux of a quantum field through ${\mathcal C}$ bounds the second time derivative of the von Neumann entropy outside of ${\mathcal C}$. We have shown \cite{hwz} that when applied to a (perturbed) Killing horizon, QNEC is equivalent to a version of the generalized second law, with the total entropy taken to $S_{\mathcal C}$ plus a dynamically modified version of the von Neumann entropy for the matter outside the black hole. On the other hand, an integrated form of QNEC with suitable future boundary conditions yields the generalized second law for the Dong-Wall entropy of the black hole plus the ordinary von Neumann entropy of the matter outside the black hole.

There are a number of issues raised by our new formula, eq.(\ref{grent}), for black hole entropy in general relativity and its generalization to arbitrary theories of gravity. First, the dynamical correction term is well defined only at leading order in perturbations about a stationary black hole. Even in the case of general relativity, the choice of origin of $V$ in eq.(\ref{grent}) implicitly depends on a choice of background horizon Killing field and would not be well defined on an arbitrary, non-stationary horizon. This contrasts with the Bekenstein-Hawking entropy, which is well defined for an arbitrary black hole. For more general theories of gravity, the are considerable additional ambiguities in the definition of entropy beyond leading order in deviations from stationarity \cite{hwz}. It does not seem unreasonable that the notion of black hole entropy should be well defined only ``near equilibrium,'' but it would be worth considering whether our new notion of black hole entropy can be generalized to the case of black holes that are far from equilibrium.

Another issue arises from the fact that, at leading order, $S_{\mathcal C}$ in general relativity is equal to the area of the apparent horizon. This suggests that, perhaps, the entropy of a black hole should be associated with a structure that is defined locally in time, such as a dynamical horizon \cite{ak}, rather than the globally defined event horizon. In general, apparent horizons are slicing dependent and dynamical horizons are similarly non-unique, so it is hard to see how to view them as fundamental structures underlying the properties of black holes. However, this non-uniqueness is greatly alleviated at leading order in perturbations of a stationary black hole, which, as discussed above, is the only context in which our entropy has been defined. Thus, the possibility of attributing black hole entropy to a locally defined structure remains open.

A further issue is the appearance of the von Neumann entropy of matter and/or its dynamical modification in the generalized second law. In the first place, it is quite surprising that there is {\em any} precise statement and argument/proof of the generalized second law for black holes and matter, since there is no corresponding precise statement or proof of the ordinary second law in the context of non-general-relativistic physics. It is even more surprising that the von Neumann entropy and/or its dynamical modification appears in the generalized second law, since the von Neumann entropy would be constant in time for any complete system in non-general-relativistic physics. In addition, it would be interesting to see if further justification can be given for the dynamical modification term to the von Neumann entropy introduced in \cite{hwz}.

Finally, the issue of the fundamental meaning of black hole entropy remains to be resolved. In ordinary quantum statistical physics, the entropy of a system at a given value of energy and other state parameters is the logarithm of the number of states of that system in a small interval around that energy and other state parameters. It is normally taken for granted that the same interpretation of entropy applies to a black hole, so that the value of the entropy of a black hole corresponds to counting states of the black hole in quantum gravity description. But where do the degrees of freedom associated with these states of a black hole reside? Are they in the deep interior of the black hole? Or, are they located on the event horizon? Or, alternatively, do they reside in the ``thermal atmosphere'' surrounding the black hole? There are difficulties with all of these possible answers. Furthermore, in the context of quantum general relativity, it should be possible for new effective degrees of freedom to be created and/or destroyed. For example, in an expanding universe, field modes that would be counted as ordinary degrees of freedom in the present universe would have been sub-Planckian in the early universe and presumably would not have been included in the effective degrees of freedom of the field at early times. Similarly, near the singularity (or whatever replaces it) inside a black hole, it is possible that effective degrees of freedom may be destroyed. Thus, it does not seem obvious that black hole entropy corresponds to counting states in any usual sense.

\bibliographystyle{utphys}

\bibliography{references}

%

\title{The Non-linear Regime of Gravity}

\author{Luis Lehner}

\institute{\textit{Perimeter Institute for Theoretical Physics, 31 Caroline St., Waterloo, ON, N2L 2Y5, Canada}}

\maketitle

\begin{abstract}
The second century of General Relativity, building upon
the exquisite foundation that analytical and perturbative 
studies have provided, the detailed understanding of the non-linear regime of gravity will
increasingly take a prominent role. Fueled in part by computational advances as well as
observational challenges, and drawing inspiration and tools from other areas in physics,
new insights will be unraveled and likely exciting surprises.

\end{abstract}

\section{The non-linear regime of gravity}
Our understanding of General Relativity and the gravity phenomena it describes has from
the beginning relied on simplifying assumptions to enable its exploration. Such efforts include
restricting to particular symmetries, focusing on linearized perturbations off special solutions, 
assessing asymptotic/late time behavior, etc. As well, remarkable understanding has been obtained
purely on geometrical grounds, which has provided insights into global aspects of solutions.
To these undoubtedly impressive body of knowledge, the turn of the century saw the ability
to explore the highly dynamical/non-linear regime through simulations which both elucidated key
questions and pointed up new and interesting questions to explore.
A remarkable example in this front is the discovery of critical phenomena~\cite{Choptuik:1992jv} which provided a violation of cosmic censorship in 4D but crucially uncovered a rich phenomena
arising as the system approached the singular solution describing it, with the possibility of
'zero-mass' black holes, a universal scaling law above and below the threshold of black hole formation and a self-similar behavior of the spacetime in its vicinity.
This groundbreaking result illustrated the power of simulations to uncover new phenomena. While much effort in the computational front has been devoted to the description of the two-body problem in General Relativity (crucial for its relevance to gravitational wave astronomy), a number of efforts have been devoted for continued exploration of fundamental aspects of the theory in diverse fronts.
As I look towards the future, I see the dynamical/highly non-linear as the new frontier in gravity
where exciting results await. In what follows, I give some representative examples.\footnote{References listed are representative to the topics/literature raised and by
no means exhaustive. They are intended for illustration rather that for a definitive and inclusive presentation of results/discussions obtained in the areas discussed.}

\subsection{Cosmology}
Proposed solutions to the horizon and flatness problems in Cosmology are at the core of
the main models for cosmology. Cosmic inflation, bouncing cosmologies to name a couple of
options both rely (in different ways) on the assumption that the non-linear regime (either
at the onset of inflation or through the bounce) proceeds in particular ways. Incipient, and
gradually increasing, work is exploring in detail the reliability of these assumptions, pointing
out both robustness and limitations depending on the particular scenarios adopted~\cite{Corman:2022alv}.
Future studies will explore models with increasing generality, helping to further constrain models
and understand the role that non-linearities play in setting relevant scales.

\subsection{Non-linear behavior of black holes and the ultimate state conjecture}
The merger of black holes in 4D show a remarkable ``simplicity''. Gravitational waves
in the inspiraling case display an uncomplicated structure, devoid of significant energy
transfer to shorter wavelengths and soon after its peak amplitude describable in
terms of quasi-normal modes in accurate terms. What mediates this behavior? When does
the non-linear to linear transition takes place? How to characterize it?
These questions and the mounting evidence for connections between gravitational
phenomena and fluid-dynamics suggests the understanding of similar questions in the latter
might help inform avenues for exploration in the former. For instance, tantalizing connections
with turbulence have been recently indicated~\cite{Yang:2014tla}.

\subsection{Energy transfer,  weak turbulence and long-term behavior of key systems}
A number of systems have been identified with instabilities at the linear regime with a
future development that needs scrutiny in a fully non-linear treatment. To name a couple of examples:
the super-radiance instability~\cite{PhysRevD.7.949}, mass-inflation instability~\cite{Poisson:1990eh}, extreme black hole instability~\cite{Aretakis:2012ei}, and the black string instability~\cite{Gregory:1993vy} are all instances where an exponential departure from the unperturbed solution is identified and understanding its future development is important for practical reasons (for instance in the context of boson-clouds around  black holes) and/or
fundamental understanding (e.g. to assess cosmic censorship and stability of Kerr black holes).
Others require a non-linear analysis from the onset for its identification, e.g. pure AdS 
instability~\cite{Bizon:2011gg} and are also important in multiple contexts (e.g. holography). The non-linear
dynamics of these problems have, at its core, non-trivial energy flow and saturation towards a regular behavior or progress towards a singular behavior. This process can take place in a  relatively short time or in very long scales depending on the problem. The former can potentially
be resolved numerically~\cite{East:2017ovw,Lehner:2010pn} while for the latter simulations might provide guidance for a judicious analytical or perturbative approach to be pursued both for long-term arguments and to understand
the intricacies of the process~\cite{Balasubramanian:2014cja}.

\section{Final words}
The above are just but a few fronts where the understanding of the non-linear behavior
of gravity is of utmost importance to unravel particularly exciting questions. These may
range from practical ones (for their relevance to gravitational wave astronomy or cosmology) to
fundamental ones. The combination of simulations in combination with traditional approaches promise
to yield new breakthroughs in our understanding of gravity.

\bibliographystyle{utphys}
\bibliography{references}

%
\part{Black holes beyond general relativity}
%
\title{Black Holes Galore in \texorpdfstring{$D>4$}{D>4}}
\author{Roberto Emparan}
\institute{\textit{Institució Catalana de Recerca i Estudis Avançats (ICREA)\\
 Passeig Lluis Companys, 23, 08010 Barcelona, Spain} \and
 \textit{Departament de Física Quàntica i Astrofísica and
  Institut de Ciències del Cosmos,\\
 Universitat de Barcelona, 08028 Barcelona, Spain}}

\maketitle

\begin{abstract}
The black hole solutions to Einstein's vacuum equations in four dimensions contain just one example: the Kerr black hole. Over the past two decades, we have understood that higher-dimensional black holes are far more plentiful. I give a straightforward account of the reasons for this abundance based on three key ideas: (i) Horizons in $D>4$ can be long. (ii) Long horizons are flexible. (iii) Long horizons are unstable. I conclude with some comments and conjectures about the classification problem.
\end{abstract}

\section{Introduction}

Not long after Kerr generalized the Schwarzschild spacetime to include rotation, it was proven that the black hole solutions to Einstein's vacuum equations in four dimensions consist of a single member: the Kerr black hole \cite{Kerr:1963ud,Robinson:1975bv}. However, over the past two decades we have discovered a bounty of higher-dimensional black holes even in the vacuum sector. My goal here is to offer a concise explanation of the reasons behind this, using three elementary ideas:
\begin{enumerate}
\item Horizons in $D>4$ can be long.
\item Long horizons are flexible.
\item Long horizons are unstable.
\end{enumerate}
%
I will mostly confine my presentation to black hole solutions of
\begin{equation}
    R_{ab}=0
\end{equation}
in $D>4$ without any matter or compact dimensions. This is the simplest and most universal setup and the three properties above are generally true in it. A non-zero $\Lambda$, positive or negative, brings in new qualitative features, as does the addition of charges under different $p$-form gauge fields. However, many of the lessons learned in the vacuum sector have implications in those cases too. 

I apologize in advance since I will entirely omit a large part of the many fascinating results in this field and I will not properly credit many (most) of the participants in it. My presentation largely reflects my personal view of what I regard as some of the main general lessons, rather than specific findings, and is far from an exhaustive review. For more detailed accounts and other perspectives, the reader is referred to \cite{Emparan:2008eg} and \cite{Horowitz:2012nnc}.

\section{Why are there black holes galore in $D>4$?}

The reason why black holes should become more abundant as $D$ grows cannot simply be the larger number of degrees of freedom in the gravitational field. Indeed, the profusion of black holes first arose in the study of solutions that only depend on one radial and one angular direction, just like the Kerr solution, with all other directions being cyclic. The explanation begins instead with a simple observation about the gravitational field in different dimensions.

\subsection{Horizons in $D>4$ can be long}

By a long horizon, I mean one that is much longer in some directions than in others.
Imagine, then, an object that is very long in $p$ directions in a $D$-dimensional spacetime. Over scales smaller than this length, gravity in the directions transverse to the object is effectively $(D-p)$-dimensional, with the Newtonian potential falling off like
\begin{equation}
    \Phi\sim \frac1{r^{D-p-3}}\,.
\end{equation}
We can say that the lines of force dilute, or spread, only in $D-1-p$ dimensions.

In particular, this means that when
\begin{equation}
    D-p=3
\end{equation}
gravity is effectively three-dimensional. But it is known that there are no black hole solutions to the gravitational equations $R_{ij}=0$ in $D=3$. The reason is a matter of scales: in three dimensions $GM$ is a dimensionless quantity, so in contrast with the situation in four dimensions, where $GM$ sets the size scale for the Schwarzschild black hole radius, in three dimensions there is no size scale for a horizon of a given mass $M$. If we want a horizon we must introduce an explicit length, e.g., through $\Lambda=-1/L^2$, which leads to the BTZ black hole \cite{Banados:1992wn}.\footnote{With $\Lambda<0$ the gravitational repulsion does not yield a black hole but a cosmological horizon instead.}

\begin{figure}[h]
\centering
\includegraphics[width=0.35\linewidth]{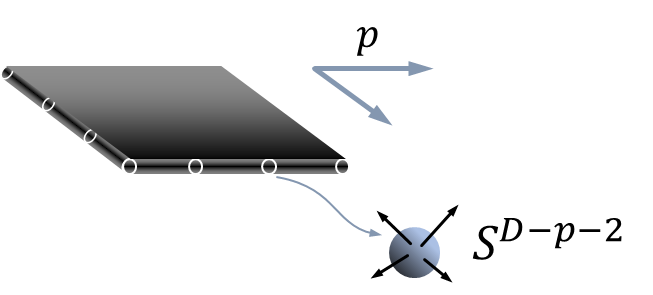}
\caption{Black $p$-branes can exist only in $D\geq p+4 $. \label{fig:blackbrane}}
\end{figure}

Therefore, to have horizon that is long in $p$ directions we need $D>p+3$. If $p\geq 1$ this means $D>4$.
That black holes with these long horizons actually exist is easily proven by simply adding $p$ flat directions to the Schwarzschild-Tangherlini solution \cite{Tangherlini:1963bw}. This results in black strings (with $p=1$) and more generally black $p$-branes (Fig.~\ref{fig:blackbrane}).

\begin{figure}[h]
\centering
\includegraphics[width=0.4\linewidth]{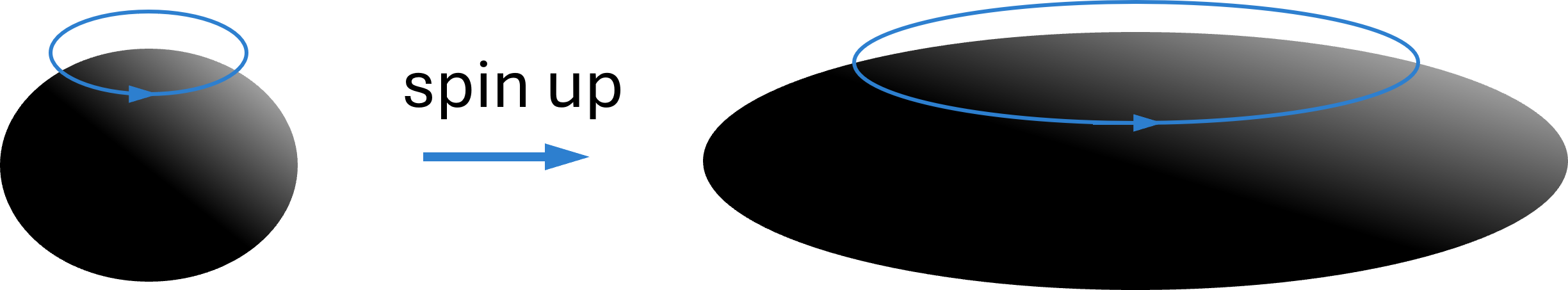}
\caption{When a black hole in $D\geq 6$ is spun-up, it spreads along the rotation plane and approaches the geometry of a black membrane. \label{fig:ultraspin}}
\end{figure}

Less trivially, one finds that ultraspinning black holes are possible in $D\geq 6$. The family of Myers-Perry black hole solutions \cite{Myers:1986un} with rotation in, say, a single plane, can be spun up to arbitrarily angular momentum $J$ for any given mass $M$. As conventional intuition suggests, such black holes spread---become long---on the rotation plane, and when $J$ grows very large they are well approximated, near the axis of rotation, by a black 2-brane \cite{Emparan:2003sy}  (Fig.~\ref{fig:ultraspin}).

\subsection{Long horizons are flexible}

Just like a black hole can be assigned a mass $M$ that is measured from the fall-off of the gravitational field near asymptotic infinity, a black $p$-brane can be assigned a stress-energy tensor---not only a mass (density) but also a pressure and possibly a momentum density---with components
\begin{equation}
    T_{ab}\,,\qquad a,b=0,1,\dots p
\end{equation}
along the $p$ directions in which the gravitational field does not decay.
 
Neutral black $p$-branes have positive energy density, $\varepsilon=T_{tt}>0$, and positive tension, $\tau=-T_{ii}>0$, $i=1,\dots p$. Viewed from a distance much larger than their thickness, the black $p$-brane looks like a flexible extended object: a long-wavelength deformation of the worldvolume in the directions transverse to it can be shown to obey the equations of an elastic deformation, and since the tension is positive, the brane is stable. Thus, black branes can bend like elastic material sheets \cite{Emparan:2009at,Camps:2012hw,Armas:2011uf}.

\begin{figure}[h]
\centering
\includegraphics[width=0.5\linewidth]{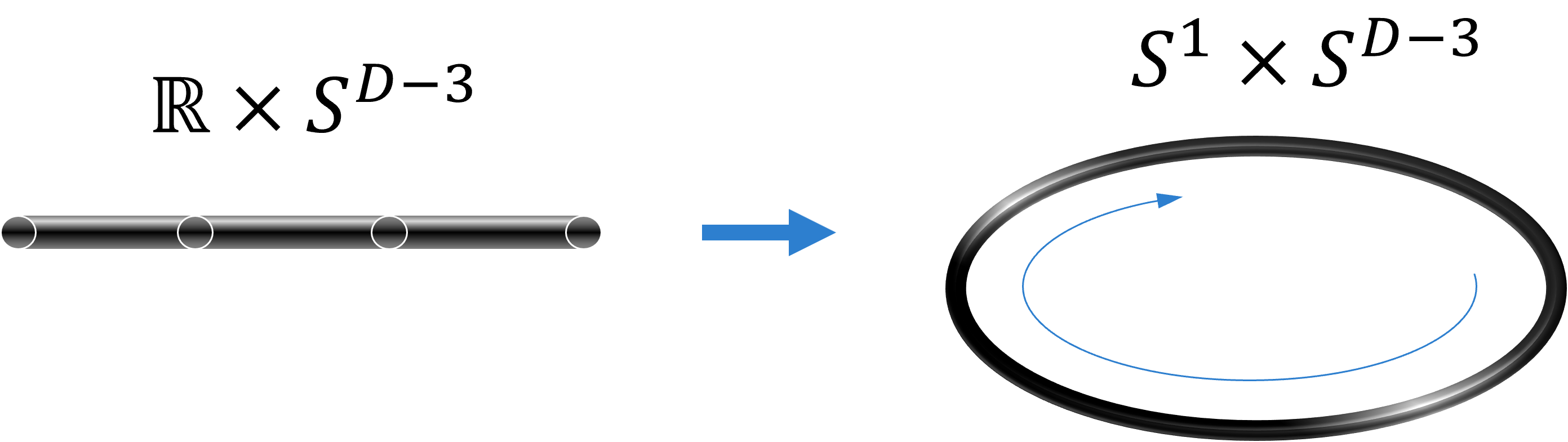}
\caption{A black string bent into a circle and spun up to balance its tension results in a black ring. \label{fig:blackring}}
\end{figure}

If we apply this idea to the simplest case of a black string, $p=1$, we can envisage bending it in a circular shape to form a \emph{black ring}. If we make it rotate along its length we can provide enough centrifugal force to balance its tendency to shrink under its tension. If the initial black string had horizon topology $\mathbb{R}\times S^{D-3}$, the final result is a black ring with $S^1\times S^{D-3}$ horizon \cite{Emparan:2001wn,Emparan:2007wm}  (Fig.~\ref{fig:blackring}).

The flexibility of the $D>4$ horizons thus allows them to have non-spherical topologies.

\subsection{Long horizons are unstable}
\begin{figure}[h]
\centering
\includegraphics[width=0.5\linewidth]{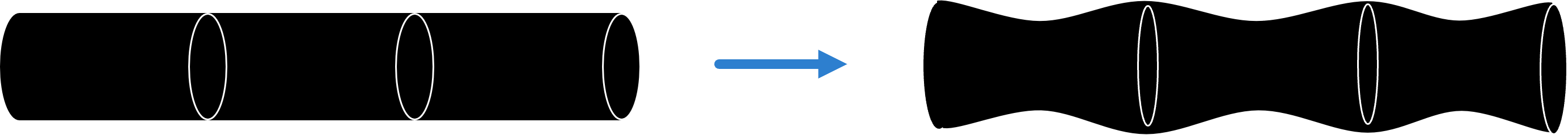}
\caption{Black strings and branes are unstable to developing ripples along their length. \label{fig:GL}}
\end{figure}

While the positive tension of black branes guarantees their stability under transverse, elastic deformations, it implies that they have negative pressure $P=-\tau<0$, which bodes ill for their stability under longitudinal fluctuations. Imagine a black string undergoing a fluctuation such that its thickness varies along the string length. In the limit where the fluctuation wavelength is very long, it can be described as a hydrodynamic pressure wave \cite{Emparan:2009at}. The soundspeed squared of these waves is $dP/d\varepsilon$, which is negative for a black string, and more generally for black branes, since $\varepsilon>0$ and $P<0$ are proportional to each other. An imaginary soundspeed means that the fluctuations do not propagate as sound, but rather they grow and make the string clump. This is the instability discovered by Gregory and Laflamme \cite{Gregory:1993vy}   (Fig.~\ref{fig:GL}).

This argument applies in the regime of small fluctuations with very long wavelength. The instability must nevertheless switch off for perturbation wavelengths sufficiently shorter than the horizon thickness, since these would be fluctuations of a geometry that is very approximately flat spacetime, which is stable. A static zero-mode deformation of the horizon appears when the wavelength is at the threshold of the instability. 

Gregory and Laflamme presented a heuristic argument based on global thermodynamics to support their result: the entropy of a black string wrapping a long enough compact circle is less than the entropy of a black hole localized in the circle. The argument above is instead based on local thermodynamics: since $dP/d\varepsilon$ is inversely proportional to the specific heat of the black string, small variations of its thickness increase the entropy. This can be made rigorous in the long wave limit.

We conclude that black-brane-like horizons are unstable. Then, ultraspinning black holes, which approach the geometry of black branes, must be unstable too \cite{Emparan:2003sy}. The conclusion also applies to thin black rings.\footnote{Higher-dimensional black holes can also be unstable to fluctuations different than the pressure waves discussed above.}

When the growth of the instability is followed into the deeply non-linear regime, a singular pinch forms on the horizon \cite{Lehner:2010pn}. Violations of cosmic censorship of this kind are a fascinating and pervasive feature of higher-dimensional black hole physics.

Another implication of these results is that, in contrast with the smoothness of four-dimensional horizons, higher-dimensional black holes can be bumpy. The reason lies in the zero modes that appear at the threshold of the instability (which are not hydrodynamic). When these linearized zero modes are extended non-linearly, they give rise to bumpy horizons: static non-uniform black strings and lattice black branes, and stationary rotating bumpy black holes \cite{Wiseman:2002zc,Dias:2017coo,Emparan:2014pra}.

\subsection{A matter of scales}

Let us recap briefly what the previous observations are telling us. In vacuum, the  only length scales of a black hole are associated with their mass and spin,
\begin{equation}
    \ell_M=(GM)^\frac1{D-3}\,,\qquad \ell_J=\frac{J}{M}\,.
\end{equation}

In $D=3$, $GM$ is dimensionless. Rotation gives a length $\ell_J$ but not the attraction that would be needed for a black hole.

In $D=4$ there cannot be two widely separate length scales on the horizon, and consistently, if we try to make $\ell_J$ larger than  $\ell_M$, the Kerr horizon ceases to exist.\footnote{Close to extremality there arises a direction $\gg \ell_M$, but this is not along the horizon but transverse to it: a throat.} The presence of a single scale $\sim \ell_M$ then helps us understand why four-dimensional black holes are scarce, plump, smooth, stable, and topologically spheres (more on the latter in the appendix).

However, in $D>4$ these constraints disappear since horizons can be long with two separate scales, $\ell_J\gg \ell_M$. This is the reason why there are so many black holes in $D>4$ that are 
\begin{itemize}
\item not always plump and smooth, but can be long and bumpy,
\item not always stable,
\item not always topologically spherical, but ring-like and much more.
\end{itemize}

\section{Classification vs.\ guidance, and a conjecture}

When approaching this subject it may seem natural to attempt a classification of all possible higher-dimensional black holes, at least in the vacuum theory. After all, the classification in $D=4$ was solved once the uniqueness theorems were proven.

In my opinion, we can only hope for a complete classification in $D=5$, where there already exist partial but substantial results \cite{Hollands:2012xy}. In particular, the integrability of the stationary spacetimes with two commuting spatial isometries allows the explicit construction and analysis of infinite families of new solutions. 

In $D\geq 6$ the possibilities for the horizons explode, as we have seen.\footnote{The lack of a complete classification of topologies for manifolds of four or more dimensions does not seem to be directly related to this problem, but it certainly does not help with it.} But even if we may not solve the mathematical problem of classification, we can aim at the physically more important question of whether a few guiding principles govern the main features of the space of possible solutions. This is what the ideas exposed above attempt to achieve.

The classification problem might nevertheless be solvable in a physically relevant manner. It seems likely that the only black hole solutions of $R_{ab}=0$ that are dynamically stable are those in the Myers-Perry family with angular momenta that are bounded above,
\begin{equation}\label{Dbound}
    |J_i|< \alpha_D M(GM)^\frac1{D-3}\,,
\end{equation}
where $\alpha_D$ is a $D$-dependent pure number that, in general, must be determined numerically. All other black holes are conjectured to be unstable, and there is considerable evidence that they are.

These stable, low-spin Myers-Perry black holes resemble in many ways the Kerr black hole, even though in general they do not become extremal when the bound \eqref{Dbound} is saturated. Stability, which played no role in the proof of the uniqueness of the Kerr black hole, may be the criterion that makes black holes in $D\geq 5$ unique again.

\section*{Acknowledgements}
I thank the organizers of `Black Holes Inside and Out' for the invitation to present these ideas in such a stimulating environment, and for prompting me to write them up. Work supported by MICINN grant PID2022-136224NB-C22 and AGAUR grant 2021 SGR 00872.

\appendix

\section{Some caveats}

\paragraph{$\Lambda<0$.} Long horizons in $D\geq 4$ can exist with $\Lambda<0$. It is left to the reader to reason along the lines above to justify why this is so.

\paragraph{$p$-brane charge.} Long horizons can be stabilized with $p$-brane charges. These black $p$-branes are elastically stable since they have positive tension, but it is not simply proportional to the energy density. The condition $dP/d\varepsilon>0$ for the stability of sound waves can then be satisfied for a sufficiently large $p$-brane charge.

\paragraph{Topological censorship.}The impossibility of long horizons in $D=4$ suggests that the black hole must be roughly spherical but does not really forbid the existence of `fat black rings' with toroidal horizons. The complete proof relies on the topological censorship theorem, which states that any causal curve $\gamma$ that extends from $\mathcal I^-$ to $\mathcal I^+$ can be smoothly pushed to lie entirely at asymptotic infinity \cite{Friedman:1993ty}. In four dimensions, if a $S^1\times S^1$ horizon existed, the curve $\gamma$ could thread it and the link with the torus would make it impossible to deform it to infinity. In $D\geq 5$ spacetime dimensions there is no obstruction since two curves never form a link. Thus $\gamma$ cannot link a ring-like object with surface $S^1\times S^{D-3}$ and horizons with this topology are allowed. Indeed we have seen that they result from the flexibility of the horizons in $D>4$. The possibilities go much beyond black rings. For instance, a black $3$-brane with horizon $\mathbb{R}^3\times S^{D-5}$ can be bent into $S^3\times S^{D-5}$ and then balanced by adjusting two independent angular momenta \cite{Emparan:2009vd}. 

\bibliographystyle{utphys}
\bibliography{references}

\title{Same as Ever: Looking for (In)variants in the Black Holes Landscape}
\author{Carlos A. R. Herdeiro}
\institute{\textit{Departamento de Matem\'atica da Universidade de Aveiro and
Centre for Research and
Development in Mathematics and Applications (CIDMA), Campus de Santiago, 3810-193
Aveiro, Portugal}}

\maketitle 

\begin{abstract}
\textit{"Is it a goose, a duck or a swan?"} -  asked the alien.\textit{"I do not know; and to \underline{know} we have to look \underline{closer}."} - said the earthling. \textit{"But even from here we can see it has webbed feet... so it is not a chicken."}
\end{abstract}


\section{Prologue: What never changes}

In science one is constantly looking for the next breakthrough, the next game changer. It then becomes easy to overlook the basic, which is what is  \textit{the same as ever}~\cite{housel}. 

The same applies in business. The Amazon founder Jeff Bezos made the following remark: \textit{I very frequently get the question: "What's going to change in the next 10 years?" And that is a very interesting question; it's a very common one. I almost never get the question: "What's not going to change in the next 10 years?" And I submit to you that that second question is actually the more important of the two -- because you can build a business strategy around the things that are stable in time.}~\cite{JB}

In this era of strong gravity data, can we build a strategy for understanding  the nature of the most compact objects in the universe around what is ``the same as ever"?

\section{Two questions and their answers. And then, their denials.}
There are paradigmatic black hole (BH) features, learned from the canonical electro-vacuum solutions ($i.e.$ the Kerr-Newman family~\cite{Townsend:1997ku}), $e.g.$: a) BHs have a spherical horizon; b) Spinning BHs have an ergo-region; c) BHs have light rings (LRs) and other bound photon orbits; d) BHs have ``no hair".  Many of these and other features have phenomenological impact. But it is natural to question the generality of such features in broader classes of models. Here, I will focus on the latter two aspects, LRs and hair, each addressed as a question.

\subsection{ “Do all black holes have LRs?”}

That is “planar”, in an appropriate sense, bound photon orbits? LRs are extreme examples of light bending and their existence and properties have relevance for BH imaging~\cite{Cunha:2017eoe} and ringdown~\cite{Cardoso:2016rao}.  There is a generic statement (theorem), which states: \textit{A stationary, axisymmetric, asymptotically flat BH in D=4, with a non-extremal, topologically spherical Killing horizon admits, at least, one standard LR outside the horizon for each rotation sense}~\cite{Cunha:2020azh}. 

This LRs theorem exemplifies a \underline{field equations independent} statement. So it is \textit{the same as ever}, regardless of the theory of gravity or the specific solution. It is topological and associates to LRs a topological charge $w$, first introduced in~\cite{Cunha:2017qtt}. 

\begin{figure}[ht!]
\centering
\includegraphics[width=0.25\linewidth]{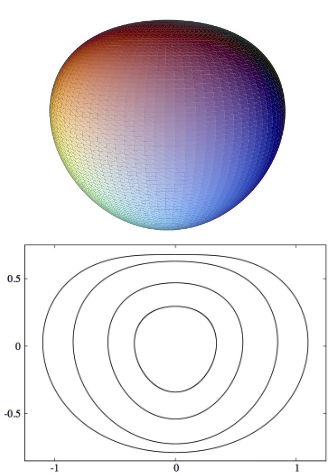}
\includegraphics[width=0.35\linewidth]{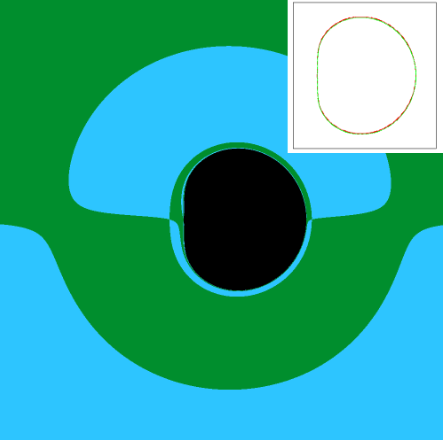}
\caption{A spinning BH with electric and magnetic charge in KK theory~\cite{Rasheed:1995zv} (Left) Isometric embedding of the spatial sections of the horizon, in 3D (top) and 2D contours for different spacetime parameters (bottom). (Right) Lensing and shadow. Extracted from~\cite{Cunha:2018uzc}.
\label{fig2}}
\end{figure}

The theorem does not specify where the LRs are, only that they are present. For instance, unlike the Kerr case, for a general BH the LRs need not be on an equatorial plane of the spacetime. In fact, the BH spacetime needs not have an equatorial plane, as the theorem does not assume that the BH has a $\mathbb{Z}_2$ north-south symmetry and hence an equatorial plane as the fixed plane of the symmetry. As curiosity, there are in fact examples of BHs (even with a connected event horizon) which do not have such a $\mathbb{Z}_2$ symmetry and still possess the LRs predicted by the theorem. Fig.~\ref{fig2} shows one such example: a spinning, electrically and magnetically charged BH in Kaluza-Klein (KK) theory~\cite{Cunha:2018uzc}. The horizon embedding (left panels) is clearly non-$\mathbb{Z}_2$ symmetric. But the so-called "shadow" - the dark region (right panel) - is $\mathbb{Z}_2$ symmetric, regardless of the observation angle; on the other hand, the lensing (right panel) is not $\mathbb{Z}_2$ symmetric, in general.

\begin{figure}[ht!]
\centering
\includegraphics[width=0.25\linewidth]{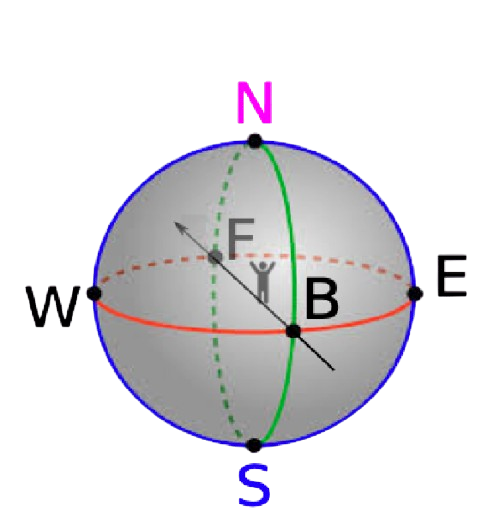}
\includegraphics[width=0.28\linewidth]{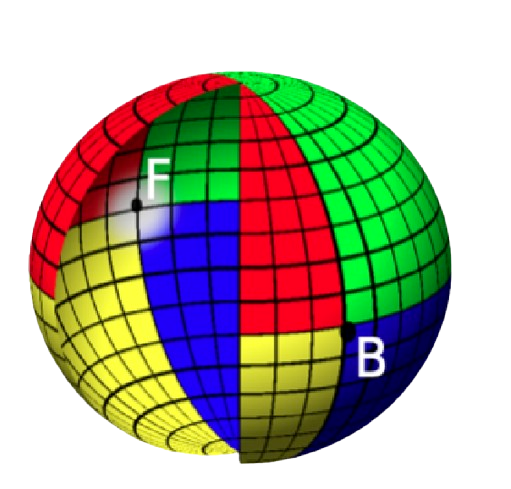}
\includegraphics[width=0.45\linewidth]{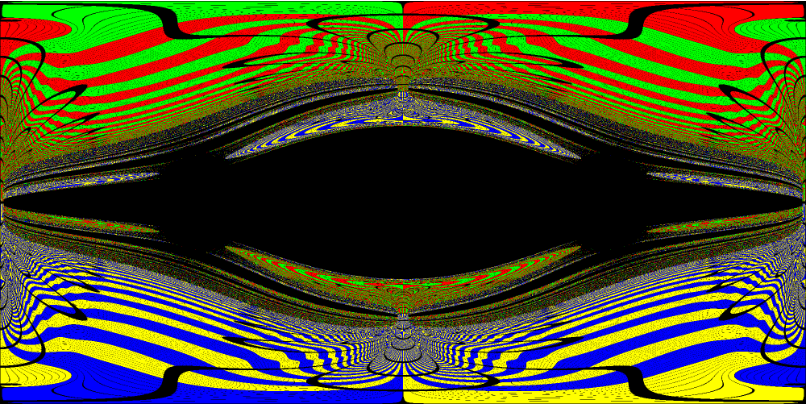}
\caption{In a family of BHs in Melvin Universes, an equatorial observer looking forward at the BH region (left), within a colourful
celestial sphere (middle), observes a panoramic “shadow” (right), in a 360$\degree$ wide observation window. Extracted from~\cite{Junior:2021svb}.
\label{fig:fig3}}
\end{figure}

Perhaps the result of the above theorem does not appear surprising. But, in fact, there are BHs  without LRs - albeit only in academic examples so far. And the theorem offers guidelines of how to find them. Let me illustrate this. The asymptotically flat Schwarzschild BH has a topological charge $w=-1$. Since the topological charge is additive, if one could immerse the latter in a background with $w=+1$ then such non-linear superposition would have $w=0$ offering the possibility of a BH spacetime without LRs.  It turns out the Melvin universe of electro-vacuum~\cite{Melvin:1963qx}, a self-gravitating uniform magnetic field, has $w=+1$; it always brings along a stable LR. Then, a Schwarzschild BH inside a Melvin Universe has a total topological charge zero~\cite{Junior:2021dyw}. For sufficiently large magnetic fields, in fact, there are no LRs outside the BH horizon. Since LRs yield critical points for light scattering, the absence of LRs leads to a curious feature: light sent from an equatorial observer, along the equatorial plane, along any azimuthal direction (but one point) falls onto the BH. Consequently, the observer sees the BH shadow (this dark region in the image) all around, along the equator: a \textit{panoramic} shadow~\cite{Junior:2021dyw,Junior:2021svb} - Fig.~\ref{fig:fig3}. This illustrates the impact on the imaging of the LRs, which in this (rather academic) case is dramatic. In fact, the analysis of LRs, and their non-planar generalisations - fundamental photon orbits~\cite{Cunha:2017eoe} - provides a simple (albeit not precise) diagnosis of how non-Kerr the phenomenology of an alternative model may be.

Studying this topological charge is an informative technique. For instance, in different asymptotics it could be shown that the topological charge is still $w=-1$~\cite{Wei:2020rbh,Moreira:2024sjq}, also implying the existence of standard LRs. This also holds for some topologically non-trivial horizonless objects such as wormholes - they always have LRs~\cite{Xavier:2024iwr}. But for smooth, topologically trivial horizonless objects, $w=0$~\cite{Cunha:2017qtt,DiFilippo:2024mnc}, meaning that LRs, if present, come in pairs with one standard and one exotic, which may impact on the stability of such objects~\cite{Cunha:2022gde}. In Einstein-Maxwell-dilaton, the topological charge for Melvin asymptotics varies from $w=0$ for Einstein-Maxwell to $w=-1$ when increasing the dilatonic coupling, with the discontinuity at the KK value~\cite{Junior:2021svb}. There are also connections between the existence of ergo-regions and LRs~\cite{Ghosh:2021txu} and various generalizations~\cite{Tavlayan:2022hzl,Cunha:2024ajc}, besides the technique inspiring some spin-offs, for instance to understand special timelike~\cite{Wei:2022mzv,Cunha:2022nyw} (rather than null)  orbits, or the topology of BH thermodynamics~\cite{Wei:2021vdx}.

\subsection{“Do all (asymptotically flat) black holes have no hair?”}

\begin{figure}[ht!]
\centering
\includegraphics[width=0.9\linewidth]{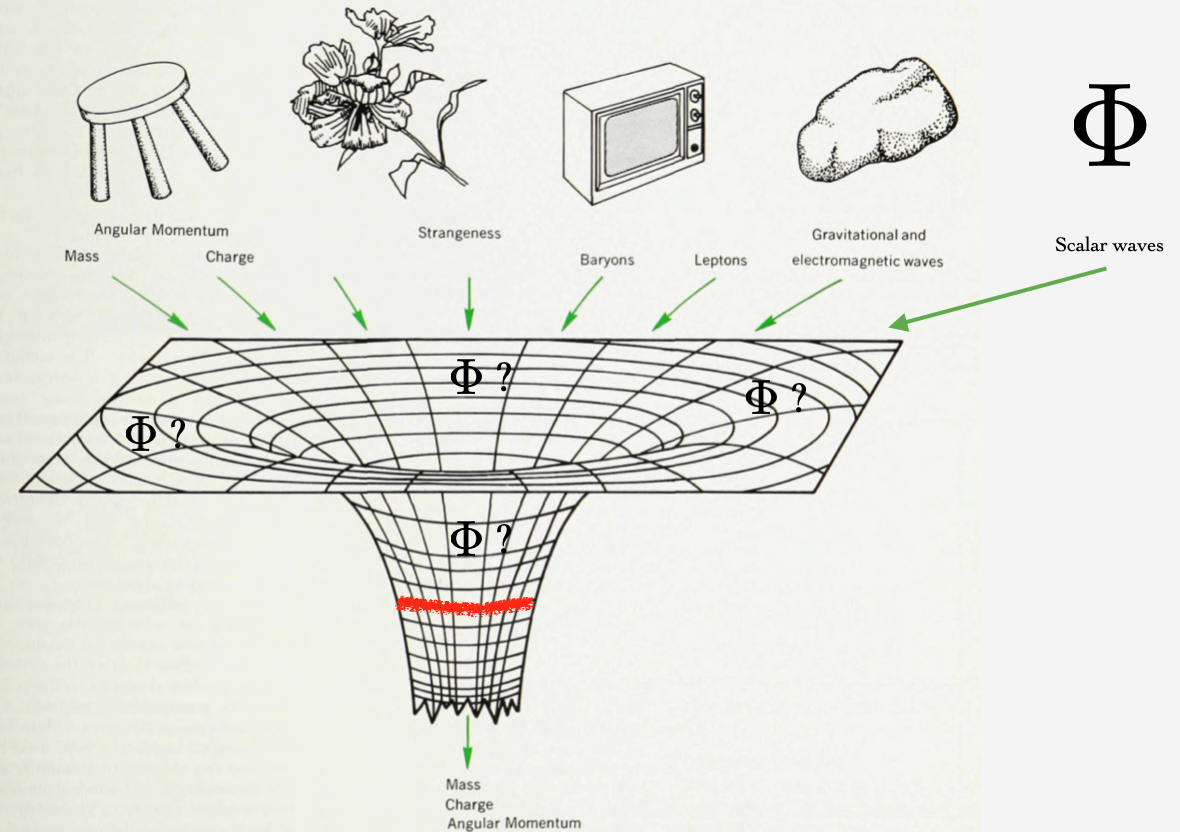}
\caption{Can BHs have "hair" of a scalar field not associated to a Gauss law? Adaptation of the classical figure in~\cite{Ruffini:1971bza}, first shown by the author in a seminar in Cambridge in March 2014~\cite{cambridge}.
\label{fig4}}
\end{figure}

That is, are BHs uniquely determined by their total mass M, total angular momentum J, and total gauge charges Q, all asymptotically measured quantities subject to a Gauss law - and no other independent characteristics (“hair”)~\cite{Ruffini:1971bza}?  Here, there are two types of generic statements (theorems). Firstly, in electro-vacuum, there is a set of \textit{uniqueness} theorems, establishing, within generic  assumptions (but see $e.g.$~\cite{Carter:1997im}), that asymptotically flat stationary BH solutions with a connected event horizon, regular on and outside an event horizon, belong to the Kerr-Newman family~\cite{Israel:1967wq,Israel:1967za,Carter:1971zc,Robinson:1975bv,Racz:1995nh,Chrusciel:2012jk}. Secondly, beyond electro-vacuum, there are “no-hair” (which are no-go) theorems; there are many such different theorems/arguments applying to different models and under different assumptions. In the case of asymptotically flat BHs, for models with scalar fields and no gauge fields, where the scalar field $\Phi$ is not subject to a Gauss law - Fig.~\ref{fig4} - a partial list can be found in~\cite{Herdeiro:2015waa} - see also~\cite{Bekenstein:1996pn,Volkov:2016ehx}.

A simple and paradigmatic no-hair theorem is due to Bekenstein~\cite{Bekenstein:1972ny}, that uses only the scalar field equation (not the Einstein equations). It is enlightening,
because its assumptions give the leads to how it can be circumvented. One considers a stationary, axi-symmetric, asymptotically flat BH spacetime and the following assumptions:
\begin{description}
    \item[1)] The scalar field is canonical in its kinetic term and is minimally
coupled to Einstein’s gravity. But one allows a scalar potential, $V(\Phi)$. 
\item[2)] The scalar field inherits the spacetime symmetries: it is invariant under the action of the Killing vectors.
\item[3)] Requirement on the scalar potential: the original Bekenstein version ($\Phi dV/d\Phi\geq 0)$ is not particularly significant and may be violated for potentials of interest; other versions assume, say, convexity ($d^2V/d\Phi^2\geq 0)$~\cite{Sotiriou:2011dz}, positivity ($V\geq 0)$~\cite{Heusler:1996ft} or the weak energy condition~\cite{Bekenstein:1995un,Heusler:1992ss,Sudarsky:1995zg}. They all rule out scalar hair of one (or many) real, massive free scalar field minimally coupled to Einstein's gravity.
\end{description} 

Whereas violating assumptions 1 and 3 lead to many classes of scalar ``hairy" BHs, with non-minimal couplings, non-canonical kinetic terms, or potential violating energy conditions - see~\cite{Herdeiro:2015waa} for examples -, a most curious trap door relates to the innocuous looking assumption 2. In fact, dropping this symmetry inheritance between the geometry and matter opens up a general mechanism, which works in GR and beyond and, moreover, relates to the wonderful physics of \textit{sync}.

\bigskip

\underline{Sync: same as ever}

\bigskip

Dynamical synchronization (sync) occurs ubiquitously in physical and biological systems. Communities of fireflies or crickets, sets of metronomes or pendulums, are illustrations wherein individual cycles converge to the same phase, if appropriate interactions are present~\cite{sync}. In the mangrove forests of Thailand, experiment shows thousands of live fireflies synchronizing their flashes due to a few computer controlled LED’s~\cite{fireflies}. Another striking example is a set of metronomes synchronizing by being placed on the same moving base~\cite{metronomes}.

Sync is also just in front of us in the sky. The Moon always shows the same face towards the Earth. This synchronous rotation of the Moon with its orbit is dynamically driven by tidal forces~(see $e.g.$~\cite{Hut:1981}) and it is known as \textit{tidal locking}. It is a common feature in binary planetary systems. For instance, most large moons of Jupiter, Saturn, Uranus and Neptune are tidally locked with their planet~\cite{wikitidal}. The Earth, however, is not \textit{yet} showing always the same face towards the Moon; but it is being driven into that configuration by tidal dynamics.

In a detached binary system, each ``planet" raises tides on the other's surface~(see $e.g.$~\cite{Hut:1981}). Dissipation -- $e.g.$ friction between the Earth's oceans and crust -- causes a misalignment between the tides and the line joining the 2 planets. This produces a gravitational torque, which exchanges angular momentum between the rotation, $\vec{S}$, and the orbit $\vec{L}$, of the planets. Focusing on the  ``planet", with rotational frequency, $\Omega_{\rm rot}$, tides due to a smaller companion  ``moon", with revolution frequency $\omega_{\rm rev}$, in a binary system with aligned orbital and equatorial planes, lead to three possibilities:
\begin{description}
\item[i)] If $\omega_{\rm rev}>\Omega_{\rm rot}$, there is an \textit{inwards} tidal acceleration, and the moon drifts \textit{towards} the planet. This occurs, $e.g.$, with Mars' moon Phoebos~\cite{phoebos} ($T_{\rm rev}\simeq 7$h$40$; $T_{\rm rot}\simeq 24$h$37$). 
\item[ii)] If $\omega_{\rm rev}<\Omega_{\rm rot}$, angular momentum transfer from $\vec{S}$ to $\vec{L}$ yields an \textit{outwards} tidal acceleration, and the moon drifts away from the planet. This is currently occurring with Earth's Moon~\cite{Touma:1994sf} ($T_{\rm rev}\simeq 27.3$ days [sideral month]~\cite{moondata}; $T_{\rm rot}\simeq 23$h$56$ m [sideral day]). 
\item[iii)] If $\omega_{\rm rev}=\Omega_{\rm rot}$, then the system is in tidal equilibrium. Angular momentum transfer ceased and there is no drift. This is the case of the Pluto-Charon system~\cite{cheng}. 
\end{description}
Thus, \textit{Nature likes synchrony} and, in Newtonian gravity, tidal dynamics enforces it, creating synchronized, balanced binaries as the endpoint of tidal evolution.

\begin{figure}[ht!]
\centering
\includegraphics[width=0.7\linewidth]{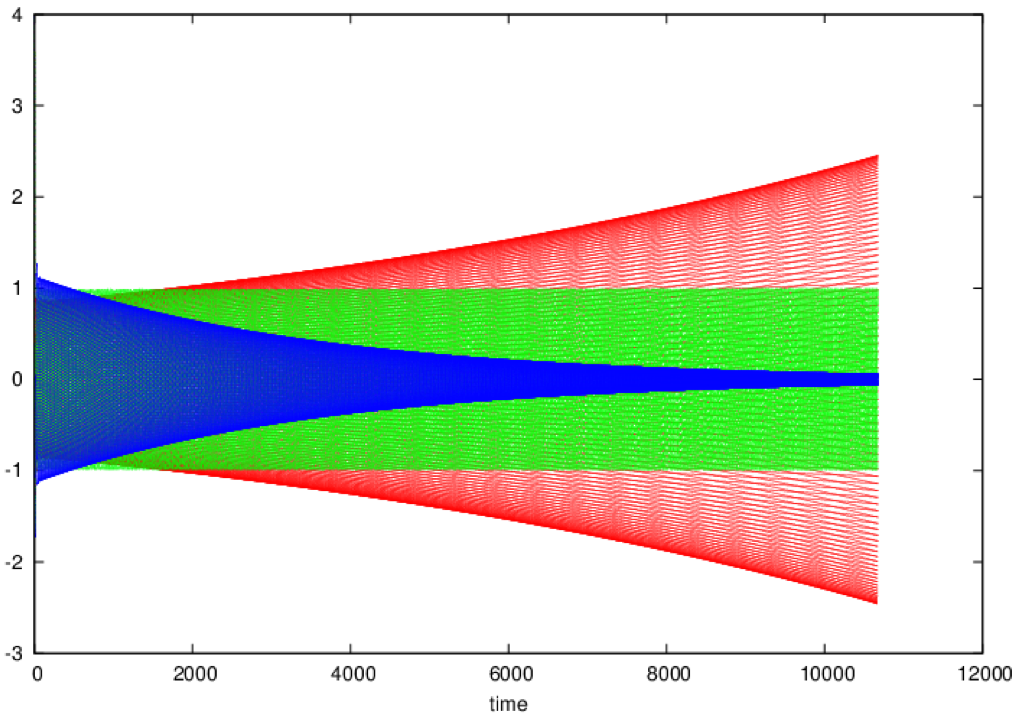}
\caption{Evolution of the amplitude of three different types of scalar modes around a BH - courtesy of J.C. Degollado (unpublished).
\label{fig5}}
\end{figure}

\begin{figure}[ht!]
\begin{center}
\includegraphics[width=0.45\textwidth]{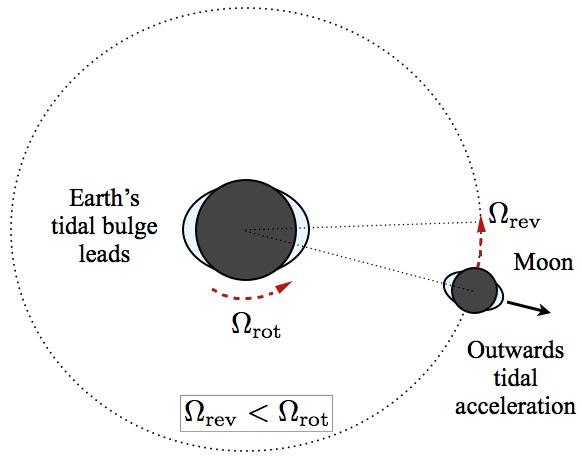}\ \ \ \ \ \ \ \ \ \ \ 
\includegraphics[width=0.35\textwidth]{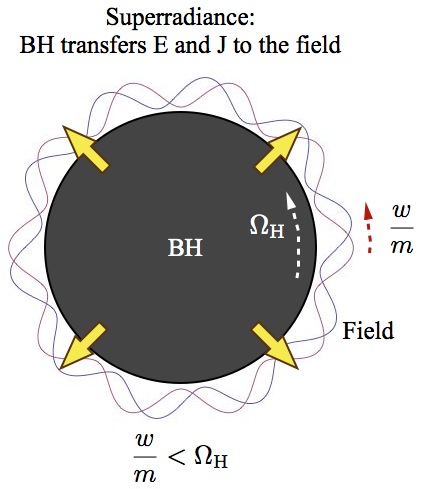} \\ \vspace{0.5cm}
\includegraphics[width=0.46\textwidth]{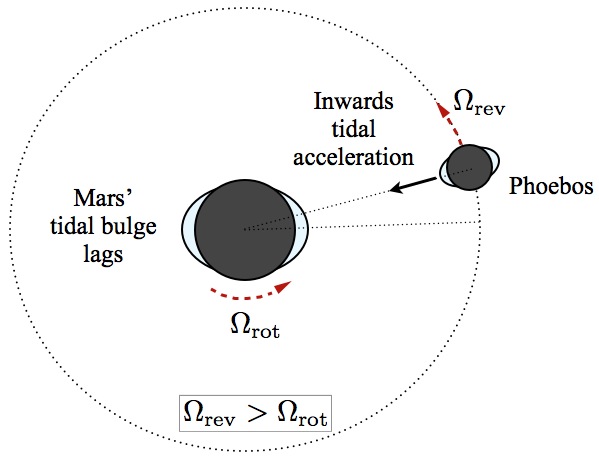} \ \ \ \ \ \ \ \ \ \ \ 
\includegraphics[width=0.35\textwidth]{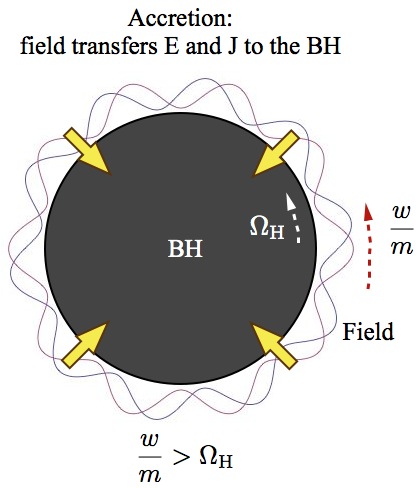}\\ \vspace{0.5cm}
\includegraphics[width=0.46\textwidth]{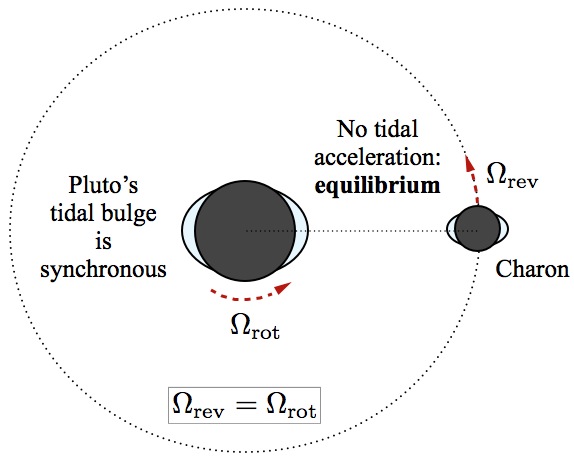} \ \ \ \ \ \ \ \ \ \ \ 
\includegraphics[width=0.35\textwidth]{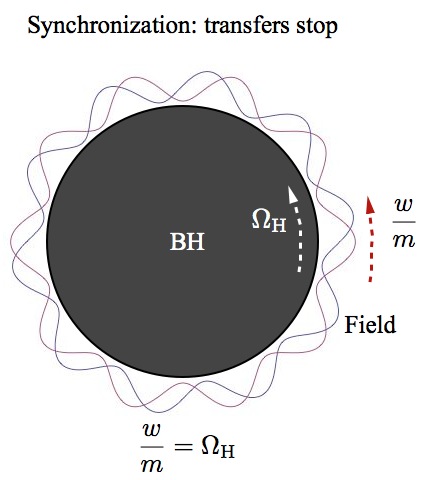}
\caption{\small{Tidal dynamics in Newtonian theory (left) and Einstein's gravity (right). }}
\label{fig6}
\end{center}
\end{figure}

In relativistic gravity, there is something alike in the interaction between a Kerr BH with horizon angular velocity $\Omega_H$ (the "planet") and a bosonic field (the "moon"). Take a scalar field as the simplest example. In the test field approximation consider the Klein-Gordon equation on Kerr in Boyer-Lindquist coordinates $(t,r,\theta,\varphi)$, as a test field. Performing the standard mode decomposition $\Phi=e^{-i\omega t+im\varphi}S_{\ell m}(\theta)R_{\ell m}(r)$~\cite{Brill:1972xj}, which introduces the scalar field frequency $\omega$ and the integer harmonic index $m$, this leads to a radial Teukolsky equation~\cite{Teukolsky:1972my}. Imposing appropriate boundary conditions, one obtains quasi-bound states, with a complex frequency, $\omega=\omega_R+i\omega_I$. The sign of the imaginary part $\omega _I$ depends on the relation between the real part and a critical frequency $\omega_c$, defined as $\omega_c/m=\Omega$. This is a \textit{sync condition} between the phase angular velocity of the scalar field and the horizon angular velocity of the BH. Then, there are three possibilities for the dynamics of the BH-scalar field ``binary":
\begin{description}
\item[i)] If $\omega_{R}>\omega_{c}$, then $\omega_I<0$  and the mode decays, being absorbed by the BH. This is the only regime for a Schwarzschild BH. It is the accretion regime - blue mode in Fig.~\ref{fig5}. 
\item[ii)] If $\omega_{R}<\omega_{c}$, then $\omega_I>0$  and the mode grows. This is the superradiant regime. Press and Teukolsky imagined a black hole bomb~\cite{Press:1972zz} sourced by it -  red mode in Fig.~\ref{fig5}.
\item[iii)]  If $\omega_{R}=\omega_{c}$, then $\omega_I=0$ and the modes are true bound states - with a real frequency - in equilibrium with the Kerr BH - green mode in Fig.~\ref{fig5}.
\end{description}
The dynamics of the planet-moon and BH-field binaries are compared in Fig.~\ref{fig6}.

The modes with $\omega_{R}=\omega_{c}$ can be continued to the full non-linear regime of Einstein's gravity minimally coupled to a \textit{complex} massive (free) scalar field. The complexity allows the above ansatz for the scalar field, which does not inherit the spacetime symmetries and circumvents all the no-hair theorems to yield a whole landscape of asymptotically flat BHs with sync bosonic hair, pioneered by~\cite{Herdeiro:2014goa}. 
Moreover, the sync state could be dynamically obtained by superradiance as shown from fully non-linear numerical simulations~\cite{East:2017ovw,Herdeiro:2017phl}.

The dynamics of these BHs with sync hair is involved. They are certainly unstable against higher $m$ superradiant modes~\cite{Herdeiro:2014jaa,Ganchev:2017uuo}. This is akin to a planet in which further moons with different orbital periods tend to destabilize the tidal locking achieved with the closest moon. In this sense, no such hairy BH is absolutely (classically) stable against small perturbations, neither is the Kerr BH in the same models. But time scales are important if one is considering possible astrophysical roles and they could be cosmologically long depending on the mass of the bosonic field and of the BH~\cite{Degollado:2018ypf}.

From the viewpoint of the landscape of BH solutions sync is a very generic mechanism to endow spinning black objects with ``hair".  One can add scalar self-interactions, $e.g.$~\cite{Herdeiro:2015tia}  or replace the scalar by Proca hair~$e.g.$~\cite{Herdeiro:2016tmi};  one can consider other spacetime dimensions and asymptotics. In fact the first example was in $AdS_5$~\cite{Dias:2011at}.  One can add gauge charges, use non-minimal couplings or non-canonical kinetic terms~$e.g.$~\cite{Delgado:2016jxq,Kleihaus:2015iea,Herdeiro:2018daq}. Different types of excited states can be solutions, $e.g.$~\cite{Wang:2018xhw}; and using such excited states, actually a scalar dipole, it is possible to put two spinning BHs in equilibrium in an asymptotically flat spacetime, without conical singularities~\cite{Herdeiro:2023roz}. There is also a purely charged version, in which  the sync condition takes the guise of a resonance condition, and one can add this type of hair to Reissner-Nordstr\"om BHs - an interesting related story~\cite{Herdeiro:2020xmb,Hong:2020miv}.

\section{Epilogue}

\textit{Jim was traveling through Omaha, Nebraska, with Warren Buffett at the end of 2003. The global economy was bad. One could see closed shops and bankrupted companies. Jim said to Warren: 
\\ - Everything is bad. How can the economy recover from such a thing?\\
- Jim, do you know what was the top-sellling candy bar in 1962? \\
- No.\\
- Snickers - said Warren - and do you know what is the top-sellling candy bar today?\\
- No.\\
- Snickers. \\
Silence followed -  
}
(Adapted from~\cite{housel}.) 

\bigskip


Some things are same as ever, or change very little, and to find them is reassuring. There are BH properties that are independent of the model as in the case of LRs; exceptions exist but they may be merely curiosities. Or, in some other cases, in the exceptions one also finds a familiar pattern - same as ever - and a whole new family of possibilities opens up, as in the case of the sync bosonic hair.

\section*{Acknowledgements}
I thank the \textit{``Black Holes Inside and Out"} organizers for a truly stimulating meeting. I thank all my collaborators throughout the years, for the amazingly fruitful scientific adventure. This work is supported by the Center for Research and Development in Mathematics and Applications (CIDMA) through the Portuguese Foundation for Science and Technology (FCT -- Fundaç\~ao para a Ci\^encia e a Tecnologia) through projects: UIDB/04106/2020 (DOI identifier \href{https://doi.org/10.54499/UIDB/04106/2020}{10.54499/UIDB/04106/2020}); UIDP/04106/2020 \linebreak(DOI identifier \href{https://doi.org/10.54499/UIDP/04106/2020}{10.54499/UIDP/04106/2020});  PTDC/FIS-AST/3041/2020 (DOI identifier \href{http://doi.org/10.54499/PTDC/FIS-AST/3041/2020}{10.54499/PTDC/FIS-AST/3041/2020}); and 2022.04560.PTDC (DOI identifier \linebreak\href{https://doi.org/10.54499/2022.04560.PTDC}{10.54499/2022.04560.PTDC}). This work has further been supported by the European Horizon Europe staff exchange (SE) programme HORIZON-MSCA-2021-SE-01 Grant No.\ NewFunFiCO-101086251.

\bibliographystyle{utphys}
\bibliography{references}

%

\title{Black Holes, Cauchy Horizons, and Mass Inflation}

\author{Matt Visser}
\institute{\textit{Victoria University of Wellington, Wellington 6140, New Zealand}}

\maketitle

\begin{abstract}
Event horizons and Cauchy horizons are highly idealized mathematical constructions that do not fully capture the key physics of either Hawking radiation or mass inflation. 
Indeed, because they are teleological, both event horizons and Cauchy horizons are (in a precise technical sense) not physically observable.
In contrast, by inspecting the quasi-local behaviour of null geodesics,  long-lived apparent horizons (or more generally long-lived quasi-local horizons) are in principle physically observable, and 
are ``good enough" for then pragmatically redefining a black hole, and ``good enough'' for generating Hawking radiation. Furthermore it is now also clear that long lived apparent horizons (quasi-local horizons) are also ``good enough" for generating mass inflation. These observations suggest that one should be somewhat careful when trying to extrapolate rigorous mathematical theorems, which often embody mathematical idealizations that do not necessarily correspond to what a  finite resource astronomer can actually measure, into the astrophysical realm. 
\end{abstract}

\label{sec:visser}

\section{Introduction}

Event horizons (absolute horizons) are teleological, (telos = causa finalis = final cause),
in the sense that one has to back-track from the infinite future in order to decide whether or not an event horizon is present right now~\cite{Ellis:1993,Hayward:1993,Hayward:1993b, Ashtekar:2004, Andersson:2005, Visser:2014, Ashtekar:2023}. Perhaps worse, if the end-point of the black hole Hawking evaporation process~\cite{Hawking:1974,Hawking:1975} is complete evaporation, (as opposed to a naked singularity or remnant), then strictly speaking no event horizon ever formed in the first place, implying the need for a more pragmatic and useful redefinition of what a black hole actually is --- a redefinition that eschews the use of event horizons. 
The key distinction to be made here is between the teleological causal horizon (event/absolute horizon) and the various quasi-local (non-causal) horizons.
These issues are well-known within the relativity community, but less well appreciated in the astronomy, astrophysical, and high energy communities~\cite{Ellis:1993,Hayward:1993, Hayward:1993b,Ashtekar:2004, Andersson:2005, Visser:2014, Ashtekar:2023}.

Twenty years ago, in July 2004 at GR17 in Dublin, Stephen Hawking really did say: \\
\centerline{``{... a true event horizon never forms, just an apparent horizon.}''}

A decade ago (2014) Stephen Hawking really did say~\cite{Hawking:2014}: \\
\centerline{``The absence of event horizons mean that there are no black holes}
\centerline{ --- in the sense of regimes from which light can't escape to infinity.}
\centerline{\qquad\qquad 
There are however apparent horizons which persist for a period of time.''}
(One can imagine how some elements of the popular press managed to mangle these quotes.)

We shall soon see that similar issues apply to mass inflation~\cite{Carballo-Rubio:2024}. 
The point is that because event horizons are teleological,
they are detectable only by omniscient super observers~\cite{Ashtekar:2004, Andersson:2005, Visser:2014, Ashtekar:2023}.
Event (absolute) horizons are good for developing abstract mathematical theorems --- but not so good for physicists and astronomers. (At the very least some effort would need to put into modifying standard theorems so that they are not crucially dependent on the existence of strict event horizons.)
Apparent horizons, in contrast, are quasi-locally detectable by finite-resource observers~\cite{Visser:2014}.
Apparent horizons (or dynamical horizons, or close-to-Killing horizons, or trapping horizons, or more generally, quasi-local horizons) are not quite as good for proving mathematical theorems, but they are at least physically observable~\cite{Visser:2014}.

Better yet, there are various ways of extending the notion of surface gravity ${\kappa}$ to apparent horizons (or dynamical horizons, or close-to-Killing horizons, or trapping horizons)~\cite{Cropp:2013,Hayward:2008,Hayward:2005,Nielsen:2005}.
This then lets one define the notion of a slowly-evolving apparent/quasi-local horizon:
\begin{equation}
|\dot \kappa| \ll \kappa^2.
\end{equation}

Once one has a handle on the notion of a slowly-evolving apparent horizon
then it becomes easy to see that this is ``good enough'' for inducing Hawking radiation~\cite{Barcelo:2010a,Barcelo:2010b,Barcelo:2006}. 
Note that the adiabaticity condition
\begin{equation}
{\frac{\mathbf{|\dot \kappa|}}{|\kappa|} \ll |\kappa|}
\end{equation}
implies that the fractional change in Hawking temperature is small --- on the oscillation timescale of a typical Hawking photon.
This is all very mainstream, quite well established material.
But I am now going to argue~\cite{Carballo-Rubio:2024} that similar considerations apply
 to both Cauchy horizons and mass inflation~\cite{Poisson:1989,Poisson:1990,Ori:1991,Balbinot:1993,Hod:1998,Barcelo:2022, DiFilippo:2022}.

\section{Cauchy horizons:}

The particular class of Cauchy horizons of interest are the inner event horizons~\cite{Poisson:1989,Poisson:1990,Ori:1991,Balbinot:1993,Hod:1998,Barcelo:2022, DiFilippo:2022}.
There is widespread agreement that inner event horizons lead to mass inflation, in the sense that 
an initially small infalling perturbation will grow exponentially as it approaches the inner horizon.
In terms of a suitable null coordinate $v$:
\begin{equation}
\delta m(v) \propto (v-v_0)^{-p} \; \exp\left(|\kappa_\mathrm{in}| (v-v_0)\right).
\end{equation}
But inner event horizons are, in particular, event horizons --- 
 so they are teleological, and hence detectable only by omniscient super observers~\cite{Ashtekar:2004, Andersson:2005, Visser:2014, Ashtekar:2023}.

In contrast inner apparent  horizons \emph{are} in principle physically observable, 
(quasi-locally detectable), 
at least to a possibly doomed finite-resource internal-to-the-black-hole  observer. 
Better yet, there are again ways of extending the notion of surface gravity ${\kappa_{inner}}$ to inner apparent horizons~\cite{Cropp:2013,Hayward:2008}, which then lets one define the notion of a slowly-evolving inner apparent horizon.
Once one has a handle on the notion of a slowly-evolving inner apparent horizon
\begin{equation}
|\dot \kappa_\mathrm{in}| \ll \kappa_\mathrm{in}^2
\end{equation}
then it becomes relatively easy to see that this is ``good enough'' for inducing mass inflation.
This adiabatic condition now has a new interpretation:
The fractional change in (inner horizon) surface gravity  is small ---
on the e-folding timescale of mass inflation.
Thence  for a slowly evolving inner apparent horizon one would now expect
\begin{equation}
\delta m(v) \propto \left(\int^v_{v_0} |\kappa_\mathrm{in}(\bar v)| \; d\bar v\right)^{-p} \; 
\exp\left(\int^v_{v_0} |\kappa_\mathrm{in}(\bar v)| \; d\bar v\right).
\end{equation}
Under suitable circumstances we can now argue that the exponential factor in mass inflation will ultimately become large but finite --- being subject to a kinematic cutoff at late times.

\begin{figure}
\begin{center}
    \includegraphics[width=0.32\textwidth]{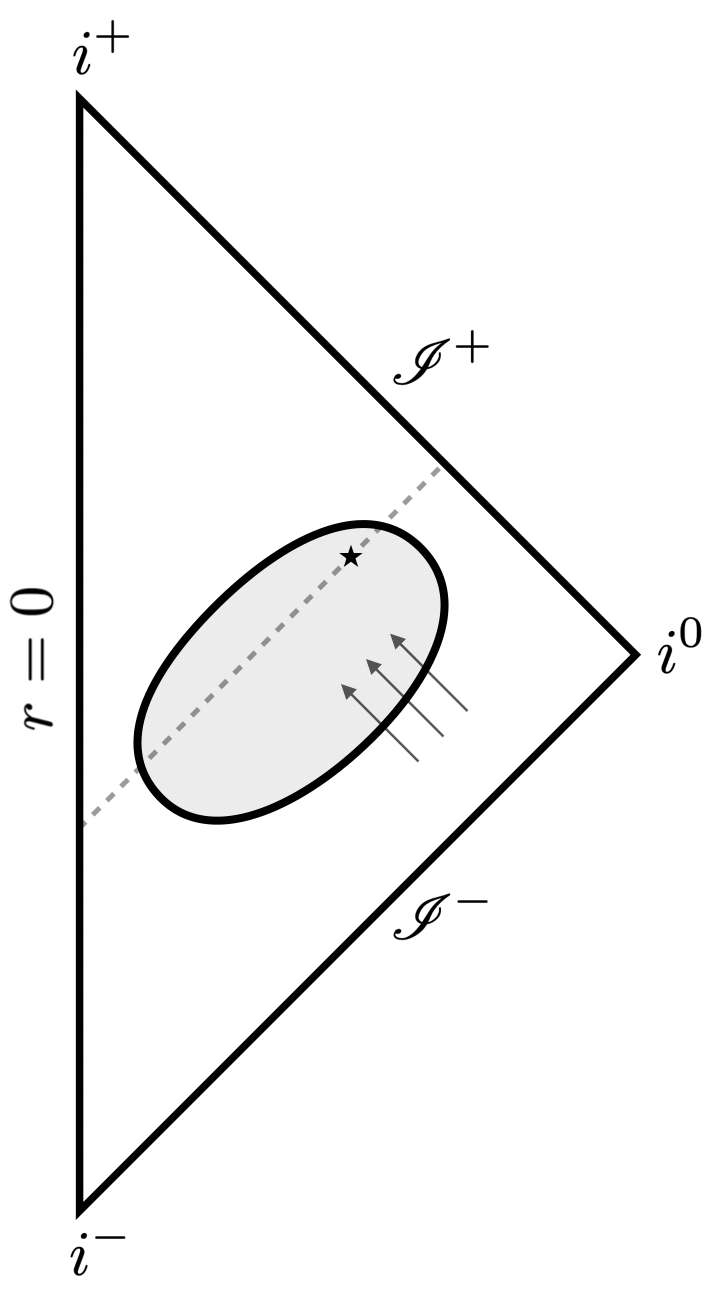}
    \caption{Penrose diagram describing the formation and disappearance of a black hole with both outer and inner horizons, but not Cauchy horizons, marking the boundary of the trapped region. (Topologically closed, as envisaged by Frolov~\cite{Frolov:2014} and others, eg~\cite{Roman:1983}. See also reference~\cite{Carballo-Rubio:2024}.)
}
\end{center}
\end{figure}

Consider a spherically symmetric model generalized Eddington-Finkelstein coordinates:
\begin{equation}
\text{d}s^2=-e^{-2\Phi(v,r)}F(v,r)\text{d}v^2+2e^{-\Phi(v,r)}\text{d}r\text{d}v+r^2\text{d}\Omega^2.
\end{equation}
The Misner--Sharp quasi-local mass is:
\begin{equation}
M(v,r)=\frac{r}{2}(1- g^{ab} \;\partial_a r\; \partial_b r)=\frac{r}{2}(1-g^{rr} )
= \frac{r}{2} (1- F(v,r)).
\end{equation}
We have inner and outer apparent horizons:
\begin{equation}
F(v,r) = e^{\Psi(v,r)} 
\left( 1- \frac{r_{\rm in}(v)}{r}\right)
\left( 1- \frac{r_{\rm out}(v)}{r}\right).
\end{equation}
(Generalizing this to multiple inner-outer horizon pairs would be possible --- but that is overkill for now.
Similarly we can also smooth out the curvature singularity at $r=0$ --- but that is again overkill for now.)
For the inner horizon quasi-local mass
\begin{equation}
M(v,r_{\rm in}(v))= \frac{r_{\rm in}(v)}{2}.
\end{equation}
Similarly for the inner horizon surface gravity:
\begin{equation}
\kappa_{\rm in}(v)=
\frac{e^{-\Phi(v,r_{\rm in}(v))}}{2}\left.\frac{\partial F}{\partial r}\right|_{(v,r_{\rm in}(v))}
\end{equation}
which evaluates to
\begin{equation}
\kappa_{\rm in}(v)=
-\frac{e^{-\Phi(v,r_{\rm in}(v))}\;e^{\Psi(v,r_{\rm in}(v))}}{2}
\left(\frac{r_{\rm out}(v)-r_{\rm in}(v)}{r_{\rm in}(v)^2}\right)
=-|\kappa_{\rm in}(v)|.
\end{equation}
(Similar constructions apply to outer apparent horizons.)
Note the kinematic constraint $r_{\rm in}(v) < r_{\rm out}(v)$. 
Thence $M{(v,r_{\rm in}(v))} < M(v,r_{\rm out}(v))$.
For the outer apparent horizon, it is well established that the
quantities $ r_{\rm out}(v)$ and $M(v,r_{\rm out}(v))$ can evolve rapidly via accretion and/or  slowly via Hawking radiation.
For the inner apparent horizon, quantities closely related to $ r_{\rm in}(v)$ and $M(v,r_{\rm in}(v))$ evolve rapidly via mass inflation.
But watch out for that kinematic constraint!
There is a cutoff!
Note  that $\kappa_{\rm in}(v)\to 0$ as $r_{\rm in}(v) \to r_{\rm out}(v)$. 
So ultimately the exponential growth switches off and the exponential factor is large but finite.
To be a little more explicit about this, (see  reference~\cite{Carballo-Rubio:2024} for details), note that outgoing null geodesics $r(v)$ satisfy
\begin{equation}
\frac{\text{d}r (v)}{\text{d}v}=\frac{e^{-\Phi(v,r)}F(v,r)}{2}.   
\end{equation}
Expand around the position of the inner horizon, $r=r_{\rm in}(v)$:
\begin{equation}
\frac{\text{d}r(v) }{\text{d}v}=\frac{e^{-\Phi(v,r_{\rm in}(v))}}{2}\left(\left.\frac{\partial F}{\partial r}\right|_{(v,r_{\rm in}(v))}
\left[r(v)-r_{\rm in}(v)\right]\right)+...
\end{equation}
Thence
\begin{equation}
\frac{\text{d}[r(v)-r_{\rm in}(v)]}{\text{d}v}=
-|\kappa_{\rm in}(v)|\left[r(v)-
r_{\rm in}(v)\right]-\frac{\text{d}r_{\rm in}(v)}{\text{d}v}+...
\end{equation}
Now suppose we satisfy two adiabaticity conditions, 
one on the surface gravity itself, and the other on the location of the inner horizon, 
namely:
\begin{itemize}
\item $\left|\frac{\text{d} r_{\rm in}(v)}{\text{d}v}\right|\ll |\kappa_{\rm in}(v)|
    \left|r(v)-r_{\rm in} (v)\right|,$
\item
    $\left|\frac{\text{d} \kappa_{\rm in}(v)}{\text{d}v}\right|\ll |\kappa_{\rm in}(v)|^2.$
\end{itemize}
Then we can write:
\begin{equation}
r(v)\approx r_{\rm in}(v)+\left[r(v_0)-r_{\rm in}(v_0)\right]e^{-|\kappa_{\rm in}(v)|(v-v_0)},
\end{equation}
subject to
$r(v_0)\in(r_{\rm in}(v_0),r_{\rm out}(v_0))$.
Thence for the motion of the inner horizon
\begin{equation}
 \left | \frac{\text{d} r_{\rm in}{(v)}}{\text{d}v}\right | \ll |\kappa_{\rm in}(v)|\left[r(v_0)-r_{\rm in}(v_0)\right]e^{-|\kappa_{\rm in}(v)|(v-v_0)}.
\end{equation}
Eventually, for sufficiently large $v$, adiabaticity of the motion of the inner horizon will {always} break down, certainly by time~\cite{Carballo-Rubio:2024}
\begin{equation}
v_\star \approx v_0 + \frac{1}{|\kappa_{\rm in}(v_\star)|} 
\ln \left\{\frac{
|\kappa_{\rm in}(v_\star)|\; \left[r(v_0)-r_{\rm in}(v_0)\right]} {\left|\text{d} r_{\rm in}/\text{d}v\right|_{v_\star}} 
\right\}.
\end{equation}
This provides a somewhat crude  estimate for the time at which inner and outer horizons merge, at which stage the exponential factor characterizing mass inflation saturates to something large but finite.

\section{Explicit Models}

Let us consider two specific models using null dust (lightlike thin shells):
\begin{itemize}
\item Analytic model (of the Dray, t'Hooft, Barabbes, Israel, Poisson variety~\cite{Dray:1985,Barrabes:1991,Poisson:1989,Poisson:1990}.
\item Numerical model (a variant of the Ori model~\cite{Ori:1991}). .
\end{itemize}

\subsection{Analytic model}

For the analytic model consider the standard setup of crossed ingoing and outgoing null shells.
(See Dray, t'Hooft, Barabbes, Israel, Poisson~\cite{Dray:1985,Barrabes:1991,Poisson:1989,Poisson:1990}.)
A brief analysis yields~\cite{Carballo-Rubio:2024}:
\begin{eqnarray}
M_{\rm f}(v_\times,r_\times) &=& M_{\rm i}(v_\times,r_\times) + M_{\rm in} (v_\times,r_\times) +M_{\rm out}(v_\times,r_\times)\nonumber\\ 
&&- \frac{2M_{\rm in} (v_\times,r_\times) M_{\rm out}(v_\times,r_\times)}{r_\times F_{\rm i}(v_\times,r_\times)},
\end{eqnarray}
where
\begin{equation}
F^\times_{\rm i}= \frac{e^{\Psi(v_\times,r(v_\times))}}{r_{\rm in}(v_\times)} 
\left( 1- \frac{r_{\rm out}(v_\times)}{r_{\rm in}(v_\times)}\right)
\left( r(v_\times)- r_{\rm in}(v_\times)\right).
\end{equation}
But allowing for the breakdown of adiabaticity, (that is, inner-outer horizon merger), we at worst have
\begin{equation}
F^\star_{\rm i} = \frac{e^{\Psi(v_\star,r_{\rm in}(v_\star))}}{r_{\rm in}(v_\star)} 
\left( 1- \frac{r_{\rm out}(v_\star)}{r_{\rm in}(v_\star)}\right)
\frac{\left|\text{d}r_{\rm in}/\text{d}v\right|_{v_\star}}{|\kappa_{\rm in}(v_\star)|}.
\end{equation}
Thence the mass $M_{\rm f}(v_\star,r_\star)$ exhibits an exponential behaviour in $v_\times$,
up to a maximum value
\begin{equation}
M_{\rm max}\approx \frac{r_{\rm in}(v_\star)\;|\kappa_{\rm in}(v_\star)|}{
\left|\left.\text{d}r_{\rm in}(v)/\text{d}v\right|_{v_\star}\right|} \; 
\frac{2 M_{\rm in} (v_\times, r_\times) M_{\rm out}(v_\times, r_\times)}{r_\times}.
\end{equation}
This can be factorized into the form $M_{\rm max}\simeq f_1(v_{\star}) \; f_2(v_\times,r_\times)$, with one function depending on the end of exponential mass inflation, and the other only on the null-shell crossing.
This is a regularized version of $M_{\rm max}=\infty$ that is obtained in the static case, where the regulator comes from the exponential approximation ceasing to be valid due to the non-zero  value of $\left.\text{d}r_{\rm in}(v)/\text{d}v\right|_{v_\star}$. 
(As a consistency check, for $\left.\text{d}r_{\rm in}(v)/\text{d}v\right|_{v_\star}\to 0$ we recover the naive $M_{\rm max}=\infty$; infinite mass inflation.)
Thus this analytic model confirms our general expectations.

\subsection{Numeric model}

\begin{figure}[!ht]
\begin{center}
\includegraphics[width=.75\linewidth]{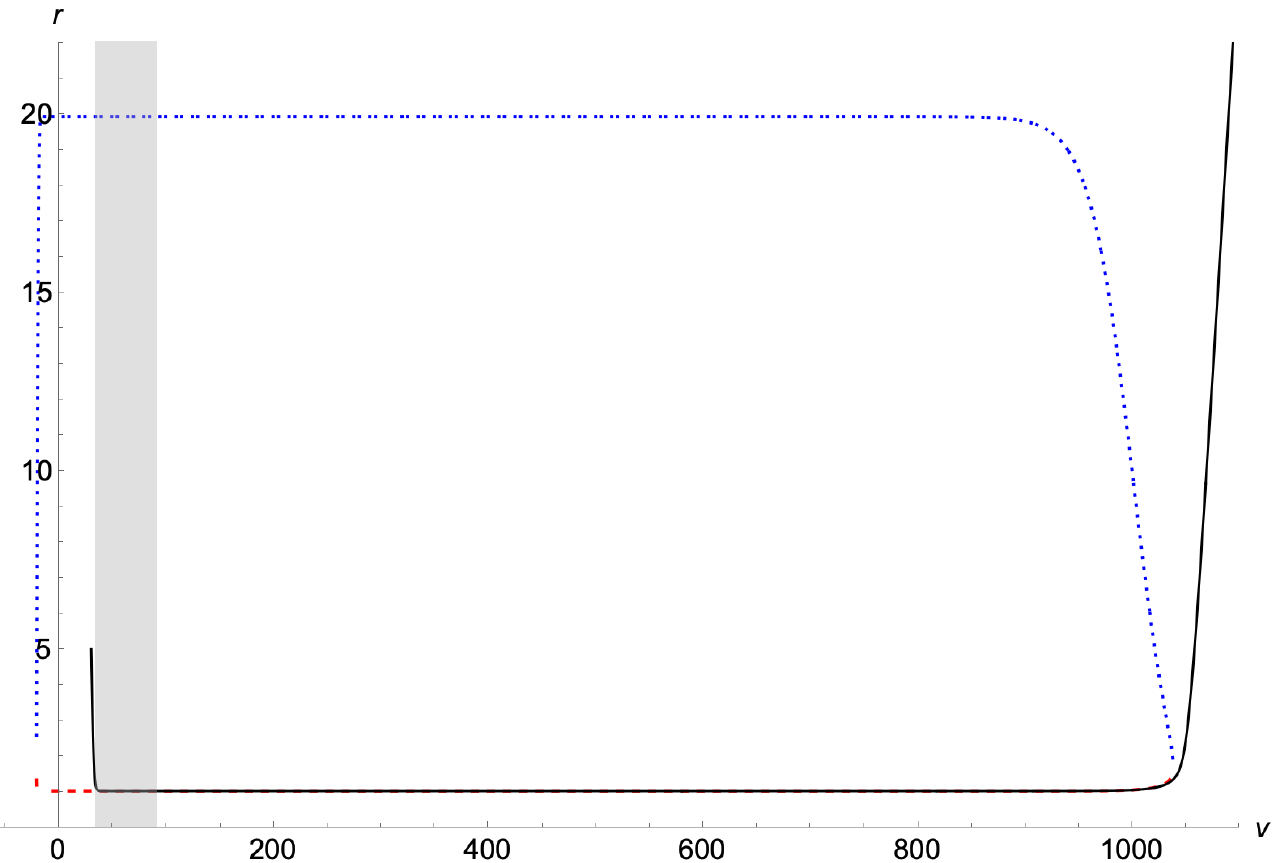}
\caption{\label{F:2} Trapped region in $(r,v)$ coordinates. See~\cite{Carballo-Rubio:2024}.}
\end{center}
\end{figure}

For the numeric model (a variant of the Ori model~\cite{Ori:1991}):
\begin{itemize}
\item 
Keep the outgoing null dust shell.
\item
Replace the ingoing null dust shell by a stream of null fluid.
\end{itemize}
For definiteness let us work with the Hayward metric: 
\begin{equation}
F_\pm(v,r)=1-\frac{2r^2m_\pm(v)}{r^3+2\ell^2m_\pm(v)}; \qquad \Phi(v,r)=0.
\end{equation}
Choose the null normal to be
\begin{equation}n^\mu=\frac{d x^\mu}{d r}=\left( \frac{2}{F},1,0,0 \right),\end{equation}
and work with pressureless shells
\begin{equation}
\left. \frac{1}{F_+}\frac{\partial M_+}{\partial v}\right|_{r=R(v)}=\left.\frac{1}{F_-}\frac{\partial M_-}{\partial v}\right|_{r=R(v)}\,.
\end{equation}
After a bit of fiddling with the numerics~\cite{Carballo-Rubio:2024}, the trapped region, in terms of $(r,v)$ coordinates, is presented in Figure \ref{F:2}.

\begin{figure}[!ht]
\begin{center}
\centering\includegraphics[width=.75\linewidth]{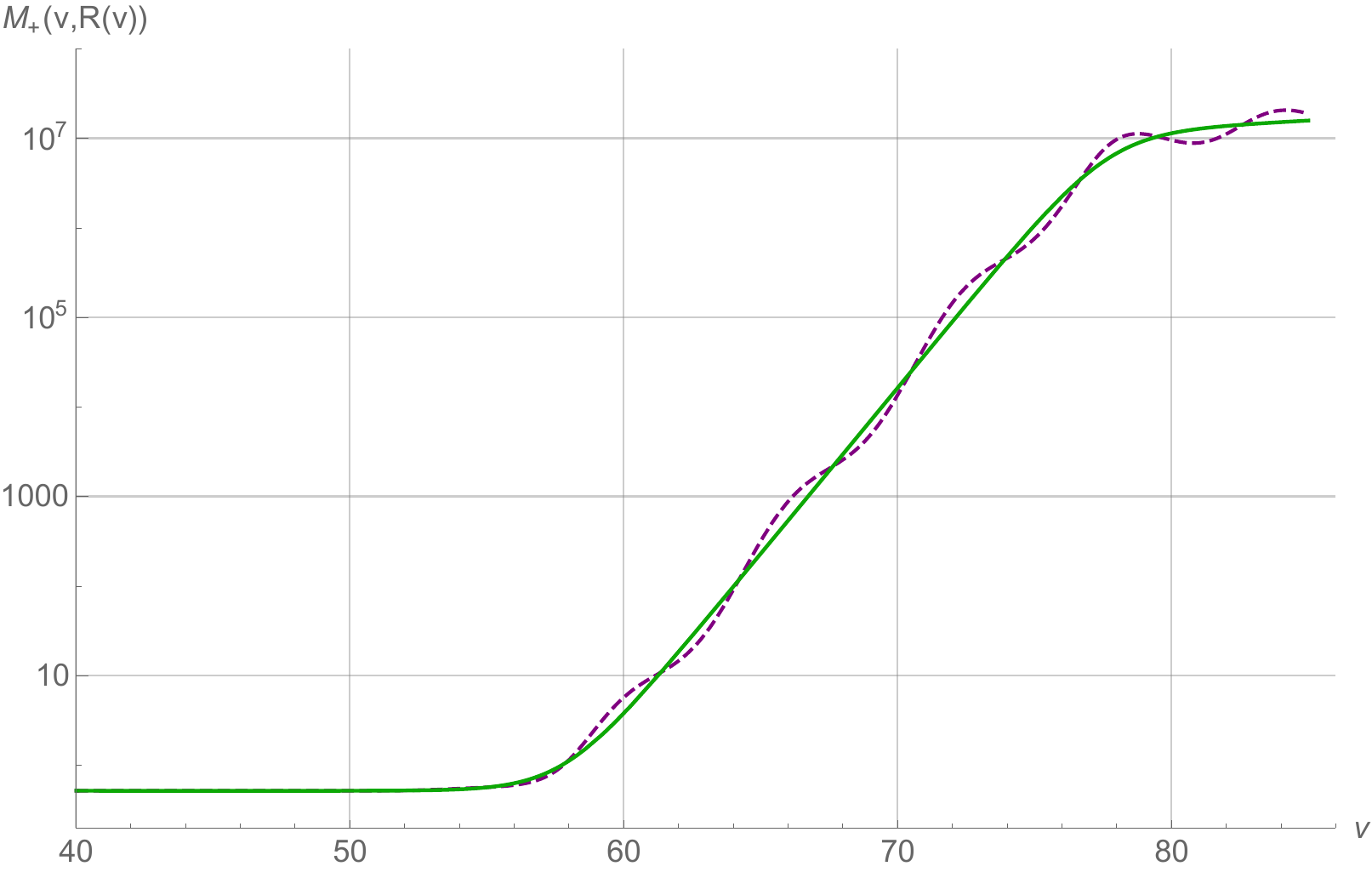}
\caption{\label{F:3}(Truncated) exponential mass inflation. Note the log-linear plot. See reference~\cite{Carballo-Rubio:2024}.}
\end{center}
\end{figure}

That is, going to $(r,v)$ coordinates preserves the topology, but not the shape. (Compare with the corresponding figures in Frolov 2014~\cite{Frolov:2014}.)
Furthermore the mass of the ingoing shell exhibits (truncated) exponential mass inflation.
For a log-linear plot  see Figure \ref{F:3}. 

\section{Fully extremal regular black holes}

\def\d{{\mathrm{d}}}

We have seen that mass inflation is a generic feature of inner apparent horizons~\cite{Carballo-Rubio:2024}, not being limited to Cauchy horizons. 
Furthermore the timescale of mass inflation is rapid, being determined by the inner horizon surface gravity $\kappa_{\rm in}$. 
This strongly suggests the rapid merger of inner and outer horizons, thereby switching off both Hawking radiation and mass inflation.

The simplest example of such behaviour is a geometry of the form~\cite{DiFilippo:2024}:
\begin{equation}
\d s^2 = - \exp[-2\Phi(r)-2\Psi(r)] \left(1-\frac{r_H}{r}\right)^2 \d t^2 + 
\frac{\exp[2\Psi(r)]}{\left(1-\frac{r_H}{r}\right)^2} \d r^2 + r^2 \d\Omega^2.
\end{equation}
This is a distortion of extremal Reissner--Nordstr\"om spacetime.  Outer and inner horizons have merged driving both $\kappa_\mathrm{out}\to 0$ and $\kappa_\mathrm{in}\to 0$, thereby switching off both Hawking radiation and mass inflation.
(Though this switches off the two most obvious of the black hole instabilities, there are several other instabilities one might also wish to take into account, see~\cite{Barcelo:2022, DiFilippo:2024b, Cunha:2022,Cunha:2024,Aretakis:2012,Aretakis:2013,Cardoso:2014}.)

To regularize the origin introduce a non-zero parameter $r_0\neq 0$~\cite{DiFilippo:2024}:
\begin{equation}
\hspace{-25pt}
\d s^2 = - \exp[-2\Phi(r)-2\Psi(r)] \left[\frac{(r-r_H)^2}{r^2+r_0^2}\right] 
\d t^2 + 
\exp[2\Psi(r)] \left[\frac{r^2+r_0^2}{(r-r_H)^2}\right] \d r^2 + r^2 \d\Omega^2.
\end{equation}
This is now a regular spacetime, with no Hawking radiation, and no mass inflation.

Even more generally one could write~\cite{DiFilippo:2024}:
\begin{eqnarray}
\d s^2 &=& - \exp[-2\Phi(r)-2\Psi(r)] \left[\frac{(r-r_H)^2}{r^2+r_0^2}\right]  \d t^2 
\nonumber\\
&& \qquad\qquad\;\; + 
\exp[2\Psi(r)] \left[\frac{r^2+r_0^2}{(r-r_H)^2}\right] \d r^2 + 
\Xi(r)^2 \;\d\Omega^2,
\end{eqnarray}
where we now entertain the possibility of a wormhole throat.
The Misner--Sharp quasi-local mass is now slightly more complicated~\cite{DiFilippo:2024}:
\begin{equation}
m(r) = \frac{\Xi(r)}{2} \left\{ 1 - [\Xi'(r)]^2 \exp[-2\Psi(r)] \left[\frac{(r-r_H)^2} {r^2+r_0^2}\right] \right\}.
\end{equation}
This particular spacetime is regular, possibly with a wormhole throat, with
no Hawking radiation, and no mass inflation.

Even more generally one could take $n\in\mathbb{Z}^+$ with $n\geq 2$. Then~\cite{DiFilippo:2024}:
\begin{eqnarray}
\d s^2 &=& - \exp[-2\Phi(r)-2\Psi(r)] \left[\frac{(r-r_H)^n}{r^n+r_0^n}\right]  \d t^2 
\nonumber\\
&& \qquad\qquad\;\;+ 
\exp[2\Psi(r)] \left[\frac{r^n+r_0^n}{(r-r_H)^n}\right] \d r^2 + \Xi(r)^2 \d\Omega^2.
\end{eqnarray}
This is now an $n$ fold degenerate horizon,
(with a regular center, the possibility of a wormhole throat, and no Hawking radiation, no mass inflation).

Ultimately take 
$n_i\in\mathbb{Z}^+$, with $n_i \geq 2$,  $i\in\{1,2,\dots, N\}$, and $r_{H_i}\in \mathbb{R}^+$,  and then set~\cite{DiFilippo:2024}:
\begin{eqnarray}
\d s^2 &=& - \exp[-2\Phi(r)-2\Psi(r)] \;
\prod_{i=1}^N\left[\frac{(r-r_{H_i})^{n_i}}{r^{n_i}+r_0^{n_i}}\right] \d t^2
\nonumber\\
&&\qquad\qquad\;\;\;+ 
\exp[2\Psi(r)] 
\prod_{i=1}^N\left[\frac{r^{n_i}+r_0^{n_i}}{(r-r_{H_i})^{n_i}}\right] \d r^2 + \Xi(r)^2 \d\Omega^2.
\end{eqnarray}
One now has 
$N$ degenerate horizons, each with $n_i$ fold degeneracy.
The geometry is possibly regular, possibly with wormhole throat, with
no Hawking radiation, and no mass inflation.
(With a bit more work one could further drive the central density and central pressure to zero, asymptoting to a ``Minkowski ore''~\cite{Simpson:2019,Berry:2020}, or explore other features of regular black holes~\cite{Roman:1983,Berry:2021,Simpson:2021a,Simpson:2021b,
Carballo-Rubio:2019a,Carballo-Rubio:2019b,Carballo-Rubio:2022}.)

One can furthermore easily generalize these constructions to rotating systems~\cite{DiFilippo:2024}.

\section{Conclusions}
One key observation from the above discussion is that both Hawking radiation and mass inflation are more robust than commonly appreciated --- strict event horizons (absolute horizons) and strict Cauchy horizons are not really necessary~\cite{Hawking:2014,Visser:2014,Ashtekar:2023,Carballo-Rubio:2024}. A slowly evolving variant of an apparent horizon (trapping horizon, dynamical horizon, quasi-local horizon) is quite sufficient~\cite{Hawking:2014,Visser:2014,Ashtekar:2023,Carballo-Rubio:2024}. 
A subtlety of the quasi-local horizons is that they are generally not causal horizons, and one cannot rely on naive causality arguments. This underlies Hawking's comments in~\cite{Hawking:2014}, and related comments in~\cite{Visser:2014,Visser:2014b,Alonso-Serrano:2015a,Alonso-Serrano:2015b,Alonso-Serrano:2017}.
(This distinction between causal and non-causal horizons is also important in cosmology~\cite{Ellis:1993}.)

A second key observation is that the timescale on which the mass inflation instability operates is rapid, being set by the surface gravity $\kappa_\mathrm{in}$ of the inner horizon~\cite{Carballo-Rubio:2024}. Since this timescale is typically of order the light-crossing-time for the black hole, 
the classical mass inflation instability potentially becomes of relevance to astrophysical black holes.

\enlargethispage{15pt}
A third key observation is that switching off the two most obvious black hole instabilities, the mass inflation and Hawking instabilities,  already puts very tight constraints on the spacetime geometry --- there are other more subtle instabilities relevant to extremal horizons~\cite{Barcelo:2022, DiFilippo:2024b,Cunha:2022,Cunha:2024,Aretakis:2012,Aretakis:2013,Cardoso:2014}, but switching off the two most obvious instabilities already tells one quite a lot regarding the general form of the spacetime geometry.

Overall, the emerging picture of black hole internal structure is considerably more subtle and nuanced than we might have expected even a few years ago~\cite{Cardoso:2019}.

\section*{Acknowledgements}
This research was supported by a Victoria University of Wellington travel grant.\\
MV particularly wishes to thank  Ra\'ul Carballo-Rubio, Francesco Di Filippo, and Stefano Liberati for pertinent comments and observations.

\bibliographystyle{utphys}

\bibliography{references}

\title{The Backreaction Problem for Black Holes in Semiclassical Gravity}
\author{Adrian del Rio}
\institute{\textit{Departamento de Matem\'aticas, Universidad Carlos III de Madrid. Avda. de la Universidad 30, 28911 Leganes, Spain}}

\maketitle

\begin{abstract}
The question of black hole evaporation is reviewed in the framework of quantum field theory in curved spacetimes and semiclassical gravity. We highlight the importance of taking backreaction effects into account to have a consistent picture of the fate of gravitational collapse in this framework. We describe the difficulties of solving the backreaction semiclassical equations due to practical complications of renormalizing the stress-energy tensor of quantum fields in general 3+1 spacetimes. We end with some personal views and plans on the subject.
\end{abstract}

\section{Hawking radiation and black hole evaporation}

Quantum field theory (QFT) predicts that any dynamical spacetime is able to spontaneously excite particle pairs out of the quantum vacuum \cite{PhysRevLett.21.562, PhysRev.183.1057, PhysRevD.3.346}. In this framework,  Hawking found that the formation of a black hole (BH) of mass $M$ by gravitational collapse will emit, at sufficiently late times,   a flux of particles with a thermal spectrum,  with temperature $T=\frac{\hbar c^3}{8\pi G M k_B}$, and irrespectively of the  details of the original star \cite{Hawking:1974rv, cmp/1103899181, cmp/1103899393, PhysRevD.12.1519, cmp/1104180137}. Immediately after, this outgoing  flux was  found to be compensated by a negative-energy flux across the horizon \cite{PhysRevD.13.2720, PhysRevD.21.2185, ELSTER1983205, PhysRevLett.117.231101} which is expected to backreact on the BH and decrease its mass. For macroscopic BHs,    quasi-stationary calculations show that this ingoing  flux makes the horizon  shrink adiabatically in time following the Stefan-Boltzmann law \cite{PhysRevLett.46.382, Balbinot_1984, PhysRevLett.73.2805, PhysRevD.52.5857}: the BH evaporates thermally.

\begin{figure}[ht]
\centering
\includegraphics[width=0.4\linewidth]{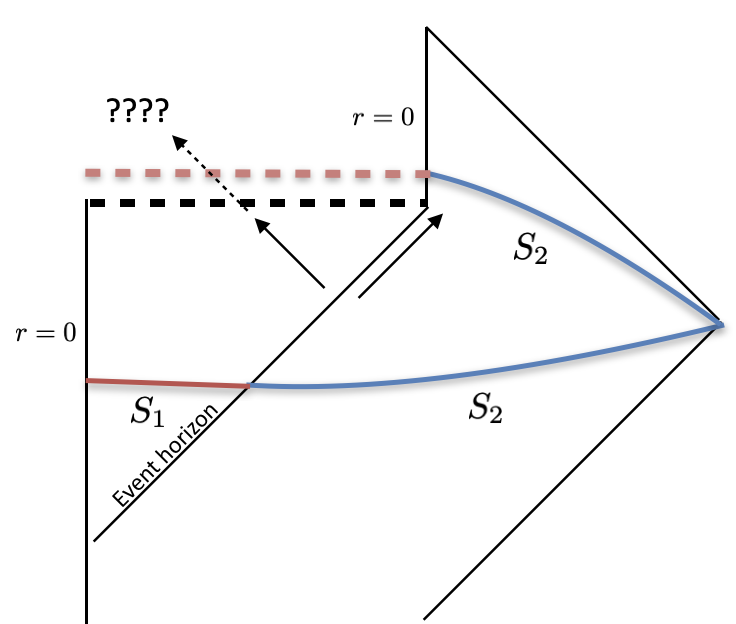}
\caption{Penrose diagram for black evaporation conjectured in \cite{cmp/1103899181, PhysRevD.14.2460}. \label{fig:1}}
\end{figure}

The quanta emitted to infinity at late times  are quantum-correlated with those particles that fall into the BH horizon \cite{cmp/1103899393, PhysRevD.12.1519}. In other words, the Hawking radiation is entangled with degrees of freedom of the quantum field inside the BH. If the result of BH evaporation leads to a spacetime with Penrose diagram given by Fig~\ref{fig:1}, where the BH interior is completely excised, as  conjectured originally in \cite{cmp/1103899181}, then these correlations are lost after the evaporation, and the quantum field would evolve from an initial pure vacuum state  to a mixed thermal state. Based on this, some people (initially Hawking  \cite{PhysRevD.14.2460}) have suggested that BH evaporation leads to a unitarity problem in QFT, and speculated on how information may be restored in quantum gravity (see \cite{Unruh_2017} for a review). 

The BH evaporation picture of  Fig~\ref{fig:1} was only conjectured based on the quasi-stationary/test field approximations, where the renormalized stress-energy tensor of the quantum fields $\langle T_{ab}\rangle$ which sources Einstein's equations is calculated on a fixed, classical BH background. To figure out the final state of gravitational collapse and address the information  issue, the problem of backreaction must be taken more seriously. Despite that the  vacuum energy of a quantum field is negligible for astrophysical (solar-mass) scales, it increases with spacetime curvature, roughly as $\hbar R^{abcd}R_{abcd}$. Therefore, in a spacelike hypersurface close to the classical curvature singularity during gravitational collapse, the effects of $\langle T_{ab}\rangle\neq 0$ are no longer negligible, and can drastically change the original causal structure of the spacetime! There is no reason to expect that Fig~\ref{fig:1} will hold. In particular, the classical trapped region may dissappear before a spacelike/null singularity or Cauchy horizon ever forms. What is, ultimately, the actual end point of gravitational collapse of stars? 

Although we lack a full theory of quantum gravity to properly answer this question, one can expect to obtain a qualitative, physically realistic solution for BH evaporation by solving the so-called semiclassical Einstein's equations. These are simply the ordinary Einstein's equations, but sourced by the quantum vacuum energy of quantum fields: $G_{ab}=8\pi G \langle 0 | T_{ab}|0 \rangle$. Interestingly,  in QFT the classical energy conditions do not hold \cite{Epstein:1965zza}. In fact, many examples of  negative energy densities are known \cite{Casimir:1948dh, Davies:1977yv, PhysRevLett.57.2520, PhysRevD.46.4566}. Consequently, singularity theorems in classical General Relativity do not apply anymore \cite{PhysRevLett.14.57, PhysRevLett.15.689, PhysRevLett.17.444, PhysRevLett.17.445, Hawking:1970zqf}  (see  \cite{Senovilla_2015} for a detailed historical treatment). Can semiclassical effects avoid the emergence of singularities in gravitational collapse, and lead to a regular BH spacetime where information is not `lost'?    

\section{The renormalized stress-energy tensor: a short review}

Unfortunately, solving the semiclassical Einstein's equations is a formidable task.  The problem resides in the difficulties for calculating the  renormalized stress-energy tensor in general curved spacetimes,  as a consequence of which $\langle 0 | T_{ab}|0 \rangle$ remains as an unknown functional of the spacetime metric and quantum state in the semiclassical field equations.

For definiteness, let $\phi(x)$ represent a massless, minimally coupled scalar field on a general curved spacetime \cite{DEWITT1975295, birrell_davies_1982,Fulling_1989, Wald:1995yp,parker_toms_2009, doi:10.1142/S0217751X13300238, c37b9fad8f104637ad2b22a76f1e1e1e}, with metric $g_{ab}$ and  Levi-Civita connection $\nabla_a$.  As is well-known, Fock quantization gives rise to  infinitely-many Hilbert space representations of the canonical commutation relations, which translates into infinitely-many possible choices of vacuum states  in the quantum theory \cite{PhysRevD.7.2850}. To each quantum state $|0\rangle$ we can assign a local notion of energy and stress for the quantum field,  by calculating the {\it expectation value} of the stress-energy tensor, $\langle 0|T_{ab}|0\rangle$.  Typically, the way to do this is to promote the classical expression, $T_{ab}=\nabla_a \phi \nabla_b \phi-\frac{1}{2}g_{ab}\nabla_c \phi\nabla^c \phi$, to  the quantum theory as an operator acting on the Fock space. Unfortunately, since the classical formula is quadratic in the fields, and fields in the quantum theory are well-defined only as operator-valued distributions \cite{osti_4606723} (not simply as operators), this formula involves taking the product of two distributions at the same spacetime point, which is mathematically ill-defined.   As a consequence, the vacuum expectation value of this formal expression gives rise to several ultraviolet (UV) divergences, i.e.  integrals and sums that are ill-defined in the high-frequency limit of the field modes.

To get a physically sensible result, a prescription to regularize and renormalize these divergences is required. Since UV divergences arise from the high-frequency component of the field modes, which only probe the geometry in the immediate vicinity of the spacetime point of interest, it seems natural to split the two points in the quadratic expression to regulate the UV divergences. Indeed, if the two field distributions in $T_{ab}$ are evaluated at slightly separated points, the product is now mathematically well-defined.  This is known as  point-splitting regularization \cite{PhysRevD.14.2490, PhysRevD.17.946}.
Now, to get a finite quantity for the stress-energy tensor, the standard prescription   is  to subtract  the  singular behaviour of the two-point function as the two point merge,  
\begin{equation}
\left\langle T_{a b}(x)\right\rangle_{\mathrm{}}:=\lim _{x^{\prime} \rightarrow x}\left\{\nabla_{a} \nabla_{b^{\prime}}-\frac{1}{2} g_{a b'} g^{cd'}\nabla_{c} \nabla_{d'}\right\}\left[\langle\{\hat{\phi}(x), \hat{\phi}(x')\}\rangle-\langle\{\hat{\phi}(x), \hat{\phi}(x') \} \rangle_{\mathrm{sing}}\right]\, , \label{Tab}
\end{equation}
where brackets denote symmetrization, and the limit is taken along a geodesic that connects the two points (which is unique for sufficiently close points \cite{oneill1983semiriemannian}). In  Minkowski space we can take $\langle\hat{\phi}(x) \hat{\phi}(x')\rangle_{\rm sing}=\langle 0_M|\hat{\phi}(x) \hat{\phi}(x')|0_M\rangle$, where $|0_M\rangle$ is the usual Minkowski vacuum.
In any curved spacetime, the short-distance  structure of the two-point function $\langle\hat{\phi}(x) \hat{\phi}(x')\rangle$ can be obtained as an asymptotic  series by solving iteratively the field equation, and taking $\langle 0_M|\hat{\phi}(x) \hat{\phi}(x')|0_M\rangle$ as a starting point. This is  the DeWitt-Schwinger expansion \cite{DEWITT1975295, PhysRev.82.664}.  

This prescription  manages to remove the UV divergences so long the two-point function has the ``Hadamard'' form: 
\begin{equation}
\langle 0 |\hat{\phi}(x) \hat{\phi}\left(x^{\prime}\right)|0\rangle\sim \frac{U\left(x, x^{\prime}\right)}{\sigma\left(x, x^{\prime}\right)}+V\left(x, x^{\prime}\right) \log \sigma\left(x, x^{\prime}\right)+W\left(x, x^{\prime}\right)\, , \label{hadamard}
\end{equation}
where $\sigma(x,x')$ represents the geodesic separation between the two points, which is zero for $x'=x$, and $U(x,x')$, $V(x,x')$, $W(x,x')$ are smooth.  For this reason, only those quantum states satisfying (\ref{hadamard}) are regarded as physically admissible. Any spacetime that evolves from a static  regime (like a collapsing star) admits such class of quantum states \cite{cmp/1103904566, FULLING1981243}.

Although the full two-point function  depends on the choice of quantum state, the short-distance  singular structure $\langle\hat{\phi}(x) \hat{\phi}(x')\rangle_{\rm sing}$ is independent of it, i.e. $U(x,x'), V(x,x')$ only depend on the local spacetime geometry, making this prescription useful  in any curved spacetime (i.e. we don't need  any fiducial state, like $|0_M\rangle$ in Minkowski). The regulated UV divergences in $\left\langle T_{a b}(x)\right\rangle$ can be further reabsorbed in the coupling constants of the semiclassical  equations (cosmological constant,  gravitational constant, etc), yielding a fully satisfactory renormalized theory \cite{PhysRevD.17.946} (see also \cite{10.1063/1.1724264}). Furthermore, it can be proven that (\ref{Tab}) is the unique prescription (up to minor ambiguities) consistent with locality, causality, and stress-energy conservation \cite{robert_m__wald_1977}. 

While conceptually successful, the practical implementation of this prescription is technically challenging. This is because, in most cases, practical calculations  require solving the field modes numerically, however the limit in (\ref{Tab}) cannot be carried out using numerical techniques. To obtain explicit expressions for $\left\langle T_{a b}(x)\right\rangle_{\mathrm{}}$ we need more efficient prescriptions which can be adapted in a computer for numerical calculations.  Several  methods are available today (mathematically equivalent to point-splitting)  that can do this for some class of fixed spacetime metrics and vacuum states. For instance, for Fridman-Lemaitre-Robertson-Walker (FLRW) spacetimes in Cosmology, there is the so-called ``adiabatic''  method \cite{PhysRevD.9.341, FULLING1974176, PhysRevD.10.3905, Bunch_1978, PhysRevD.18.1844, 300b6d7c-10d8-3655-99b9-25256f6839ee, Bunch_1980, PhysRevD.36.2963, PhysRevD.41.3101, PhysRevD.89.044030, PhysRevD.90.084017, PhysRevD.91.064031, PhysRevD.95.105003}. On the other hand, for different stationary BHs we have the   ``pragmatic mode-sum'' method \cite{PhysRevD.91.104028, PhysRevD.94.044054, PhysRevD.95.025007, PhysRevLett.117.231101, PhysRevLett.118.141102, PhysRevD.97.024033, PhysRevLett.124.171302, PhysRevD.103.105002,  PhysRevD.104.024066, PhysRevLett.129.261102};   different ``euclidean space'' techniques \cite{PhysRevLett.53.403, PhysRevD.30.2532, PhysRevLett.70.1739, PhysRevD.51.4337, PhysRevD.66.124017, PhysRevD.94.125024,  PhysRevD.96.105020, PhysRevD.106.065023}; and a generalization of the ``adiabatic'' method for  the interior of the Schwarzschild BH \cite{dRS}. 

These alternative prescriptions produce explicit expressions of the renormalized stress-energy tensor in particular spacetimes with high degree of symmetry. However, solving the semiclassical Einstein's equations requires knowledge of $\langle 0|T_{ab}|0\rangle$ as a function of the spacetime metric, but we still lack a general formula.

In view of these complications, and because  the causal structure of a spherically-symmetric gravitational collapse is  two-dimensional, a number of effective 1+1  models have alternatively been explored to gain further insights into the  dynamics of BH evaporation (see \cite{doi:10.1142/p378} and references therein). In sharp contrast with the four-dimensional case,  in 1+1 dimensions it is possible to derive an exact formula for the renormalized stress-energy tensor of conformal fields in any curved spacetime. Namely, if the spacetime metric is written as $ds_{(2)}^2=e^{2\rho}ds_{(2),{\rm flat}}^2$ for some conformal factor $\rho$ (recall that every 2-dimensional spacetime is conformally flat) then, for a {\it conformally static vacuum} $|0_{\rho}\rangle$ we can write \cite{ceb8dc1c-d11b-3e0b-9c15-54c356f2bc68, 0fb308e0-b7a9-3288-b0cc-1f935580a086, PhysRevD.13.2720, PhysRevD.16.1712, Bunch_1978, WALD1978472, doi:10.1142/p378} 
\begin{equation}
\langle 0_{\rho}|T_{a b}^{(2)}| 0_{\rho}\rangle=\frac{\hbar R_{(2)}}{48 \pi} g^{(2)}_{a b}+\frac{\hbar}{12 \pi}\left[\nabla_{a} \nabla_{b} \rho+\nabla_{a} \rho \nabla_{b} \rho -\frac{1}{2}g^{(2)}_{ab}(\nabla_c \nabla^c \rho+\nabla_c \rho \nabla^c\rho)\right]\, , \label{T2} 
\end{equation}
where $R_{(2)}$ is the 2-dimensional Ricci scalar.
This observation can motivate us to reduce a QFT on a spherically-symmetric 3+1 spacetime  down to an effective 1+1 theory, so as to take advantage of equation (\ref{T2}) or similar. This is known as {\it dimensional reduction}. Starting from the Einstein-Hilbert action and a minimally coupled  scalar field $f$, working with a spherically-symmetric metric $ds^2=ds^2_{(2)}+r^2d\Omega^2$ with radial variable $r\equiv \kappa^{-1} e^{-\phi}$, and expanding the scalar field $f$ in spherical harmonics, then the integration of the angular degrees of freedom in the action results in  \cite{MUKHANOV1994283, doi:10.1142/p378}  
\begin{equation}
S=\frac{1}{2\kappa^2}\int d^{2} x \sqrt{-g_{(2)}} e^{-2 \phi}\left[\frac{R_{(2)}}{2}+ \nabla_{a} \phi \nabla^{a} \phi+\kappa^2 e^{2 \phi}- \nabla_{a} f_{0} \nabla^{a} f_{0}\right]+\ldots \label{2dtheory}\, ,
\end{equation}
where dots denote contributions  $\ell\geq 1$ of the  field's harmonic modes $f_{\ell}$. The restriction to the $\ell=0$  sector yields an effective  2-dimensional theory which is called the {\it $s$-wave approximation}. With this truncation, the 4-dimensional stress-energy tensor for $f$ is  given by $T_{ab}  =\frac{ T_{a b}^{(\ell=0)} }{4\pi r^2}$. If $f_0$ is now quantized,  we may use  (\ref{T2}) to obtain an $s$-wave approximation for $\langle  T_{ab} \rangle$. 
(This is called the Polyakov approximation; a more accurate $s$-wave approximation which generalizes formula (\ref{T2}) by incorporating  the   coupling  with $\phi$ in (\ref{2dtheory}) is also available \cite{doi:10.1142/p378}). 
This strategy has been adopted to gain some  insights in semiclassical gravity \cite{PhysRevD.73.104023, PhysRevLett.120.061102, PhysRevD.104.084071, Barceloetal, PhysRevD.106.124006}.

Unfortunately, the backreaction equations derived from this effective 2-dimensional model are still challenging to solve due to the explicit coupling of the scalar field $f_0$ with  $\phi$. To study BH formation and evaporation, slightly different toy models have been analyzed instead,  which can be related to (\ref{2dtheory}) under a ``{\it near-horizon}'' approximation. The most popular one is  the Callan-Giddings-Harvey-Strominger ``stringy'' model (see \cite{doi:10.1142/p378} and references therein),  
\begin{equation}
S_{C G H S}=\int d^{2} x \sqrt{-g_{(2)}}\left\{e^{-2 \phi}\left[\frac{R_{(2)}}{2}+2 \nabla_{a} \phi \nabla^{a} \phi+2 \kappa^{2}\right]-\frac{1}{2} \nabla_{a} f_{0} \nabla^{a} f_{0}\right\}\, . \label{cghs}
\end{equation}
This theory can now be solved exactly.  Most up-to-date analytical and numerical calculations  indicate that, despite the formation of a singularity, BH evaporation in this framework does not involve  information loss, in the sense that the $S$-matrix is unitary \cite{PhysRevLett.106.161303, PhysRevD.83.044040}.

Despite the interest of  these results, the accumulation of so many hypothesis / approximations in this 2-dimensional approach raises some questions.  To which extent does the 1+1 framework provide a reliable picture of gravitational collapse in 3+1 semiclassical gravity? 

The reliability of the $s$-wave approximation (\ref{2dtheory}) and the corresponding stress-energy formula $T_{ab}  =\frac{ T_{a b}^{(\ell=0)} }{4\pi r^2}$ lies in its ability to reproduce the $\ell=0$ contribution of the Hawking luminosity at future null infinity, as well as the $\ell=0$ part of the late-time negative-energy flux across the horizon \cite{doi:10.1142/p378}. However, it is known that dimensional reduction  does not commute with  quantization, not even in flat space \cite{PhysRevD.61.024021, PhysRevD.62.044033}. This is,  although we can  regard a classical scalar field $f$ as an infinite collection of 2-d fields $\{f_{\ell} (t,r)\}_{\ell=0}^{\infty}$ by expanding it in spherical harmonics, ${f}=\sum_{\ell=0}^{\infty} \sum_{m=-\ell}^{\ell} f_{\ell} (t,r) \frac{Y_{\ell m}(\theta, \phi)}{r}$,  the quantization of $f$ on a 3+1 spacetime is not equivalent to the quantization of the  modes $f_\ell$ on the reduced 1+1 spacetime. In particular, using (\ref{Tab}) separately for both $f$ and $f_{\ell}$, it can be shown that \cite{PhysRevD.61.024021, PhysRevD.62.044033} 
\begin{equation}
\langle{ T_{ab}}(t,r,\theta,\phi)\rangle \neq \sum_{\ell=0}^{\infty} \frac{(2 \ell+1)}{4 \pi r^{2}}\langle{ T_{ab}}_{,\ell}(t,r)\rangle= \frac{1}{4 \pi r^{2}}\langle{ T_{ab}}_{,0}(t,r)\rangle+\dots \, ,
\end{equation}
where dots denote $\ell\geq 1$ contributions. This  result shows that dimensional reduction only provides a {\it naive} $s$-wave approximation of the actual renormalized stress-energy tensor of a quantum field. Therefore, although it does recover the $s$-wave part of the Hawking luminosity at infinity, it can dramatically fail to describe the $s$-wave part of $\langle T_{ab}\rangle$ in other spacetime regions where gravity is strong, like in a neighborhood of  curvature singularities. Besides this issue,  higher $\ell$ contributions become relevant near the curvature singularity, so even the actual $s$-wave approximation can fail to correctly capture semiclassical effects.  Not to talk about the reliability of ``near-horizon approximations'' that are used to justify toy models like (\ref{cghs}). Overall, my opinion is that  any conclusion extracted or inspired from these 1+1 approximations should perhaps be taken with a grain of salt.

\section{Some personal ideas and future prospects}

The problematic discussed in this paper about the renormalized stress-energy tensor in 3+1 spacetimes is a long-standing one. Unfortunately, it does not look that a  fully satisfactory answer is going to appear in the near future with current methodology  (mostly concerned in obtaining numerical results for concrete gravitational backgrounds). If we aim/hope to make some progress in the backreaction problem, I believe that radically new and fresh ideas need to be discussed and proposed.  

To my view, one  interesting idea   is to look for strategies that may allow us to generalize the well-known formula  (\ref{T2}) to 3+1 spacetimes.  All current derivations of this result  rely crucially on the conformal flatness of 1+1 spacetimes \cite{ceb8dc1c-d11b-3e0b-9c15-54c356f2bc68, 0fb308e0-b7a9-3288-b0cc-1f935580a086, PhysRevD.13.2720, PhysRevD.16.1712, Bunch_1978, WALD1978472} , and the lack of it in 3+1 backgrounds has prevented, perhaps, a generalization of these calculations. We have recently deduced an explicit analytical expression for the renormalized stress-energy tensor in 1+1 dimensions which does not assume the hypothesis of conformal flatness, and which recovers equation (\ref{T2}) in  conformal coordinates \cite{dRpepejavier2}. This calculation can in principle be generalized to 3+1 spacetimes, although it is technically much more involved. We are currently exploring this computation and expect to reach a conclusion in the near future. 

Another possible way to proceed is to reformulate the standard renormalization prescription for (\ref{Tab}) in a manifest 3+1 covariant language. The choice of quantum state might be  connected to a choice of spacetime foliation. If this is possible, it may be used to solve the semiclassical Einstein's equations using standard techniques in numerical relativity. I am not aware of any previous attempt to develop  this idea. 

 To address the backreaction problem it can also be useful to deviate the focus from the explicit calculation of the renormalized stress-energy tensor. For instance, instead of focusing in getting $\langle T_{ab}\rangle$ as a functional of the metric, we can perhaps leave it as an unknown in the backreaction equations, and try to solve the full set of variables imposing some physically motivated constraints between them. This idea has  been explored and developed for static and spherically-symmetric spacetimes in \cite{PhysRevD.107.085023} using the trace anomaly \cite{DESER197645, PhysRevD.15.1469, PhysRevD.15.2810, PhysRevD.16.3390, PhysRevD.17.1477} and an isotropy condition for the renormalized stress-energy components \cite{PhysRevD.21.2185}.

\begin{figure}[ht]
\centering
\includegraphics[width=0.4\linewidth]{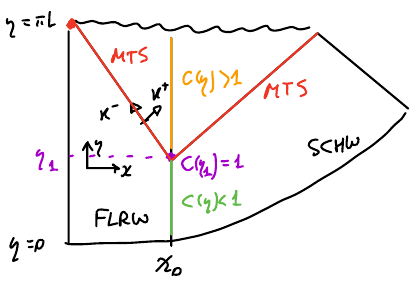}
\includegraphics[width=0.35\linewidth]{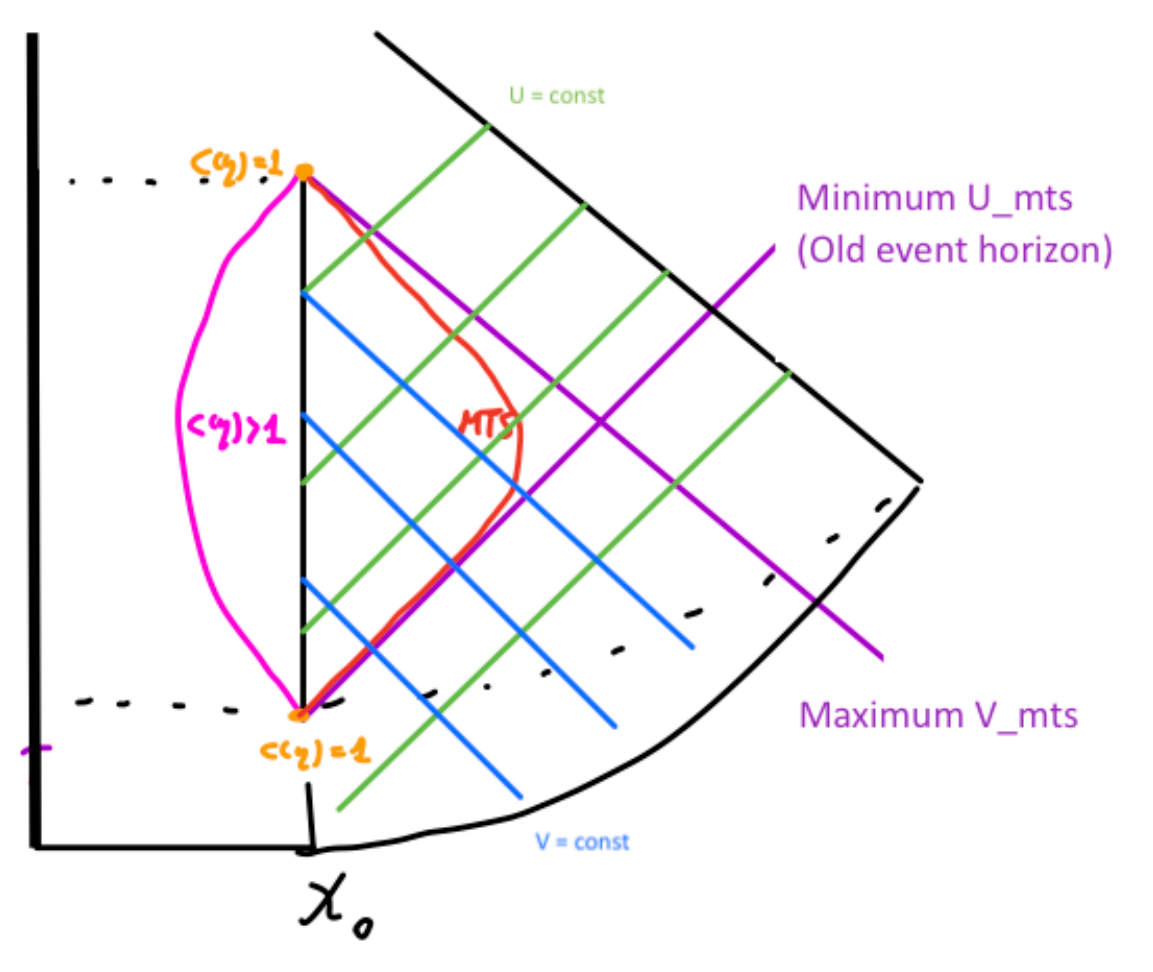}
\caption{Left: Penrose diagram for the causal structure of the Oppenheimer-Snyder gravitational collapse. Right: Penrose diagram inferred in \cite{dRpepejavier2} for a gravitational collapse in semiclassical gravity.  \label{fig:2}}
\end{figure}

In connection with the above, we have also attempted to study a semiclassical version of the dynamical Oppenheimer-Snyder (OS) collapse model without solving explicitly the field equations \cite{dRpepejavier}.  Let me recall the causal structure of the classical OS gravitational collapse, given in the left panel of Fig \ref{fig:2}. This model consists of two solutions of the Einstein's equations that are matched   through the surface of the star. On the one hand, the interior is modeled by a  homogeneous and isotropic dust fluid, giving rise to a FLRW metric, whose scale factor describes the  collapse of the star. On the other hand, the exterior vacuum is described by the Schwarzschild line element.   The causal diagram is characterized by two null apparent horizons, one in the interior and another one in the exterior (event horizon),  which enclose a trapped region and a curvature singularity in the future. The first marginally trapped surface forms when the radius of the star reaches the Schwarzschild radius. As time evolves further the compactness of the star increases monotonically in time, until the singularity is formed in finite time.

 In semiclassical gravity we have to solve again the field equations with the addition of the renormalized stress-energy tensor of a quantum field. In sharp contrast to the classical solution, the interior spacetime will fail to be homogeneous. Even though the classical metric  in the interior is homogeneous and isotropic, to specify the vacuum state we need to impose some boundary conditions for the field modes on the surface of the star, which will in general introduce inhomogeneities through backreaction. This is, unlike in cosmological studies, the most natural vacuum state of the quantum field inside the star will only be isotropic, but not homogeneous. This make things significantly harder.

 By working out the semiclassical Einstein's equations (using some results for $\langle T_{ab}\rangle$ in specific inhomogeneous FLRW metrics, and some approximations) we find that some solutions may admit a Penrose diagram which is topologically equivalent to the right panel of Fig \ref{fig:2}. In this diagram, the two apparent horizons of the classical OS model are no longer null but timelike, they are bounded from below and from above, and eventually reconnect again in the future, enclosing a compact and regular trapped region. Physically, the radius of the star  (compactness) no longer decreases (increases) monotonically, but rather reaches a minimum (maximum) value, when the star undergoes a bounce. Interestingly, this bounce occurs at a time when the compactness of the star is again below $1$. Thus, the trapped region must have fully vanished or ``evaporated'' before, when the radius of the star reaches back the Schwarzschild radius from below.  Since the formation of trapped surfaces in this picture is bounded in a compact region of spacetime, in this solution information falling in the trapped region is not lost, it is only kept captive during a large amount of time, and it eventually goes out. 
 
This global causal structure has been widely expected and conjectured for a long time in different frameworks of quantum gravity \cite{FROLOV1981307, PhysRevD.28.1265, Ashtekar_2006, PhysRevLett.96.031103, doi:10.1142/S0218271814420267} (for a review see \cite{Frolov:2014jva, universe3020048} and references therein). So far, our findings indicate that a spacetime bounce after a  trapped region ``evaporates'' can  be predicted at the semiclassical level, during the late stages of gravitational collapse of stars. We hope to confirm these expectations with more accurate calculations.

{\bf Acknowledgements.} I am grateful to J. Navarro-Salas and A. Fabbri for useful comments while preparing this manuscript. 
I acknowledge financial support via  ``{\it Atraccion de Talento Cesar Nombela}'', grant No 2023-T1/TEC-29023, funded by Comunidad de Madrid, Spain. This work is also supported by  the Spanish Grant PID2023-149560NB-C21, funded by MICIU/AEI/10.13039/501100011033.

\bibliographystyle{utphys}
\bibliography{referencesRio}
\title{Black Holes Beyond General Relativity}

\author{Enrico Barausse\inst{1,2} \and Jutta Kunz\inst{3}}

\institute{\textit{SISSA, Via Bonomea 265, 34136 Trieste, Italy and INFN Sezione di Trieste} \and \textit{IFPU - Institute for Fundamental Physics of the Universe, Via Beirut 2, 34014 Trieste, Italy} \and \textit{Institute of Physics, University of Oldenburg, Oldenburg, Germany}}

\maketitle

\begin{abstract}
Here, we will discuss some ideas for possible classical/semi-classical modifications of the black hole solutions in General Relativity (GR). These modifications/extensions include black holes in higher dimensions; black holes with additional gravitational fields, or fields beyond the Standard Model of Particle Physics; black holes in alternative classical theories of gravity and in semiclassical gravity; phenomenological models that extend the GR black hole solutions.
\end{abstract}

\section{Black holes in higher dimensions}

Black holes in higher dimensions naturally arise  from string theory.
Treating the $D-1$ spatial dimensions on  equal footing, the Myers-Perry black holes~\cite{Myers:1986un} generalize the Kerr solution to rotation in $\lfloor (D-1)/2 \rfloor$ independent planes, while retaining a spherical horizon topology. 
The horizon topology of black holes/objects in higher dimensions is, however, no longer necessarily spherical, as seen, for instance, from the presence of black rings, black Saturns, etc., i.e., a whole new zoo of black objects arises with non-trivial horizon topologies \cite{Emparan:2008eg}.
\textit{Open questions: Would it be possible to obtain black rings or other black objects with non-spherical horizon topology in four dimensions in some theory beyond GR?}

The Anti-de Sitter/conformal field theory (AdS/CFT) correspondence~\cite{Maldacena:1997re} can be exploited to study strongly coupled theories in the non-perturbative regime. The celestial holography program aims to realize this duality in asymptotically flat spacetimes (AFS) \cite{Raclariu:2021zjz}.
By translating dynamical problems in $D$ dimensions into gravitational problems in $D+1$ dimensions, the study of black holes (and other black objects) in higher dimensions allows for new insights and predictions.
\textit{Open questions: Can higher-dimensional black holes teach us about the underlying theory of quantum gravity, thanks to the AdS/CFT duality?}

\section{GR with known/unkown fields}

The electrovacuum black hole solutions of GR are well-studied and possess intriguing properties, e.g., they only possess three ``hairs'' (mass, spin and electric charge).
However, in the presence of non-Abelian fields, as is the case in the Standard Model of Particle Physics, this changes, i.e. black holes can carry non-Abelian hair \cite{Volkov:1998cc,Kleihaus:2016rgf}.
Whereas earlier studies assumed that such hairy black holes would all be microscopic and therefore not of astrophysical interest, it was recently shown that hairy black holes of planetary size can be obtained in the electroweak sector of the Standard Model coupled to GR \cite{Gervalle:2024yxj}.
\textit{Open questions: What further surprises are lurking in the Standard Model coupled to GR, waiting to be unveiled?}

Retaining GR but allowing for fields beyond the Standard Model opens further avenues to obtain potentially astrophysically relevant black holes \cite{Herdeiro:2014goa,Herdeiro:2015waa,Herdeiro:2016tmi} and black hole ``mimickers''~\cite{Boskovic:2021nfs}.
These objects need massive (light) complex scalar or vector fields with harmonic time-dependence for their existence.
Their astrophysical features, e.g., their shadows, or their gravitational waveforms~\cite{Toubiana:2020lzd,Boskovic:2021nfs},
can be sufficiently distinct from those of Kerr black holes to be detectable in future observations.
However, the allowed mass range for such complex fields is continuously narrowing, also due to constraints coming from superradiance and from pulsar timing array observations~\cite{Brito:2017wnc,Brito:2017zvb,EuropeanPulsarTimingArray:2023egv}.
\textit{Open questions: Will the mass parameter space for these light bosons  close in due time? Are  additional hairy rotating black hole solutions still conceivable and relevant astrophysically?}

\section{Alternative gravity theories}

Among the vast set of alternative theories of gravity \cite{Berti:2015itd,CANTATA:2021asi}, much work  has focused on black holes in quadratic gravity.
Coupling a scalar field to the Gauss-Bonnet or Chern-Simons invariants allows for scalarized rotating black holes of various types, depending on the coupling function~\cite{Silva:2017uqg,Doneva:2017bvd,Dima:2020yac,Herdeiro:2020wei}.
The additional scalar degree of freedom is expected to betray itself in binaries due to its associated dipole radiation, not present in GR.
Comparison of calculations and observations of gravitational waves during the inspiral phase can therefore put strong bounds on such theories \cite{Barausse:2016eii,Julie:2024fwy}. 
The scalar degree of freedom can also significantly modify the spectrum of quasi-normal modes arising in the ringdown of the merger remnant~\cite{Volkel:2022aca}.
First calculations of the spectrum of rapidly rotating black holes are now available in this type of theories for small couplings \cite{Chung:2024vaf} and for large couplings \cite{Blazquez-Salcedo:2024oek}.

Another class of theories in which black hole solutions have been 
systematically investigated is Lorentz-violating gravity. The latter has 
received significant attention over the past 15 years,
because Lorentz violations may allow for
 developing a renormalizable theory of gravity at the power-counting and potentially perturbative levels~\cite{Jacobson:2000xp,Horava:2009uw,Blas:2009qj,Blas:2010hb,Barvinsky:2015kil,Barvinsky:2019rwn,Bellorin:2022qeu}. However, defining black holes in such theories is challenging, due to the presence of multiple propagation speeds for the gravitational degrees of freedom, including superluminal speeds diverging at high energies. This has led to intense investigation of the black hole causal structure~\cite{Eling:2006ec,Blas:2011ni,Barausse:2011pu,Barausse:2013nwa,Lara:2021jul,Barausse:2015frm,Adam:2021vsk}, revealing multiple event horizons and, at least in spherical symmetry, a universal horizon, i.e. a causal boundary for arbitrarily fast signals~\cite{Blas:2011ni,Barausse:2011pu}.

\textit{Open questions: Will it be possible to 
significantly shrink the ``theory space'' of GR alternatives with astrophysical probes? Will future observations of the ringdown phase of black hole mergers be sufficiently precise to distinguish GR quasinormal modes from those of alternative theories? Will we be able to disentangle deviations from GR from environmental effects in astrophysical observations?}

\section{Semiclassical gravity}

Approaches to quantum gravity often focus on their implications for black holes, like Hawking radiation, the information loss problem, or the central curvature singularity.
Guidance from observations to scrutinize these approaches would be invaluable.
This motivates, for instance, the study of quantum gravity signatures from gravitational waves during the inspiral, merger and ringdown phases  of binaries of black holes or exotic compact objects \cite{Barausse:2018vdb,Konoplya:2019xmn,Agullo:2020hxe}.
\textit{Open questions: Will it be possible to discern quantum gravity effects from other effects modifying gravitational signals?}

\section{Phenomenological modelling}

An interesting alternative route is to explore (expected) quantum gravity effects based on a phenomenological modelling of classical spacetimes, e.g., by excising the central curvature singularity of a black hole \cite{Torres:2022twv,Lan:2023cvz}.
Employing such ``engineered'' spacetimes, their astrophysical viability can then be tested \cite{Simpson:2021zfl}, for instance, by gravitational wave observations or by observations of their light ring images.
\textit{Open questions: Will observations provide sufficient hints regarding the nature of the underlying theory?}

\bibliographystyle{utphys}
\bibliography{references}
%
\part{Black holes as laboratories: Tests of general relativity}
%
\title{Black Holes as Laboratories: Searching for Ultralight Fields}

\author{Richard Brito}

\institute{\textit{CENTRA, Departamento de F\'{\i}sica, Instituto Superior T\'ecnico -- IST, Universidade de Lisboa -- UL, Avenida Rovisco Pais 1, 1049-001 Lisboa, Portugal}}

\maketitle
\begin{abstract}
The interaction of black holes with classical fields can lead to many interesting phenomena such as black-hole superradiance and the superradiant instability. The existence of these effects has been shown to have implications for beyond Standard Model particles that could explain dark matter, namely ultralight bosonic fields. In this note I give a historical account of this topic and briefly go through some recent developments. I conclude with some personal reflections on the importance of constraining simple dark matter models, even if such models may ultimately be the incorrect description of nature.
\end{abstract}

\section{Black holes and classical fields}
The study of how classical fields interact with black hole (BH) spacetimes has a long story that can be traced back to the early studies regarding the mathematical properties of the Kerr metric that followed its discovery in 1963~\cite{Kerr:1963ud}. Not long after Roy Kerr first wrote down the metric that bears its name, it was quickly understood that the Kerr family of solutions features a number of important mathematical properties that make many calculations tractable\footnote{As Subrahmanyan Chandrasekhar noted in his 1983 Nobel Lecture ``contrary to every prior expectation, all the standard equations of mathematical physics can be solved exactly in the Kerr space-time''~\cite{ChandraNobel}.} (to this day the best and most remarkable review on the topic continues to be Chandrasekhar's influential book~\cite{Chandrasekharbook}). Most notably, as first noted by Carter in 1968, the Hamilton-Jacobi and the scalar wave equation are separable in the Kerr metric~\cite{Carter:1968ks,Carter:1968rr} (see also Ref.~\cite{Brill:1972xj} were the details of this separability for the Klein-Gordon equation were first given explicitly). A no less remarkable result is the fact that scalar, neutrino,
electromagnetic, and gravitational perturbations of the Kerr metric can all be described by a single \emph{separable} master equation, as was first shown by Teukolsky in 1972 and 1973~\cite{Teukolsky:1972my,Teukolsky:1973ha}. These results were fundamental to explicitly show that by scattering massless bosonic waves (i.e. scalar, electromagnetic and gravitational waves) off a Kerr BH, one can extract energy and angular momentum from a BH through a process now known as \emph{superradiance}~\cite{Press:1972zz,Starobinsky:1973aij,Starobinskil:1974nkd,Teukolsky:1974yv}, a wave analogue to the Penrose process~\cite{Penrose:1971uk} that had been first predicted by Zel'Dovich in 1971~\cite{1971JETPL..14..180Z}. The existence of superradiance was used to devise a \emph{gedankenexperiment} in which one can make a Kerr BH unstable by placing a mirror around it that reflects radiation back towards the hole multiple times, the ``\emph{black-hole bomb}''~\cite{Press:1972zz} idea (this effect was latter studied in more detail in Ref.~\cite{Cardoso:2004nk}). This was followed by works showing that a \emph{massive} scalar field can become unstable in a Kerr BH~\cite{Damour:1976kh,Ternov:1978gq,Zouros:1979iw,Detweiler:1980uk} (the existence of this instability has also been rigorously demonstrated in Ref.~\cite{Shlapentokh-Rothman:2013ysa}). The physical mechanism behind this instability can be traced back to superradiance: massive fields admit quasi-bound state modes in BH spacetimes which can be superradiantly amplified multiple times inside the ergoregion, leading to an instability in analogy with the BH bomb effect.

These early studies regarding classical fields in BH spacetimes were mostly concerned about understanding the properties of Kerr BHs, such as their stability under small perturbations. This was crucial to establish the astrophysical relevance of the Kerr solution, given that the superradiant instability, the only known (classical) instability for (sub-extremal) Kerr BHs, seemed to be irrelevant for astrophysical purposes.\footnote{I am not including mass inflation as a relevant instability here since in principle it should not affect the spacetime outside the event horizon. Also note that the full nonlinear stability of the (sub-extremal) Kerr metric under gravitational perturbations has not yet been formally established. However, the sub-extremal Kerr metric has been shown to be linearly mode stable~\cite{Whiting:1988vc} against massless perturbations, whereas significant steps towards proving the full linear stability of the sub-extremal Kerr metric under massless perturbations were recently made in Refs.~\cite{Shlapentokh-Rothman:2020vpj,Shlapentokh-Rothman:2023bwo}.} Specifically, the superradiant instability was shown to be maximized when the Compton wavelength $\lambda_c$ of the massive field is comparable to the BH's horizon radius $r_H$.\footnote{Early works on the subject computed the superradiant instability rates only in the limits $r_H/\lambda_c\gg 1$~\cite{Zouros:1979iw} and $r_H/\lambda_c\ll 1$~~\cite{Detweiler:1980uk}. As far as I am aware, the first studies looking at the most interesting regime $r_H/\lambda_c\sim 1$ were actually only done in the 2000's through time-domain~\cite{Strafuss:2004qc} and frequency-domain computations~\cite{Furuhashi:2004jk,Cardoso:2005vk,Dolan:2007mj}.} On the other hand, if $r_H/\lambda_c\gg 1$, the instability growth rate $\Gamma$ is exponentially suppressed $\Gamma\propto e^{-r_H/\lambda_c}$~\cite{Zouros:1979iw}. This then means that for a BH of mass $M$, the instability is maximized for bosons with mass  
\begin{equation}
m_b\sim 5.6\times 10^{-11}\left(\frac{M_\odot}{M}\right)\,{\rm eV}\,.\label{eq:mass_super}
\end{equation}
The conclusion was then that no particles known within the Standard Model can turn astrophysical BHs unstable on a relevant timescale. For example, a simple calculation shows that for a pion field (which can be described as a massive scalar field) around a solar-mass BH one has $r_H/\lambda_c\sim 10^{18}$ and the instability is extremely suppressed. In order to have  $r_H/\lambda_c\sim 1$ for pions one would need to have BHs with masses $M\sim 10^{12}$ kg, i.e., if one only considers Standard Model particles, such instability is only relevant for primordial BHs (if at all relevant, given that the maximum instability growth rate for the specific case of the pion field is of the same order of magnitude as its lifetime~\cite{Detweiler:1980uk,Dolan:2007mj}).
 
\section{Black holes as particle detectors for ultralight fields}

Although many applications of the superradiant instability were found through the years (for extensive reviews on superradiance and its applications see Refs.~\cite{Bekenstein:1998nt,Brito:2015oca}), its relevance for searches of physics beyond the Standard Model seems to have remained mostly under the radar at least until 2009, when Ref.~\cite{Arvanitaki:2009fg} proposed the \emph{string axiverse} scenario. In particular they argued that ``string theory suggests the simultaneous presence of many ultralight axion-like particles, possibly populating each decade of mass down to the Hubble scale $10^{-33}$ eV''~\cite{Arvanitaki:2009fg}. They then realized that axion-like particles\footnote{Axion-like particles are pseudo-scalar particles with properties similar to the QCD axion~\cite{Peccei:1977hh,Peccei:1977ur,Weinberg:1977ma,Wilczek:1977pj}, hence their name.} in the mass range $\sim 10^{-21}-10^{-10}$ eV [see Eq.~\eqref{eq:mass_super}], as the ones possibly arising from the string axiverse scenario, could significantly affect the dynamics of rapidly rotating astrophysical BHs\footnote{I should note however that Ref.~\cite{Zouros:1979iw} did notice that the superradiant instability would be relevant for a BH with mass $M\sim M_{\odot}$ if fields with masses around $m_b \sim 10^{-10}$ eV existed, noticing also that the observations of a rapidly rotating black hole could rule out the existence of these fields. Similarly Ref.~\cite{Dolan:2007mj} noted that ``the instability is insignificant for astrophysical
black holes, unless there exists an unknown particle with a tiny but non-zero rest mass''. However, apart from these short remarks, Ref.~\cite{Arvanitaki:2009fg} seems to have been the first paper noticing that the BH superradiant instability could have implications for physics beyond the Standard Model in a well-motivated model.} precisely due to the superradiant instability of massive scalar fields that had been discovered back in the 1970's. 

The realization that BH superradiance has implications for a well motivated beyond Standard Model scenario, as well as the fact that ultralight bosons are known to be possible candidates for cold dark matter (see Ref.~\cite{Ferreira:2020fam} for an extensive review on ultralight dark matter) have been one of the main motivations behind the flurry of developments that have occurred since 2010. For example, it was only in the last $\sim 10$ years that it was shown for the first time that massive vector fields fields~\cite{Witek:2012tr,Pani:2012bp,Pani:2012vp,Baryakhtar:2017ngi,East:2017mrj,Cardoso:2018tly,Frolov:2018ezx,Dolan:2018dqv}, as well as massive tensor fields~\cite{Brito:2013wya,Brito:2020lup,Dias:2023ynv} are also prone to the superradiant instability.\footnote{The relevance of the superradiant instability for massive tensor fields is still not fully clear though, since massive tensors are also prone to an axisymmetric instability around Kerr BHs~\cite{Babichev:2013una,Brito:2013wya,East:2023nsk} that is unrelated to superradiance and that dominates over the superradiant instability in much of the parameter space.} This was somehow expected given that superradiance occurs for any bosonic field, however the more complex structure of the field equations (as well as perhaps a lack of motivation to study this problem before 2010) meant that it took nearly 40 years to go from the first works showing that massive scalar fields were superradiantly unstable to similar works for higher-spin fields.

The newly-found motivation to study the superradiant instability also prompted the need to understand in more detail how the instability evolves over time and what its astrophysical consequences are. Studies of the evolution of the instability have been done using both quasi-adiabatic approximations~\cite{Brito:2014wla,Ficarra:2018rfu,Hui:2022sri} and through numerical relativity simulations~\cite{East:2017ovw,East:2018glu} (this later case only for massive vector fields) and have explicitly shown that the instability leads to an exponentially growing bosonic field around the BH, while the BH spins down in the process. This process continues until the BH and the bosonic environment synchronize their rotation, a process that is analogous to tidal locking in binary systems~\cite{Cardoso:2012zn}. This scenario provides a natural mechanism to form long-lived \emph{boson clouds} around BHs, also known as \emph{gravitational atoms} due to the similarity of these configurations with the hydrogen atom (see e.g.~\cite{Detweiler:1980uk,Dolan:2007mj,Arvanitaki:2009fg,Baumann:2019eav}). The long term evolution of these clouds depends in general on whether the field is complex (i.e. the field admits a global $U(1)$ symmetry) or real. For complex fields, the stress-energy tensor of the cloud can be stationary and axisymmetric (even though the bosonic field itself might not be) and therefore the instability can evolve towards truly stationary and axisymmetric BH solutions known as ``Kerr BHs with bosonic hair''~\cite{Herdeiro:2014goa,Herdeiro:2016tmi,Herdeiro:2017phl}.\footnote{Kerr BHs with bosonic hair are themselves unstable against the superradiant instability of higher-order modes~\cite{Ganchev:2017uuo}, and therefore they are only metastable configurations with a lifetime that depends on the specific parameters of the system. However, since the instability rates are smaller for higher-order modes, effectively only a couple of modes will be relevant over the lifetime of a given BH.} On the other hand, for real fields, the bosonic cloud will necessarily dissipate through the emission of nearly monochromatic gravitational-waves (GWs) over long timescales~\cite{Arvanitaki:2010sy,Yoshino:2013ofa,Arvanitaki:2014wva,Baryakhtar:2017ngi,Brito:2017zvb,Siemonsen:2019ebd}.

A large number of observational consequences of the superradiant instability have also been uncovered in the last fifteen years. In particular, purely gravitational signatures that have been studied over the years include: the prediction that the existence of ultralight bosons in nature would imply a lack of highly spinning BHs in certain mass ranges, with the specific range being dependent on the putative boson mass~\cite{Arvanitaki:2009fg,Arvanitaki:2010sy,Arvanitaki:2016qwi,Brito:2017zvb,Baryakhtar:2017ngi,Cardoso:2018tly,Stott:2018opm,Ng:2019jsx,Ng:2020ruv,Stott:2020gjj,Hoof:2024quk}; the possibility to detect long-lived nearly monochromatic GWs emitted by boson clouds~\cite{Arvanitaki:2009fg,Arvanitaki:2010sy,Yoshino:2013ofa,Arvanitaki:2014wva,Baryakhtar:2017ngi,Brito:2017zvb,Siemonsen:2019ebd,Siemonsen:2022yyf} either through all-sky~\cite{Arvanitaki:2014wva,Brito:2017zvb,DAntonio:2018sff,Zhu:2020tht,LIGOScientific:2021rnv}, targeted~\cite{Arvanitaki:2016qwi,Isi:2018pzk,Ghosh:2018gaw,Sun:2019mqb,Jones:2023fzz} or stochastic background searches~\cite{Brito:2017wnc,Tsukada:2018mbp,Tsukada:2020lgt,Yuan:2021ebu,Yuan:2022bem}; the prediction that the dynamics of binary BH systems can be strongly affected by boson clouds if one or both BHs is surrounded by a cloud~\cite{Ferreira:2017pth,Hannuksela:2018izj,Baumann:2018vus,Baumann:2019ztm,Zhang:2019eid,Cardoso:2020hca,Baumann:2021fkf,Tomaselli:2023ysb,Brito:2023pyl,Duque:2023seg,Cannizzaro:2023jle,Tomaselli:2024bdd,Boskovic:2024fga}; or even signatures in the motion of S-stars orbiting around Sgr A$^*$ which could be perturbed by the presence of a cloud if it exists around the BH at the center of our galaxy~\cite{Ferreira:2017pth,GRAVITY:2019tuf,GRAVITY:2023cjt,GRAVITY:2023azi}. Some of these observables have in fact already been used to impose constraints on ultralight bosons in parts of the parameter space (I refer the reader to Ref.~\cite{Brito:2015oca} for a review of these constraints).

One of the most important caveats of the works I mentioned so far, is the fact that, in general, they assume ultralight fields described by a simple massive bosonic field that only interacts with other particles through gravity. As such, especially in the last five years, several studies have started to explore in more detail how self-interactions or couplings of the bosonic field with Standard Model particles affect the superradiant instability and its observables (see e.g.~\cite{Ikeda:2018nhb,Fukuda:2019ewf,Baryakhtar:2020gao,Omiya:2022gwu,Omiya:2022mwv,East:2022ppo,Siemonsen:2022ivj,Spieksma:2023vwl,Chen:2023vkq,Takahashi:2024fyq}). From these studies it is now clear that non-gravitational interactions can significantly affect the evolution of the superradiant instability, mainly by limiting the growth of the cloud and slowing down the rate at which the BH spins down during the instability phase, which can then affect the purely gravitational-wave observables I mentioned before. Moreover, perhaps not surprisingly, novel signatures that do not occur in the free field case, have also been uncovered. To give some recent examples: in clouds formed by self-interacting scalar fields, GW signals due to level transitions in the cloud can be more common and longer-lived than in the absence of self-interactions~\cite{Baryakhtar:2020gao,Omiya:2024xlz,Collaviti:2024mvh}; binary tidal interactions can facilitate the ``bosenova'' collapse of the cloud if self-interactions are sufficiently strong~\cite{Takahashi:2024fyq} (see also~\cite{Aurrekoetxea:2024cqd} where a similar effect was seen in numerical simulations of binary BHs surrounded by self-interacting scalar dark matter); couplings of the bosons in the cloud with photons could lead to electromagnetic signatures~\cite{Siemonsen:2022ivj,Spieksma:2023vwl} or birefringence effects that can be measured through polarimetric measurements of the radiation emitted near supermassive BHs~\cite{Chen:2021lvo,Ayzenberg:2023hfw}.

The (incomplete) overview I gave above highlights the fact that most of what has been learned regarding the implications of BH superradiance for beyond Standard Model physics, happened just in the last fifteen years. I will not try to speculate where this line of research will be in fifteen years from now, but I believe there is still plenty of work to be done. A specific example regards the modelling of BH binaries moving inside ultralight bosonic environments (such as boson clouds). While I would say that the general picture of how a bosonic environment can affect the dynamics of BH binaries is now understood (see e.g.~\cite{Ferreira:2017pth,Hannuksela:2018izj,Baumann:2018vus,Baumann:2019ztm,Zhang:2019eid,Cardoso:2020hca,Traykova:2021dua,Baumann:2021fkf,Bamber:2022pbs,Tomaselli:2023ysb,Brito:2023pyl,Duque:2023seg,Cannizzaro:2023jle,Aurrekoetxea:2023jwk,Traykova:2023qyv,Tomaselli:2024bdd,Boskovic:2024fga,Aurrekoetxea:2024cqd}), there is still a lot of work to be done in order to eventually produce \emph{generic} and \emph{accurate} gravitational waveforms from binary BH systems moving in such environments, both in the comparable-mass-ratio regime and in the extreme-mass-ratio regime. This problem is especially interesting given that future GW detectors will most likely have the capability to detect the impact of astrophysical environments, such as accretion disks, dark matter spikes or boson clouds, on the evolution of binary BH systems (see e.g.~\cite{Cardoso:2019rou,Cole:2022yzw,Speri:2022upm,CanevaSantoro:2023aol}). Within this context, attempting at modeling BH binaries in ultralight bosonic environments offers a perfect playground, much simpler to model than other astrophysical environments. Therefore I am certain there is a lot to be learned from this exercise. 

\section{Concluding remarks}
Particles such as the QCD axion, axion-like particles or dark photons remain elusive so far, with laboratory and astrophysical data constraining them to have a mass below the eV scale and to couple ``weakly'' enough with Standard Model particles. A very interesting aspect of BH superradiance is the fact that it is a pure gravitational effect. Therefore the phenomenology associated with this effect is particularly suited to constrain ultralight bosons in regions that are difficult to constrain through usual laboratory searches, in which some type of non-gravitational interaction of the bosons with Standard Model particles is assumed (see e.g.~\cite{AxionLimits} for the latest limits on the mass and couplings of axion-like particles and dark photons).
Therefore, constraints coming from BH superradiance are in general highly complementary to other type of searches, making this a particularly interesting topic to explore.

In the most pessimistic (or optimistic, depending on the point-of-view) scenario one can envision that in the not so far future the joint constraints coming from BH superradiance (and associated phenomenology) and other types of searches will exclude most of the parameter space for ultralight bosons in the mass range $\sim [10^{-21}, 10^{-10}]$ eV. Perhaps a more exciting possibility would be to actually detect some signatures signaling the existence of bosons with such masses. The main difficulty then, will be to convince ourselves that what we are observing are in fact ultralight bosons. In that respect, another particularly interesting aspect of BH superradiance, is that it predicts a wide array of possible observables. Therefore, a convincing way to pinpoint the existence of an ultralight boson would be to find convincing evidence for the same type of boson through more than one observational channel (for example through both continuous GW searches and BH spin measurements). 

\subsubsection*{Constraining dark matter models: an endless game?}
One of the motivations for the line of research discussed in this document is the fact that ultralight bosons provide a possible solution to the dark matter problem~\cite{Ferreira:2020fam}. This is in fact one the simplest solutions to the problem given that it only requires the existence of an ultralight scalar field. As with other dark matter candidates, such as WIMPs or primordial BHs, one can also motivate the model from the fact that such particles arise naturally in some well-motivated theoretical scenarios (in this particular case, from the string axiverse scenario for example~\cite{Arvanitaki:2009fg}). However, theoretical motivations alone are not really enough to constrain the nature of dark matter. The main difficulty in devising dark matter searches is precisely the fact that based on what we know currently from observations of dark matter, a wide range of theoretical scenarios provide viable dark matter candidates (see e.g.~\cite{Marsh:2024ury}). Therefore, one of the standard paradigms in dark matter searches has been to remain as agnostic as possible, considering simple dark matter candidates that can be parameterized by very few parameters (for example the mass of the particles, their interaction to photons, etc...), and ultimately to come up with methods to constrain the parameter space of these candidates as much as possible. Some of the research discussed in this document falls precisely within this line of reasoning. 

There is of course a possibility that this might become an endless game of putting stronger and stronger constraints without ever actually detecting new particles that could explain dark matter. The solution to the dark matter problem may require more complex models than just hypothesizing a new particle and could even require a dramatic paradigm shift in physics. However, even with this possibility in mind, I do not think that playing the game of constraining simple models is a worthless effort (although perhaps discouraging at times). In fact, I would say it is almost an obligation. On the one hand, if dark matter can be explained within the simplest models we currently know, then attempting at constraining them as best as we can might eventually result in a positive detection. On the other hand, in a scenario where explaining dark matter might require a more complex solution, then such solution will only be accepted once all the simplest explanations have been excluded. Therefore I see the current paradigm of trying to constrain simple dark matter models as being part of a long journey that might take decades or even centuries to reach its conclusion. Whether at the end of this (most likely very long) journey some future humans will find a proper solution to the dark matter problem I do not know, but I am confident that by attempting at constraining all the simplest explanations we can think of, we will eventually either find positive evidence for a dark matter candidate or reach the conclusion that a drastic paradigm shift might actually be needed.

\section*{Acknowledgments}
I thank the organizers of the \emph{Black Holes Inside and Out} conference for the invitation to write this short ``vision'' document. 
I also thank all my collaborators with whom I worked on topics related to black-hole superradiance throughout the years. In particular, I am especially indebted to Vitor Cardoso and Paolo Pani for long-lasting collaborations on this topic.
I acknowledge financial support provided by FCT – Fundação para a Ciência e a Tecnologia, I.P., under the Scientific Employment Stimulus -- Individual Call -- Grant No. 2020.00470.CEECIND (DOI: \href{https://doi.org/10.54499/2020.00470.CEECIND/CP1587/CT0010}{10.54499/2020.00470.CEECIND/CP1587/CT0010}). This work is also supported by FCT's project No. 2022.01324.PTDC (DOI: \linebreak\href{https://doi.org/10.54499/2022.01324.PTDC}{10.54499/2022.01324.PTDC}) and by the ``GravNewFields'' project, funded through \linebreak FCT's ERC-Portugal program.

\bibliographystyle{utphys}
\bibliography{references}

%





\title{Primordial Black Holes from Inflation}
\author{Misao Sasaki}

\institute{\textit{Kavli Institute for the Physics and Mathematics of the Universe (WPI), The University of Tokyo, Kashiwa, Chiba 277-8583, Japan} \and \textit{Center for Gravitational Physics and Quantum Information,
	Yukawa Institute for Theoretical Physics, Kyoto University, Kyoto 606-8502, Japan} \and \textit{Leung Center for Cosmology and Particle Astrophysics, National Taiwan University, Taipei 10617, Taiwan}}

\maketitle 
\begin{abstract}
It has now been widely accepted that many inflation models can account for the formation of primordial black holes (PBHs). 
In particular, it has been fully realized that the PBH formation depends crucially on the tails of the probability distribution function of the curvature perturbation. 
I will review some models of inflation that give rise to the PBH formation, and their observational implications, in particular, for the blossoming field of
gravitational wave cosmology.
\end{abstract}

\section{Introduction}

The possibility that black holes may be produced in the very early universe was pointed out more than half a century ago~\cite{Hawking:1971ei}. An intriguing property of primordial black holes (PBHs) is that there is no limitation in the range of masses.
In particular, if a sub-solar mass black hole is found, one can regard it as strong evidence for the existence of PBHs because there is no known mechanism to produce such a small mass black hole astrophysically.
But PBHs had never been one of the major topics in cosmology until recently. 

The situation changed when LIGO detected gravitational waves from the coalescence of binary black holes in 2015, the event GW20150914~\cite{Abbott:2016blz}.
Because the mass of these black holes were unexpectedly large, $\sim 30 M_\odot$, it lead to the idea that these black holes may be primordial~\cite{ Bird:2016dcv,Sasaki:2016jop}.
This idea sparked research efforts in the physics of PBHs.

Now after almost a decade of intensive studies, our understanding of PBHs has improved considerably; from the formation mechanisms and criteria to the cosmological and observational implications, particularly in the light of the new window to explore the universe, gravitational waves (For a comprehensive overview on PBHs, see \cite{pbhtextbook:2024}).
For example, compared to a simple criterion by the threshold amplitude of the density perturbation, more sophisticated criteria for the PBH formation have been proposed. 
However, here we do not discuss formation criteria but focus on the formation mechanisms and their implications.

\section{Conventional scenario}
The conventional scenario is that a PBH is formed from a rare, large positive curvature perturbation ${}^{(3}R\sim\Delta\calR/a^2$ during radiation dominance 
that becomes $O(1)$ when the comoving scale crosses the Hubble horizon scale,
where $a$ is the scale factor of the universe. 
Such a large curvature perturbation is too rare to produce any appreciable amount of PBHs if the rms amplitude is as small as that on the CMB scale,
 ie, $\sim 10^{-5}$ on Gpc scale. However, it can be much larger on small scales, say, $\sim 10^{-2}-10^{-1}$ on $\lesssim 1{\rm kpc}$ scales. 

\begin{figure}[ht]
\centering
\includegraphics[width=0.5\linewidth]{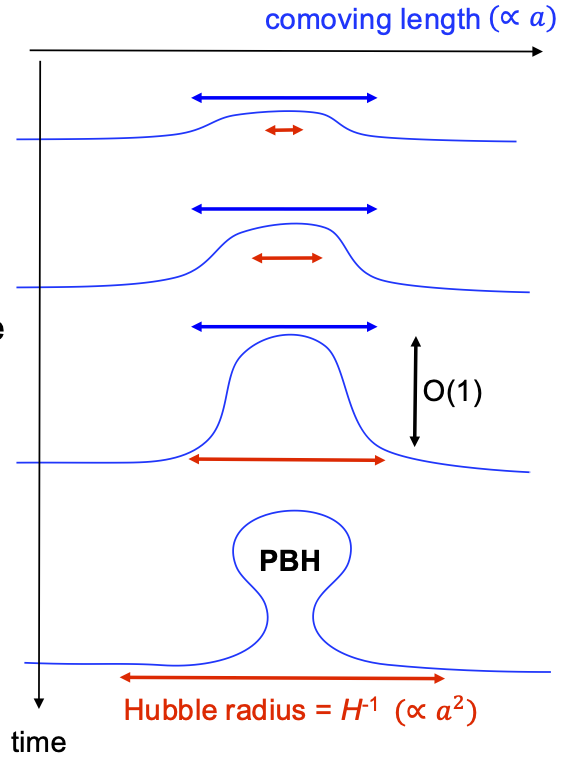}
\caption{An illustration of the PBH formation during radiation dominance.} \label{fig:pbhform}
\end{figure}

As illustrated in Fig.~\ref{fig:pbhform}, a region of positive spatial curvature collapses to form a PBH if the amplitude of the curvature perturbation is $O(1)$ when the comoving scale becomes equal to the Hubble horizon size,  just like the case of a closed, positive curvature universe that stops expanding when the curvature begins to dominate and collapse to a big crunch singularity. 
This tells us that the characteristic mass of PBHs is determined by the Hubble horizon size at the time when the PBH scale crosses it. Hence,
\begin{equation}
    M_{PBH}\sim M_{H}\sim \left(\frac{100{\rm MeV}}{T}\right)^2M_\odot
    \sim \left(\frac{\ell}{1{\rm pc}}\right)^2M_\odot\,,
\end{equation}
where $T$ is the temperature when the comoving scale $\ell$ crosses the Hubble horizon.

Here the important point is that the PBH formation is rare, 
hence sensitive to the tail behavior of the probability distribution function (PDF) of the curvature perturbation. This implies that any primordial non-Gaussianities will affect the PBH formation significantly.

\section{Inflation}

Conventionally single-field slow-roll inflation is assumed, and it fits the observational data such as the Planck CMB data very well~\cite{Planck:2018jri}.
Nevertheless, the range of the inflaton potential probed by CMB is very narrow; corresponding to the duration of at most $10$ e-folds around $50$-$60$ e-folds 
from the end of inflation. Thus on very small scales, say $10$-$20$ e-folds from the end of inflation, non-trivial features may appear in the potential or a different inflaton field may dominate the late stage of inflation. In such cases, the power spectrum of the curvature perturbation, $P_{\calR}(k)$,
may be enhanced considerably. A number of such models have been proposed.  

Here let us focus on a model of inflation in which the potential has a small upward step~\cite{Cai:2021zsp}. Even if the step is extremely small, it may cause a substantial effect on the curvature perturbation. First, $P_{\calR}(k)$
 be significantly enhanced to $\sim10^{-2}$. But most importantly, the tail of the PDF of the curvature perturbation has a highly non-perturbative behavior; it has a cutoff at a certain value of $calR$ because, if the fluctuation just before the step is such that it pulls back the inflaton field to make the kinetic energy of the inflaton too small, it cannot climb the step. Thus, in this model, depending on the choice of parameters, it can either enhance the PBH formation or completely suppress it.

\begin{figure}[ht]
\centering
\includegraphics[width=0.8\linewidth]{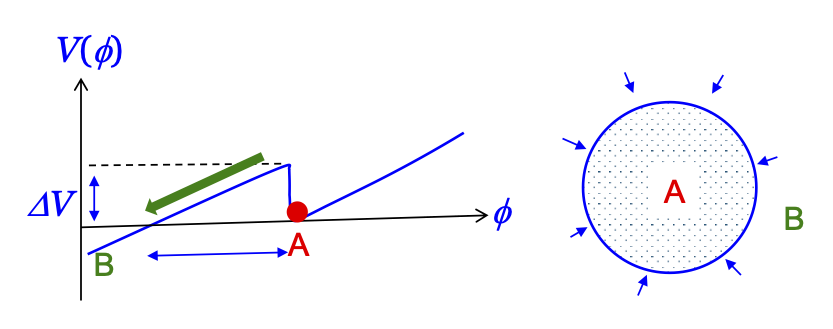}
\caption{An illustration of a Hubble size region that becomes a PBH. }
\label{fig:deadend}
\end{figure}

The above picture, however, happens to be too naive. 
The intriguing point is that a Hubble size region that couldn't climb the step becomes a black hole. As illustrated in the left panel of Fig.~\ref{fig:deadend}, if the region A is left behind due to a vacuum fluctuation, it will be surrounded by the region B with the potential higher than A right after the inflaton has climbed the step.
Then the region A behaves like a true vacuum bubble in the sea of false vacuum, and the radius keeps expanding until the potential energy of the region B becomes equal to that of the region A. After that epoch, the situation reverses. The region A begins to behave like a false vacuum bubble. The wall is pushed inward and eventually turns into a black hole.
This is similar to the PBH formation due to quantum tunneling during inflation~\cite{Deng:2017uwc}. Thus in this model, PBHs may be produced both by a conventional mechanism and from left-over regions during inflation.

\section{Implications and Conclusions}

One of the most important implications of the PBH formation due to large positive curvature perturbations is that it will be associated with the production of gravitational waves (GWs) at second order, the so-called induced GWs.
As we discussed before, the PBH formation occurs only in regions with rare, large curvature perturbations. Thus most of the space will not collapse into a black hole. Instead, as the scale enters the Hubble horizon, they begin to behave like sound waves, and induce oscillations in the energy momentum tensor at second order in perturbation. 
This leads to the production of GWs. Since the comoving scales of these two events are the same, one concludes that the PBH function is tightly related to the spectrum of the induced GWs, hence can be used to test PBH scenarios~\cite{Saito_2009}.

Since the PBH formation is sensitive to the tail behavior of the PDF, the existence of the induced GWs may not necessarily imply that of PBHs. However, conversely, if we obtain an upper bound on the amount of the induced GWs by future experiments/observations, it will either tightly constrain or totally exclude the corresponding PBH scenario, as one would need a power spectral amplitude of $O(10^{-2})$ in order to produce any appreciable amount of PBHs.

This idea was first put forward by Saito and Yokoyama~\cite{Saito_2009}, 
in which they pointed out that the scenario of PBHs of mass $\sim10^{20}{\rm g}$ as dark matter of the universe can be strongly constrained by future space GW detectors such as LISA, where the corresponding induced GW frequency is $\sim10^{-3}{\rm Hz}$, while the scenario of 
binary black holes found by ground-based detectors such as LIGO being PBHs of $10$-$100\,M_\odot$ is strongly constrained by pulsar timing array experiments. 

Another interesting implication is that small PBHs which have completely evaporated due to Hawking radiation in the early universe may give rise to a large amount of induced GWs that may be detected in the near future, provided that the universe was once dominated by those PBHs~\cite{Inomata:2019ivs, Papanikolaou:2020qtd,Domenech:2020ssp}.
This is because of spatial inhomogeneities in the distribution of PBHs which behave as a primordial isocurvature perturbation turn into an adiabatic perturbation when the universe became PBH dominated. 

In this respect, it is worth mentioning that the role of primordial isocurvature perturbations in PBH formation scenarios, as well as in GW cosmology,
has been little explored. 

In conclusion, thanks to the discovery of GWs from binary black holes by LIGO, PBHs have become one of the major topics in cosmology, particularly in GW cosmology. 
In this sense, we may say that we have entered an era of GW cosmology where PBHs play a central role.
\\

This work is supported in part by JSPS KAKENHI Nos. 20H05853 and 24K00624.

\bibliographystyle{utphys}
\bibliography{PBH}


%

\title{Tests of General Relativity with Future Detectors}

\author{Emanuele Berti}

\institute{\textit{William H. Miller III Department of Physics and Astronomy, Johns Hopkins University, Baltimore, Maryland 21218, USA}}

\maketitle

\begin{abstract}
This ``vision document'' is about what the future has in store for
tests of general relativity with gravitational wave detectors. I will
make an honest attempt to answer this question, but I may well prove
to be wrong. To quote Yogi Berra: ``It's hard to make predictions,
especially about the future.''
\end{abstract}

\noindent
{\bf Why should we test general relativity?}
First of all, it is important to ask why we are doing this. Is there
value in testing general relativity with higher and higher precision?
I obviously believe that the answer is
``yes''~\cite{Will:2014kxa,Berti:2015itd}, but your mileage may
vary. While Clifford Will was a postdoc in Chicago, Subrahmanyan
Chandrasekhar once asked him: ``Why do you bother testing general
relativity? We know that the theory is right.''  I recently heard over
dinner with Bob Wald and Sean Carroll that Chandra must have made that
remark more than once. When someone pointed out that Einstein was
elated when the 1919 eclipse expedition confirmed the validity of his
theory (in fact, he famously said that he had ``palpitations'' when he
first computed the perihelion advance of Mercury), Chandra quipped:
``That’s because Einstein did not understand his own theory very
well.''

There is, of course, great value in testing the limits of validity of
any of our physical theories. In his 1894 dedication of Ryerson
Physical Laboratory in Chicago (quoted also in Weinberg's ``Dreams of
a final theory''), Albert Michelson stated: ``While it is never safe
to affirm that the future of physical science has no marvels in store
even more astonishing than those of the past, it seems probable that
most of the grand underlying principles have been firmly established
and that further advances are to be sought chiefly in the rigorous
application of these principles to all the phenomena which come under
our notice. It is here that the science of measurement shows its
importance — where quantitative work is more to be desired than
qualitative work. An eminent physicist [probably Lord Kelvin?]
remarked that {\em the future truths of physical science are to be looked
for in the sixth place of decimals}.'' By building better detectors, we
will be better equipped to look for those ``future truths.'' Such is
the nature of any experimental science. 

\noindent
{\bf What future detectors?}
Ongoing efforts to build more sensitive gravitational-wave detectors
on Earth include the Einstein Telescope~\cite{Punturo:2010zz}, Cosmic
Explorer~\cite{Evans:2023euw,Evans:2021gyd} and
NEMO~\cite{Ackley:2020atn}, among others.  The reach of these
detectors can be extended to lower frequencies through detectors on
the Moon~\cite{Cozzumbo:2023gzs} and in space: besides
LISA~\cite{Colpi:2024xhw}, TianQin~\cite{TianQin:2015yph} and
Taiji~\cite{Hu:2017mde}, there are proposals for follow-up space
missions on a longer
timescale~\cite{Sedda:2019uro,Baibhav:2019rsa,Sesana:2019vho}. There
are also major experimental efforts to develop detectors based on atom
interferometry~\cite{Badurina:2019hst,MAGIS-100:2021etm,AEDGE:2019nxb,Canuel:2017rrp},
and ideas to extend the range of observable gravitational waves
to high frequencies~\cite{Aggarwal:2020olq}. Each of these future
detectors will observe different astrophysical sources and probe
different physics.

What are we going to learn about theories beyond general relativity,
or about general relativity itself, from these observations?
Here I will focus on the science achievable with the Einstein
Telescope, Cosmic Explorer and LISA -- detectors for which the main
astrophysical sources (inspiraling compact binaries) are relatively
well understood. The science case for these detectors is clearly much
broader and stronger than just tests of general relativity (see
e.g.~\cite{Sathyaprakash:2012jk,Evans:2021gyd,Gupta:2023lga,Branchesi:2023mws,Colpi:2024xhw}).
For single events, the increased sensitivity of ground-based detectors
implies higher probability of detecting ``exotic'' binary merger
events with unexpected parameters; better constraints on the nuclear
equation of state; more accurate studies of binary dynamics, including
spin precession and eccentricity; and a broad range of applications to
(and of) multimessenger astronomy~\cite{Evans:2023euw}. Detecting
hundreds or thousands of events per day, these observatories will
drastically improve our understanding of compact binary populations
and formation channels. In particular, they will probe sources in the
high-redshift universe, such as Population III stars, primordial black
holes, and intermediate-mass black holes. One of the Holy Grails of
future interferometers is the detection of stochastic backgrounds of
astrophysical and cosmological origin. When viewed in this broader
context, testing general relativity is only a cherry on top of the
cake -- but of course, the observation of a real violation of general
relativity (as opposed to one of the many possible false violations
due to noise systematics, waveform systematics or
astrophysics~\cite{Gupta:2024gun}) could lead to a paradigm shift in
physics, and for this reason alone it is well worth pursuing.


\noindent
{\bf Pushing the boundaries: inspiral radiation.}
A flexible, theory-agnostic framework to explore the theoretical
physics implications of binary merger observations, particularly in
the inspiral phase, is the so-called parametrized post-Einsteinian
formalism~\cite{Arun:2006yw,Yunes:2016jcc,Berti:2018cxi,Yunes:2024lzm}. The
idea is that any specific theory will induce modifications in the
waveform phasing starting at some post-Newtonian order. If those
modifications are consistent with zero, they can be turned into bounds
on specific classes of theories. Perkins et al.~\cite{Perkins:2020tra}
used astrophysical models to estimate how these bounds will improve in
the future. The conclusions are affected by astrophysical
uncertainties on compact binary population, and they also depend on
the sensitivity upgrades of the instruments. For theory-agnostic
bounds, ground-based observations of stellar-mass black holes and LISA
observations of massive black holes can each lead to improvements
ranging between 2 and 4 orders of magnitude with respect to present
constraints. Multiband observations of sources inspiraling in the LISA
band and merging in the band of ground-based interferometers can yield
improvements between 1 and 6 orders of magnitude.

These predictions must be taken with a grain of salt for two main
reasons.  The first, and most important, is that they ignore
systematics due to detector noise, waveform modeling and
astrophysics~\cite{Gupta:2024gun}. Our understanding of waveforms
within general relativity must improve by orders of magnitude if we
want to perform tests of gravity at high signal-to-noise ratios. In
fact, some events display apparent violations of general relativity
due either to waveform systematics or to data-quality issues even at
current sensitivities~\cite{Maggio:2022hre}.
The second reason is that any predictions based on the population
observed so far are conservative, in the sense that a single
``special'' event can be sufficient to kill many theories in one
stroke. We already have two excellent examples of this: (i) the
simultaneous detection of gravitational and electromagnetic waves in
the binary neutron star event GW170817 ruled out broad classes of
theories that predict modifications in the propagation properties of
gravitational
waves~\cite{Creminelli:2017sry,Ezquiaga:2017ekz,Sakstein:2017xjx};
(ii) the recent observation of the merger of a neutron star with a
compact object in the lower mass-gap (GW200115\_042309) placed
stringent bounds on high-order curvature theories with a
mass-dependent coupling~\cite{Gao:2024rel,Sanger:2024axs}.



\noindent {\bf Pushing the boundaries: ringdown radiation.} The
LIGO-Virgo-KAGRA collaboration routinely performs ``black hole
spectroscopy'' tests of the Kerr nature of a merger
remnant~\cite{Detweiler:1980gk,Dreyer:2003bv,Berti:2005ys,Berti:2009kk,Berti:2018vdi,Cardoso:2019rvt}
on all detected
events~\cite{LIGOScientific:2016lio,LIGOScientific:2021sio}. There are
claims that an overtone of the dominant ($\ell=m=2$) component of the
radiation may be present in GW150914~\cite{Isi:2019aib} and that
higher multipole modes can be seen in GW190521~\cite{Capano:2021etf},
but these conclusions depend on subtle assumptions about waveform
modeling and data analysis. The evidence for a second mode in the data
is affected by uncertainties in the ringdown starting time, choices of
sampling rate, noise modeling, and data analysis methods (e.g., whether
the analysis is performed in the time or frequency domain, or whether
one considers only the ringdown portion of the waveform or models the
whole inspiral-merger-ringdown
signal)~\cite{Brito:2018rfr,Ghosh:2021mrv,Finch:2022ynt,Cotesta:2022pci,Crisostomi:2023tle,Gennari:2023gmx,Wang:2023xsy}.

From a theoretical point of view, it is important to perform an
agnostic analysis of numerical relativity waveforms that takes into
account all of the relevant physics (including a proper choice of the
BMS frame, the early transient, late-time tails, spherical-spheroidal
mode mixing, counterrotating and nonlinear modes) and not just a
simple superposition of linear modes~\cite{Baibhav:2023clw}. High
overtones do not contribute much to the remnant's mass and spin
estimates, and (unlike the fundamental mode) they can be spectrally
unstable~\cite{Nollert:1996rf,Jaramillo:2020tuu,Cheung:2021bol,Berti:2022xfj}.
A quasinormal mode analysis within linear perturbation theory can only be
trusted $\sim 10M$ after the waveform peak, where the ringdown
signal-to-noise ratio in current detectors is low.
Many of these issues will be resolved in the future, when ``golden''
events will have ringdown signal-to-noise ratios in the hundreds for
next-generation ground-based detectors, and in the thousands for
LISA~\cite{Berti:2016lat,Baibhav:2018rfk,Baibhav:2020tma,Ota:2021ypb,Bhagwat:2021kwv}.
However, even at such high signal-to-noise ratios there is a
compromise between the complexity of the signal model and the
information that can be extracted because of Occam
penalties~\cite{CalderonBustillo:2020rmh}.


\noindent {\bf Can ringdown radiation be used to look for smoking guns
  of modified gravity?} Finding an answer to this question is
technically challenging.  It is well known that the separability of
the Teukolsky equation relies on the Petrov Type D of the Kerr metric
and on the existence of ``hidden symmetries'' (i.e., of a Killing
tensor)~\cite{Teukolsky:2014vca}.  These special properties are lost,
in general, in modified theories of gravity: it is difficult (if not
impossible) to find closed-form analytical expressions for the
background metric, the perturbation equations are usually not
separable, and the equations of motion are often of higher
order. Despite these difficulties, there has been remarkable progress
in quasinormal mode calculations in modified theories of gravity using
three main approaches:

\noindent {\em (1) Metric perturbations} (similar to the classic
Regge-Wheeler/Zerilli approach) have been used to compute quasinormal
modes for slowly rotating black holes in specific theories (such as
Einstein-scalar-Gauss-Bonnet and dynamical Chern Simons gravity). This
approach is limited to low spins, but it can deal with large
couplings~\cite{Molina:2010fb,Blazquez-Salcedo:2016enn,Pierini:2021jxd,Pierini:2022eim}.

\noindent
{\em (2) Generalized Teukolsky equations}
can be derived under the assumption of small coupling. This technique
can be used to compute shifts in the quasinormal mode frequencies that
are linear in the coupling using either variants of Leaver's
method~\cite{Leaver:1985ax} or eigenvalue perturbation techniques.
While the technique is based on a small-rotation expansion, it has
been pushed (in some cases) up to order $a^{18}$ in the small rotation
parameter
$a$~\cite{Cano:2021myl,Li:2022pcy,Hussain:2022ins,Li:2023ulk,Wagle:2023fwl,Cano:2023tmv,Cano:2023jbk,Cano:2024ezp}.

\noindent
{\em (3) Spectral methods}
can be used to compute quasinormal mode frequencies of black holes in
modified gravity for arbitrary coupling and (in principle, but not in
practice) for arbitrarily large rotation
parameters~\cite{Chung:2024ira,Chung:2024vaf,Blazquez-Salcedo:2024oek}.

If modified gravity effects can be treated perturbatively, one can
incorporate theory-specific predictions within a ``parametrized
ringdown'' formalism~\cite{Cardoso:2019mqo,McManus:2019ulj} that has
recently been extended to the rotating
case~\cite{Cano:2024jkd}. Parametrized ringdown templates (similar to
the ``parametrized post-Einsteinian'' formalism) can also be
implemented in data analysis~\cite{Maselli:2019mjd,Carullo:2021dui}.
This idea can be used to estimate how well future ringdown detections
can constrain modified gravity theories. A recent study considers
agnostic (null) tests, as well as theory-specific tests for two
quadratic gravity theories (Einstein-scalar-Gauss-Bonnet and dynamical
Chern-Simons gravity) and various classes of effective field
theories~\cite{Maselli:2023khq}. It finds that high-order terms in the
slow-rotation expansion are necessary for robust inference of
hypothetical corrections to general relativity. However, even when
high-order expansions are available, ringdown observations alone may
not be sufficient to measure deviations from the Kerr spectrum for
theories with dimensionful coupling constants. This is because the
constraints are dominated by ``light'' black hole merger remnants
(where curvature-dependent corrections are largest), but even for
next-generation ground-based detectors, only a few of these events
have sufficiently high signal-to-noise ratio in the ringdown.


\noindent {\bf Light ring tests are complementary to ringdown tests.}
The well-known connection between light ring geodesics and ringdown
physics~\cite{Cardoso:2008bp} means that experiments probing light
rings, such as the Event Horizon
Telescope~\cite{EventHorizonTelescope:2019ggy,EventHorizonTelescope:2022xqj}
and its next-generation follow-ups (ngEHT~\cite{Tiede:2022grp} and the
Black Hole Explorer~\cite{Lupsasca:2024xhq}) probe similar physics
and yield complementary tests~\cite{Volkel:2022khh}. Much work is
needed to explore the complementarity of these experiments and
gravitational wave detectors in testing general relativity and black
hole dynamics.

\noindent {\bf ``Null tests'' of general relativity in the strong
  gravity regime: nonlinear quasinormal modes and gravitational
  memory.}
The high signal-to-noise ratio of future detectors holds promise to
better test the nonlinear nature of gravity in at least two ways:

\noindent {\em Nonlinear quasinormal modes} are present in binary
black hole
waveforms~\cite{London:2014cma,Cheung:2022rbm,Mitman:2022qdl} and
potentially detectable: optimistic astrophysical scenarios predict
thousands of LISA binary black hole merger events with a detectable
quadratic mode~\cite{Yi:2024elj}. The ratio between the nonlinear mode
amplitudes and the linear amplitudes driving them can be computed
within Einstein's theory (see e.g.~\cite{Gleiser:1998rw,Campanelli:1998jv,Loutrel:2020wbw,Ripley:2020xby,Redondo-Yuste:2023seq,Perrone:2023jzq,Bucciotti:2023ets,Cheung:2023vki,Redondo-Yuste:2023ipg,Zhu:2024dyl,Zhu:2024rej,Ma:2024qcv,Bucciotti:2024zyp,Bourg:2024jme,Bucciotti:2024jrv,May:2024rrg}), and used as an additional ``null'' test of
the nonlinear nature of general relativity.

\noindent {\em Gravitational wave memory effects} arise from
non-oscillatory components of gravitational wave signals. They are
predictions of general relativity in the nonlinear regime that have
close connections to the asymptotic properties of isolated gravitating
systems, and can shed light on the infrared structure of
gravity~\cite{Christodoulou:1991cr,Strominger:2017zoo}.  The largest
of these effects from binary black hole mergers are the “displacement”
and “spin” memories -- a change in the relative separation of two
initially comoving observers due to a burst of gravitational waves,
and a portion of the change in relative separation of observers with
initial relative velocity, respectively. Both of these effects are
small, but by combining data from multiple events, the displacement
memory could be detected by a LIGO-Virgo-KAGRA network operating at
the (current) O4 sensitivity for 1.5 years, and then operating at the
O5 sensitivity for an additional year. By contrast, a next-generation
interferometer such as Cosmic Explorer could detect the displacement
memory for {\em individual} loud events, and detect the spin memory in a
population within 2 years of observation time~\cite{Grant:2022bla}.

\noindent {\bf Should we care about testing general relativity?} The
answer is obviously ``yes'' if it turns out that general relativity is
indeed modified {\em ``in the sixth place of decimals,''} but I would
argue that we should care anyway. Some of the biggest mysteries in
modern physics (dark matter, dark energy, the information paradox,
strong cosmic censorship) are related to the behavior of gravity at
large scales and near black holes. Any progress in understanding the
gravitational interaction can only lead us closer to unveiling those
mysteries.

%

%
%
%
%
%

\bibliographystyle{utphys}
\bibliography{references}

\title{Black Holes as Laboratories: Tests of General Relativity}

\author{Ruth Gregory\inst{1,2} \and Samaya Nissanke\inst{3,4}}

\institute{\textit{Department of Physics, King’s College London, University of London, Strand, London, WC2R 2LS, UK} \and
\textit{Perimeter Institute, 31 Caroline Street North, Waterloo, ON, N2L 2Y5, Canada} \and \textit{GRAPPA, Anton Pannekoek Institute for Astronomy and Institute of High-Energy Physics, University of Amsterdam, Science Park 904, 1098 XH Amsterdam, The Netherlands} \and
\textit{Nikhef, Science Park 105, 1098 XG Amsterdam, The Netherlands}}

\maketitle

\begin{abstract}
We briefly overview the case for using black holes as a discriminator for theories of gravity.
The opportunities and challenges for the various observational experiments are outlined, and
key questions for the community identified. This note summarises the discussion from the roundtable on the third day 
of \emph{Black Holes Inside and Out 2024}.
\end{abstract}

The past decade has seen a complete transformation in the way black holes are regarded.
A decade ago, we were pretty sure that black holes existed, but it was via deductive
reasoning: It is small, heavy, and non-luminous, so what else can it be? Direct
evidence was missing. However, with first the gravitational wave signal from merging
black holes \cite{LIGOScientific:2016aoc}, then the imaging of the black hole shadow \cite{EventHorizonTelescope:2019dse}, we are now in the position
of having direct evidence of strongly curved near, or at, horizon geometry.

Having this data has opened the exciting possibility that we can now test General Relativity (GR),
not as a correction to Newtonian Gravity, or in the very special case of the high degree of
symmetry of a cosmological background, but in the highly nonlinear regime
of black hole horizon. 
While many of us often assume that GR is right, and certainly weak field
GR has been well verified, the conflict between GR and Quantum Theory is
so profound that it is essential we step back and ask: is GR really the right
theory of gravity? Much of the theoretical conflict between gravity and quantum theory arises
when we are at extreme scales, however, even the relatively mild environment
of an event horizon leads to quantum puzzles: do we have a horizon, a firewall,
a fuzzball? It is therefore essential to test GR in these strong gravity, but
weakly quantum, regimes.

We are currently experiencing a golden age in strong-field gravity, marked by groundbreaking observations and measurements from a range of experiments, including the LIGO \cite{Aasi2015det}, Virgo \cite{Virgo}, and KAGRA \cite{KAGRA} detectors, the GRAVITY instrument \cite{GRAVITY}, and ongoing Event Horizon Telescope (EHT; e.g., \cite{EHT}) campaigns. This boom period is set to expand in the 2030s and 2040s with the advent of next-generation observatories, such as funded telescopes and instruments such as the Extremely Large Telescope (ELT, e.g., GRAVITY+, \cite{GRAVITY+}), the Square Kilometer Array, and the Laser Interferometer Space Antenna (LISA; \cite{Colpi:2024xhw}). Additionally, proposed projects like the next-generation Event Horizon Telescope (ngEHT; e.g., \cite{ayzenberg2023fundamentalphysicsopportunitiesnextgeneration}), Cosmic Explorer (CE; \cite{evans2021horizonstudycosmicexplorer,CE}), and the Einstein Telescope (ET; \cite{Maggiore_2020,Branchesi_2023}) promise further advancements. These new observatories represent major investments and a multitude of ambitious experimental advances, encompassing decades of engineering, instrumentation, and cutting-edge technological breakthroughs.

Given these developments, it is crucial that the scientific community prepares to maximize the opportunities they present for testing general relativity, particularly through the analysis of black hole observations across diverse data sets. For example, as discussed below, this includes exploring theoretical, computational, and analytical approaches to gravitational wave data, along with future GRAVITY observations.

Firstly, for gravitational waves (GWs), compared to the few hundred stellar-mass binary black hole and neutron star mergers observed since 2015, we can expect to observe several hundred to thousands of binary black hole mergers per year at redshifts up to 1, starting in the early 2030s with the proposed LIGO A$\#$ and Virgo\_nEXT upgrades (see e.g., \cite{ligoasharp}). With entirely new GW facilities like ET and CE, we could observe several hundred thousand binary black hole mergers per year, reaching redshifts of up to 30 and beyond \cite{evans2021horizonstudycosmicexplorer,CE,Maggiore_2020,Branchesi_2023}. In the case of LISA, we anticipate detecting numerous supermassive black hole mergers per year, along with extreme mass-ratio inspirals \cite{Colpi:2024xhw}.

These upgrades and next-generation GW observatories represent a paradigm shift in our ability to use black hole observations to test General Relativity. Notably, they will enable: (1) exquisite precision in high signal-to-noise ratio (SNR) events, with precision physics becoming possible for the tens of binary black hole mergers per year that will have SNRs greater than 1000. (2) greater sensitivity to more parts of the gravitational waveform, whether it be the inspiral phase of heavier binary black hole mergers or the post-merger remnant for low-mass neutron star binaries. For instance, we may observe thousands of post-inspiral binary black hole mergers with SNRs exceeding 100. (3) the discovery of a population of intermediate-mass black holes. (4) a dramatic increase in the number of binary black hole mergers due to the larger volume of space being probed. (5). The potential detection of primordial black holes, particularly from binary black hole mergers observed at redshifts greater than 20.

All of these advances will enable a plethora of tests of GR. These include improving graviton mass constraints by three orders of magnitude, conducting ringdown tests of the no-hair theorem with SNRs potentially reaching 11,000 for the root-sum-square post-inspiral of binary black holes, detecting gravitational wave memory, and testing for beyond-GR polarizations during the earlier inspiral phase, e.g., \cite{CE}. We will also be able to perform parameterized consistency tests of GR, such as those using the post-Einsteinian (PPE) formalism, e.g., \cite{Yunes:2009ke}.

To meet these challenges, significant and coordinated investment is required in developing complete and accurate approximants for binary black hole (BBH) mergers, as well as other binary sources. This is urgently needed as we approach the era of next-generation GW detectors, where SNRs could reach as high as thousands. A combination of analytical relativity, numerical relativity, effective field theory and machine learning is expected to be employed for source modeling and the production of gravitational waveforms. Additionally, advances in amplitude scattering methods are likely to play an increasingly important role in theoretical modeling, e.g., \cite{2021PhRvL.126q1601B}.

Over the past several years, it has also become clear that environmental effects\linebreak—whether from beyond-standard-model physics, ultra-light boson clouds, or astrophysical environments like accretion disks—can be degenerate with one another and may mimic the effects of non-GR compact object mergers. During our roundtable discussions, it was emphasised that such degeneracies could pose a significant challenge in identifying non-standard sources or environments (e.g., \cite{evstafyeva2024gravitationalwavedataanalysishighprecision,2023NatAs...7..943C}). Therefore, it is crucial to develop a systematic and comprehensive framework for producing and analysing gravitational waveform models within the context of GR, beyond-GR theories, and beyond standard model physics, while incorporating physical effects from astrophysics, such as accretion disks and finite-size effects such as the information encoded about the ultra-cold, dense matter in the case of neutron stars. To achieve this goal, we stress the need for active collaboration across multiple disciplines including theoretical and observational astrophysics, general relativity, high energy theory and phenomenology, and nuclear physics.

Moreover, this multi-disciplinary community should also focus on the identification of clear observational signatures, or ``smoking guns," for new types of GW sources. These include primordial black holes, specific resonances from binary black holes in unique environments, and the tidal deformabilities of exotic compact objects.

In this context, we stress the necessity of advancing our analysis methods to handle the multitude of GW signals from both standard and exotic sources across various beyond-GR theories. For transient events like binary black hole and neutron star mergers observed with future detectors like the ET and CE, many signals will overlap, each signal with durations potentially lasting several hours — much longer than the fraction of a second to minutes that we currently observe. Moreover, new algorithms will be essential to address the significant computational challenges involved in analyzing continuous sources, such as axisymmetric rotating neutron stars and stochastic astrophysical or cosmological backgrounds. These methods must also ensure the detection of anomalous signals, which could originate from either new sources or new physics. This will be critical in the era of next-generation GW observatories, where strong synergies will exist between space-based detectors like LISA and ground-based detectors like ET and CE. For example, in the case of LISA, preparations are already underway to implement a ``global fit" for analyzing synthetic datasets spanning years (e.g., \cite{Littenberg_2023}), which will include signals from several supermassive black hole mergers, millions of Galactic white dwarf binary inspirals, stellar-mass black hole inspirals, and extreme mass-ratio inspirals. We are already witnessing the deployment of machine learning and artificial intelligence methods in both GW detection and parameter inference—such as simulation-based inference using convolutional neural networks (e.g., \cite{2021PhRvL.127x1103D,2023PhRvD.108d2004B})—as well as in the forward modeling of GW emission from various sources. These techniques are becoming increasingly critical as we prepare for the data deluge expected from future detectors, and also for successful follow-up by other electromagnetic and astroparticle telescopes for nearby golden events.

Secondly, in the case of the GRAVITY+ experiment, before addressing beyond-GR effects, there is an urgent need to account for the stronger foreground of Newtonian perturbations from other classical objects, such as stellar-mass black holes, neutron stars, and white dwarfs, near Sagittarius A*. How many of these classical objects is tied to the star formation history of the universe, as well as to how Sagittarius A* acquired its own mass. Fully modeling the astrophysical environment, while also incorporating beyond-GR effects, remains a significant challenge. Disentangling these effects will require considerable effort in both modeling and simulations, collaborations with observers and instrumentalists, especially as the precision of our data increases dramatically with the arrival of the ELT. 

Finally, we must be prepared for the possibility that our current and next-generation detectors may not detect or reveal any hints of ``new" physics. In that case, we must ask ourselves: what are the implications of black hole measurements for testing GR and beyond-standard-model physics? How do we interpret the absence of new physics in these observations, and what would this mean for future facilities and for our understanding of fundamental physics?

\bibliographystyle{utphys}
\bibliography{references}

%
\part{Testing the black hole paradigm}

%





\title{Simulating Black Hole Imposters}

\author{Frans Pretorius}
\institute{\textit{Department of Physics and Princeton Gravity Initiative, Princeton University, Princeton, New Jersey 08544, USA}}

\maketitle

\begin{abstract}
I briefly describe motivation for, and the current state of research into 
understanding the structure and dynamics of black hole ``imposters'' : 
objects that could be misidentified as Kerr black holes given the current precision 
of LIGO/Virgo gravitational wave observations, or EHT accretion disk measurements.
I use the term ``weak imposter'' to describe an object which is a black hole,
i.e. it has an event horizon, but whose structure and dynamics is governed
by a modified gravity theory. At the other end of the spectrum are ``strong imposters'' :
hypothetical horizonless, ultra-compact objects conjectured to form instead of black holes during gravitational collapse. To discover or rule-out imposters will require a quantitative understanding of their merger dynamics. This is hampered
at present by a dearth of well-posed theoretical frameworks to describe imposters beyond perturbations of Kerr black holes and their general relativistic binary dynamics. That so little is known about non-perturbative modifications to dynamical, strongfield gravity is, I argue, due to a lamppost effect.

\end{abstract}

\section{Observing the Dynamical, Strongfield Regime of Gravity}
Einstein's theory of general relativity makes several remarkable
predictions about the nature of spacetime in the strongfield regime.
First is the fact that there {\em is} a strongfield regime
in a theory without any (geometric) dimensionful scales or parameters.
This is unlike any of the other fundamental forces we understand
today, where new physics arises at distinct energy scales
associated with fundamental constants in the corresponding theories. 
For general relativity, this begs the question of what property
of the theory governs the onset of the qualitatively distinctive
strong versus weakfield dynamics of spacetime. An answer,
which is difficult to make precise due to the coordinate invariance
of the theory, is strongfield behavior is manifest when
the nonlinear terms in the field equations become important or dominant
in determining its solution. That there is no fundamental scale in the
theory means that the nonlinear regime can, in principle,
be reached at {\em any} length or energy scale. 

A second remarkable fact about general relativity is how
strikingly different its predictions of weak versus
strongfield spacetime structure is. The former
is given by Minkowski spacetime and its perturbations,
well described on Earth and in the solar system by Newtonian gravity with
small relativistic corrections. The strongfield most notably includes black hole
solutions, where within their corresponding Schwarzschild radii
spacetime undergoes {\em gravitational collapse}, becoming intrinsically 
dynamical, flowing to some kind of singularity or Cauchy horizon
where general relativity breaks down as a sensible theory. Presumably
some new theory, call it {\em quantum gravity}, will supercede
Einstein's theory in providing a better description of nature then.
General relativity putatively says this kind of breakdown occurs
hidden behind an event horizon, and thus the best we can
do to test this astonishing prediction of the theory
is to observe the exterior structure and dynamics of black holes.
Doing so has become possible in the last decade with LIGO/Virgo
observations of binary black hole mergers~\cite{LIGOScientific:2016aoc}, and the EHT
observation of accretion disks about the supermassive 
black holes in M87~\cite{EventHorizonTelescope:2019dse} and the Milky Way~\cite{EventHorizonTelescope:2022wkp}. 

Given the uncertainties in modeling accretion disk astrophysics,
there has been much debate in the literature about
whether present day EHT observations can place quantitative
bounds on how well the Kerr metric describes the corresponding
spacetimes (see e.g.~\cite{Lupsasca:2024wkp} and the citations therein). Moreover, given how massive M87* and SgrA* are 
relative to their accretion disks, outside the horizons these spacetimes are essentially
stationary, and cannot be used to test the radiative
degrees of freedom of strongfield gravity. 
Black hole mergers observed by LIGO/Virgo are not (as far as we
know) meaningfully influenced by a circumbinary environment,
and certainly probe dynamical gravity, giving us at present
the only direct tests of this regime of general relativity.
Current observations of black hole mergers are consistent
with the predictions of general relativity (GR), though
are not yet at the level of ``precision'' tests : lack of any
detectable residual following subtraction of ``best-fit'' 
GR waveforms give consistency at the O($10\%$) level~\cite{LIGOScientific:2021sio};
of this order or slightly better constraints can be placed for parameters
governing various specific modified gravity theories (see e.g.~\cite{Yunes:2016jcc}).

This then still leaves some room for the candidate
black holes we have observed to be ``imposters'', where their structure
and dynamics is governed by some modified or beyond-GR theory.
Most proposed examples of imposters can be called ``weak'' imposters,
for example as coming from Einstein-dilaton-Gauss-Bonnet (EdGB) or
Chern-Simons (CS) modified gravity, where the solutions are still
black holes but with only perturbative corrections compared
to the structure and dynamics predicted by GR. Proposed ``strong''
imposters, such as gravastars, fuzzballs, or AdS black shells,
do not have horizons. The purpose of this note is not 
to describe any of these imposters -- see e.g. ~\cite{Will:2014kxa,Cardoso:2019rvt} for more comprehensive
reviews --- rather, it is to briefly discuss some motivation
for studying them, and what the present difficulties are in
simulating them with the intent of making quantitative
merger waveform predictions. 

One motivation is as already discussed : testing general relativity
in the dynamical strongfield regime. Though the residual test
is theory agnostic, somewhat better constraints for a specific
imposter can be made if we have waveform models for it. Moreover,
if an anomaly is ever detected, the residual by itself will give
little information on the physics behind it; a better understanding
of the gamut of theoretically viable beyond-GR phenomenology
will be crucial to decipher the underlying cause.  

A second motivation, discussed more after describing
the difficulties in simulating imposters, is what can be
called the ``lamppost'' problem in beyond-GR theories. Essentially
it is that, at present, we are limited to studying theories that are perturbatively
close to GR, not because observational data is telling us this is adequate,
but rather because we do not know how to go beyond the perturbative
description. Of course, since there is no detected waveform
anomaly yet, the data {\em is}
telling us we do not need anything but GR at present, let alone
perturbative corrections, and solving the lamppost problem might
only ever be of pure theoretical interest.

\section{Difficulties simulating black hole imposters}

The main difficulty at present in simulating many black hole
imposters is coming up with a well-posed set of partial
differential equations (PDEs) governing their structure and
dynamics. For weak imposters, the issues stem from the fact that most beyond-GR theories 
introduce their modifications by adding higher order
curvature scalars to the Einstein-Hilbert action, which typically
leads to the principle parts of the PDEs having higher than
second derivatives, and also complicates the nonlinear structure of
the PDEs. One approach people have taken to deal with the 
problems this causes is to only treat these beyond-GR terms
perturbatively, e.g.~\cite{Witek:2018dmd,Okounkova:2019dfo,Okounkova:2020rqw}. The equations then
retain the mathematical structure of GR, and all the modern
methods for numerically solving the Einstein equations
for black hole mergers can be applied essentially verbatim.
One problem with this approach is it is susceptible to spurious
secularly growing solutions that require care to mitigate~\cite{Okounkova:2019zjf}.

For the Horndeski class of theories~\cite{Horndeski:1974wa}, including EdGB gravity,
the equations of motion are still second order, though have more complicated
non-linear structure than GR, and one needs to introduce an additional
scalar degree of freedom to give non-trivial modifications to GR
in four spacetime dimensions. Using standard generalized harmonic coordinates, EdGB gravity
is not well posed in generic settings~\cite{Papallo:2017qvl,Papallo:2017ddx}, 
however recently Kovacs and Reall
introduced a ``modified harmonic'' scheme that is well posed~\cite{Kovacs:2020ywu}, and was applied
to binary merger simulations in~\cite{East:2020hgw,Corman:2022xqg,AresteSalo:2023mmd}. 
A problem here though arises
when the non-linear corrections to GR become too large (which for
binary mergers is achieved by making the black holes smaller than
the length scale set by the coupling constant of the EdGB correction). Then
the PDEs cease being hyperbolic, and the predictability of the theory
fails form the perspective of the initial value problem~\cite{Ripley:2019hxt}.

A novel approach to deal with ill-posedness introduced by the addition of derivatives
beyond second order, yet still be able to study non-perturbative
corrections, is to elevate the beyond-GR terms to new fields, and introduce
new equations of motion that ``relax'' them to their desired
values on some chosen timescale~\cite{Cayuso:2017iqc,Cayuso:2020lca}.\footnote{This method was inspired
by the Muller-Israel-Steward approach of fixing similar ill-posedness
problems in the relativistic Navier-Stokes equations.} This approach
is not coordinate invariant, and somewhat ad-hoc, though early
studies show it does work, both in the sense of mitigating the ill-posedness problem,
and that when in the perturbative regime the ad-hoc relaxation
parameters do not significantly affect the solution~\cite{Corman:2024cdr}. 
How far solutions
can be pushed beyond the perturbative regime remains to be seen, questions
aside as to how such solutions should be interpreted, or what
putative well-posed theory they may correspond to.

A very recent work by Figueras et al.~\cite{Figueras:2024bba} seems to show that, 
remarkably, a broad class of higher-curvature metric Lagrangians
can, through the introduction of regularizing terms and field redefinitions, have their equations of motion
cast into a well-posed hierarchy of coupled, nonlinear wave equations.\footnote{This is a generalization of the previously known case of
quadratic gravity~\cite{Noakes:1983xd,Stelle:1977ry}.} This introduces a corresponding
set of new massive degrees of freedom, but they can be chosen
to not affect the dynamics below some desired cutoff scale.

In all then, quite significant progress has been made toward
taming the structure of beyond-GR theories, making it feasible
to solve them numerically for binary merger processes. This should
allow for a more thorough understanding of weak black hole imposters,
and set more stringent constraints on specific theories with current
GW data than agnostic tests can.

The theoretical health of strong imposters, i.e. ultra compact,
horizonless alternatives to black holes, is in a more precarious
state. Motivations for proposing these objects are largely theoretical,
including to avoid the information loss problem in semi-classical GR,
and/or singularity formation in classical GR. 
However, arguably none of the proposed black hole alternatives have a complete
theory behind them, in particular from which solvable equations
of motion predicting their dynamics can be derived. One possible exception
are AdS black shells (or black bubbles)~\cite{Danielsson:2017riq}. They are string theory inspired objects, argued to form from a tunneling process prior to formation of a classical
horizon. Once formed, for macroscopic shells, the effective four dimensional system
is argued to be well described by classical physics : a thin shell of matter
separating an interior AdS spacetime from an exterior asymptotically flat spacetime, 
where the spacetime dynamics is governed by GR, and the matter by a set of (possibly charged)
dissipative fluids. To date, the spherically symmetry case with a Schwarzschild exterior
has been shown to be stable to radial perturbations and accretion with
an appropriate equation of state~\cite{Danielsson:2021ykm}, slowly rotating solutions have
been found~\cite{Danielsson:2017pvl,Danielsson:2023onu}, and certain electromagnetic properties studied~\cite{Giri:2024cks} that suggest
AdS black shells could be compatible with EHT observations despite the shell
being a macroscopically large distance outside of the would be black hole horizon
(it is at the Buchdahl Radius, $9M/4$, for a non-rotating black shell).
It remains to be seen if they are stable to non-radial perturbations, and it is
unclear how their collision in a merger should be treated. If the latter
is a new tunneling process creating a larger shell, then this model also
suffers from lack of a theoretical framework to calculate a merger
waveform in the coalescence regime. If it can be treated as the collision of two classical
shells of matter, a waveform calculation should be possible, though it will be challenging 
to simulate due to the effectively distributional source of matter.

\section{The Lamppost problem in theoretical, strongfield gravity}

In principle, non-GR metric theories of gravity should allow for black hole imposters
that are {\em significantly} different from Kerr black holes, however none are presently known. For example, stable geons, black holes
with different horizon topologies (as occurs 
in higher dimensional GR), or black holes that can support ``true hair''.
What I mean by the latter is a horizon with (say) spherical topology, but with a stable,
non-trivial geometric structure that depends on its {\em formation history}. Many current
examples of modified GR theories are often claimed to imbue a black hole with hair; however given that said hair has a structure uniquely set by the parameters of the underlying black hole and coupling constant of the theory, this does not violate the spirit of the GR no-hair theorems that do not allow a black hole horizon to support non-decaying, formation dependent structure. 

This begs the question of why all known 4D, asymptotically flat black holes in modified-GR theories are Kerr-like. I suspect the answer is because we have not been able
to successfully look outside of the {\em lamppost} of GR. This is partly because of the conceptual and mathematical difficulties in doing so, that I will discuss in the following paragraphs, but also that for black holes there is not yet an observational anomaly that requires we do so. It would be surprising if the answer
were instead that there is some as-of-yet undiscovered property of strongfield gravity that, regardless of the nature of modifications to the Einstein Hilbert action, the only viable black hole solutions are Kerr with small perturbations,
and we do not need to rely on observation to vet amongst possible theories describing the leading order phenomenology of horizons in our universe. Contrast for example the analogous question
of what radiative degrees of freedom are in principle allowed in the weak field 
in a metric based theory of gravity : here we know the two transverse/traceless polarizations of GR is {\em not} the unique answer, but in general {\em six} linearly independent modes are possible~\cite{Will:2014kxa}. 

One aspect of the lamppost problem is simply the difficulty of finding solutions to the more complicated, non-linear PDEs arising from curvature based modifications to the Einstein-Hilbert action. This is not surprising given that in GR itself almost 50 years of research occurred between discovery of the Schwarzschild~\cite{Schwarzschild:1916uq} and Kerr solutions~\cite{Kerr:1963ud}, the former satisfying considerably simpler differential equations compared to the latter (in appropriate symmetry adapted coordinates). That we have analytical forms for Schwarzschild and Kerr, and that most modified-GR theories reduced to GR in some well-defined manner, implies it is straight-forward to seek perturbative corrections to these solutions going beyond-GR. Looking for a completely different classes of black hole solutions, or proving that they do not exist, is a considerably more challenging endeavor, and is difficult to justify
undertaking if the only reason to do so is theoretical curiosity.

A second aspect of the lamppost problem relates to a common expectation that new physics will manifest at some new scale governed by a corresponding dimensionful constant. For example, that the putative quantum nature of spacetime will be revealed at a Planck or string scale, or that a Hubble scale correction to classical gravity could explain the mystery of dark energy. The lamppost here is in part this philosophical bias that modified gravity should behave in this manner, but also in part the difficulty in contemplating how to alter strongfield gravity to retain the scale-free nature of GR without breaking the well tested weakfield regime. Given the many similarities
between cosmological and black hole horizons in classical gravity, most importantly that both reside in the strong-field regime of GR (see for e.g. the discussion in ~\cite{Pretorius:2023hwx}), and the unknown nature of dark energy, there is significantly more justification to spend effort to solve this aspect of the lamppost problem. In particular, this would allow testing such theories with GW observations of stellar mass black hole mergers, either constraining such theories or discovering an anomaly that could then be argued to also pertain to dark energy. 

\bibliographystyle{utphys}
\bibliography{references_FP}

%





\title{Black Hole Spectroscopy: Status Report}
\author{Gregorio Carullo}

\institute{\textit{Niels Bohr International Academy, Niels Bohr Institute, Blegdamsvej 17, 2100 Copenhagen, Denmark}}
\institute{\textit{School of Physics and Astronomy and Institute for Gravitational Wave Astronomy, University of Birmingham, Edgbaston, Birmingham, B15 2TT, United Kingdom}}

\maketitle

\begin{abstract}
A brief overview of the ``Black Hole Spectroscopy program'' status is presented. 
Albeit given from a personal angle, it constitutes an attempt to convey the impressive progress achieved within the field in the last few years.
Modeling and observational aspects are touched upon, although both from an observationally-oriented perspective.
Particular emphasis is given to recent advancements within General Relativity and challenging open problems.
\end{abstract}

\section{Introduction}

The idea that precise measurements of vibrational spectra can provide convincing evidence for the black hole (BH) nature of dark compact objects, and for the accuracy of a general relativistic (GR) description of large curvature dynamics is now decades old~\cite{Detweiler:1980gk,Dreyer:2003bv, Cardoso_QNM_LISA}.
Interest in the mathematical description of black hole perturbations was initially fueled by the fundamental question of their stability, namely physical viability~\cite{Regge:1957td, Press_Teuk, Dias:2015wqa}, and since then has evolved towards including BH perturbations in connection with astrophysical phenomena.
For excellent reviews of early developments we refer to~\cite{Kokkotas_QNM,Ferrari_QNM,Cardoso:2016ryw,Nollert_QNM,QNM_review_BCS}, while for pointers to the mathematical literature see~\cite{Klainerman:2017nrb,TeixeiradaCosta:2019skg,Dafermos:2021cbw,Klainerman:2022ric, Shlapentokh-Rothman:2023bwo}.

More recently, focus of the physics community shifted towards a specific setting exciting BH perturbations: the coalescence of two compact objects, particularly in the form of a merger from a binary at the end of a long-lived bounded orbit.
A robust understanding of this process fueled the opportunity of exploiting this type of signals to search for new physics\footnote{These efforts are also commonly interpreted as \textit{testing General Relativity}, through the lenses of a framework inherited from previous experimental investigations~\cite{Will:2014kxa}.} either in the form of non-BH dark compact objects~\cite{Cardoso:2019rvt}, or modifications to the Einstein-Hilbert action~\cite{Berti:2015itd,Perkins:2020tra}.

Consequently, the modeling of the resulting post-merger signal and the experimental extraction of characteristic BH vibrational frequencies has been a leading topic of gravitational wave (GW) physics.
Significant milestones in these directions were achieved on the theoretical side through the first fully general relativistic simulation of a binary merger~\cite{Pretorius:2005gq,Campanelli:2005dd,Baker:2005vv} and the subsequent extraction of the remnant BH vibrational properties~\cite{Buonanno:2006ui,Berti:2007fi}, while on the experimental one through the groundbreaking observations of GWs from a binary merger, GW150914~\cite{LIGOScientific:2016aoc}.
Beyond marking the first detection of a GW signal, the observation of GW150914 simultaneously happened to be the first time that a BH relaxation process has ever been observed~\cite{LIGOScientific:2016lio}.
The combination of these two advancements opened the era of modern \textit{observational} BH spectroscopy.

Another intriguing possibility worth mentioning is the usage of this type of signals to verify if the BH background describing the process can be well-approximated by the Kerr solution, or if corrections due to dark-matter halos, accretion disks or other ``environmental'' effects might play a role.
The latter question is clearly of paramount importance before being able to draw any conclusions about ``fundamental'' gravity properties.
Current investigations point to the fact that such environmental contaminations are expected to be negligible in the ringdown phase, for signals observed by current and planned observatory~\cite{Barausse:2014tra, Carullo:2021oxn, Spieksma:2024voy}, albeit full numerical evolution are required to consolidate this picture~\cite{Aurrekoetxea:2024cqd}. 

Below, we concisely review past progress, current efforts and open problems of the physical description of BH spectra excitations in the aftermath of a binary merger, and the data analysis challenges and techniques required to extract information on such phenomena from interferometric data.
For brevity, the analysis will focus on the (yet incomplete) understanding on the process within GR, albeit computation of QNM spectra relaxing the GR or Kerr assumptions have witnessed remarkable progress in recent times~\cite{Blazquez-Salcedo:2016enn,Blazquez-Salcedo:2017txk,Blazquez-Salcedo:2020caw,Cardoso:2019mqo,McManus:2019ulj,Pierini:2021jxd,Pierini:2022eim,Srivastava:2021imr,Wagle:2021tam,Cano:2019ore,Adair:2020vso,Cano:2020cao,Cano:2021myl,Cano:2023tmv,Cano:2023jbk,Cano:2023qqm,Li:2023ulk,Tattersall:2017erk,Franciolini:2018uyq,Chung:2024vaf,Chung:2024ira,Chung:2023wkd,Cano:2024jkd}.

\section{The physical process}

\subsection{Introduction}

The aftermath of a binary coalescence of two Kerr BHs in GR is an initially highly perturbed object, rapidly relaxing towards a stationary Kerr state.
A large body of analytical results and numerical experiments indicate that at intermediate times (after $\simeq 20 M$ for a typical binary merger, where geometric units are used and $M$ is the mass of the binary) the signal is well-approximated by a linear superposition of quasi-normal modes (QNMs), namely damped sinusoids carrying the characteristic frequencies of the background.
This can be appreciated e.g. by looking at Fig.~\ref{fig:0305}, displaying the dominant quadrupolar mode of the asymptotic gravitational wave strain from the highly accurate \texttt{SXS:0305} numerical simulation at highest resolution, representing a binary with parameters close to GW150914~\cite{Mroue:2013xna,Boyle:2019kee}.
Around $\simeq 20 M$, the frequency has approached a constant (equal to the corresponding longest-lived QNM frequency), and the amplitude scales exponentially according to the $A \cdot e^{-t/\tau},$\footnote{At first order only a single mode dominates. While mode-mixing effects and higher-order modes are present even at these intermediate times, they don't affect the conceptual picture discussed here.} with $A$ a constant number, see top left of Fig.3 in~\cite{Baibhav:2023clw}.
This is what will be referred to as the ``stationary QNM regime'',  which has historically been the focus of spectroscopic analyses.
The late-time $\sim 100 M$ signal (not included in the above figure, since we won't be discussing it) is known to be dominated by power-law tails~\cite{Price:1971fb}, which can be signifcantly more excited than previously expected when the binary eccentricity is sufficiently high~\cite{Albanesi:2023bgi,Carullo:2023tff,DeAmicis:2024not}.

During the stationary regime, the waveform observed far from the source will be of the form:

\begin{figure}[t!]
\centering
\includegraphics[width=\linewidth]{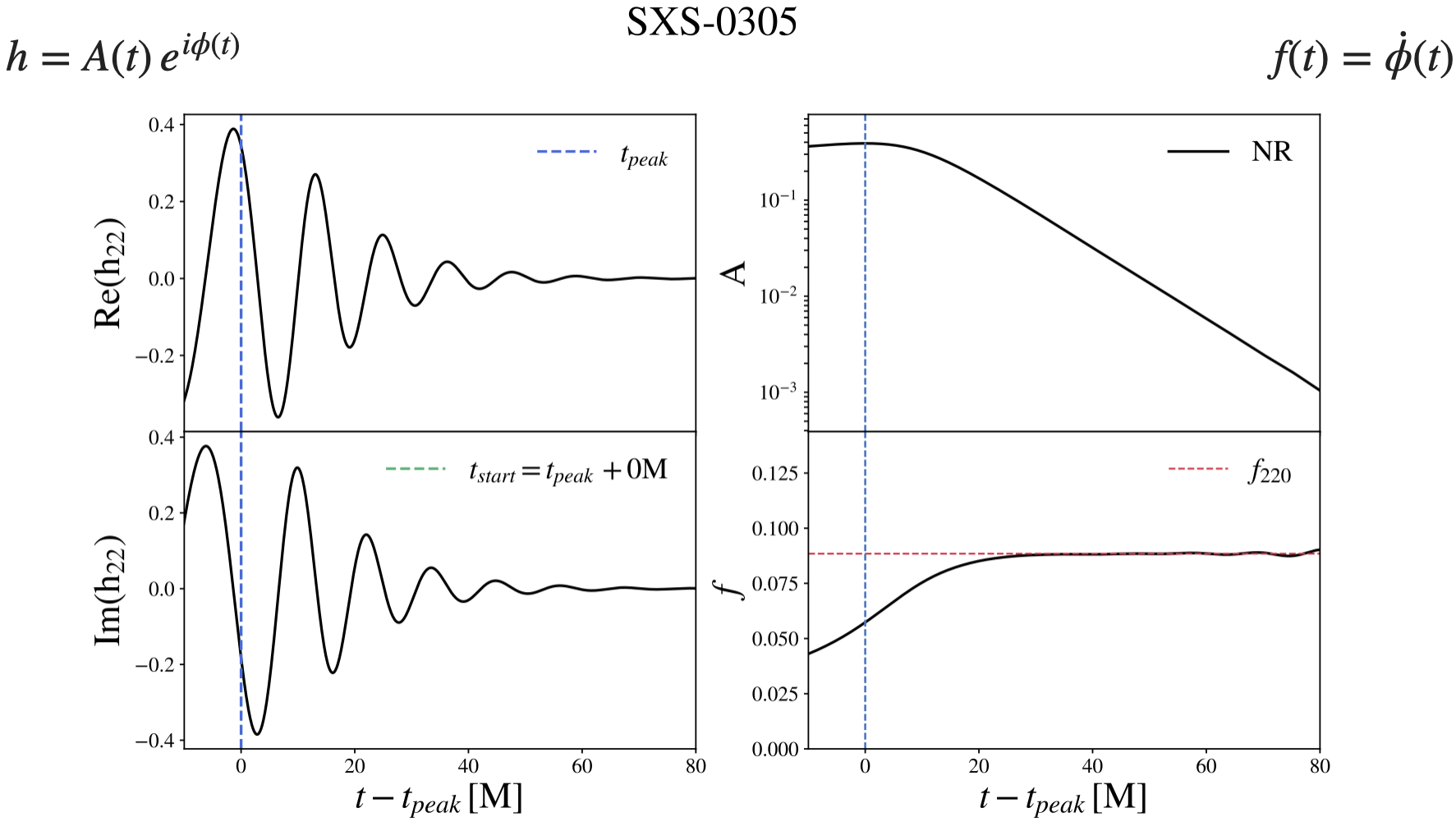}
\caption{Dominant quadrupolar mode (left: real and imaginary part; right: amplitude and frequency split, following the definitions given at the top of the figure) of the strain extracted from the \texttt{SXS:0305} highly-accurate simulation. The horizontal line marks the corresponding QNM frequency computed using the remnant mass and spin extracted at late times. \label{fig:0305}}
\end{figure}

\begin{eqnarray}\label{eq:template}
    h^{\rm Kerr}_{\ell  m} (t) &=& \sum_{\ell^\prime=2}^{\infty} \sum_{n=0}^{\infty} \left[ A^{+}_{\ell^\prime m n} \, e^{i \left( \omega^{+}_{\ell^\prime m n}(t-t_{\rm ref})+\phi^{+}_{\ell^\prime m n} \right) } + A^{-}_{\ell^\prime m n} \, e^{i \left(\omega^{-}_{\ell^\prime m n}(t-t_{\rm ref})+\phi^{-}_{\ell^\prime m n} \right)} \right] 
\end{eqnarray}

with $\ell>2$, $m\geq0$, where $\omega^{\pm}_{\ell m n}$ correspond to Kerr complex QNM frequencies fixed by the final BH mass and angular momentum (spin).
The ${\pm}$ labels refer to the two family of modes contributing to each harmonic, physically corresponding to perturbations that respectively co-rotate (``+'') or counter-rotate (``-'') with respect to the remnant BH spin.
Albeit the complex QNMs frequencies can be predicted from known perturbative results in terms of the BH parameters, the amplitudes and phases are initial-data dependent.
Chiefly, in this regime the amplitudes and phases are \textit{constant} to a very good approximation.

Instead, the early stage of the process is (in principle) expected to be contaminated by all sorts of nonlinear effects, such as non-linear couplings, non-modal contributions, mass and spin evolution, mode-spreading etc. (see below for details). 
Beyond these, other features contributing to the early-time relaxation are instead connected to the dynamical nature of the QNMs excitation.
Albeit less studied, this latter phenomenon is predicted even in linear theory~\cite{Lagos:2022otp}, and is simply due to the fact that QNM excitation doesn't happen instantaneously.
Instead, the modes will experience a transient ``activation'' regime, driven by the dynamical multipolar structure of the background.
All these effects prevent an a-priori straightforward extension of the above description to the near-peak regime.
Such extensions could possibly be achieved by including the time-dependence of the BH mass and spin, of the QNM amplitudes, together with other non-modal contributions.

The lack of understanding of the early ringdown regime, and possibility of contamination to the QNM extraction were circumvented by initial BH spectroscopy strategies by assuming to start the analysis during the intermediate stationary regime.
This comes at the price of renouncing to the largest portion of the signal-to-noise ratio (SNR), but ensures a clean interpretation of the measurement, and an unbiased extraction of the physical QNM BH spectra.
Indeed, these investigations were mainly targeted at high-SNR signals observed by future detectors~\cite{Cardoso_QNM_LISA, Gossan:2011ha}. 

\subsection{Past work}

We now briefly review the current status of QNM waveform modeling, with sparse pointers to key early developments.

\subsubsection{Early models and linear attempts}

While early studies had to rely on ``educated guesses''~\cite{Baibhav:2023clw} concerning modes excitations and validity of perturbative results~\cite{Buonanno:2000ef}, as soon as the first numerical simulations of binary mergers became available, the question of the domain of validity of linear perturbation theory was investigated, which sparked significant interest also in the development of second-order perturbation theory~\cite{Gleiser:1995gx,Campanelli:1998jv,Baker:2001sf,Brizuela:2009qd,Ioka:2007ak,Nakano:2007cj,Pazos:2010xf,Loutrel:2020wbw,Ripley:2020xby,Pound:2021qin,Spiers:2023mor}.
Later, when simulations of mergers of quasi-circular binaries were achieved~\cite{Pretorius:2005gq,Campanelli:2005dd,Baker:2005vv}, studies along these lines were conducted in~\cite{Buonanno:2006ui,Berti:2007fi}.
An immediate attempt underwent to extend the QNM description up to the waveform peak by including multiple overtones in a fit. 
Indeed, early effective-one-body (EOB) models calibrated to numerical relativity featured a post-peak signal completion in terms of a superposition of up to $n=7$ overtones.
Additionally, they included "pseudo-QNM" contributions, found to be necessary to obtain a more stable fit that could ``bridge'' the post-merger frequency to the pre-merger orbital one~\cite{Buonanno:2009qa}.
However, these attempts were explicitly recognised as phenomenological (see e.g.~\cite{Berti:2007fi}), since they were not accounting for the wealth of beyond-linear effects previously mentioned and described in more detail below.
It is interesting to note that similar considerations were already presented in~\cite{Andersson:1996cm}, as recounted in~\cite{Maggiore:2018sht}.
Later on, pure QNM superpositions were abandoned in the construction of phenomenological post-peak models employed in inspiral-merger-ringdown (IMR) waveforms, since the (resummation-inspired) strategy of~\cite{Damour_Nagar_ring} proved to be more accurate and stable. 
See also~\cite{Baker:2008mj} for earlier proposals.
Such models include a phenomenological time-dependence of QNM complex amplitudes, chosen to mimick the waveform morphology observed in Fig.~\ref{fig:0305}.

More recently, considerable interest was again devoted to the problem following the attempts of~\cite{Giesler:2019uxc} to propose a picture based once more on a superposition of a large number of overtones, again up to $n=7$, carrying 16 free parameters when constructing a fit.
This proposal suffered from the same conceptual and practical shortcomings of earlier attempts discussed above, and a critical assessment of this newer study was investigated in the by~\cite{Baibhav:2023clw,Nee:2023osy}, see also~\cite{Forteza:2020hbw}.
A fruitful lesson from~\cite{Baibhav:2023clw} (surprisingly ignored in the many subsequent studies inspired by~\cite{Giesler:2019uxc}, which were thus following similar assumptions), is that models based on linear superpositions of constant-amplitude QNMs should be tested for self-consistency, especially before physically interpreting any improvement based on fits with large numbers of parameters.
The simplest of such self-consistency tests is the complex amplitude \textit{independence} on the fit starting time (within the stationary model validity regime), as already internally applied in e.g.~\cite{MMRDNS,MMRDNP}.
It is of fundamental importance to stress that, under the assumptions through which the model is constructed, such consistency requirement is implied for \textit{all} modes, not just a subset of them~\cite{Clarke:2024lwi, Zhu:2023mzv}. 
In the opposite scenario, it may be the case that ``inconsistent modes'' fit away non-QNM or non-linear features, preventing a clean interpretation of the remaining modes.
The second requirement is a smooth variation of the extracted modes amplitudes upon changes of the initial conditions (e.g. binary mass ratio).
Again this was well-known in earlier works~\cite{Buonanno:2009qa, MMRDNS, MMRDNP}, but surprisingly ignored by a large body of literature in more recent overtones investigations.
Simple superpositions of a large number of constant-amplitude overtones extended up to the signal peak do not respect neither of these consistency criteria, as shown in~\cite{Forteza:2020hbw,Baibhav:2023clw}.

Before moving on, a word of caution is needed regarding the interpretation of these latter studies.
A superficial reading could lead to think that such latter studies argue for the complete absence of overtones in the post-merger binary signal.
Such conclusion would be obviously flawed, as some of the same authors have analytically computed the physical overtones excitations up to $N=3$ in simplified post-merger perturbative settings~\cite{Zhang:2013ksa}.
It might well be the case that from a certain point onwards the post-peak signal is indeed well-described by a (time-dependent) linear description, for example because most of the non-linear features are cloacked behind the newly-formed horizon~\cite{Okounkova:2020vwu}.
The key message of~\cite{Baibhav:2023clw} is that in such a dynamical and complex evolution many effects are at play, and before being able to confidently identify short-lived overtones components, these effects must be investigated in detail before being possibly excluded.
Only after achieving a robust modeling of beyond-linear effects and QNM time dependence can a physical interpretation of early-time modes superpositions be assessed.

\subsubsection{Beyond the stationary linear picture}

Indeed, recent years have witnessed an impressive effort by the community to model non-linear and time-dependent effects, leading to a wealth of new results, with rapid progress still undergoing.
Towards this goal, novel results in the definition of a ``QNM scalar-product''~\cite{Green:2022htq,London:2023aeo} hold great promise in advancing the understanding of such process.

Key effects that have been targeted include:
\begin{itemize}
	\item \textbf{Dynamical background:} from the moment a common horizon forms, to the one a stationary state is reached, the BH background parameters (mass and spin, for an astrophysical Kerr black hole) will be changing with time~\cite{May:2024rrg}, see also~\cite{Buonanno:2006ui,Berti:2007fi,Redondo-Yuste:2023ipg,Zhu:2024dyl,Capuano:2024qhv}.
	This will be driving a complex frequency drift, but also induce ``mode-spreading''~\cite{Sberna:2021eui, May:2024rrg}.
	Namely, a given QNM generated around the initial background will experience an amplitude variation due to the background dynamical evolution, while also giving rise to a family of additional modes with respect to the final background.
	Albeit this effect will certainly be present, it seems unlikely that the entire frequency increase in the waveform might be captured by it, given the percentage-level variation of mass and spin, which should account for the order-unity variation displayed by the waveform frequency;
	\item \textbf{Modes coupling:} higher-order perturbative contributions imply that two "parent" modes can interact, giving rise to "child" modes with frequencies composed of linear combinations of parent frequencies, and an angular structure dictated by Clebsch-Gordan rules~\cite{Lagos:2022otp}. 
	The most-studied among these couplings has been the $(2,2,0) \times (2,2,0) \rightarrow (4,4,0)$ one, given that the $(\ell, m, n) = (2,2,0)$ mode is typically the dominant one for such systems.
	After the first extraction of such coupling from nonlinear simulations~\cite{Cheung:2022rbm,Mitman:2022qdl} (see earlier work in~\cite{MMRDNS}), a wealth of perturbative studies investigated the quadratic amplitude generation dependence on initial conditions and BH spin, both in numerical~\cite{Baibhav:2023clw,Cheung:2023vki,Redondo-Yuste:2023seq,Zhu:2024rej} and analytical~\cite{Bucciotti:2023ets,Perrone:2023jzq,Ma:2024qcv,Bucciotti:2024zyp,Bourg:2024jme} settings.
	While this effect has a significant impact on non-quadrupolar harmonics, and its inclusion will certainly be required to achieve high-fidelity templates for e.g. $(\ell, m, n) = (4,4,0)$, nonlinear couplings relevance to model the early ringdown regime seems to be limited.
	In fact, angular selection rules imply that in order for nonlinear couplings involving the highly excited $(\ell, m, n) = (2,2,0)$ to enter back the dominant $(\ell, m) = (2,2)$ spherical mode, interactions with typical sub-dominant modes such as the $(\ell, m) = (2,0)$ or counter-rotating modes are required, which are expected to deliver an overall small contribution.
	These arguments apply as well to third-order couplings, which are expected to deliver an even larger suppression.
\end{itemize}

\subsubsection{Numerical extraction of QNM excitation amplitudes}

Another aspect actively pursued in recent years is the numerical extraction of QNM constant amplitude values from the stationary portion of the ringdown signal for comparable-mass binaries, and their modeling in terms of the merger initial conditions (e.g. the binary masses and spins).
The quasi-circular case with binary spins aligned to the orbital angular momentum has now been tackled, reaching very high accuracy, with a variety of methodologies~\cite{Kamaretsos2012,MMRDNS,MMRDNP,Cheung:2023vki,Pacilio:2024tdl,MaganaZertuche:2024ajz}.

The mis-aligned spins case (determing spin-induced precession, hence dubbed ``precessing case'' below) has witnessed less attention from the QNM amplitudes modeling side,\footnote{Phenomenological descriptions entering inspiral-merger-ringdown models that do not focus on the modeling of QNM components are not discussed here, see references in~\cite{Zhu:2023fnf}.} albeit significant progress has been recently achieved~\cite{Hamilton:2023znn,Zhu:2023fnf}.
In particular, the methodology leveraged in~\cite{Pacilio:2024tdl,MaganaZertuche:2024ajz} holds great promise in obtaining a robust extraction of these quantities.
Extensions of the phenomenological models akin to~\cite{Damour_Nagar_ring} to spin-precessing configurations were instead included in e.g.~\cite{Estelles:2021gvs,Ramos-Buades:2023ehm}, with~\cite{Pompili:2023tna} including for the first time mode-mixing contributions in these type of templates.

Due to the fact that gravitational radiation is highly efficient in circularising binaries evolving in bounded orbits, the non-circular case has also received little attention in the past.
Recently, interest was revived by the possibility that a non-negligible fraction of binaries observed by the LIGO-Virgo-Kagra (LVK) detectors might have formed dynamically or that BH captures could be observed directly in the detectors band~\cite{Romero-Shaw:2020thy,Gamba:2021gap,Gupte:2024jfe,Gayathri:2020coq,CalderonBustillo:2020xms}.
In Ref.~\cite{Carullo:2024smg}, the extraction of QNM amplitudes beyond the quasi-circular case was achieved in the case of a non-spinning progenitors binary.
The methodology employed stems from~\cite{Albanesi:2023bgi, Carullo:2023kvj}, which have introduced a gauge-invariant parameterisation of non-circular orbits based on combinations of energy and angular momentum parameters, applicable during the entire coalescence process.

Recent advancements have shown that although the initial main motivation for QNM studies was to search for new physics, modeling efforts arising within the BH spectroscopy program can also have direct astrophysical implications, informing IMR models used in LVK searches and binary parameters extraction.
Even further, physically-motivated models of QNM excitations and prompt ringdown, required to advance the BH spectroscopy program, could deliver a first-principles model with much higher interpretability properties. This feature would be key to allow for easier extensions to include environmental effects or beyond-GR/Kerr signatures, saving the need for costly numerical simulations that become hardly feasible when considering such host of additional effects.

\subsection{Future developments and open problems}

Future milestones and open problems include:

\begin{itemize}
	\item \textbf{QNM amplitudes in generic orbits:} the extension of closed-form QNM amplitudes models to the quasi-circular precessing case, the spin-aligned noncircular case, and finally to the generic noncircular-precessing one.
	Noncircular extensions of post-merger phenomenological models similar to~\cite{Damour_Nagar_ring,Albanesi:2023bgi}, together with investigations on whether standard twisting-up techniques~\cite{Schmidt:2012rh,Estelles:2021gvs} are sufficient to accurately extend these noncircular templates to the spin-precessing case are also future key milestones.
	\item \textbf{Amplitude growth:} as detailed in e.g.~\cite{Lagos:2022otp}, transient effects induced by initial data impart a time-dependent amplitude growth even within the context of linear perturbation theory. Extensions of these models to the realistic case of comparable mass binaries appears challenging, but would be necessary to achieve physically motivated templates of the prompt ringdown emission.
	\item \textbf{Orbital imprint:} already from the first investigation concerning QNM amplitudes models as a function of binary parameters~\cite{Kamaretsos2012}, it was understood that for many QNM modes such dependence closely resembles the leading-order post-Newtonian scaling governing inspiral amplitudes.
	This observation was key in the construction of subsequent high-accuracy closed-form models, and was investigated numerically in a systematic way in~\cite{Borhanian:2019kxt}.
	Analytical models predicting this dependence are currently missing, but as argued in~\cite{MMRDNP}, the simplicity of these expressions gives hope that such computation may be achieved in the near future.

\end{itemize}


\section{The data analysis problem}

\subsection{Introduction and past work}

\subsubsection{Frequency-domain vs. time-domain formulations}

The extraction of BH vibrational spectra from GWs data has been proposed long ago with the key motivation of investigating the Kerr nature of the emitting remnant compact object.
Initial analyses were working in an idealised frequency-domain framework~\cite{Cardoso_QNM_LISA, Gossan:2011ha, Meidam:2014jpa}, where the simulated data consisted of a Fourier transformed QNM-only template with constant amplitudes.
This approach ignored altogether contaminations from the pre-stationary ringdown signal and the pre-merger signal.
Such setup is sufficient to obtain a reasonable estimate of measurable ringdown properties.
However, for real observational scenarios in which a pre-stationary (or even pre-ringdown) signal is always present, these portions will pollute the QNM analysis if not properly taken into account.

A way to get around this problem when analysing complete signals is to multiply the data and the template by a ``windowing'' function, manually setting to zero previous contributions, as done in~\cite{Carullo:2018sfu}.
This method is valid but cumbersome, since the steepness of the window needs to be optimised to strike a balance between signal loss (decreased by a steeply-rising window) and spurious Gibbs-phenomena (increased by a steeply-rising window).
Formulating the test directly in the time domain can overcome this problem, and this methodology has been applied in the white-noise case (process with equal power over all frequencies, thus uncorrelated in time) in~\cite{DelPozzo:2016kmd}, then generalised to the coloured noise (correlated over time) weakly-sense stationary case in~\cite{Carullo:2019flw}, when assuming small correlation between ringdown and inspiral components.
The full solution for a fixed analysis start time was given in ~\cite{Isi:2019aib} through a truncated likelihood formulation, implemented in the \texttt{ringdown} package~\cite{Isi:2021iql}.
An alternative, but mathematically equivalent formulation in the frequency domain was presented in~\cite{Capano:2021etf}, building on~\cite{Zackay:2019kkv}.
For a comparison of the latter with time-domain methods see~\cite{Isi:2021iql}.

\subsubsection{The LVK search in a nutshell}

These works led to the time-domain ringdown analysis pipeline built around the \texttt{pyRing} package~\cite{Carullo:2019flw,pyRing}, employed  by the LVK collaboration to isolate and analyse ringdown signals observed throughout the first, second and third observing runs~\cite{LIGOScientific:2020iuh,LIGOScientific:2020ufj,LIGOScientific:2020tif, LIGOScientific:2021sio}.
The pipeline explored a wide range of starting times, and employed models of increasing complexity, ranging from agnostic superpositions of damped sinusoids to templates calibrated against binary BH numerical simulations~\cite{MMRDNP}.
All these analyses are complementary to each other.
The most agnostic ones are supposed to be able to capture very generic beyond-Kerr/GR signatures, such as additional non-tensorial modes~\cite{Crescimbeni:2024sam}, but also to be less sensitive to small deviations from Kerr GR templates (i.e. will require a much larger SNR to flag possible deviations).
Conversely, highly-informed templates such as~\cite{MMRDNP, Gennari:2023gmx} are expected to have a harder time faithfully characterising features that strongly deviate from the GR predicted behaviour, but are expected to be more sensitive to minute deviations (also because less affected by Occam's penalty).
Beyond the most-informed analyses performed through \texttt{pyRing}, an even more informed analysis is the \texttt{pSEOB} one~\cite{Ghosh:2021mrv,Brito:2018rfr,Maggio:2022hre}, where now the pre-merger signal is assumed to correspond to the GR one (unlike time-domain spectroscopic analyses that do not make such assumption, and remove contributions from the pre-merger).
The analysis proceeds by retaining the entire NR calibration of a standard full-signal waveform, and adding deviations to QNM parameters.
Thus, it will be more sensitive than spectroscopic analyses to small GR deviations, although the advantages of interpretability, and flexibility of damped-sinusoids superpositions in capturing more exotic signals are lost.
Details of the actual sensitivity to new physics signatures will likely depend on the details of the modification searched for.
This extensive LVK search, together with the subsequent extension~\cite{Gennari:2023gmx}, did not find robust evidence for the presence of subdominant modes, within the physical regime of applicability of respective models, nor evidence for new physics signatures.

\subsubsection{Multiple overtones claims}

Evidence for multiple modes was instead reported in~\cite{Isi:2019aib} regarding an additional overtone ($n=1$) in GW150914.
The overtone detection was challenged in~\cite{Cotesta:2022pci}, with subsequent responses in~\cite{Isi:2022mhy,Isi:comment,Carullo:2023gtf} and analyses from independent groups in~\cite{Crisostomi:2023tle,Finch:2022ynt,wang2023frequencydomain,Correia:2023bfn}.
Leaving aside the (previously described) fundamental issues with the underlying physical model, through which detection was claimed, many of these analyses initially seemed to return different results regarding the statistical significance of an overtone component starting at the peak of the waveform.

Part of this descrepancy was due to some analyses neglecting certain sources of uncertainty or using an incorrect labeling of the time axis.
For example, in~\cite{Cotesta:2022pci} a mislabeling of the time stamps induced a systematic error around $0.06 \, ms$, while the initial $3.6 \sigma$ detection claim was ignoring the statistical uncertainty on the determination of $t_{\rm start}$, amounting to neglect a statistical uncertainty of around $2.5 \, ms$.
For reference, the damping time of the putative $n=1$ overtone is $\tau_{221} \simeq 1.4 \, ms$, while for the fundamental mode $\tau_{220} \simeq 4.2 \, ms$.
Once all these elements had been fixed and start time uncertainty was kept into account, analyses converged towards a lower overtone significance, slightly smaller than $2 \sigma$ with a negative Bayes Factor for the overtone presence, for most of the independent analyses conducted by groups not part of the initial debate~\cite{Crisostomi:2023tle,Finch:2022ynt}.
Initially the authors of~\cite{wang2023frequencydomain} found higher significance, but later some of the same authors obtained low evidence for an overtone presence with similar methods, but now marginalising over the time uncertainty~\cite{Correia:2023bfn}.

Albeit different techniques have been applied by different groups, hence it is not always immediate to compare all these results, some of these analyses did yield different results even when using mathematically equivalent methods.
The reason behind this apparent inconsistency is that for the type of signal analysed (very short overtone component, low signal-to-noise ratio) even small differences in data conditioning (sampling rate, bandpassing) and noise estimation methods (namely the computation of the autocorrelation function setting the variance of the gaussian likelihood employed) can lead to non-negligible differences.
Indeed these inputs were not always uniform across analyses, as recollected e.g. in~\cite{Carullo:2023gtf}, which motivates these discrepancies.
Instead, in~\cite{Siegel:2023lxl} is it shown how similar analyses do obtain compatible outputs when employing similar inputs.

A clear lesson from the ensuing debate is that more attention needs to be paid to develop robust and commonly agreed procedures to include all statistical and systematic uncertainties, together with robust detection thresholds (see below).
The recounted debate has certainly been extremely useful in investigating the details of data-analysis methods in spectroscopic settings, hence useful in the context of future detections. 
However, the view of the author is that for what concerns the physical interpretation of the analysis result, the outcome is moot, independently of the claimed significance.
This is because, as extensively motivated in~\cite{Baibhav:2023clw} and summarised above, the underlying model ignores a vast array of physical effects that are known to come into play when analysing data close to the peak (time-varying background, nonlinearities, dynamical QNM excitation, ...), hence simply cannot be interpreted as a spectroscopic measurement.

\subsubsection{Multiple fundamental modes claims}

For what concerns detection claims of fundamental modes corresponding to harmonics different than $(\ell,m)=(2,2)$, the $(\ell,m)=(3,3)$ observability was claimed in GW190521 data in~\cite{Capano:2021etf}.
Hints of higher harmonics were uncovered in~\cite{Siegel:2023lxl} for the $(\ell,m)=(2,1)$ mode in GW190521, and in several events for both $(\ell,m)= [(3,3), (2,1)]$ modes  by~\cite{Gennari:2023gmx}. 
Both these latter analyses did not make any strong detection claim.

Focusing on GW190521, most of the aforementioned problems affecting overtones detection have a significantly smaller impact, given the longer duration of the $n=0$ components with respect to an overtone.
However, many more complications arise in this case due to the morphology of the investigated signal.
In fact, GW190521 was originated by a highly massive system (remnant mass $ M_f \simeq 250 M_{\odot}$ in the detector frame), thus spending a very short duration in the detector band.
For this reason, it is not possible to discriminate whether this event had measurable spin-induced precession or orbital eccentricity, or a combination of both~\cite{CalderonBustillo:2020xms, Romero-Shaw:2022fbf}.\footnote{Further, the gaussian-noise simulation study conducted in~\cite{Gamba:2021gap} indicates that some amount of non-circular behaviour was present, although the same exercise should be repeated using real interferometric noise to validate this conclusion.} 
However, all inspiral-merger-ringdown analyses conducted by the LVK and external groups~\cite{LIGOScientific:2020iuh,LIGOScientific:2020ufj,Olsen:2021qin} agree on the presence of at least one of the two effects, and on the incompatibility of such signal with a spin-aligned quasi-circular binary.

The spectroscopic analysis of~\cite{Capano:2021etf}, claiming detection of the $(\ell,m,n)=(3,3,0)$ mode, suffered from two major issues related to the aforementioned signal morphology:
 
\begin{itemize}
	\item the amplitude model, employed to reported Bayes Factor for multiple Kerr harmonics, assumed an upper cut on the amplitudes ratio valid for spin-aligned quasi-circular systems only. Additionally, almost all results are quoted assuming reflection symmetry around the orbital plane, broken in generic spin-precessing binaries;
	\item the chosen $t_{\rm start}$ was based on a frequency-domain approximant family~\cite{Pratten:2020ceb} known to provide a sub-optimal signal description than the \texttt{NRSur7dq4} model~\cite{Varma:2019csw}.
	The latter model was employed by the LVK collaboration when quoting main results~\cite{LIGOScientific:2020iuh} precisely because of its higher accuracy, especially in the region of parameter space probed by this signal.
	When translating the $t_{\rm start}$ at which detection is claimed to the peaktime $t_{\rm peak}$ of $h_+^2 + h_{\times}^2$ as reconstructed using \texttt{NRSur7dq4}, this corresponds to  $t_{\rm start} \simeq t_{\rm peak} + [2-5] M_f$, where a stationary QNM model based on fundamental harmonics with constant amplitudes is not valid.
\end{itemize}

Albeit the investigations presented in~\cite{Capano:2021etf} are certainly tantalising and indicative of a non-trivial signal structure, both these factors prevent a rigorous interpretation of the measurement as a multi-modal spectroscopic detection.
Indeed, when keeping these two elements into account, the analysis of~\cite{Siegel:2023lxl} finds preference not for the $(\ell,m,n)=(3,3,0)$ mode, but for the $(\ell,m,n)=(2,1,0)$ mode, but at a much lower significance for which no strong detection claim was made.

In summary, the early ringdown regime bears a significant prize in terms of signal-to-noise ratio (hence resolving power) gained, and it is only natural that analyses will attempt at leveraging this increase.
However, as extensively argued in the previous section, the risk is that the outcome of such analyes becomes similar to phenomenological fits already performed at the time of the first detection~\cite{LIGOScientific:2016lio}, constituting a valid consistency test of GR, but not a spectroscopic measurement.

\subsubsection{Start time marginalisation}

In~\cite{Carullo:2019flw} the ringdown analysis starting time $t_{\rm start}$ was recognised as the parameter controlling the largest systematic uncertainty of these type of analyses, and its marginalisation attempted, producing a posterior on this quantity.
Unfortunately, when adopting the unbiased truncated formulation, this marginalisation does not appear to be possible in a straightforward way.
The intuitive reason behind this is that $t_{\rm start}$ does not only control when a QNM-superposition template starts to be non-zero, but also the amount of data included in the likelihood calculation. 
Hence, the analysis output will tend to push $t_{\rm start}$ as early as possible, to include as much signal as possible, increasing the likelihood function.

Note that this behaviour is dependent on the signal/template features.
In fact, it is partly due to the flexibility of damped-sinusoids templates being used, and on the smooth morphology of the binary BH waveform signals (similar to what predicted by effective-one-body arguments~\cite{Buonanno:2000ef}), for which no abrupt changes between the merger and ringdown portions can be easily discerned using feature-detection algorithms.
Such complication would have instead been avoided, if merger waveforms would have given rise to high-frequency complicated emission patterns, as sometimes foreseen before numerical simulations of binary mergers were achieved~\cite{Schutz:2004uj}.
Also note that depending on the problem formulation, the opposite behaviour might arise, with $t_{\rm start}$ being pushed as late as possible to have a close-to-zero signal, consistent with gaussian noise~\cite{Correia:2023bfn}.

Current strategies that do not assume pre-merger information simply involve repeating the analysis at multiple times, veryfing the robustness of a given result within the a certain confidence region (typically $90\%$) of $t_{\rm start}$ as inferred from IMR analyses.
This effectively ``discretises'' the $t_{\rm start}$ support, providing some sort of ``marginalisation'' of this parameter.
However, combining the result of these discrete inference steps is not immediate.
This is because each inference with a given $t_{\rm start}$ removes all data before $t_{\rm start}$, implying that different runs are performed on different subsets of the same \textit{data}.
Hence, computing a meaningful Bayes Factor between runs at different times (and producing averaged posteriors), is not straightforward.
Note, however, that Bayes Factors between different models at a fixed start time retain a standard Bayesian interpretation.
A conservative way to avoid this problem when interested in e.g. the detection of a mode (or a posterior of a certain parameter attaining a given value), is to require that such Bayes Factor is larger than a given threshold throughout the uncertainty band of $t_{\rm start}$, as done e.g. in~\cite{Gennari:2023gmx}.
This is clearly sub-optimal, but valid. 

Another way to resolve the issue is to avoid ignoring the pre-ringdown data, making an assumption on the pre-merger signal, in which case $t_{\rm start}$ becomes again a ``standard parameter'' (i.e. it does not control the amount of data included). 
This was the strategy employed in~\cite{Correia:2023bfn} by assuming a GR template.
This strategy is valid, but loses some of the agnostic character of the search, going closer to \texttt{pSEOB}-like searches.
In~\cite{Finch:2022ynt} an agnostic superposition of wavelets for the pre-ringdown data was considered.
However, without imposing any GR prediction on when the QNM-driven signal starts, the authors had to resort to a series of narrow Gaussian priors over $t_{\rm start}$, which is more computationally efficient than the discretisation strategy employed in time-domain analyses, but ultimately equivalent.

\subsection{Open issues}

From the above discussion, open issues and challenges that should be addressed in the future to advance spectroscopic analyses are:

\begin{itemize}

	\item Agreement on a mode detection threshold, dependent on the binary parameter space and instrumental background;

	\item Marginalisation of $t_{\rm start}$ and sky-position parameters with as little pre-ringdown information as possible;
	
	\item Inclusion of systematics in the determination of $t_{\rm start}$, induced by systematic uncertainties in IMR models;
	
	\item Identification of the (SNR-dependent) validity regime of the different classes of stationary QNM-like models, as a function of the binary parameter space;
	
	\item Autocorrelation methods comparisons, and inclusion of systematic uncertainties induced by different methods or off-source computations~\cite{Littenberg:2014oda,Plunkett:2022zmx};
	
\end{itemize}

Albeit all these items need to be thoroughly addressed, the prospects of a multi-modal spectroscopic detection in the stationary QNM regime, applied to loud ``golden'' signals sourced by well-understood binary mergers  remain very optimistic~\cite{Cabero:2019zyt,Bhagwat:2019dtm}.
A typical example is the determination of the $(\ell,m,n)=(3,3,0)$ mode from a slowly-spinning binary with $m_1/m_2 = [2,3]$.
Past literature detection debates were instead mostly sparkled by attempts to apply spectroscopic analyses to the least understood signals available (GW190521), for which many uncertainties remain even on the IMR side, or to extend QNM templates to the pre-stationary regime, for which robust models are currently lacking.
All these complications are instead expected to fade away for aforementioned golden signals, albeit past debates have clearly shown that robust spectroscopic analyses require more work than initially thought.
A study rigorously re-evaluating spectroscopic detections on golden binaries, now applying all the analyses developments stemmed from recent debates, appears timely and necessary.

Concerning the analysis of high-SNR signals that are going to be observed by planned detectors~\cite{Bhagwat:2021kwv,Bhagwat:2023jwv}, many complications that normally affect future detectors measurements (noise non-stationarities, signals overlap, ...) are typically greatly reduced by the short ringdown signal duration.
However, challenges are still present, and a discussion of all the complications of high-accuracy measurements achieved next-generation detectors, such as waveform systematics (including remnant recoil~\cite{Gerosa:2016vip,Varma:2018aht} and hereditary effects~\cite{Mitman:2024uss,DeAmicis:2024not}) and the simultaneous modelling of many QNM contributions, is beyond the scope of this overview.

\bibliographystyle{utphys}
\bibliography{references}

\title{VLBI as a Precision Strong Gravity Instrument}
\author{Paul Tiede}
\institute{\textit{Black Hole Initiative at Harvard University, 20 Garden Street, Cambridge, MA 02138, USA}\and \textit{Center for Astrophysics | Harvard \& Smithsonian, 60 Garden Street, Cambridge, MA 02138, USA}}

\maketitle
\begin{abstract}
We are entering the era of precision black hole astronomy. Thanks to gravitational wave interferometers like LIGO, and potential next-generation detectors like Baker:2019nia the next few decades will provide direct gravitational probes of black holes on horizon scales. However, gravitational wave detectors are not alone. Thanks to the unprecedented resolution of very-long-baseline interferometry (VLBI) instruments like the Event Horizon Telescope and potential space extensions like the Black Hole Explorer, black hole horizons will come into focus within the next decade. In this letter I will highlight the potential for precision measurements of black holes using VLBI and how it can provide an independent probe of black holes. 
\end{abstract}

\section{Introduction}

Black holes were first predicted over a century from Einstein's theory of general relativity (GR). Black holes are both exceptionally simple \cite{Israel:1967wq, IsraelChargeNoHair, Nohair}, and exquisite laboratories for exploring quantum gravity effects for over 50 years \cite{Bekenstein, Hawking:1974rv, Hawking}. Furthermore, black holes have several theoretically unique properties interior to the horizon, such as the singularity, and exterior, such as the ergosphere and photon sphere, that are responsible for some of the more exotic predictions of GR. 

Astrophysically, it is now established that the center of all active galactic nuclei is a supermassive black hole (SMBH) \cite{ReeseBH} powering some of the most consistently luminous objects in the night sky. The impact of black holes on their surroundings, from galactic to intergalactic scales, has been studied for over half a century. Only recently have black holes started to be directly studied on horizon scales.

In 2016, the LIGO collaboration made the first discovery of gravitational waves from two colliding stellar mass black holes \cite{LIGO}, establishing the existence of gravitational waves and confirming the existence of black holes. In the coming year, gravitational interferometry will continue to expand, and telescopes like Baker:2019nia \cite{Baker:2019nia} will open an entirely new window into the horizons of massive black holes.

While much of the gravitational community has focused on measurements of gravity on horizon scales using LIGO, this article will argue that precision VLBI astronomy measurements can provide an additional probe of black hole horizons \cite{Ayzenberg:2023hfw}. In April 2019, the Event Horizon Telescope (EHT) provided the first-ever resolved images of the supermassive black hole M87*. Two years later, the first images of the black hole at the center of the Milky Way, Sgr A*, were published. These images provided direct visual evidence of the nature of compact objects sitting at the centers of galaxies, providing evidence for a high redshift region of spacetime similar in scale to the predicted size of a black hole's event horizon \cite{EventHorizonTelescope:2019ggy, EventHorizonTelescope:2022xqj}. Furthermore, future VLBI arrays will provide unprecedented resolution, and future space missions will be able to visually probe key predictions from GR around SMBHs directly, such as signatures of the photon sphere, ergosphere, and the horizon. Section 2 will review several detectable features of black hole horizons. In section 3, I will describe their observability in the coming decade.

\section{Electromagnetic Signatures of Strong Gravity}

One of the difficulties of measuring gravity using electromagnetic signatures is that the emission from accretion material around the black hole can obfuscate gravitational signatures. This ``astrophysical noise'' can make measuring properties of gravitational sources strongly dependent on the assumptions of accretion around black holes. Unfortunately, accretion physics on these scales is poorly understood, making measurements of gravity prone to significant systematic errors and highly model-dependent. Fortunately, strong lensing provides an avenue to asymptotically astrophysical invariant observables in the limit that the emission region near a black hole horizon is optically thin, namely photon rings \cite{Bardeen:1973tla, universal}.  

Photon rings arise from the presence of the region of spacetime where light can travel on close orbits called the photon sphere. For optically thin accretion flows, these photon spheres will generically produce images of a nested set of bring rings, often called ``photon rings'' for a variety of geometries and simulations \cite{EventHorizonTelescope:2022urf, EventHorizonTelescope:2022urf}. These photon rings are indexed by the number of equatorial crossings \cite{universal} or length of the Mino time \cite{Chang}. As the number of crossing increases, the rings become exponentially sharper and brighter, eventually converging to the asymptotic critical curve \cite{Bardeen:1973tla, Luminet:1979nyg}. The properties of the photon ring provide a sensitive probe of the black hole's mass, spin, and inclination. Furthermore, photon rings are related to the quasi-normal mode spectrum of both Schwarzschild and Kerr black holes \cite{SchwarzschildQuasi, KerrQuasi}, suggesting that while LIGO and EHT may be looking at different sources, they are probing similar physics. Furthermore, unlike the critical curve, which is not observable, the photon rings are observable, and the $n=1$ photon ring could potentially be detected in the coming decade, as I describe below.

Only recently have a photon ring's polarization properties been researched \cite{Himwich, Palumbo:2022pzj}. While the polarization properties of the photon ring are more dependent on astrophysical assumptions, like the Faraday rotation depth, they provide a largely independent avenue to measure black hole spin \cite{Palumbo_2023, Shavelle_2024}. This spin dependence is partly thanks to the impact of frame dragging and the ergosphere on magnetic fields. Relatedly, by tracing magnetic field lines in the accretion disk to the horizon, it may also be possible to directly see the impact of the ergosphere on matter, especially for systems where the accretion flow is counter-rotating relative to the black hole's angular momentum \cite{Ricarte_2022}.

High dynamic range images of black holes may be able to directly see the imprint of the horizon on the accretion flow, known as the \textit{inner shadow}. For accretion flows with subdominant emission in the interior funnel region of a black hole's jet, the black hole's horizon will create an inner central dark region in the image \cite{Dokuchaev2019, Dokuchaev2020a, Dokuchaev2020b}.

\section{Near-Future Detectability of Electromagnetic Signatures of Strong Gravity}

The $n=1$ photon ring is expected to account for $10-30\%$ of the total flux of the image \cite{EventHorizonTelescope:2022urf}. This amount of flux already implies that, at EHT resolution, the $n=1$ photon ring has a measurable signal, even in the 2017 EHT data. However, the expected width of the ring is predicted to be $\sim 1\, \mu{\rm as}$, well below the nominal resolution of the EHT ($\sim 20 \mu{\rm as}$). \cite{Broderick:2022tfu} claimed detection of the $n=1$ photon ring from the EHT M87* 2017 data. Using an astrophysically motivated model, specific properties of the $n=1$ photon ring, such as its diameter and azimuthal brightness profile, were measured. However, follow-up work \cite{TiedePhotonRings} demonstrated that the analysis used in \cite{Broderick:2022tfu} is prone to false positives and biased measurements. Furthermore, \cite{TiedePhotonRings} argued that improving the ground array does not increase the robustness of the results for this class of models. To make a robust measurement of the $n=1$ photon ring without strong astrophysical assumptions requires moving to space.

The Black Hole Explorer~\cite{Johnson:2024ttr} proposal aims to measure the $n=1$ photon ring conclusively with a $3.5\,{\rm m}$ space radio dish, connected with existing earth-based radio telescopes and, if selected, would be launch in the early 2030s. Unlike ground-based VLBI arrays, the Black Hole Explorer will provide measurements at scales where, after time-averaging, will be dominated by the $n=1$ photon ring \cite{PRCardenas, Johnson:2024ttr}. The Black Hole Explorer's unprecedented resolution will measure the $n=1$ photon ring's size, position, and brightness profile relative to the $n=0$ or direct emission. Theoretically, recent studies have explored the relation of $n=1$ photon ring properties to spacetime, demonstrating that measurements of spin, inclination, and and should be feasible \cite{PhotonRingSizes, PRShape, BroderickShadowSize, CardenasHeld, HePR, Salehi,Johnson:2024ttr}. However, what measurable low-order photon ring quantities are most sensitive to, e.g., black hole spin, mass, and inclination, is still under active research.

While photon rings provide a direct avenue to precision black hole measurements largely invariant of astrophysics, other promising observables exist. As mentioned above, the \textit{inner shadow} is the projected image of the black hole horizon as seen by an observer at infinity. While the presence of an inner shadow does depend on the particular emission geometry, studies of M87* suggest that its emission geometry may give rise to this feature \cite{EventHorizonTelescope:2021srq}. As demonstrated in \cite{ChaelInnerShadow}, the inner shadow may be observable by future VLBI arrays like the next-generation EHT \cite{Doeleman:2023kzg, ngehtscience}. Furthermore, in combination with the Black Hole Explorer, a measurement of the location of the inner shadow relative to the photon rings would potentially be another precise measurement of black hole spacetime \cite{ChaelInnerShadow}. Finally, we should mention that much of the research in the field is in a very early stage, and other sensitive probes of spacetime are likely to be discovered.

\section{Conclusion}

LIGO, the EHT, and Gravity have provided unprecedented views of black holes on horizon scales and can both probe the quasi-normal modes of black holes from entirely different perspectives. However, in the coming decade, the precision of the measurements will improve dramatically. Potential space telescopes like the Black Hole Explorer will provide the first direct measurement of the $n=1$ photon ring and a direct view into the nature of gravity around SMBHs. More than a century after their discovery, physics enters the era of multi-messenger precision tests of black holes on horizon scales.

\bibliographystyle{utphys}
\bibliography{references}

\title{Testing the Nature of Compact Objects and the Black Hole Paradigm}

\author{Mariafelicia De Laurentis\inst{1} \and Paolo Pani\inst{2}}

\institute{\textit{Dipartimento di Fisica, Universitá di Napoli and INFN Sezione di Napoli, 
Compl. Univ. di Monte S. Angelo, Edificio G, Via Cinthia,
I-80126, Napoli, Italy} \and \textit{Dipartimento di Fisica, Sapienza Università 
	di Roma and INFN Sezione di Roma, Piazzale Aldo Moro 5, 00185, Roma, Italy}}

\maketitle

\begin{abstract}
\emph{Do compact objects other than black holes and neutron stars exist in the universe? Do all black holes conform with the predictions of Einstein's General Relativity? Do classical black holes exist at all?}
Future gravitational-wave observations and black-hole imaging might shed light on these foundational questions and deepen our understanding of the dark cosmos.
\end{abstract}

\section{Motivation \& Current Status}

A common mantra in astrophysics is that \emph{compact objects} --~self-gravitating bodies with mass $M$ and radius $R$ such that their compactness $GM/(c^2 R)\sim 1$~-- must be either neutron stars or black holes~(BHs). 
This is predicted by stellar evolution and by the universality of gravitational collapse in Einstein's General Relativity~(GR)~\cite{Penrose:1964wq,Hawking:1965mf,Hawking:1966sx,Hawking:1966jv,Hawking:1967ju,Christodoulou:2008nj}.

There are strong motivations to challenge this paradigm. 
On the one hand, the behavior of matter inside ultradense stars is unknown and might involve new degrees of freedom. At least some of them might be associated with the dark matter comprising about $85\%$ of the total matter content of the universe, of which little is known, aside from its gravitational interactions.
On the other hand, BHs, once considered bizarre solutions to GR, are now pillars of high-energy astrophysics, cosmology, and gravitational-wave~(GW) astronomy. Their importance has grown to the point where the initial issues tied to these unique solutions might be easily overlooked.
BHs connect gravity, thermodynamics, quantum mechanics and astrophysics, and the paradoxes they reveal are deep and far-reaching: their Hawking evaporation is incompatible with quantum unitarity, the microstates underlying their enormous entropy are unknown, and they conceal singularities where Einstein’s theory breaks down. 

Given this state of affair, and the ever-growing wealth of data made accessible by GW observations, BH imaging, and other electromagnetic probes, it is compulsory to \emph{test the nature of compact objects} with an agnostic attitude. Our reflections will revolve around three related questions, which become progressively more radical:
\begin{itemize}
    \item \emph{Do compact objects other than BHs and neutron stars exist in the universe?}
    \item \emph{Do all BHs in the universe conform with the GR prediction?}
    \item \emph{Do classical BHs exist at all?}
\end{itemize}
We will provide a bird's-eye view on these problems and discuss the  many opportunities for this field in the years to come.


\subsection{Do compact objects other than BHs and neutron stars exist in the universe?}
In GR, any compact object with a mass exceeding a few times that of the Sun must be a BH. Observations contradicting this would imply
either new physics beyond GR or new exotic matter fields beyond the Standard Model.
Such discoveries could also offer insights into the mysterious properties of BHs (see~\cite{Cardoso:2019rvt,Maggio:2021ans} for some reviews).
 
From a phenomenological standpoint, BHs and neutron stars might just be two “species” within a broader category including \emph{exotic compact objects}~(ECOs). These objects could possess unique characteristics that enable precision searches using current and future experiments. 
Notably, some GW events are compatible with objects in the so-called lower-mass and upper-mass gap forbidden for standard stellar-origin BHs~\cite{LIGOScientific:2020zkf,LIGOScientific:2024elc,LIGOScientific:2020iuh}, and do not rule out more exotic interpretations~\cite{CalderonBustillo:2020fyi}.

A helpful guide for exploring the ECO landscape comes from Buchdhal’s theorem~\cite{Buchdahl:1959zz}, which states that, under certain conditions, the maximum compactness of a self-gravitating object is $GM/(c^2R) = 4/9$. 
This result rules out the existence of ECOs with compactness arbitrarily close to that of a BH. However, relaxing some of these conditions offers a way around the theorem and provides a framework for classifying ECOs~\cite{Cardoso:2019rvt}. In addition to some technical assumptions, Buchdahl’s theorem relies on GR, spherical symmetry, and the assumption that the matter sector consists of a single fluid that is, at most, mildly anisotropic.
A quite common feature of ECOs is the presence of strong tangential stresses to support very compact self-gravitating configurations. This is the case, for instance, of boson stars~\cite{Kaup:1968zz,Ruffini:1969qy}, gravastars~\cite{Mazur:2001fv,Mazur:2004fk,Mottola:2006ew,Cattoen:2005he}, fermion-boson stars~\cite{DelGrosso:2023trq}, ultracompact anisotropic stars~\cite{1974ApJ...188..657B,Raposo:2018rjn}, elastic stars~\cite{Alho:2022bki,Alho:2023mfc,Alho:2023ris}, and wormholes~\cite{Morris:1988cz,Visser:1995cc,Lemos:2003jb}.
Only some of the ECOs models are embedded in consistent field theories coupled to gravity (although the number of consistent models is steadily growing), in which case they are prone to be studied in their full glory, including their formation and nonlinear dynamics. 

\subsection{Do all BHs in the universe conform with the GR prediction?}
If GR deviations are relevant at the scale of compact objects (namely when $GM/(c^2R)\sim1$), the properties of BHs (and neutron stars) could differ from those predicted by GR.
Specifically, GR uniqueness theorems suggest that all properties of a BH are ultimately determined solely in terms of its mass and angular momentum, offering multiple complementary approaches to test and potentially falsify the GR BH hypothesis.
Indeed, the entire quasinormal mode~(QNM) spectrum, ringdown, multipolar structure, tidal deformability, but also all geodesics properties such as the innermost stable circular orbits and photon rings, are known functions of the BH mass and spin.

Strong-field tests of gravity largely rely on (dis)proving these predictions (see~\cite{Berti:2015itd,Yunes:2024lzm} for some reviews). In the context of testing the BH paradigm, it is useful to highlight some analogies:
\begin{itemize}
    \item several tests of GR (especially the model-agnostic ones~\cite{LIGOScientific:2021sio,Yunes:2009ke}) can be directly mapped into --~or easily adapted to~-- tests for ECOs, since also in this case one expects a deviations from the standard GR BH predictions;
    \item most gravity theories beyond GR introduce extra scales and dimensionful coupling constants, so they might introduce observable deviations only for BHs in given mass ranges. This means that supermassive BHs can be practically indistinguishable from GR ones while stellar-mass are not (as in the case of theories with higher-curvature corrections) or viceversa (as in the case of massive gravity and other theories with ultralight bosonic degrees of freedom);
    \item Some models of ECOs require modified gravity either for their existence or to have a stronger theoretical motivation. For example, ordinary matter supporting wormhole solutions beyond GR might not need to violate energy conditions~\cite{Kanti:2011jz}. In these models testing ECOs is another flavor of testing GR.
\end{itemize}

Finally, it is worth noting that standard tests of gravity typically assume \emph{vacuum}. Dirty astrophysics and environmental effects can cause false GR violations~\cite{Gupta:2024gun}. For the ordinary environments expected around BHs, however, such violations are small~\cite{Barausse:2014tra,Barausse:2014pra}. Exceptions might be orbital migration in accretion disks~\cite{Yunes:2011ws,Kocsis:2011dr,Barausse:2014tra}, or more exotic scenarios such as large dark-matter spikes around intermediate-mass BHs~\cite{Cole:2022yzw} and ultralight-boson condensates grown around BHs due to superradiance~\cite{Brito:2015oca}. 
As for all environmental effects, deviations would be source dependent. Therefore, by detecting multiple sources with similar mass and spin, it should be relatively easy to disentangle environmental effects from beyond-GR ones~\cite{Barausse:2014tra,Gupta:2024gun}. On the other hand, distinguishing whether a given source is a BH with strong environmental effects or an ECO can be more challenging and likely requires a detailed model selection.

\subsection{Do classical BHs exist at all?}
The BH hosted in M87, featured in the very first image taken by the Event Horizon Telescope~(EHT)~\cite{EventHorizonTelescope:2019dse}, and the remnant of GW150814, the first binary BH merger ever detected by LIGO/Virgo~\cite{GW150914}, have entropies of the order $10^{95}$ and $10^{80}$, respectively, and hence an enormous number of states. 
The same holds true for SgrA*, the supermassive BH at the center of our galaxy, observed with the EHT~\cite{EventHorizonTelescope:2022wkp} and by the GRAVITY collaboration~\cite{GRAVITY:2020gka}, which provided unprecedented precision in tracking the orbits of stars around this object. Like M87*, SgrA* provides crucial insights into the nature of event horizons and strengthens the case for the classical BH paradigm as predicted by GR~\cite{deLaurentis:2022oqa}. The consistency of both images, despite the difference in scale between M87* and SgrA*, offers compelling observational evidence that GR successfully describes supermassive BHs across various mass ranges~\cite{EventHorizonTelescope:2022xqj}.
GR explains all BH observations so far, but does not offer any clue about the microscopic origin of the gigantic entropy of astrophysical BHs. On the contrary, it predicts a huge discrepancy: for a given mass, angular momentum and  charge, a BH is unique~\cite{Robinson}. Furthermore, GR predicts that every BH hosts a singularity in its interior, where classical physics breaks down. Finally, BHs evaporate by emitting entangled Hawking particles in a process that violates quantum unitarity~\cite{Hawking:1976ra,Polchinski:2016hrw}.
Thus, despite the recent observational breakthroughs in GW astronomy and BH images, these \emph{entropy, singularity, and information-loss problems} remain dramatic manifestations of the incompleteness of our fundamental laws of nature.
A very general argument in quantum information theory shows that if BHs have vacuum at the horizon and ``normal'' local physics (as GR predicts), quantum unitarity is violated~\cite{Mathur:2009hf,Almheiri:2012rt,Hayden:2020vyo}. 
No small deviation from GR or effective-field-theory arguments can come to rescue~\cite{Mathur:2009hf}, so a resolution to this theoretical problem might require a radical change of the GR paradigms.
In particular, the information loss paradox might be resolved by postulating the existence of structure at the scale of the horizon~\cite{Mathur:2009hf,Almheiri:2012rt}. This finds a concrete and particularly appealing realization in the \emph{fuzzball} program of string theory, which aims at describing the classical BH horizon as a coarse-grained description of a superposition of regular quantum states~\cite{Mathur:2009hf,Bena:2022rna,Bena:2022ldq}.
The horizon-scale structure is provided by \emph{microstate geometries}: solitons with the same mass and angular momentum as a BH, but where the horizon is replaced by a smooth horizonless cap~\cite{Bena:2006kb,Bena:2016ypk,Bena:2017xbt,Bah:2021owp,Bah:2022yji}.
Besides potentially resolving all theoretical problems associated with BHs, microstate geometries are possibly the only motivated BH mimicker because, just like BHs, they: i)~have a general (albeit complicated~\cite{Kraus:2015zda,Bena:2015dpt}) formation mechanism, ii)~can exist with any mass, iii)~can be more compact than neutron stars without collapsing.
The very existence of these solutions and their unique properties hinge on nonperturbative string theory effects, the presence of nontrivial topologies and fluxes, and on the fact that these solutions are intrinsically higher dimensional. Although they are perfectly regular in higher dimensions, from a four-dimensional perspective they appear to have pathologies.

The lesson to learn here is that, in order to solve the paradoxes associated to BHs, one needs \emph{radically new effects} that cannot be captured by educated phenomenological models or ordinary parametrizations. Testing the BH paradigm would then require computing observables in these highly nontrivial theories from scratch. 


\section{Outlook}
The previous considerations suggest that, as is often the case, to test the nature of compact objects it is advisable developing \emph{both} model-agnostic searches and top-down predictions in well-motivated theories.
The former approach is broad but potentially inaccurate, while the latter is tailored to a specific theory and hence more precise.
Either ways, it is essential to model the signatures in these scenarios as accurately as possible, since current detection strategies rely heavily on matched-filtering techniques and Bayesian model comparison.

Below we list the most promising observables and smoking guns to test the nature of compact objects:

\begin{itemize}
    \item {\bf Kerr bound:}~While the mass $M$ of a BH in GR is arbitrary, its angular momentum $J$ is not and in fact limited by the Kerr bound, $J< GM^2/c$. This bound is easily (and in fact enormously!) exceeded in everyday life, e.g. by a spinning top. It would therefore be interesting to devise agnostic tests of the Kerr bound in compact objects. This is a challenging task in the strong-field regime, since quantities associated to $J$ are typically model dependent and oftentimes models, templates, and parameter estimation pipelines \emph{assume} the Kerr bound. Extreme mass-ratio inspirals detectable by LISA provide an appealing opportunity to measure the spin of the secondary object in a model-agnostic fashion~\cite{Piovano:2020ooe}, but such measurement will be very challenging due to parameter correlations~\cite{Piovano:2021iwv}.
    \item {\bf Multipolar structure:}~Due to their axial and equatorial symmetry, GR BHs have a very rigid and highly constrained multipolar structure. The difference between model-agnostic and model-specific approaches is particularly clear in this case. The model-agnostic approach is based on assuming the same symmetries as Kerr, introducing parametrized deviations from the Kerr multipole moments, in particular the mass quadrupole~\cite{Krishnendu:2017shb,Krishnendu:2019tjp,LIGOScientific:2021sio}. However, motivated by certain concrete models of microstate geometries~\cite{Bena:2020see,Bianchi:2020bxa,Bena:2020uup,Bianchi:2020miz}, one might also consider multipole moments that are vanishing in the Kerr case (e.g., current dipoles~\cite{Fransen:2022jtw}), and even breaking axial symmetry~\cite{Loutrel:2022ant}. 
    Multipole moments affect the inspiral waveforms at different post-Newtonian order~\cite{Blanchet:2013haa}. While the model-agnostic approach can be directly implemented in existing waveforms~\cite{Krishnendu:2017shb,LIGOScientific:2021sio}, the absence of symmetries in specific models makes the computation much more involved due to intrinsic precession~\cite{Loutrel:2022ant}. However, these effects break parameter degeneracies and can drastically improve measurability~\cite{Loutrel:2023boq}.    
    \item {\bf Extra charges/dipoles:}~No-hair theorems~\cite{Robinson} guarantee that Kerr BHs in GR cannot be endowed with a long list of extra matter fields. Some exceptions exist, for example in case of oscillating bosonic fields~\cite{Herdeiro:2014goa,Herdeiro:2016tmi} or electroweak hair~\cite{Gervalle:2024yxj} (see~\cite{Herdeiro:2015waa} for a review). Furthermore, various extensions of GR predict BHs with extra massless hair. Finally, although astrophysical BHs are expected to be electrically neutral, they can have extra $U(1)$ charges in beyond-Standard Model theories, e.g. in the presence of dark photons~\cite{Cardoso:2016olt}.
    In addition to modifying the BH GR solution, if compact objects are endowed with light degrees of freedom one expects extra dissipative channels during the inspiral, and non-gravitational modes during the 
    ringdown. This can be studied in a model-agnostic way within post-Newtonian theory~\cite{Barausse:2016eii}, in extreme mass-ratio inspirals for a large class of theories~\cite{Maselli:2021men}, and for ringdown~\cite{Crescimbeni:2024sam}.
    Finally, even if globally neutral, in various scenarios compact objects can have a nonvanishing dipole moment, which can also be modelled generically in some scenarios~\cite{Lestingi:2023ovn}. 
    \item {\bf Geodesics \& integrability:}~Due to special symmetries of the Kerr metric, geodesic motion around a Kerr BH is fully integrable. This property is broken by essentially any other object, even if axisymmetric. Examples include neutron stars, boson stars, BHs in modified gravity. In fact, beside GR BHs, non-integrable geodesic motion is the rule rather than the exception. 
    This fact would impact imaging, accretion, and extreme mass-ratio inspirals. In general, geodesic motion beyond GR BHs can also be chaotic~\cite{Destounis:2020kss,Destounis:2021mqv,Destounis:2021rko,Destounis:2023khj}. This is hard to model generically, but a qualitative signature would be the existence of resonances, glitches, or islands were orbital dephasing is particularly large.
    \item {\bf Tidal properties:}~If at least one of the two objects in a binary system does not have a horizon, the GW emission will be different, due to finite-size effects~\cite{Cardoso:2017cfl,Datta:2019epe,Maggio:2021uge}.
    Compact objects without an event horizon, such as neutron stars or ECOs like boson stars, exhibit different tidal characteristics compared to classical BHs. Unlike BHs, objects with a solid surface or internal structure can undergo significant tidal deformation. This effect would cause deviations in the waveform during the inspiral phase, especially in the late stages when the objects are in close proximity. Such deviations could provide key signatures that distinguish horizonless objects from standard BHs~\cite{Cardoso:2017cfl}. In addition, horizonless objects would exhibit negligible tidal heating or dissipation, significantly altering the GW signal from comparable-mass binaries~\cite{Chia:2024bwc} and extreme mass-ratio inspirals~\cite{Datta:2019epe, Maggio:2021uge,Datta:2024vll}. The detection of these tidal imprints is particularly relevant for future GW observatories like LISA and the next generation of ground-based detectors, which will have the sensitivity to probe these subtle effects. 
    \item {\bf Ringdown, QNMs, echoes:}
    The ringdown phase marks the final stage of a BH merger, during which the newly formed BH settles into a stable state by emitting GWs mostly in the form of QNMs. While the amplitudes and phases of these modes are set by the binary parameters, their frequencies and damping times are uniquely determined by the remnant's mass and angular momentum, as predicted by GR. These modes allow for precise tests of the no-hair theorem (see~\cite{Baibhav:2023clw} for a recent detailed analysis). If the remnant is a horizonless compact object or one that carries additional degrees of freedom, the ringdown signal will exhibit deviations from the GR prediction. These could manifest as deviations to the BH QNMs, different QNM excitation factors~\cite{Forteza:2022tgq} or excitation of new non-gravitational modes~\cite{Crescimbeni:2024sam}, or also as post-merger GW echoes, delayed and weaker signals caused by wave reflections near the surface or by the internal structure of the object~\cite{Cardoso:2017cqb}. The presence of such echoes would indicate the breakdown of the classical event horizon and provide strong evidence for new physics beyond GR. Testing the nature of the remnant through a multitude of (both agnostic and theory-specific) ringdown tests is a major focus of current theoretical work and observational efforts~\cite{Baibhav:2023clw,LIGOScientific:2021sio}, as they could offer key insights into the nature of compact objects, quantum gravity effects~\cite{Abedi:2020ujo}, or the existence of new fundamental fields and interactions.
    \item {\bf Accretion:}~ Accretion flows around compact objects offer critical observational opportunities to study their nature (see, e.g.,~\cite{Zulianello:2020cmx}). In the case of horizonless objects, infalling material would not disappear across an event horizon but could instead interact with a surface, leading to distinct signatures in the dynamics of the accretion disk. These interactions could manifest as increased luminosity from the surface~\cite{Broderick:2007ek}, different thermal emission properties, or variability in the radiation due to the accumulation of matter~\cite{EventHorizonTelescope:2022xqj}. Accretion-induced signals in such scenarios might reveal deviations from the classical BH predictions, where matter is absorbed without observable effects near the horizon, due to the infinite redshift. The presence of a surface would alter the energy dissipation process, potentially producing additional observational features, such as enhanced emission from the innermost regions of the accretion disk or even secondary signals reflecting the interaction of the infalling matter with a solid boundary. 
    The recent multi-wavelength campaigns, including the EHT's continued observation of SgrA* and M87*, are already beginning to constrain accretion models, offering a direct test of whether the observed objects conform to classical BHs or exhibit signs of some structure at the horizon scale.
    In this context, it would be important to improve accretion models onto ECOs, to include realistic dissipation effects.
    \item {\bf Merger counterparts:}~The absence of an event horizon or the presence of additional degrees of freedom can lead to the generation of observable electromagnetic or neutrino counterparts during a merger event. This contrasts sharply with a classical BH, where the event horizon prevents any significant emission of such radiation during or after the merger. For horizonless objects, instead, matter that would otherwise be absorbed by a BH could interact with the object's surface or internal structure. The interaction between the merger debris and the surface of these exotic objects could lead to a variety of post-merger emissions, potentially akin to the case of neutron-star mergers~\cite{Metzger:2011bv,Metzger:2019zeh} but occurring also for heavier mergers and potentially probing higher-curvature and higher-redshift regions. These emissions might range from prompt flashes of electromagnetic radiation to more sustained afterglows, depending on the object's specific properties, such as surface composition, compactness, and temperature. In addition, shock waves generated by the collision of the compact objects might accelerate particles to relativistic speeds, producing detectable high-energy emissions, including X-rays or gamma rays. Similarly, neutrinos could be produced in significant quantities if the dense matter interacts with the horizonless surface during or after the merger, offering a unique multi-messenger signal.
    Finally, novel degrees of freedom might be present or excited, resulting in dark radiation.
    The observation of such counterparts (or lack thereof) would provide critical insights into the nature of the merging objects. A detection of electromagnetic or neutrino signals in coincidence with GW events would strongly suggest the presence of a horizonless compact object or some other exotic physics beyond the standard BH paradigm. On the other hand, the absence of these signals would reinforce the classical BH interpretation, in line with the predictions of GR, or at least constrain ECO models even further. Future multi-messenger observations, combining GWs, electromagnetic radiation, and neutrino detections, will play a key role in distinguishing between these possibilities and testing the fundamental nature of compact objects in the universe.
\end{itemize}
%

For all these observables, strong efforts are ongoing to develop model-agnostic tests, improve existing ones to make them more accurate, and in parallel improve the theoretical understanding of well-grounded models to compute the above observables in a motivated top-down approach, possibly looking 
``outside the lamppost'' for more radical deviations.
Eventually, for selected well-motivated theories, it is imperative to develop models and templates that are as accurate as the current ones for GR BHs, in order to perform a robust data-driven model comparison, possibly using a synergy of multiwavelength/multimessenger probes of BH candidates across various mass ranges.
Finally, in case some tension with the standard paradigm is detected in future experiments, one would need to carefully examine degeneracies with model systematics, experimental errors, and astrophysical uncertainties~\cite{Gupta:2024gun}.

\vspace{1cm}

{\bf Acknowledgments.}
PP is partially supported by the MUR PRIN Grant 2020KR4KN2 ``String Theory as a bridge between Gauge Theories and Quantum Gravity'' and by the MUR FARE programme (GW-NEXT, CUP:~B84I20000100001).  M.D.L. acknowledges INFN Sez. di Napoli (Iniziativa Specifica TEONGRAV.

\bibliographystyle{utphys}
\bibliography{references}

%
\part{Quantum black holes}

\title{Some Thoughts about Black Holes in Asymptotic Safety}

\author{Alessia Platania}

\institute{\textit{Niels Bohr International Academy, The Niels Bohr Institute, Blegdamsvej 17, DK-2100 Copenhagen \O, DENMARK}}

\maketitle

\begin{abstract}
We discuss the current status of black holes within quantum gravity approaches based on quantum field theory, and particularly asymptotic safety. We highlight key open questions in the field, including the difficulty of deriving global solutions from the effective action, the role of matter in constraining the set of all possible alternatives to classical black holes, the importance of the formation probability, and its implications for both conceptual and practical aspects.
These challenges highlight some necessary developments in the field and single out a few future directions, which could be promising to progress our understanding of black holes in quantum gravity.
\end{abstract}

\section{Introduction}

\textbf{Quantum gravity and asymptotic safety}---The search for a consistent theory of quantum gravity (QG) remains one of the most challenging problems in modern physics. While General Relativity (GR) successfully describes gravity at large scales, spacetime singularities and the consequent loss of predictivity indicate that the classical description breaks down at short distances. Singularities thus signal the need for a (quantum) theory capable of describing gravity across all scales. Among various approaches to QG, in the asymptotic safety (AS) scenario~\cite{Knorr:2022dsx, Eichhorn:2022gku, Morris:2022btf, Martini:2022sll, Wetterich:2022ncl, Platania:2023srt, Saueressig:2023irs, Pawlowski:2023gym,Bonanno:2024xne} gravity is described as a quantum field theory (QFT) whose ultraviolet (UV) completion is an interacting theory---a non-trivial fixed point of the gravitational renormalization group (RG) flow. This fixed point provides a mechanism for making the theory finite and predictive at all energy scales, addressing the issue of renormalizability posed by traditional perturbative approaches. The strength of the AS approach lies in its ability to describe gravity as a QFT across all scales and its inherent predictive power: the specific UV completion is not a choice, rather a property of the RG flow; the infrared (IR) limit from such a completion is a restricted set of effective field theories (EFTs), parametrized by two or three free parameters; the resulting landscape of EFTs is compatible with the Standard Model, with some of its masses and couplings being either predicted or postdicted by the theory~\cite{Eichhorn:2022gku,Pawlowski:2023gym}. While investigating the implications of AS for particle physics has a clear path and is an area of active research, its predictions in the context of black holes and cosmology remain less well understood (see~\cite{Bonanno:2017pkg,Platania:2020lqb,Eichhorn:2022bgu,Platania:2023srt,Bonanno:2024xne} for reviews). 

\textbf{Black holes and asymptotic safety}---One of the central challenges in QG is understanding the nature of black holes. Vice versa, black holes offer an arena for exploring the implications of QG theories. QG is believed to modify  classical black holes, smoothing out the singularities that plague the classical description, and providing a top-down explanation for their thermodynamical properties. Even in the absence of a complete solution, we may still be able to extract important insights about black holes based on the general properties of the underlying theory. 
In AS, the existence of a UV fixed point, akin to the case of Quantum Chromodynamics, indicates that gravity exhibits anti-screening behavior---its strength diminishes at high energies. This hallmark of AS suggests that the singularities predicted by GR could be avoided. For example, one might argue that if gravity weakens at high energies, the gravitational collapse of a star would be halted~\cite{Bonanno:2017zen,Bonanno:2023rzk,Delaporte:2024and}, potentially preventing the formation of a singularity or at least mitigating its severity~\cite{Tipler:1977zzb}. Gravitational anti-screening can be qualitatively captured through the variation of the Newton coupling with respect to the RG scale. Although this running is unphysical~\cite{Bonanno:2020bil}, if a decoupling mechanism is in effect~\cite{Borissova:2022mgd}, it may provide qualitative insights into the consequences of quantum effects via the so-called RG improvement.
If the mechanism works, an effective, coordinate-dependent Newton coupling in some generalized Einstein equations encodes (an approximation to) the effects of higher derivative corrections generated by quantum fluctuations. Such corrections are expected to manifest in the effective action on general grounds, irrespective of the specific UV completion, while the physical Newton coupling is constant in the latter. Due to computational constraints, RG improvement has been widely employed over the past few decades to explore the potential implications of AS in black hole physics and cosmology. However, direct derivations are essential for making conclusive, quantitative statements. Such derivations necessitate solving the quantum field equations associated with the effective action and deriving the effective action from the ground up in the first place. Recent progress in the field has made these top-down approaches increasingly viable. Although definitive answers regarding the nature and realization of quantum black holes within the framework of AS are still far from reach, the field has made significant strides in this direction~\cite{Knorr:2022kqp,Pawlowski:2023dda,Daas:2023axu}. 

\section{Critical thoughts and key open questions}

\textbf{Possible solutions and their realization---}In fundamental QG approaches that are based on QFT, deriving black hole solutions requires the effective action, which encodes quantum corrections in the form of higher-derivative terms. However, finding \emph{global} solutions to the corresponding field equations is notoriously difficult. A case in point is the example of black holes in Stelle gravity, where even with four derivatives only, solving the field equations is numerically very demanding~\cite{Lu:2015cqa,Bonanno:2019rsq,Daas:2023axu}. This highlights the need for new mathematical tools and methods to solve field equations with higher derivatives within reliable approximations.

Focusing on QFT-based fundamental QGs---which entails disregarding non-geometric configurations---there are some general considerations one can start with. The geometrical realizations of the alternatives to classical black holes are likely only four~\cite{Carballo-Rubio:2019nel}, but not all of them can be consistently realized in QFT-based approaches:
\begin{itemize}
    \item Regular black holes avoid singularities at the core by the formation of an internal horizon. However, at least in vacuum, they lead to problematic remnants~\cite{Giddings:1992hh, Susskind:1995da}, and to mass inflation instabilities.
    \item Quasi-black holes, i.e., ultra-compact objects with a surface, avoid singularities by remaining horizonless. However, EFT suggests that horizons always form from large collapsing objects. Hence, to replace the entire black hole population with compact objects of arbitrary masses one would need non-perturbative effects at large distance scales or even IR non-localities. Alternatively, ultra-compact objects could only offer a viable alternative to Kerr black holes when their masses are Planckian~\cite{Bonanno:2019ilz,Arrechea:2021xkp}.
    \item Wormholes are a consistent alternative to classical black holes, avoiding singularities by the presence of a throat, but their dynamical formation from a gravitational collapse might require
    topology changes or non-smooth geometries~\cite{Carballo-Rubio:2019nel}. These features are unlikely realized in QFT-based QGs~\cite{ASST-to-appear}.
    \item Bouncing solutions, e.g., black-to-white hole spacetimes, typically replace GR singularities with integrable ones and have a single apparent horizon~\cite{Bonanno:2017zen,Han:2023wxg}. These solutions seem the most plausible alternative from the standpoint of the issues presented above. Yet, their rotating incarnation could develop a Cauchy horizon, and hence some of the issues of regular black holes (see however~\cite{Rignon-Bret:2021jch}).
\end{itemize}
Should none of the above solutions work in reality, we would be left with two options: either non-geometric configurations (quantum states not corresponding to classical spacetime geometries) have to be considered, which are typically realized in approaches to QG that go beyond QFT, or the \textit{role of matter in describing black holes} is more essential than one might think. This question has not been investigated thoroughly and has the potential to lead to breakthroughs in the field. 

\textbf{Dynamical formation, formation probability, and singularity resolution---}A common argument is that if a particular structure does not form dynamically---such as from a gravitational collapse---then it might not be physically relevant. However, in the context of QG, \emph{this perspective may need to be broadened: as long as a configuration is a solution of the theory, it has the potential to form}, no matter how small the formation probability is. For instance, even if a particular black hole solution is not typically produced through classical processes, quantum effects could still lead to its formation, e.g., via quantum tunneling. 

Looking at the formation probability, rather than the dynamical formation via a gravitational collapse also has a more practical advantage: it may be easier to understand whether a particular configuration is suppressed by quantum fluctuations (e.g., by  path-integral arguments~\cite{Borissova:2020knn,Borissova:2023kzq}), than numerically evolving field equations with different initial conditions to see if a given black-hole-mimicker candidate can result from a collapse. 

Moreover, in this context, it is plausible that QG allows for a whole spectrum of black-hole-like solutions. A relevant question then becomes whether all these spacetimes must be fully consistent, or if it is acceptable for some solutions to be inconsistent, provided their realization is highly suppressed. This adds another open conceptual question into the game: \textit{whether the theory needs a mechanism to strictly exclude all inconsistent solutions or if their rare occurrence could be acceptable}. 
If this is the case, the dynamical suppression of certain spacetimes, coupled with the requirement of consistency of the emerging effective metric, may open up new ways to indirectly constrain QG in the UV. A prime example is the constraints on QFT-based approaches to QG that are induced by requiring singularity resolution~\cite{Borissova:2020knn,Borissova:2023kzq,Koshelev:2024wfk}: an infinite number of higher derivatives---aka, UV non-locality (or semi-locality), is necessary at the level of the bare action in order to fully suppress singular spacetimes in the path integral, and thus to avoid their dynamical formation. In this sense, singularity resolution may disfavor approaches to QG whose bare action is local, e.g., quadratic gravity. 
These are example of how the consistency of black holes can provide non-trivial constraints on the fundamental theory.

\section{Summary: Future directions and necessary developments}

Black holes are crucial testing ground for QG, especially from a theoretical standpoint: QG has to resolve singularities, tell us how black holes (or their alternatives) look like, and eventually explain black hole thermodynamics from the top-down. Within AS, gravitational anti-screening weakens gravity at high energies, offering a simple mechanism to prevent singularities from forming in nature. Going beyond singularity resolution, however, is quite challenging: the direct computation of black hole solutions in QFT-approaches to QG involves deriving the effective action from the ground up, and solving the corresponding field equations, which, in general, involve an infinite tower of higher-derivative corrections. This highlights the need to develop new mathematical and numerical tools to construct \emph{global} solutions to the field equation within reliable approximations~\cite{Figueras:2024bba}. Partially because of these challenges, several alternative black hole models, such as regular black holes, compact objects, wormholes, and models with integrable singularities, have been proposed. We remarked that each of these presents different open questions and degrees of feasibility within QFT-based approaches to QG. At the same time, as the UV theory includes matter and other fundamental interactions, these should be taken into account to investigate structure and features of quantum black holes. It is plausible that the addition of matter into the game could eliminate (or suppress) some possibilities and introduce (or enhance) others. The role of matter in black hole models, albeit important, has not been thoroughly explored in the context of QG: this is a research direction that needs to be pursued. 

One other central question is the extent to which QG allows for multiple black hole solutions and whether all these solutions must be physically consistent. A typical argument in the literature of black hole alternatives is that if a solution does not form dynamically, for instance through gravitational collapse, it might be irrelevant. However, in a quantum theory, the question is more delicate: if a configuration is a solution of the theory, it can in principle form with some non-zero probability. A related open question is whether QG must ensure that all solutions are fully consistent, or if it is acceptable for some solutions to be inconsistent, provided they are formed with very low probability. This uncertainty suggests that the formation probability of different solutions, rather than the effective dynamics alone, may play a critical role in determining their relevance within the theory.

The questions above highlight how the interplay between QG and black hole physics is crucial to constrain both: the consistency of QG constrain the type of black hole alternatives that can be possibility realized in nature, while the consistency of black holes constrain the structure of the UV physics, e.g., its (non-)locality~\cite{Borissova:2020knn,Borissova:2023kzq,Koshelev:2024wfk}. Much work remains to be done: many of the key questions are under-explored, while the resolution of others requires developing new tools to construct quantum black holes beyond the EFT regime. Connecting theory and observations remains the biggest challenge; we hope that, in view of the increasing amount of data from astrophysics and cosmology, this gap will shrink in coming years.

\section*{Acknowledgements}

The research of A.P. is supported by a research grant (VIL60819) from VILLUM FONDEN.

\bibliographystyle{utphys}
\bibliography{references}

\title{Black Hole Evaporation in Loop Quantum Gravity}
\author{Abhay Ashtekar}

\institute{\textit{Physics Department, Penn State, University Park, PA 16802, USA} \and \textit{Perimeter Institute for Theoretical Physics, 31 Caroline St. N, Waterloo, ON N2L 2Y5, Canada}}

\maketitle

\begin{abstract}

The conference \emph{Black Holes Inside and Out} marked the 50th anniversary of Hawking's seminal paper on black hole radiance. It was clear already from Hawking's analysis that a proper quantum gravity theory would be essential for a more complete understanding of the evaporation process. This task was undertaken in Loop Quantum Gravity (LQG) two decades ago and by now the literature on the subject is quite rich. The goal of this contribution is to summarize a mainstream perspective that has emerged. The intended audience is the broad gravitational physics community, rather than quantum gravity experts. Therefore, the emphasis is on conceptual issues, especially on the key features that distinguish the LQG approach, and on concrete results that underlie the paradigm that has emerged.  This is \emph{not} meant to be an exhaustive review. Rather, it is a broad-brush stroke portrait of the present status. Further details can be found in the references listed.
\end{abstract}

\section{Introduction}
\label{s1}
Hawking's discovery \cite{swh1} that isolated black holes emit quantum radiation which is approximately thermal at late times used  quantum field theory on a classical, background space-time of a collapsing star (see the left panel of Fig. 1). The calculation involved three key approximations: \emph{(i)} Space-time geometry can be treated classically; \emph{(ii)} Quantum fields can be regarded as `test fields' so that their back reaction on space-time geometry can be ignored; and, \emph{(iii)} The matter field which collapses is classical, \emph{distinct} from the test quantum field considered.%
\footnote{\label{fn1} The third assumption is not emphasized in the literature probably because it is harmless in the external field approximation. However, it is a rather severe limitation in the subsequent discussion of `information loss', or, of `unitarity of quantum dynamics'. For these considerations, we need a closed system: we need to able to specify the full incoming \emph{quantum} state --including that of collapsing matter-- on $\scrim$ (or on $\scrim \cup \inot$).}
A truly remarkable calculation then showed that, if the incoming state on $\scri^{-}$ for the quantum field is the vacuum, the outgoing state at $\scri^{+}$ is a mixed state which, at late times, is thermal. 
 \begin{figure} 
  \begin{center}
    \begin{minipage}{1.5in}
      \begin{center}
        \includegraphics[width=1.3in,height=2.2in,angle=0]{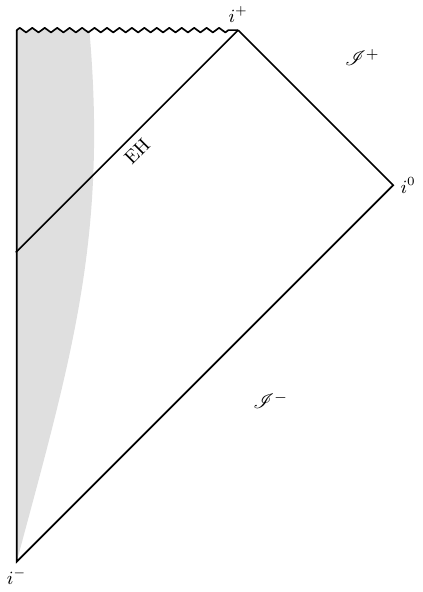} \\ {(a)}
         \end{center}
    \end{minipage}
   \hspace{.5in}
    \begin{minipage}{1.5in}
      \begin{center}
       \includegraphics[width=1.3in,height=2.4in,angle=0]{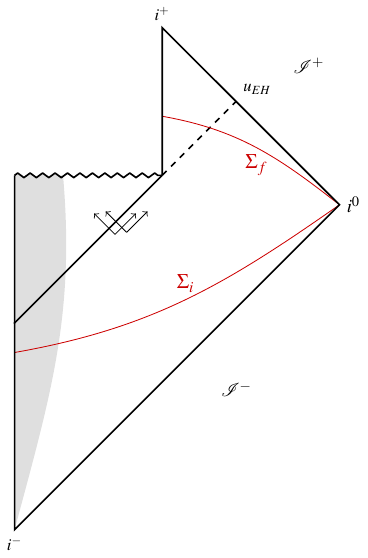} \\ {(b)}
        \end{center} 
   \end{minipage}
      \hspace{.5in}
       \begin{minipage}{1.5in}
        \begin{center}
         \includegraphics[width=1.3in,height=2.2in,angle=0]{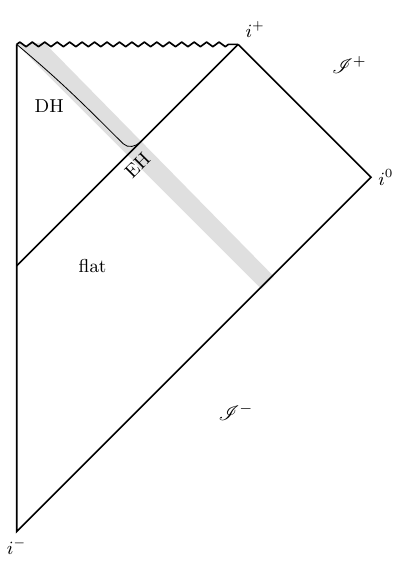} \\ {(c)}
          \end{center}
  \end{minipage}
\caption{\footnotesize{ (a) \emph{Left Panel}: The Penrose diagram of a collapsing star, used by Hawking in his calculation in the external potential approximation.
(b) \emph{Middle Panel}: The Penrose diagram proposed by Hawking in 1975 to incorporate the back reaction of an evaporation black hole.
(c) \emph{Right Panel}: The classical Vaidya space-time depicting a null-fluid collapse.}}
\vskip-1cm
\label{fig:1Ashtekar}
\end{center}
\end{figure}
To remove assumption \emph{(ii)}, one has to include the back reaction of the quantum radiation on the classical space-time geometry. A natural set of heuristics led Hawking to propose that when the back reaction is included, the space-time diagram in the  Panel (a) of Fig. 1 would be replaced by the one in Panel (b): To compensate for the mass lost through quantum radiation, the black hole would lose mass and the singularity would not extend all the way to $\scrip$ but terminate in the space-time interior, leaving us with a Minkowski metric in the upper triangular region. In this depiction, a singularity persists and serves as a `sink of information'. Put differently, because of the singularity, $\scrip$ would clearly be not the entire future boundary of space-time and therefore the S-matrix from $\scrim$ to $\scrip$ would not be unitary but would have to be replaced by the so-called \$ matrix \cite{swh2}. Even though 5 decades have passed since Hawking's discovery, we still do not have a detailed calculation of the back reaction even today. But the Panel (b) of Fig.1 continues to be widely used. Interestingly, this is done both to argue information is lost \cite{uw} and to construct mechanisms for it not to be lost (but to appear on the portion of $\scrip$ to the past of the retarded time $u_{EH}$ in the figure) \cite{marolf,amps}! 

The LQG perspective is that this Penrose diagram is incorrect. In its place, a new paradigm was introduced about 2 decades ago \cite{aamb} and most of the LQG work on black hole evaporation has been devoted to develop it systematically through innovative constructions and detailed calculations. The paradigm is based on two key observations:\vskip0.05cm
\emph{(1)} The emphasis on event horizons (EHs) that the Penrose diagram in Panel (b) places is not justified in quantum gravity. To determine if a space-time admits an EH, one needs complete knowledge of space-time geometry of the evaporating black hole to infinite future and only full quantum gravity can provide this information. One should not \emph{presuppose} the nature of this geometry; one must let evolution equations of the appropriate theory determine it. As Matt Visser reminded us in his talk, in the GR-17 conference in Dublin, Hawking emphasized that ``a true event horizon never forms". This is precisely the LQG viewpoint: there is no event horizon either in the semi-classical space-time nor in the expected space-time geometry in full quantum gravity.\vskip0.05cm  
\emph{(2)} Singularities of classical general relativity are signposts of the limitation of classical general relativity, and gates to physics beyond Einstein. In LQG there are strong reasons to expect that all space-like, strong curvature singularities will be naturally resolved by quantum geometry effects. Consequently, the quantum space-time would extend beyond the putative classical singularity, providing a path way for information recovery on the larger $\scrip$  of the quantum extended space-time.\\
It is interesting to note that Hawking changed his mind about the Panel (b) of Fig.1 in 2016  and replaced it with a Penrose diagram that has neither a singularity not an event horizon \cite{hps}!

This article is organized as follows. In section \ref{s2} we discuss in some detail main results that motivate the LQG perspective on black hole horizons and singularities. Specifically, we will address the following questions:  (i) What should replace EHs in the description of black holes in classical and quantum gravity? What is it that forms in a gravitational collapse and evaporates due to quantum emission? and, (ii) How does quantum Riemannian geometry of LQG lead to the resolution of the black hole singularity? What is the nature of the geometry in the quantum extension of space-time in the Planck regime? In section \ref{s3}, we will use these concrete results as pointers to suggest a pathway to address apparent paradoxes that arise in the analysis of the issue of information loss, first in the semi-classical theory and then in full quantum gravity. In section 4 we will summarize the main results and discuss the issues that remain. 

There is some  overlap with review articles \cite{aa-eva,aosrev,aaeb} where further details can be found.

\section{The new elements of the LQG Perspective} 
\label{s2}

As explained in section \ref{s1}, there are two key points on which LQG investigations differ from the  commonly used narrative enshrined in Hawking's original paradigm depicted in the Panel (b) of Fig. \ref{fig:1Ashtekar}. In this section we provide further details. In the first part, \ref{s2.1}, we discuss dynamical horizons (DHs) that can be used to represent black holes both in the formation and evaporation process. In the second part, \ref{s2.2}, we discuss the resolution of the Schwarzschild singularity through quantum geometry effects of LQG and summarize the key features of the resulting quantum corrected geometry.

\subsection{From EHs to Quasi-local Horizons}
\label{s2.1}
In the 1970s when black holes were first analyzed using global techniques, EHs played a seminal role and led to several interesting results. Perhaps the most important among them is the area law established by Hawking \cite{swh3} that made the similarity between the laws of black home mechanics and laws of thermodynamics compelling. Therefore subsequently, black holes were generally regarded as synonymous with EHS. 

However, as we already noted, the notion is very global: EH is the future boundary of the space-time region $J^-(\scrip)$ from which future directed causal signals can be sent to $\scrip$. Therefore to determine if a space-time admits an EH, and then to locate it, one needs the knowledge of space-time geometry to the infinite future. Already in classical general relativity, this feature prevents one from using  EHs \emph{during} numerical simulations of binary black hole mergers: One cannot use EHs to locate the progenitor black holes, nor to say when the merger happens, nor to locate and study properties of the final black hole remnant. In these simulations the EH can only be reconstructed only \emph{at the end}, as an afterthought thought. In quantum gravity, the role of EHs is even more dicey: since as of now we do not have even the adequate equations to construct the complete space-time of an evaporating black hole, we cannot look for it evan as an afterthought! 

It has long since been realized that EHs have an even more severe limitation: they are teleological and therefore inherit some spooky features. For example, an EH may well have just formed and grown in size in the very room where you are sitting, in \emph{anticipation} of a gravitational collapse in this region of our galaxy a billion years from now.  This feature is concretely realized in the null fluid collapse of a Vaidya space-time shown in the  Panel (c) of  Fig. 1, where the event horizon forms and grows in the flat region of space-time, in anticipation of the null fluid collapse, even though nothing at all is happening in the flat part of space-time. And null fluids are not necessary to illustrate this limitation of EHs. Recent results of Kahle and Ungar show that the same phenomenon occurs if one uses Vlasov fluids (that emerge from $i^-$ rather than $\scrim$, see Fig. 1 of \cite{kh}). 

These limitation motivated the introduction of \emph{quasi-local horizons} (QLHs) to describe black holes and their dynamics in a manner that not only avoids teleology but leads to a framework that is directly useful in simulations to locate black holes and to extract physics from numerical outputs. (See in particular \cite{hayward1,ih-prl,ak1,ak2,akrev,boothrev,jjrev}.) In the black hole evaporation process, one can locate the QLHs knowing only the part of space-time where semi-classical approximation holds (i.e., the the black hole continues to be macroscopic) and relate the changes in their geometry to physical processes. One does not need to know the space-time to infinite future. And, as we will see, some qualitatively new features arise already at the semi-classical level. 

\subsubsection{QLHs in classical general relativity}
\label{s2.1.1}

One begins with the notion of a \emph{marginally trapped surface}(MTs) $S$: a closed 2-manifold (which we will take to be topologically $\mathbb{S}^2$) for which one of the (future-directed) null normals, say $\ell^a$, is expansion-free: $\theta_{(\ell)} =0$. The notion is quasi-local rather than local because the expansion has to vanish on the entire 2-sphere $S$. A QLH is a 3-manifold $\mathfrak{H}$ that is foliated by MTTs. (Therefore QLHs have also been referred to as marginally trapped tubes (MTTs) \cite{aagg}.) In striking contrast to EHs, the definition of QLHs refers only to the space-time geometry in an infinitesimal neighborhood of $\mathfrak{H}$ and all their properties can be deduced just from this local geometry. Therefore there is no teleology. In particular, you can rest assured that there is no QLH contained in your room!

\begin{figure} 
  \begin{center}
     \begin{minipage}{1.5in}
      \begin{center}
       \includegraphics[width=1.7in,height=1.9in,angle=90]{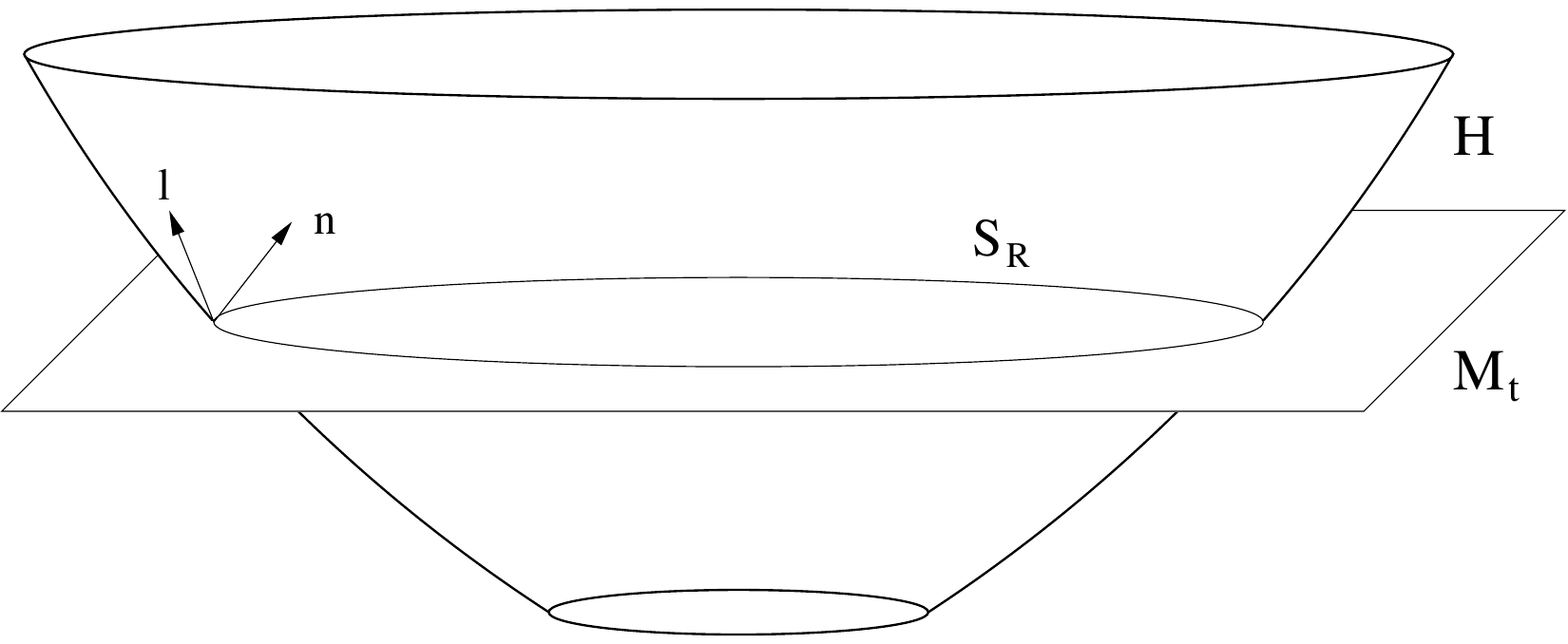}\\ (a)
         \end{center}
    \end{minipage}
   \hspace{.5in}
    \begin{minipage}{1.7in}
      \begin{center}
       \includegraphics[width=1.7in,height=1.7in,angle=0]{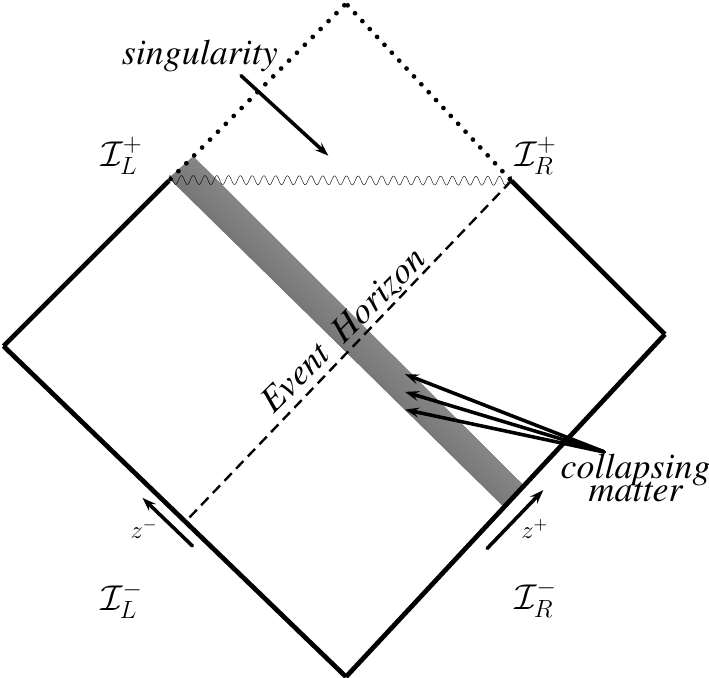} \\ {(b)}
        \end{center} 
   \end{minipage}
      \hspace{.5in}
       \begin{minipage}{1.7in}
        \begin{center}
         \includegraphics[width=1.7in,height=1.7in,angle=0]{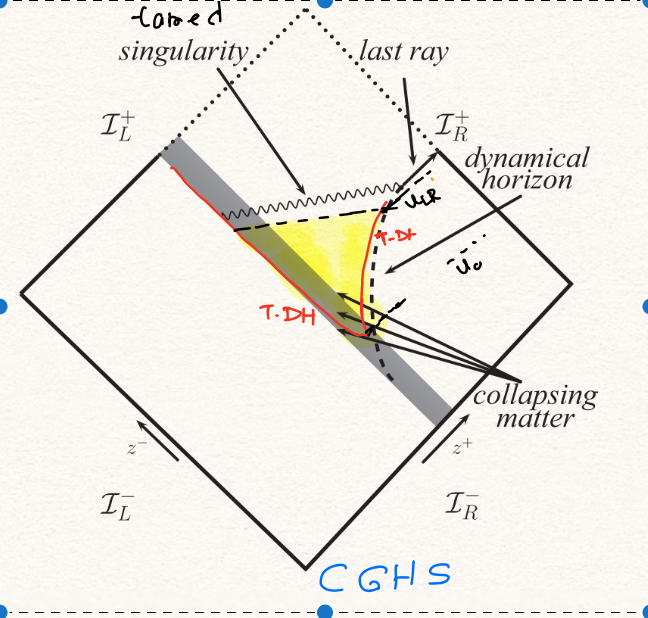} \\ {(c)}
          \end{center}
  \end{minipage}
\caption{\footnotesize{ (a) \emph{Left Panel}: A dynamical horizon $H$, foliated by marginally trapped surfaces. A typical leaf is marked $S_R$ and its two null normals are denoted by $\ell^a$ and $n^a$.  (b) \emph{Middle Panel}: The Penrose diagram of a classical 2-d black hole formed by the collapse of a massless scalar field coming in from $\scrim$ depicted by the shaded region. Space-time metric is flat to the past of the shaded region and represents a static black hole to the future. Again, the event horizon forms and evolves in the flat region of space-time.  
(c) \emph{Right Panel}: The semi-classical extension of space-time. A trapping dynamical horizon T-DH (depicted in red) forms, is space-like, and grows as the matter collapses. At the end of this process, T-DH becomes time-like and starts shrinking in area due to the influx of negative energy. The shaded yellow region is semi-classical and trapped. To its future one needs full quantum gravity. }}
\vskip-0.5cm
\label{fig:2Ashtekar}
\end{center}
\end{figure}

Of particular interest to the discussion of black hole evaporation is a sub-case of QLHs called \emph{dynamical horizons} (DHs) $H$ that are nowhere null and on which the expansion $\theta_{(n)}$ of the other null normal $n^a$ to the MTSs is either everywhere positive or negative. (See the Panel (a) of Fig. \ref{fig:2Ashtekar}).  
\footnote{This notion is a generalization of the original one \cite{ak1,ak2} which was geared to the study of QLHs that are most useful to numerical simulations of black hole mergers. The one used in this review accommodates evaporating black holes as well. It can be further generalized by weakening the `nowhere null' condition \cite{akrev2}. We use the stronger notion just to keep the discussion brief.}
It follows immediately that the area of MTSs is a monotonic function on $H$. Therefore we can use the areal radius $R$ as a coordinate on $H$ so that the MTSs are given by $R= {\rm const}$. Interestingly, one can now show that the change in area is directly related to the physical processes \emph{occurring at the DH}. For example, in the case when the DH is space-like, area increases along the projection of $\ell^a$ into the DH and the difference in the areal radius $R_1$ and $R_2$ of two MTSs is given by the flux of energy into the portion $\Delta H$ of the DH bounded by the two cross-sections \cite{ak1,ak2}:
\begin{equation}\label{balance} \f{1}{2G}\, (R_2 - R_1) =  \underbrace{\int_{\Delta H} N\, T_{ab} \ell^a \hat{\tau}^b \, \rmd^3V}_{\hbox{{\rm Matter energy flux}}} + \underbrace{\f{1}{16\pi G}\, \int_{\Delta H} N\, \left(
|\sigma|^2 + 2 |\zeta|^2 \right)\, \rmd^3V}_{\hbox{\rm GW\, energy flux}} \end{equation} 
\vskip0.1cm
\noindent Here `Energy' is defined relative to the causal vector field $N\ell^a$, where $N = |\partial_R|$ (recall: $\ell^a$ is the distinguished causal vector field on $H$). There is a similar formula in the case when $H$ is time-like, again only involving physical processes on the portion $\Delta H$ of $H$,  but now the area \emph{decreases} along the projection of $\ell^a$ into $H$ \cite{ak2}. Thus, not only does the key `area law' that elevated the status of EHs in the 1970s continue to hold for DHs, \emph{but it holds in a stronger sense}. For EHs we only have a qualitative statement that the area cannot decrease. As the example of Vaidya space-time vividly brings out, the change in not related to any physical processes near the EH since the area can grow even in flat space-time. This does not happen with DHs: Now we have a \emph{quantitative} statement (\ref{balance}) that relates the change in area to local physical processes occurring on the relevant portion $\Delta H$ of $H$. In particular, there are no DHs whose 2-sphere cross-sections lie in flat space-time, in striking contrast to EHs.
\smallskip

\emph{Remark:} Note that DHs  are 3-dimensional sub-manifolds $H$ in space-time in their own right, 
defined using structures that refer just to these sub-manifolds; in particular, one does not need a 3+1 slicing of space-time by Cauchy surfaces. The foliation by MTSs is intrinsic to $H$. Furthermore, the foliation on any given DH is unique --$H$ does not admit two distinct foliations by MTSs \cite{aagg}.  Thus the notion of a DH is distinct from that of an \emph{apparent horizon} --the `outermost' MTS on a Cauchy surface. Unfortunately, the term `apparent horizon' is often employed in the contemporary literature to emphasize that the structure used is different from an EH or a Killing horizon, when what one actually has is either an MTS or a DH, without any reference to a Cauchy surface! While the authors are well aware of which structures are actually used, this misuse of terminology can cause unnecessary confusion. Indeed, this occurred in the BHIO conference in 2 talks (see chapters~\ref{sec:wald} and ~\ref{sec:visser}). 
\smallskip

For completeness, let us consider the complementary QLHs $\mathfrak{H}$ where the underlying 
{\smash{3-manifold}} is null. These QLHs are called  \emph{non-expanding horizons} (NEHs)  and denoted by $\Delta$. In this case, the null normal $\ell^a$ is also tangential to $\Delta$; \emph{every} cross-section of $\Delta$ is expansion-free, and (if the null energy condition holds, then) the intrinsic (degenerate) metric on $\Delta$ is Lie-dragged by $\ell^a$. Of particular interest is the sub-case of NEHs is \emph{isolated horizons} on which $\ell^a$ also Lie-drags the intrinsic derivative operator on $\Delta$ \cite{ih-prl}. As the name suggests, in the context of black holes, one can think of IHs as depicting boundaries of black holes that are in equilibrium. They are more general than Killing horizons. For example, the Robinson-Trautman and Kastor-Traschen solutions admit IHs although there is no Killing field with respect to which they can be Killing horizons \cite{pc,kt}. While it is only the DHs that are directly relevant to the analysis of the black hole evaporation process, in the long adiabatic phase of evaporation during which the process is \emph{very} slow, the exact DH is well-approximated by a perturbed NEH \cite{akkl2}.

\subsubsection{DHs in semi-classical gravity: An example}
\label{s2.1.2}

As remarked in footnote \ref{fn1}, a proper analysis of the issue of `information loss' and `unitarity of the S-matrix', requires a closed system. The widely considered case of a stellar collapse does not  satisfy this criterion because it is very difficult, if not impossible, to specify the incoming \emph{quantum} state of the star in the analysis. However, this can be rectified by considering a massless scalar field collapse and using, say a coherent state peaked at an appropriate infalling classical scalar field that collapses to form a black hole. Therefore from now on we will focus on this case. 

Now, already in classical general relativity, the problem of studying this collapse is quite complicated especially because of the associated critical phenomena. But there are some slightly simplified models that are in fact exactly soluble. Perhaps the most well-known among these is the Callen, Giddings, Harvey, Strominger (CGHS) model \cite{cghs}. It describes the spherical collapse of a scalar field in a 4-d theory of gravity that differs from general relativity. However, as is well-known, one can reduce the spherical symmetric sectors of such 4-d theories to  2-d theories in which the dynamical variables are the 2-d metric $g_{ab}$, a dilation $\phi$ (where $R = e^{-\phi}$ is the radius of 2-spheres in the 4-d theory), and massless scalar fields $f$; the first two capturing the information in the 4-d metric, and the third denoting the 4-d scalar field. In the spherically symmetric sector, the CGHS action differs from that of the spherical reduction of general relativity in a seemingly minor way --two of the terms have different coefficients. But technically these differences make a huge difference: in the classical theory, calculations trivialize in the CGHS model because one can decouple dynamics of the scalar field $f$ from that of geometric fields, the metric $g_{ab}$ and the dilation $\phi$. 

Further simplifications occur because the 2-d metric $g_{ab}$ is conformally flat and the 2-d massless Klein Gordon equation is conformally invariant. Therefore one can first trivially solve the wave equation for $f$, say for a left moving wave packet in 2-dimensional Minkowski space $(\mathring{M}, \mathring{g}_{ab})$ (shown as a diamond, including the dashed lines in the Panel (b) of Fig. \ref{fig:2Ashtekar}). Given this solution $f$, one can just write down a dilation field $\phi$ and a metric $g_{ab}$, both constructed using certain integrals involving $f$. One can then verify that $f$ solves the wave equation also w.r.t. $g_{ab}$ and $g_{ab}, \phi$ satisfy the correct field equations with $f$ as a scalar field source. However, the metric $g_{ab}$ is now such that it has a space-like singularity (shown in the Panel (b) of Fig. \ref{fig:2Ashtekar}). Therefore, the physical space-time $(M, g_{ab})$ is only a portion of the Minkowski space $(\mathring{M}, \mathring{g}_{ab})$ we began with! In particular, the $\scrip$ of $g_{ab}$ is a \emph{proper subset} of the $\mathring{\scrip}$ of the underlying Minkowski space $(\mathring{M}, \mathring{g}_{ab})$. Yet, $\scrip$ is complete in the geometry endowed on $M$ by the physical metric $g_{ab}$, just as $\mathring{\scrip}$ is complete w.r.t. the Minkowski metric $\mathring{g}_{ab}$! Therefore one can examine the causal past $J^-(\scrip)$ to check if the physical space-time $(M, g_{ab})$ admits an event horizon. As the figure brings out, it does; the singularity is hidden behind this horizon. Thus, the solution $(g_{ab}, \phi, f)$ we have constructed, starting just with a solution to the wave equation in 2-d Minkowski space,  represents a black hole in this 2-d theory! (For a pedagogical self-contained discussion, see e.g. section I.A of \cite{apr}.)

Using the facts that the model is exactly soluble in the classical theory, and we are in 2 space-time dimensions, it is possible to write down the semi-classical equations as well. The key approximation in the semi-classical theory is that the quantum metric operator can be replaced by its expectation value. Thus in this theory one has a smooth, `c-number'  metric (but with coefficients that depend on $\hbar$) sourced by the quantum matter. This approximation can be justified if one has a large number $N$ of scalar fields in the problem, so that the quantum fluctuations in the geometric fields (the metric and the dilation) can be neglected compared to the total quantum fluctuations in $N$ scalar fields in a $1/N$ expansion \cite{cghs}. One then just has a set of partial differential equations for the metric coefficients. However they are non-linear and rather complicated for an analytical treatment (although in certain regimes it is possible to obtain approximate analytical expressions \cite{ori2}). There is a long history of using numerical methods to solve them but, as we will discuss below, they led to some incorrect results because (i) the numerical precision was not sufficiently high; (ii) the treatment of a key conceptual issue (the definition of Bondi energy in the semi-classical theory) was flawed; and, (iii) an important scaling symmetry of the fundamental equations was not recognized. These limitations were addressed in \cite{aprlett,apr, ori2} and the careful analysis led to several qualitatively new results that can guide us in the investigation of black hole evaporation in 4 dimensions. These results can be summarized as follows:

\emph{(1)} Simulations assumed that the quantum state of the infalling scalar field is a coherent state on $\scrim$, peaked at as classical scalar field $f$ of finite duration, that undergoes a prompt collapse. As the scalar field falls in, a dynamical horizon is formed. It is space-like and grows in area in the support of the scalar field in space-time, in response to the infalling energy flux. Immediately the infall ends, the DH turns around and becomes time-like and area of its MTSs starts shrinking due to the influx of negative energy across it that balances the outgoing positive energy flux across $\scrip$. Thus, \emph{what forms in the gravitational collapse and shrinks in the evaporation process is a DH}. (In this analysis, `area' and `expansions of null normals' used to identify the DH refer to the 4-d space-time from which the 2-d space-time is obtained using spherical symmetry reduction. In the 2-d picture their expressions involve the metric $g_{ab}$ as well as the dilation $\phi$ which encodes the radius of 2-spheres in the 4-d theory via $R= e^{-\phi}$.) 
\vskip0.05cm
\emph{(2)} \emph{There is no event horizon} in the semi-classical space-time. Numerics show that $\scrip$ of the semi-classical space-time is incomplete so one cannot use the future boundary of $J^-(\scrip)$ as the event horizon. As shown in the figure, this future boundary represents `the last ray' in  the semi-classical space-time.
\vskip0.05cm
\emph{(3)} The two branches of the DH, shown in red in the Panel (c) of Fig. \ref{fig:2Ashtekar}, enclose a trapped region, shown in yellow. The outer boundary of this region is time-like. Therefore one may be tempted to say that `information can leak out' across it and purify the quantum state at $\scrip$ before the last ray \cite{hayward-conf}. However this possibility is not realized. There is no outward flux across the time-like piece of the DH. Very recently, much more general analysis was carried out to argue that this possibility cannot be realized also in the spherical sector of general relativity (without having to use the CGHS-type simplifications) \cite{agulloetal}.
\vskip0.05cm
\emph{(4)} Both, analytical approximations \cite{ori2} and high precision numerics \cite{apr} show that the singularity is softened. The quantum corrected, semi-classical metric is continuous everywhere. However, one cannot trust the validity of the semi-classical approximation (i.e. the truncation of the $1/N$  expansion to leading order) once the curvature becomes Planckian. That is why the yellow semi-classical region in the figure does not extend all the way to the singularity.
\vskip0.05cm
\emph{(5)} While the singularity extends all the way to $\scrip$ in the classical theory (see Panel (b)  in Fig. \ref{fig:2Ashtekar}), it ends well inside the physical space-time in the semi-classical theory (see Panel (c) in Fig. \ref{fig:2Ashtekar}). There is a last ray that goes from the end of the singularity to $\scrip$ (marked $u_{LR}$ in Panel (c)). Hawking and Stewart had also analyzed this semi-classical space-time  numerically and concluded that there is a `thunderbolt singularity' along this last ray \cite{swhjs}. However, this conclusion appears to be a result of a simulations that did not have adequate accuracy. The high precision simulations carried out in \cite{apr} show that the metric is regular along this last ray; \emph{there is no thunderbolt}. There is also no trace of a `firewall' that was conjectured from string theory considerations \cite{amps}. Finally, using some plausible assumptions, it has been argued that the S-matrix would be unitary in full quantum theory \cite{atv}.
\vskip0.05cm
\emph{(6)} The retarded time at which the space-like branch of the DH ends and meets the time-like branch marks the onset of quantum radiation at $\scrip$. Furthermore, once the correct notion of Bondi-energy at $\scrip$ is identified for the semi-classical theory \cite{atv}, there is a direct correlation between the (shrinking) area of any given MTS on the time-like branch of the DH and the Bondi energy at the corresponding retarded time at $\scrip$. This `balance law' reinforces the physical significance of the DH. \vskip0.05cm

On the whole, then, these investigations have sharpened our expectations on what to expect in semi-classical limit and helped us in weeding out untenable ideas that had suggested that semi-classical gravity may fail already in rather tame situations. There are also several other striking results obtained in the CGHS model (such as a scaling symmetry and universality of the Bondi mass at the end of the semi-classical evaporation process) that may be useful to the analysis of the evaporation in the spherically symmetric sector of general relativity. 

However, the CGHS model suffers from a fundamental limitation vis a vis this more realistic system: It is a genuinely 2-d model. First, the left and right $\scri^{\pm}$ are distinct. The collapsing scalar field originates on right $\scrim$ and moves left while the quantum radiation appears on right $\scrip$, originating in the vacuum state of right moving quantum fields on left $\scrim$. Second, in this 2-d model the temperature associated with the quantum radiation on right $\scrip$ is a constant, independent of the mass of the black hole (although there is an approach to improve on this situation using approximation methods \cite{ori3}). 

To rectify this situation, very recently Varadarajan has proposed a new model that is free from these limitations but is still manageable because the scalar field sector again decouples from gravity \cite{mv}. There is a single $\scrim$ and a single $\scrip$ and a symmetry axis on which the rotational Killing fields all vanish. The classical Penrose diagram is the same as in Panel (c) of Fig. \ref{fig:1Ashtekar}, rather than the Panel (b) of Fig. \ref{fig:2Ashtekar}. What is neglected vis a via spherical symmetric reduction of the collapse of a massless scalar field $f$ in general relativity is the \emph{back scattering} of $f$ by curvature. While this is a limitation, the model can be solved exactly in the classical theory and one can write down manageable semi-classical equations. So far, ignoring back scattering does not appear to be a severe drawback. Numerical simulations (that will be soon undertaken) should shed considerable light on this issue and also provide more reliable ways to test current scenarios of quantum evaporation.

\subsection{Singularity resolution in LQG}
\label{s2.2}

Let us now turn to the second issue on which LQG approaches depart decisively from the scenario advocated  in Hawking's original proposal, depicted in Panel (b) of Fig.\ref{fig:1Ashtekar}: fate of the Schwarzschild singularity in quantum gravity. As we discussed in section \ref{s1}, if the singularity persists as a part of the future boundary, information will be lost, i.e. the evolution of the quantum state from $\scrim$ to $\scrip$ would not be unitary. However, LQG results to date strongly suggest that this singularity would be resolved because of quantum geometry effects that lie at its foundations.

Let us  make a detour to describe of the nature of quantum geometry in LQG. To begin with, let us recall the central idea behind general relativity: Gravity is not a force as in Newton's theory but a manifestation of curvature of space-time. Therefore, to develop general relativity, Einstein had to use a new syntax to describe all of classical physics: that provided differential geometry. Thus, space-time geometry is described by a metric, its derivatives operator and curvature, matter is represented by tensor fields that obey hyperbolic differential equations with respect to the metric. The LQG viewpoint is that one now needs a new syntax to formulate quantum gravity. Since gravity is encoded in geometry, a quantum theory of gravity should also be a quantum theory of geometry. Therefore, the new syntax is to be provided by \emph{quantum Riemannian geometry} in which basic geometric observables like areas of surfaces and volumes of regions and curvature of space-time are all represented by suitable operators. 

This syntax was created by a large number of researchers in the 1990s (for reviews, see, e.g., \cite{alrev,crbook,ttbook}. The syntax is based on two key ideas: \emph{(i)} A reformulation of general relativity  (with matter) in the language of gauge theories  --that successfully describe the other three basis forces of Nature--  but now \emph{without reference to any background field}, not even a spacetime metric; and, \emph{(ii)} Subsequent passage to quantum theory using non-perturbative techniques from gauge theories --such as holonomies of the gravitational connection--  again without reference to  background fields. Consequently, the emphasis is shifted from metrics to connections. Background independence implies diffeomorphism covariance, which was heavily used together with non-perturbative methods. One was then naturally led to a fundamental, in-built discreteness in geometry that foreshadows ultraviolet finiteness. The familiar spacetime continuum of general relativity is \emph{emergent} in two senses. First, it is built out of certain fields that feature naturally in gauge theories, without any reference to  a spacetime metric. Second,  it emerges only \emph{on coarse graining} of the fundamental discrete structures --the `atoms of geometry'-- of the quantum Riemannian framework. (For a recent short overview addressed to non-experts, see \cite{aaeb}.)

At a fundamental level, observables such as areas of surfaces and volumes of regions have a purely discrete spectrum. Thus geometry is `quantized' in the same sense that energy and angular momentum of the hydrogen atom are quantized. In particular, there is the smallest non-zero eigenvalue of the area operator, called the \emph{area gap} and denoted by ${\uDelta}$ whose value turns out to be $\sim  5.17$  in Planck units. It plays a key role in quantum dynamics. For, the curvature operator is defined by  considering the holonomy of the gravitational connection around a closed loop, dividing by the physical area enclosed by the loop, and then shrinking the loop until it encloses area ${\uDelta}$. Consequently, the curvature operator inherits a Planck scale non-locality from ${\uDelta}$ which in turn provides a natural ultra-violet regulator in quantum dynamics.

While full LQG is still being developed, its cosmological sector --\emph{Loop Quantum Cosmology} (LQC)-- has been investigated in great detail using non-perturbative methods of Hamiltonian LQG, the corresponding path integrals, as well as the consistent histories approach (for reviews, see, e.g., \cite{asrev,iapsrev}). These investigations have shown that the Big Bang Bang and the Big Crunch singularities of the homogeneous cosmologies  are naturally resolved by the quantum geometry effects. A key feature of the LQC dynamics is that corrections to general relativity are negligible until the matter density or curvature are $\sim 10^{-4}$ in Planck units, but then they grow very rapidly, creating an `effective repulsion' that completely overwhelm the classical attraction and causes the universe to bounce. In this singularity resolution, matter does not violate the standard energy conditions. Yet the singularity theorems in classical GR are bypassed because the quantum corrections modify Einstein's equations themselves. 

Many of the consequences of the LQC dynamics can be readily understood using the so-called \emph{effective equations} that capture the evolution of the peaks of sharply peaked quantum states (which, interestingly, remain sharply peaked also in the Planck regime). They encapsulate the leading order corrections to the classical Einstein's equation everywhere, \emph{including the Planck regime}. There is a streamlined procedure to arrive at the effective equations starting from full quantum dynamics that governs the quantum states in LQC \cite{vt,asrev}. The full dynamics of quantum states has much more detailed information. Effective equations extract from quantum states smooth metrics with coefficients that depend on $\hbar$, capturing the most important quantum corrections to dynamics. Therefore  they have been  heavily used to gain  valuable physical insights into the nature of quantum corrected geometry in the Planck regime. Note that the term `effective' is used in LQC in a sense that is different from the common usage in quantum field theory. In particular, there is no `integration of the ultraviolet degrees of freedom'; the LQC effective equations hold even in the Planck regime.
 \begin{figure} 
 
  \begin{center}
    \begin{minipage}{1.5in}
      \begin{center}
        \includegraphics[width=1.9in,height=2.2in,angle=270]{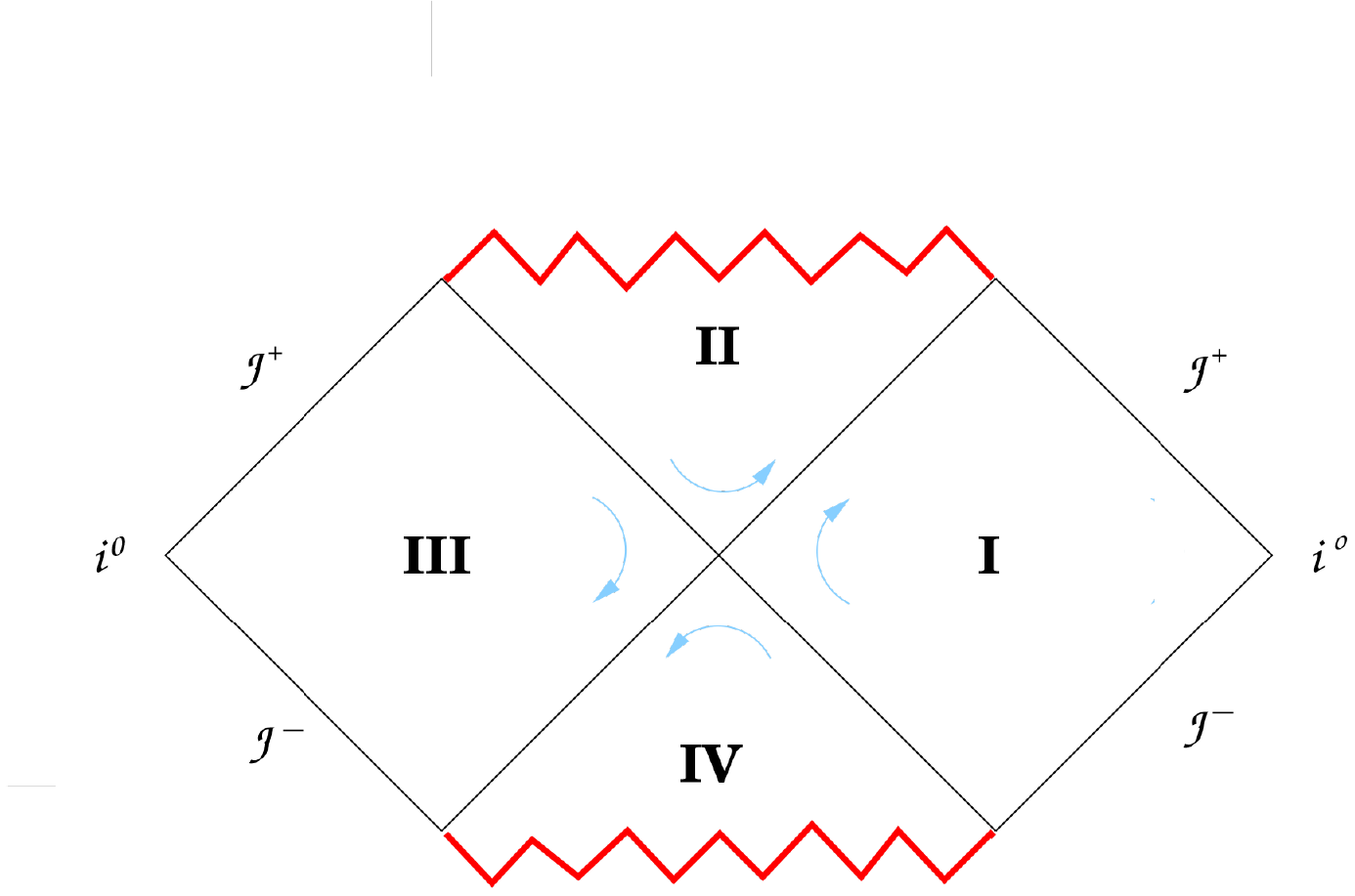} \\ (a)
         \end{center}
    \end{minipage}
   \hspace{.7in}
    \begin{minipage}{1.7in}
      \begin{center}
      \includegraphics[width=1.6in,height=1.6in]{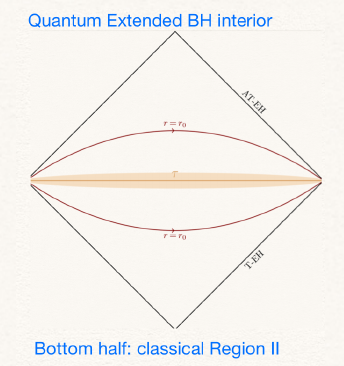} \\ {(b)}
       \end{center} 
   \end{minipage}
      \hspace{.5in}
       \begin{minipage}{1.7in}
        \begin{center}
         \includegraphics[width=1.6in,height=1.6in]{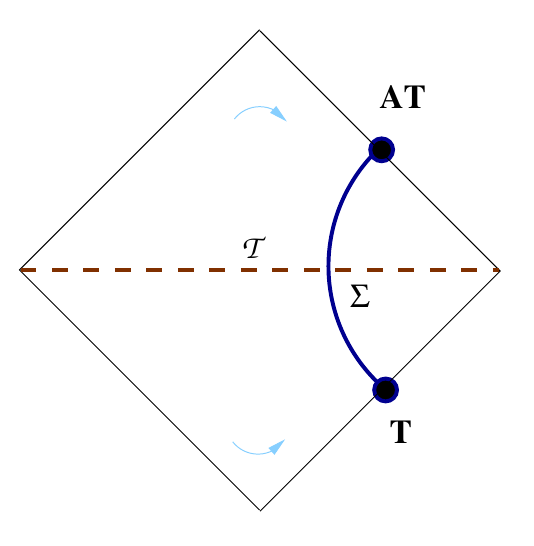} \\ {(c)}
          \end{center}
  \end{minipage}
\caption{\footnotesize{ (a) \emph{Left Panel}: Penrose diagram of Kruskal space-time. Only region II is of interest for the issue of singularity resolution.
(b) \emph{Middle Panel}: The LQG extension of region II of Kruskal space-time via singularity resolution. Singularity is replaced by a regular 3-manifold $\tau$ that separates the trapped and anti-trapped regions.
(c) \emph{Right Panel}: A time-like 3-surface $\Sigma$ joins two MTSs (depicted by blobs), one on the Trapping horizon that constitutes the past boundary of the trapped region and the other on the Anti-Trapping horizon that constitutes the future boundary of the anti-trapped region.}}
\vskip-0.5cm
\label{fig:3}
\end{center}
\end{figure}

This long detour into LQC may seem like a digression in the present context of black holes. But in fact it is directly relevant to the issue of what happens to the black hole singularity. For, the region II of Kruskal space-time that contains this singularity is isometric to a homogeneous cosmology: the vacuum Kantowski-Sachs model (see Panel (a) in Fig. \ref{fig:3}). This region is foliated by the $r={\rm const}$ space-like surfaces each of which constitutes an orbit of the 4 Killing fields of the Kruskal metric. The coordinate $r$ plays the role of time and runs from the $r=2GM \equiv 2m$ at the past boundary that constitutes an isolated horizon (IH) to $r=0$ at the future boundary that represents a Big Crunch singularity of the Kantowski-Sachs model. Therefore, one can use methods used in LQC. There is rich literature on the subject and results based on effective equations have provided significant insights into what replaces the singularity and on the nature of quantum corrected geometry in the Planck regime, especially over the last $\sim\, 5$ years. For definiteness, we will focus on the first of these recent investigations \cite{aoslett,aos} that set the stage. In more recent investigations \cite{gop,mhhl} the effective equations are obtained using a more systematic approach but the final results are very similar. Recall that, although the Kruskal space-time represents an idealized `eternal black hole', in the early investigations of the Hawking effect it provided valuable guidance on what to expect in more interesting collapsing situations. 
The situation is the same in LQG. Specifically, the following key results on the singularity resolution in Kruskal space-time have shaped our expectations in more realistic situations in which a macroscopic black hole  (i.e. one with 
$m \gg \lp$) black hole forms through gravitational collapse and then evaporates due to quantum radiation:
\vskip0.05cm

\noindent \emph{(i)} For these Kruskal black holes, the quantum geometry corrections are completely negligible near the IH that constitutes the past boundary of region II in Panel (a) of Fig. \ref{fig:3}. For example, for a solar mass black hole, they are of the order of 1 part in $10^{115}$! However, they grow as $r$ decreases, leading to the resolution of the classical singularity at $r=0$. In the quantum corrected effective geometry, the singularity is replaced by a regular \emph{transition surface} $\tau$. This a space-like 3-manifold which is again an orbital the 4 Killing fields and therefore a homogeneous 3-manifold foliated by round 2-spheres. The radius of these 2-spheres is given by $\rm{Rad}_{\tau} \approx (\uDelta^2 m)^{\f{1}{3}}\times 10^{-2}$. (By inspection, $\rm{Rad}_{\tau}$ goes to zero in the limit the area gap goes to zero, i.e., the limit in which quantum geometry effects vanish and $\tau$ coincides with the classical singularity, just as one would expect.) In the quantum corrected geometry, then,  $r$ ranges $r \approx 2m$ on the past IH to $r= \rm{Rad}_{\tau}$. \vskip0.05cm
\emph{(ii)} The quantum corrected geometry is smooth in this region and, as shown in the Panel (b) of Fig. \ref{fig:3}, effective equations extend it to a smooth metric the future of the transition surface $\tau$, all the way till a future IH.  As one would expect from Kruskal geometry, to the past of $\tau$ one has a trapped region: the expansion of both null normals to round 2-spheres is negative. Interestingly, \emph{they both vanish on $\tau$}! Thus the geometry at the transition surface is truly exceptional. To the future of $\tau$ both expansions are positive. Thus, $\tau$ is a boundary that separates a trapped region that lies to its past and an anti-trapped region that lies to its future. The radii of the round 2-spheres decrease monotonically as one moves from the past boundary of the Panel (b) of Fig. \ref{fig:3} to the transition surface $\tau$, and then increase monotonically as one moves to the future boundary. They acquire their minimum radius $\rm{Rad}_{\tau} \approx ({\uDelta}^2 m)^{\f{1}{3}}\times 10^{-2}$ on $\tau$ which grows as $m^{1/3}$ as the black hole mass grows. The past boundary is a trapping IH, while future boundary, an anti-trapping IH.
\vskip0.05cm
\emph{(iii)} Curvature of the quantum corrected metric is bounded throughout the diamond shown in Panel (b) of Fig. \ref{fig:3}. As one would expect, it reaches its maximum at the transition surface. Specifically, for the Kretschmann scalar:
\begin{equation}
\label{bounds}
 \mathcal{K}\mid_{\T}\,\,  \equiv \,\, R_{abcd}R^{abcd}\mid_{\T}\,\,=\, \frac{k_\circ}{\uDelta^2}\ + {\cal O}\big(({\uDelta}/{m^2})^{1/3} \,\ln\, ({m^2}/{\uDelta})\big)
\end{equation}
where $k_\circ$ is a constant.  Thus, the classical singularity is naturally resolved, thanks to quantum geometry that provide us with a non-zero area gap $\uDelta$: In the classical limit, $\uDelta \to 0$,\, $\tau$ is replaced by the $r=0$ surface and we return to the Kruskal singularity. Note that the leading term in (\ref{bounds}) is \emph{universal}: It does not depend on the black hole mass. This is also what happens at the big bounce (that replaces the Big Bang) in quantum cosmology: the curvature at the bounce provides a universal upper bound in Friedmann-Lema\^itre models.  \vskip0.05cm
\emph{(iv)} As one moves away from $\T$, these curvature scalars rapidly approach their classical values even for very small microscopic black holes. For instance, while the horizon radius of the effective solution is always larger than that of its classical counterpart, even for $m=10^4 \lp$, the relative difference is $\sim 10^{-15}$ and, as remarked already, for a solar mass black hole, it is $\sim 10^{-115}$! Finally one can ask for the relation between the radius $r_{{}_{\rm T}}$ of the trapping horizon, and the radius  $r_{\rm AT}$ of the anti-trapping horizon. Are they approximately the same? The answer is in the affirmative for macroscopic black holes, even though the `bounce' is not exactly symmetric. For a stellar mass black hole for example, $r_{{}_{\rm T}} = 3$km and $r_{\rm AT} = 3\,(1 + {O}(10^{-25}))$km. \vskip0.05cm
\emph{(v)} What happens to the singularity theorems? Indeed, this question was raised in the context of a singularity resolution after Adrian del Rio's talk at the BHIO conference. The answer is that, as in LQC, they are bypassed because quantum corrections to  Einstein's equations become significant as one approaches $\tau$. More precisely, the Ricci tensor of the quantum corrected metric is non-zero and one can use it to define an `effective stress-energy tensor' $T^{\rm eff}_{ab}$ via $G_{ab} = 8\pi G T^{\rm eff}_{ab}$. As one would expect, $T^{\rm eff}_{ab}$ fails to satisfy even the weak energy condition. Except very near the two horizons, the energy density is negative. For a black hole with $m = 10^6 \lp$, it  becomes $\mathcal{O}(10^{-1})$ in Planck units at the transition surface $\tau$ but  decays very rapidly as moves away from $\tau$ and is $\mathcal{O}(10^{-20})$ near the two horizons. The square root of the Kretchmann scalar (that has the same dimensions as the energy density) is $\mathcal{O}(10^{-12})$; thus the contribution of the Ricci curvature to the total curvature is negligible even for these very small macroscopic black holes. 
\vskip0.05cm
\emph{(vi)} Several non-trivial checks have been made on this geometry to verify overall consistency. A conceptually interesting one comes from the Komar mass associated with the translation Killing field $\partial/\partial t$. Consider the values of the Komar mass evaluated on a 2-sphere on the trapping horizon and another 2-sphere on anti-trapping horizon, the two being connected by a 3-manifold $\Sigma$ as in Panel (c) of Fig. \ref{fig:3}. Now, in the classical theory, the Komar mass $M_{\rm K}$ defined by the translational Killing field is given by (half the) horizon radius. As we saw, for macroscopic black holes the radii $r_{{}_{\rm T}}$ and  $r_{{}_{\rm AT}}$ are essentially the same. On the other hand, the difference between the Komar mass evaluated at the anti-trapping horizon and the trapping horizon is given by the integral involving stress-energy tensor over a 3-manifold $\Sigma$ joining cross-sections of the two horizons (see Panel (c) of Fig.~\ref{fig:3}),
\begin{equation}\label{balance2} M_{\rm K}^{\rm AT} - M_{\rm K}^{\,\rm T}\, =\, 2\,\int_{\Sigma}\! \big(T_{ab}^{\rm eff}\, -\, \f{1}{2} T^{\rm eff}\,\, g_{ab}^{\rm eff}\big)\, X^a \rmd \Sigma^b \, ,\end{equation}
and for macroscopic black holes the integrand of the right is \emph{large and negative} near $\T$ (because it represents the effective energy density). How can the two Komar masses be the same, then?  It turns out that the integrand of (\ref{balance2}) is indeed large and negative for macroscopic black holes, but its numerical value is very close to $-2M_{K}^{\rm T}$. Therefore the Komar mass associated with the anti-trapping horizon is given by $M_{\rm K}^{\rm AT} \approx M_{\rm K}^{T} - 2M_{K}^{\rm T} = -M_{\rm K}^{\rm T}$, and the minus sign is just right because while the translational Killing field is {\rm future} directed on the trapping horizon {T}, it is \emph{past} directed on the anti-trapping horizon {AT}! This resolution is an example of the conceptually subtle nature of the quantum geometry in the diamond bounded by the two horizons.\vskip0.1cm

We will conclude this subsection with two remarks:

1. Since the black hole singularity lies in region II of Kruskal space-time, in the above discussion we focused on the quantum extension of this region. The resulting quantum corrected (or effective) metric is well-defined on the boundaries: the trapping IH and the anti-trapping IH. It is then natural to ask if we can extend the metric beyond these boundaries in a systematic manner to asymptotic regions. This is indeed possible. An extension was given already in the first set of investigations \cite{aoslett,aos}. More satisfactory extensions can be found in \cite{mhhl,gop}.   

2. The quantum extensions of the type depicted in Panel (b) of Fig. \ref{fig:3} is often referred to as representing a ``black hole to white hole transition" because one has a trapped region to the past of $\tau$ and an anti-trapped region to the future. We have avoided this terminology because it has other connotations that are not realized. In particular, one loses predictivity in presence of white holes since anything can come out of their singularity. In the LQG transition from trapped to anti-trapped regions, on the other hand, there is no singularity and physics is completely deterministic across $\tau$ that replaces the singularity.   
\vskip0.1cm

Together with the discussion of DHs of section \ref{s2.1}, results summarized in this subsection have provided considerable intuition, streamlining possibilities both for permissible quantum geometries and for pathways to the recovery of information in the black hole evaporation process. These two sets of concrete results are used as stepping stones in current LQG investigations aimed at obtaining a complete description of the evaporation process. In the next section, we will summarize a mainstream perspective that has resulted.

\section{Black hole evaporation in LQG}
\label{s3}

Let us now turn to the issues related to the dynamics of black hole formation and evaporation in quantum theory using results of section \ref{s2} as guidelines. We will divide this discussion into two parts. In the first, we will focus on the semi-classical sector that excludes the Planck regime, and in the second we will discuss evolution to its future through the Planck regime. We will discuss the two interrelated but rather distinct issues: (i) Nature to the quantum corrected space-time geometry, and, (ii) issues related to entanglement, von-Neumann entropy and purity of the final state at $\scrip$. Of course, important issues remain, especially in the second part. Nonetheless, the hope is that this streamlining possibilities will lead to focused efforts to weed out ideas and concentrate on the viable paths that remain.

\subsection{The semi-classical regime}
\label{s3.1}
 
To ensure the validity of the semi-classical approximation, in most of this sub-section we will consider a solar mass $M_\odot$ blackhole that is formed by a gravitational collapse and let it evaporate till it reaches the lunar mass $M_{\text{\leftmoon}}$. This process takes some $10^{64}$ years. But even at the end of this long evaporation time, the final black hole has a macroscopic mass. Therefore, during this entire phase of evaporation, the process should be well approximated by semi-classical gravity. While there have been proposals advocating large deviations from the semi-classical theory even for astrophysical black holes (e.g. due to `firewalls'), in light of the results of the LIGO-Virgo collaboration, these proposals are no longer regarded as viable by most of the community. (For a general discussion on implausibility of the failure of semi-classical gravity well away from the Planck regime, see in particular \cite{ori1}.)

As we explained in footnote \ref{fn1}, to discuss the issue of information in a meaningful way, one needs to work with a closed system. Therefore, we will focus on the system consisting of massless scalar fields $f$ coupled to gravity in 4-dimensions. At the classical level, we will use Einstein's equations (which, however, will be appropriately modified in the quantum theory). We will restrict ourselves to spherical symmetry and, for simplicity ignore back scattering as in Varadarajan's model \cite{mv} (see the last para in section \ref{s2.1}). Finally, to justify the semi-classical approximation --in which one treats matter quantum mechanically but ignores the quantum fluctuations of geometry-- we will use a large number $N$ of scalar fields and work with the $1/N$ expansion as in section \ref{s2.2}. 

\begin{figure} 
 
  \begin{center}
    \begin{minipage}{1.5in}
      \begin{center}
        \includegraphics[width=1.5in,height=2.7in,angle=0]{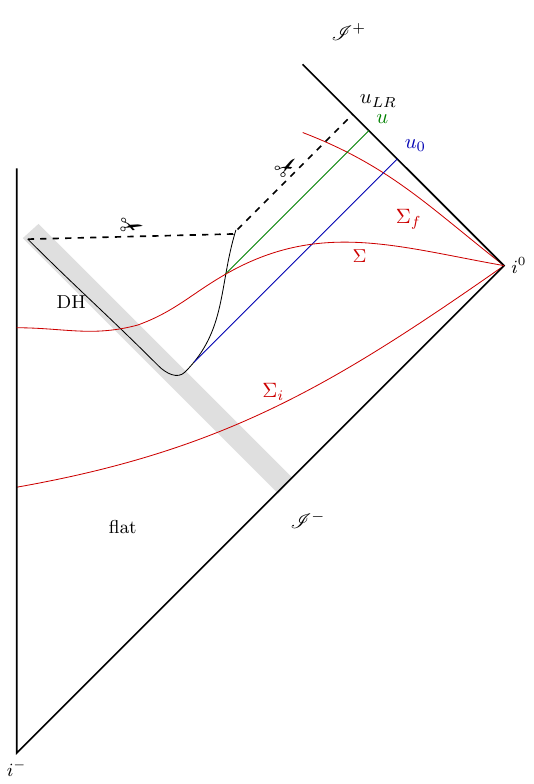} \\ (a)
         \end{center}
    \end{minipage}
   \hspace{.7in}
    \begin{minipage}{1.7in}
      \begin{center}
\includegraphics[width=1.3in,height=2.7in,angle=0]{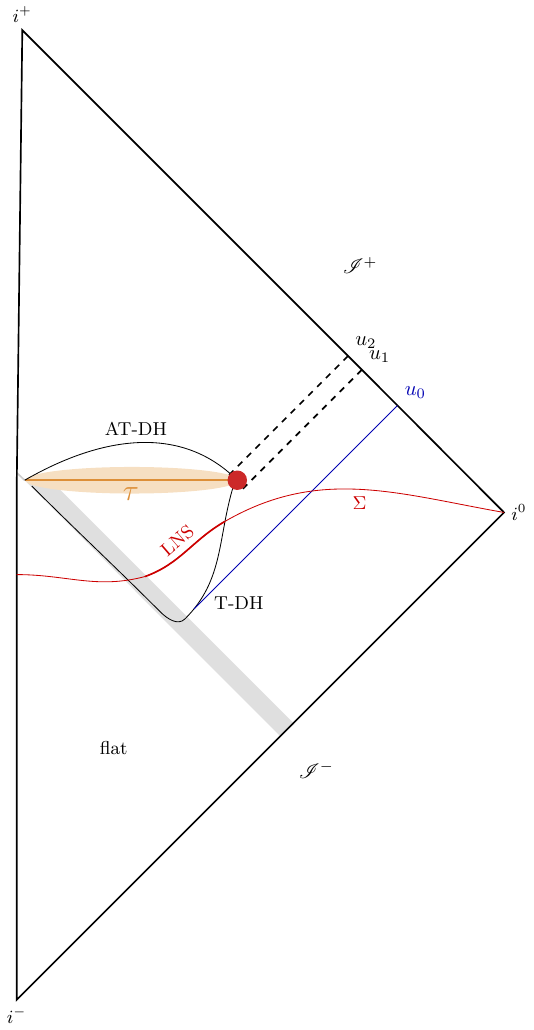} \\ {(b)}
       \end{center} 
   \end{minipage}
      \hspace{.5in}
       \begin{minipage}{1.7in}
        \begin{center}
\includegraphics[width=1.4in,height=2.7in,angle=0]{Contributions/Ashtekar/TDL-traditional-eva.pdf} \\ {(c)}
          \end{center}
  \end{minipage}
\caption{\footnotesize{ (a) \emph{Left Panel}: The expected semi-classical space-time in LQG. Planck regime has been cut out. The DH has two branches, the expanding, space-like, left branch is formed as a result of the collapse and the contracting, time-like, right branch that replaces the event horizon of the Panel (c) of Fig. \ref{fig:1Ashtekar}. Together, they bound a trapped region.\,\,
(b) \emph{Middle Panel}: The conjectured full space-time of LQG. Curvature is Planck scale in the shaded pink region that contains the transition surface $\tau$. To the past of $\tau$ we have a 
trapped region, bounded by the trapping DHs \TDH and to the future of which we have an anti-trapped region, bounded to the future by an anti-trapping DH \ATDH\!\!. The well-developed approximation methods of LQG are inapplicable to the prink blob where the fluctuations of geometry could be very large. Because their influence has not been explored, the space-time portion to the future of the anti-trapping DH is left blank.\,\,
(c) \emph{Right Panel}: Hawking's original Penrose diagram for an evaporating black hole is reproduced here for ready comparison with the LQG proposals.}}
\vskip-0.5cm
\label{fig:4}
\end{center}
\end{figure}

In the incoming state on $\scrim$, one of the quantum scalar fields, say $\hat{f}_1$, will be assumed to be in a coherent state that is peaked at a classical scalar field $f_1^\circ$  (of compact support on $\scrim$) that undergoes a prompt collapse to form a black hole. The remaining $N-1$ quantum fields $\h{f}_i$ will assumed to be in their vacuum state (as in \cite{apr,mv}). The semi-classical Einstein's equations governing this system are:
\begin{equation}\label{semi-class} G_{ab}^{\rm (sc)} = 8\pi G_{\rm N}\,  \langle\, \hat{T}_{ab}\, \rangle_{\rm ren} \quad {\rm and} \quad  \Box\, \hat{f_i} = 0\, , \hbox{\rm{with} $I = 1,\ldots N$} \end{equation}
where $G_{ab}^{\rm (sc)}$ is the Einstein tensor of the semi-classical metric $g_{ab}^{\rm (sc)}$ and the expectation value of the renormalized stress-energy tensor is computed using the Heisenberg state $\Psi$ and the space-time metric $g_{ab}^{\rm (sc)}$. The metric $g_{ab}^{\rm (sc)}$ does include quantum corrections but they are induced only by quantum matter (since the quantum geometry terms  induced by the area gap are completely negligible away from the Planck regime). These corrections to geometry are adiabatic and small. But the infalling negative energy flux has a non-trivial effect on the horizon structure already at the start of the evaporation process. In the classical theory, the space-like dynamical horizon DH would have continuously joined on to an isolated horizon --the future part of the event horizon (see the Panel (c) of Fig. \ref{fig:1Ashtekar}). Now, this space-like branch of the DH turns around and becomes time-like (see Panel (a) of Fig. \ref{fig:4}). Together, the two branches of the DH enclose a trapped region: in this region expansions of both bull normals to the MTSs are negative. 

During the evaporation process, modes are created in pairs. One escapes to $\scrip$ and its partner is trapped in this region. Therefore, as the black hole evaporates the total state on a Cauchy surface (such as $\Sigma$ in Panel (a) of Fig.\ref{fig:4}) is increasingly entangled. Now, because the right branch of the DH is time-like, light \emph{can} escape the trapped region (in sharp contrast to the situation where the trapped region is bounded by an event horizon). Therefore, one might imagine that information could leak out from the trapped region, leading to purification of the state at $\scrip$ already in the semi-classical regime \cite{hayward-pop,hayward-conf}. But, as we already indicated in section \ref{s2.1.2}, a careful analysis of the partners modes that go out to $\scrip$ shows that correlation will not be restored at $\scrip$ during the semi-classical phase \cite{agulloetal}. Thus, even at the end of the evaporation process now under consideration, when the black hole has shrunk from solar $M_\odot$ to lunar mass $M_{\text{\leftmoon}}$, the quantum state remains entangled.

But then there is an apparent paradox already in the semi-classical regime. Since $M_{\text{\leftmoon}} \sim  10^{-7}\, M_\odot$, at the end of this long evaporation process most of the initial mass is carried away to $\scrip$ by the outdoing modes. A back of the envelope calculation shows that a very large number $\mathcal{N}$\, ($\sim 10^{75}$)\, of quanta escape to $\scrip$ and all of them are correlated with the ones that are trapped in the region enclosed by \TDH\!\!\!. Therefore, at the end of the semi-classical process under consideration, one would have a huge number $\mathcal{N}$ of quanta both at $\scrip$ and in the trapped region. But the mass associated with the trapped region is only $10^{-7}$ times that carried away to $\scrip$. Furthermore, the radius of the outer part of \TDH has shrunk to only $0.1$mm -- the Schwarzschild radius of a lunar mass black hole. How can a \TDH with just a $0.1$mm radius accommodate all these $\mathcal{N}$ modes? Even if we allowed each mode to have the (apparently maximum) wavelength of $0.1$mm, one would need the horizon to have a huge mass --some $10^{22}$ times the lunar mass!  While these considerations are quite heuristic, one needs to face the conceptual tension: At the end of the semi-classical phase under consideration, the trapped region has simply too many quanta to accommodate, with a tiny energy budget. 

The resolution of this apparent paradox lies in the fact that the geometry of the trapped region has some rather extraordinary features that had not been noticed until relatively recently. Recall that the evaporation process is extremely slow, and the \TDH mass of the time-like branch decreases in response to the outgoing flux by the the standard Hawking formula: $\rmd M_{\TDH}/\rmd v = -  \hslash/(GM_{\TDH})^{2}$, where $v$ is the retarded time coordinate as in Hawking's calculation. Using these two features,  one can argue that the metric in the trapped region should be well-approximated by that of the Vaidya space-time in which the mass $M(v)$ decreases from $M_\odot$ to $M_{\text{\leftmoon}}$. To probe this geometry, let us foliate the trapped region by 3-dimensional space-like surfaces, each with topology $\mathbb{S}^2\times \mathbb{R}$. There are two particularly natural choices: $\mathcal{K} = {\rm const}$ surfaces, and $r = {\rm const}$ surfaces, where $\mathcal{K}$ denotes the Kretchmann scalar and $r$ denotes the radius of the round 2-spheres $\mathbb{S}^2$. The radius of the time-like portion of \TDH decreases from $3$km to $0.1$mm. But, as a simple calculation shows, the 3-dimensional surfaces develop \emph{astronomically long `necks'} along the $\mathbb{R}$-direction. By the time we have reached lunar mass $M_{\text{\leftmoon}}$, the lengths of these necks $\ell_{N}$ are given by:  $\ell_{N} \approx 10^{64}$ \emph{light years} for the first foliation, and $\ell_{N} \approx 10^{62}$ \emph{light years} for the second \cite{aaao,aa-ilqg}!%
\footnote{One can also consider ${\rm{tr}} K = {\rm const}$ slices \cite{cdl}. But they do not provide a foliation of the entire trapped region. But the main phenomenon of developing enormously long necks,  occurs also on these 3-surfaces.} 
These astronomically large lengths can result because the process has a really huge time at its disposal; $10^{64}$ years corresponds to  $\sim\,10^{53}$ times our cosmic history. 

This enormous stretching is analogous to the expansion that the universe undergoes in (an anisotropic) cosmology. Recall that during the cosmic expansion --e.g. during inflation-- the wavelengths of modes get stretched enormously. This suggests that partner modes in trapped region will also get enormously stretched during evolution and become infrared. Can this phenomenon resolve the quandary of `so many quanta with so little energy'? The answer is yes. With such infrared wavelengths, it is easy to accommodate $\mathcal{N}$ modes in the trapped region with the energy budget only of $M_{\text{\leftmoon}}$. Thus, even though the  outgoing modes carry away almost all of the initial mass  $M_{\odot}$ to $\scrip$, there is no obstruction to housing all their partners in the trapped region on a slice $\Sigma$ of Panel (a) of Fig.~\ref{fig:4} with the small energy budget of just $10^{-7}M_{\odot}$.  This argument removes the necessity of starting purification by Page time. In the LQG perspective, purification occurs at a much later stage.

To summarize, in the long semi-classical phase the outgoing modes that register on $\scrip$ are entangled with their partners, confined to the trapped region. But the outgoing quanta carry away most of the total energy $M_\odot$ leaving only a small remainder $M_{\text{\leftmoon}} \approx 10^{-7} M_\odot$ in the trapped region. This occurs because there is a steady negative energy flux going into the trapped region that erases most of the $M_\odot$ of infalling energy, just as one would expect from energy conservation. But energy considerations are distinct from the entanglement issues and this raises an apparent paradox already in the semi-classical phase: How can the small energy budget of the trapped partner modes suffice to hold `as many' of them as those went out to $\scrip$ carrying huge total energy $(1-10^{-7})\, M_\odot$? The resolution lies in the fact that the space-time geometry of the trapped region is highly non-trivial: As evaporation proceeds these modes get stretched \emph{enormously} and become infrared. Therefore entanglement between the modes that have registered at $\scrip$ and those in the trapped region can persist, even though the energy associated with the two sets is vastly different. Thus, at the end of the semi-classical phase, if one were to trace over the trapped partner modes, the state at $\scrip$ would be mixed. This but to be expected because the semi-classical $\scrip$ is not complete. Whether purity is finally resorted at $\scrip$ depends on what happens to the future of the semi-classical region. 

\subsection{Evolution through the Plank regime}
\label{s3.2}

In the last subsection we considered the phase during which a solar mass black hole shrinks to lunar mass to make it obvious that we are in the semi-classical regime. But, as is widely expected, the semi-classical approximation should continue to be viable until the black hole has shrunk to, say, $10^3$ Planck mass so the curvature at the \TDH is $\sim\, 10^{-6}$ in Planck units. However, then one enters the Planck regime --shown in pink in Panel (b) of Fig. \ref{fig:4}-- where one has to use the full quantum state $\Psi$ of matter \emph{and} geometry and study its evolution. The key question that remains is: What are the predictions of this evolution for space-time geometry \emph{and} for the infrared partner modes that  are confined to the trapped region of the semi-classical phase? Since they are entangled with the modes that went out to $\scrip$ throughout the very long evaporation process in which the $M_\odot$ black holes shrinks to $\sim\,10^3$ Planck mass, the associated von-Neumann entanglement entropy is very large. Nonetheless,  there \emph{is a pathway to restoration of purity} of the full quantum state at $\scrip$: If the partner modes were to evolve across the Planck regime that lies to the future of the semi-classical phase, they would arrive at $\scrip$ where they would be correlated with the modes that arrived at early times, just as they were while in the trapped region. Thus, at the end of the process, the complete quantum state at $\scrip$ should be pure, just as the complete quantum state was pure on a Cauchy surface $\Sigma$ passing through the semi-classical region (see Panel (b) of Fig.\ref{fig:4}).

To find out what happens to the partner modes in the Planck regime, we have the very difficult task of 
evolving quantum fields $\h{f}_i$ on the quantum geometry in this region. There are two aspects to the difficulty: (i) one cannot ignore the quantum fluctuations of geometry because we have Planck scale curvature, and, (ii) dynamical time scales for changes in the matter field and geometry can be Planckian, making the process highly non-adiabatic. Now in the long shaded, pink region of Panel (b) depicting the Planck regime, we do face the first issue. However, for a very long time, the evaporation  process is in fact adiabatic. Therefore, we can divide this region from left end to right into large intervals in each of which the geometry does not change significantly. This approximation fails at the right end of the region, depicted by the red blob. Here, the time-like branch of \TDH  has mass less than $\sim 10^{3}$ in Planck units, whence one expects dynamical time scales to be also Planck scale.

Let us therefore postpone the discussion pertaining to this red blob and first consider the rest of the Planck regime to its left, where we have Planck scale curvature but the adiabatic approximation holds. From our discussion of section \ref{s3.1}, one expects this region to be very long but foliated by very small 2-spheres. Fortunately, prior experience in LQC --in particular the detailed investigation of the  propagation of cosmological perturbations on the \emph{quantum} FLRW geometries-- suggests a strategy to analyze dynamics in this region (see, e.g., section II.C of \cite{aan3}). Specifically, although the quantum state of geometry does have large fluctuations, dynamics of the scalar fields $\h{f}$ is not sensitive to all of them. Consequently, one can construct from the quantum state of geometry a smooth metric $\t{g}_{ab}$ that knows not only the expectation value of the metric operator but also those fluctuations in geometry that dynamics of quantum fields $\hat{f}_i$ is sensitive to. 
$\t{g}_{ab}$ is called the \emph{dressed metric} and by construction its coefficients depend on $\hbar$. The difficult task of evolving quantum fields $\hat{f}_i$ on quantum geometry is reduced to that of evolving them on the space-time of the dressed metric $\t{g}_{ab}$. We will assume that the  $\t{g}_{ab}$ can be found in the adiabatic phase of the Planck regime in the present case as well.

Results reported in Section \ref{s2.2} on geometry in the Planck regime suggest that the shaded (pink) region will contain a transition surface $\tau$  (w.r.t. $\t{g}_{ab}$) that replaces the classical singularity and separates the trapped region that lies to its past from the untapped region that lies to its future. The metric $\t{g}_{ab}$ is expected to capture three distinct effects: (i) those that originate from quantum geometry, originating in the area gap $\uDelta$ (as in Section \ref{s2.2}); (ii) those that are induced on $\t{g}_{ab}$ by the falling quantum matter in the incident pulse of the scalar field at the left end of the (pink) shaded region, and, (iii) those associated with the negative energy flux into the trapped region across the time-like part of the \TDH\!\!\!. Results reported in \ref{s2.2} strongly suggest that the first set of effects will decay rapidly as we move away from Planck curvature into the semi-classical region. Therefore in the semi-classical region, $\t{g}_{ab}$ will be well approximated by $g^{\rm (sc)}_{ab}$ used there. As we move to the future of the (pink) shaded region, one would encounter an anti-trapping dynamical horizon \ATDH (see Panel (b) of Fig.~\ref{fig:4}). The region enclosed by the transition surface $\tau$ to the past and \ATDH to the future would be anti-trapped. Thus, while geometry in the region bounded by \TDH to the past and \ATDH to the future is qualitatively similar to that of the quantum extension of Kruskal space-time (Panel (b) of Fig.~\ref{fig:3}), there is a key difference that arises because we are now in a dynamical situation: While the boundaries in the Kruskal extension are null IHs, now the past boundary is a \TDH with a space-like and a time-like branch, while the future boundary is an \ATDH that is space-like.

Finally, one would expect that the region to the future of \ATDH would be well approximated by an approximately flat metric with an outgoing flux with a small total energy ($\sim 10^{3}$ in Planck units $\sim 10^{-2} gm$ spread over astronomical scales. It will describe the propagation of the infrared modes that will emerge from the \ATDH and arrive at $\scrip$ at very late times.  Recall that these are the partner modes which were entangled with the outgoing modes that carried away most of the initial ADM mass to $\scrip$. In the LQG scenario, then, correlations are finally restored at $\scrip$ where, in the end, the partner modes also arrive. The total energy carried by the two sets of modes is very different. But this is not an obstruction for restoring correlations, i.e., for the `purification' to occur since, as emphasized before, there is no direct correlation between energy flux and entanglement.

A commonly held notion that purification should occur before Page time (when the black hole has lost only half its mass through quantum radiation) implies that correlations have to be restored already in the semi-classical phase. As discussed earlier, recent investigations provide strong arguments against this possibility \cite{agulloetal}. In the LQG scenario purification occurs \emph{much later} because singularity resolution allows the partner modes to emerge from the trapped region and reach $\scrip$, traveling across the transition surface $\tau$ that replaces the singularity. The timescale of this purification process would be very long, $\mathcal{O}(M^{4})$ \cite{ori4,ebtdlms,bcdhr}. But there is a pathway to restoration of correlations at $\scrip$, thereby making the total quantum state at $\scrip$ pure. This is a mainstream viewpoint in LQG. 

However, a rather glaring open issue remains: the red blob in Panel (b) of Fig. \ref{fig:4}. In this region,  not only is the curvature of Planck scale, but it is varying extremely rapidly because it lies at the end point of the evaporation process. Together, these two effects make the known approximation methods inapplicable. There \emph{are} approaches to evolve across this region using full quantum gravity both in the Hamiltonian \cite{ori5} and the path integral (spinfoam) approaches  \cite{dhrv}. But they have limitations. So far they do not account for the very non-trivial features that arise in the pathway to restoration of purity outlined above. Let me use Panel (b) of Fig. \ref{fig:4} to illustrate the type of challenges that remain. If this scenario of information recovery is correct, then one would expect that, as one approaches $u=u_1$ along $\scrip$, the temperature of radiation would grow since the time-like potion of \TDH is rapidly shrinking to Planck size. Therefore, the emitted quanta would be ultra-violet at $\scrip$. On the other hand, to the future of $u=u_2$ on $\scrip$, one would find infra-red modes. How does this dramatic transition from ultraviolet to infrared come about? Presumably this is because of  non-trivial effects originating in the red blob. But so far we do not have a systematic understanding of the possibilities. Similarly, there are issues concerning the anti-trapping horizon \ATDH\!\!\!.  Is the \ATDH stable w.r.t. small perturbations?%
\footnote{The trapping horizon \TDH should be stable because both its branches are so-called FOTHs --future outer trapping horizons \cite{hayward,ak2}. The \ATDH on the other hand is not; it is `inner' in this terminology.}
One could presumably address this issue using the metric $\t{g}_{ab}$ but it is not clear what the class of physically relevant perturbations would be. A second issue concerns the geometry in its future neighborhood of \ATDH away from the red blob. In our scenario, the \ATDH  is foliated by 2-spheres with radii less than $\sim\, 10^3 \lp$ but the length in the transverse direction is enormous, $> \, 10^{60}$lyrs (both measured using $\t{g}_{ab}$). Since the total energy density in the infra-red modes is small and spread out over astronomical length scales, the geometry to the future of \ATDH should be approximately flat (assuming the red blob has no effect on it). Therefore if one chose an approximately flat 3-surface $\Sigma$ in a neighborhood of \ATDH to its future and adapted to spherical symmetry, it would be foliated by 2-spheres whose are radii increase monotonically as we move to right. Therefore, say, half way between the left end and the red blob, the radius of these 2-spheres on $\Sigma$ would have to be $\sim 10^{60}$lyrs, while the radius of the `corresponding' 2-sphere on \ATDH would be $\sim 10^3 \lp$. Is there an admissible nearly flat metric that admits such dramatic growths in the size of 2-spheres as we pass from \ATDH to a nearby $\Sigma$?  I should add that I do not know of any concrete obstructions to constructing the required nearly flat 4-geometry, or showing stability of the \ATDH. In fact I think that general ideas underlying the pathway are likely to be correct. But such issues have not been adequately addressed yet and so there is considerable food for thought.

To summarize, LQG does provide a pathway to obtain a coherent space-time description of the black hole evaporation process in which correlations are restored at late time on $\scrip$, restoring the purity of the final state. The value of the pathway lies in the fact that one has a concrete scenario that one can try to prove or falsify. Along the way one may find genuine surprises. In my view --not widely shared in the LQG community-- the possibility that genuine quantum gravity effects associated with the red blob will lead to information loss, e.g., by creation of a baby universe is not ruled out. But prior experience with the Planck scale regime --especially in LQC-- leads me to believe that this is unlikely.

\section{Discussion}
\label{s4}

The last three sections summarized a mainstream LQG viewpoint on the process of black hole evaporation. As emphasized in sections \ref{s1} and \ref{s2}, it departs from commonly held views based on Hawking's original proposal in that there is neither an EH nor a singularity in the final picture \cite{aamb}. These features constitute corner stones of most of the LQG work on black hole evaporation. The detailed scenario for the evaporation process was then built using four sets of concrete results: (i) properties of DHs in classical and semi-classical gravity; (ii) the natural resolution of space-like singularities due to quantum geometry effects; (iii) properties of the quantum extension of the part II of Kruskal space-time that contains the black hole singularity; and, (iv) a strategy to handle quantum fields propagating on quantum space-times in the Planck regime when dynamics is adiabatic. These results were obtained by a very large number of researchers, and even with a rather long bibliography I could include only a sample of this rich literature. 

In the LQG community, there is a general agreement on the description semi-classical phase summarized in section \ref{s3.1}, although we need more detailed calculations to arrive at the space-time geometry in the trapped region directly from the semi-classical equations. Currently, much of our understanding is shaped by the detailed analyses of the CGHS model \cite{ori2,apr,ori5}. While it captures several features of the 4-d spherically symmetric gravitational collapse and subsequent evaporation, as discussed at the end of section \ref{s2.1.2}, it also differs from the 4-d model in certain important respects. However, a recently proposed model \cite{mv} does not have these limitations, even though it is also exactly soluble classically. Its semi-classical equations (in the $1/N$ expansion) have been written down and they are similar to those of the CGHS model. A high precision numerical study will soon be undertaken by Fethi Ramazanoglu and Semith Tuna. They should provide a much more reliable description of the semi-classical phase of the 4-d evaporation process. The model does make an approximation: it ignores the back scattering effects. If one can make the same approximation in the derivation of the Hawking effect to begin with, one misses the gray body factors which can be added subsequently. The hope is that the situation would not be different also at the semi-classical level when the back reaction is included. 

A concrete strategy to go beyond the semi-classical approximation is sketched in section \ref{s3.2}.
It provides a plausible pathway to restoring correlations on $\scrip$, thereby ensuring the purity of the final quantum state there. A visual comparison between the LQG proposal depicted in Panel (b) of Fig. \ref{fig:4} and the commonly used proposal shown in Panel (c) brings out the fact that this pathway is possible precisely because the EH of (c) is replaced by DHs in (b) and the singularity in (c) by a regular transition surface $\tau$. However, difficult issues remain in the quantum evolution beyond the semi-classical regime, and variations on the strategy presented in section \ref{s3.2} are also being pursued in the LQG community. 
But in all these approaches a key open issue remains: The well-developed approximation methods of LQG are inapplicable to the red blob in the Panel (b) (where the trapping and anti-trapping regions meet) because not only does it have Planck scale curvature but it is also highly dynamical. The space-time portion to the future of the \ATDH is purposely left blank in panel (b) because of the uncertainties on the influence of the red blob on the geometry in this region. The value of having a concrete paradigm --such as the one sketched in section \ref{s3.2} lies primarily in the fact that it raises specific interesting questions that one may not have envisaged, thereby providing directions for further work that can confirm or falsify expectations. 

We will conclude with a few general remarks:\vskip0.05cm

1. In section \ref{s2.2}, we saw that region II of Kruskal space-time admits an LQG extension in which the singularity is replaced by a transition surface (see Panels (a) and (b) of Fig. \ref{fig:3}). As we remarked at the end of that section, this quantum corrected geometry has been also been extended to include asymptotic regions \cite{aos,gop,mhhl}. One can therefore ask if the quantum corrected metric gives rise to effects that would be relevant astrophysically. Given the quantum corrections to the metric near the horizon are \emph{extremely} small for astrophysical black holes, one would expect that the answer to be in the negative. This has been borne out in detailed analyses of quasi-normal modes (see. e.g.,\cite{kunstatter,dco}).

2. The LQG literature on collapsing models is very rich (see, e.g., recent discussions in \cite{hhcr,bcdhr,pmdcr,pmd,lmyz,hrs,fhwe,cfwe}) and often draws on earlier works on regular black holes \cite{hayward,frolov,bardeen}. These  models have provided us with concrete possibilities for the quantum corrected geometries. However, generally they do not discuss issues related to entanglement between the modes radiated to $\scrip$ and their partner modes that are initially trapped, nor to pathways to purification. At times, regular black hole models have suggested incorrect avenues for information recovery \cite{hayward-conf}. Finally, most of these models focus on stellar collapse which does not constitute a closed system that is necessary to the discussion of purification. That is why these models were not discussed in detail in this brief report, even though they have provided interesting insights. 

3. The discussion of section \ref{s3.1} shows that, in the LQG perspective, there is a major difference between a \emph{young} lunar mass black hole that just formed due to gravitational collapse, and an isolated, \emph{old} lunar mass black hole that has resulted due to quantum radiation, starting from a solar mass black hole that was formed some $10^{64}$ years ago. While their dynamical horizons will have the same radius, $\sim\,0.1$mm, and mass $M_{\TDH} = M_{\text{\leftmoon}}$, their  external environment as well as internal structure will be \emph{very} different. In the case of an old black hole, a very large number of quanta would have been emitted to $\scrip$ and their partner modes would be trapped in the region enclosed by \TDH\!\!\!.  Therefore, the area of the time-like branch of the \TDH would not be a good measure of the von-Neumann entanglement entropy for the old black hole. In the LQG perspective, for both black holes, area is a measure of the number of horizon degrees of freedom that can interact with those in the trapped region as well as those that are outside. 

4. It is often argued that there is a potential problem with the notion old black holes (and hence with the discussion of the semi-classical sector in section \ref{s3.1}): Because old black holes can have small energy but an enormous number of modes, it should be easy to produce them copiously in particle accelerators. But these arguments use only the conservation laws normally used in computing scattering amplitudes in particle physics. Old black holes, on the other hand,  have astronomically long necks with a very large number of infrared modes. They are hardly particle like remnants! It is hard to imagine how such configurations can be created on time scales of accelerator physics \cite{ori1}.

\section*{Acknowledgments}

Over the years, I have profited greatly from stimulating discussions on black hole evaporation with a large number of colleagues. For the material included in this article, I would especially like to thank I. Agullo, E. Bianchi, T. De Lorenzo, M. Han, B. Krishnan, J. Olmedo, A. Ori, F. Pretorius, F. Ramazanoglu, P. Singh and M. Varadarajan. I would also like to thank Tommaso De Lorenzo and Fethi Ramazanoglu for preparing most of the figures. This work was supported in part by the Eberly and Atherton funds of Penn State, USA and the  Distinguished Visiting Research Chair program of the Perimeter Institute, Canada.

\bibliographystyle{utphys}

\bibliography{Ashtekar_refs}

\title{How the Black Hole Puzzles are Resolved in String Theory}
\author{Samir D. Mathur}

\institute{\textit{Department of Physics\\The Ohio State University\\Columbus, OH 43210, USA}}

\maketitle
\begin{abstract}
String theory has provided a resolution of the puzzles that arise in the quantum theory of black holes. The emerging picture of the hole, encoded in the `fuzzball paradigm', offers deep lessons about the role of quantum gravity on macroscopic length scales. In this article we list these puzzles and explain how they get resolved. We extract the lessons of this resolution in a form that does not involve the technical details of string theory; it is hoped that this form will allow the lessons to be absorbed into other approaches to quantum gravity.
\end{abstract}

\section{Introduction}

How should we quantize gravity? Should we just write the classical Hamiltonian  and impose  canonical commutation relations? Or should we pass to loop variables and the novel Hilbert space structure that emerges? Or should we accept string theory with all its complicated structure of higher dimensions and extended objects?

Black hole puzzles have long posed a challenge to any theory of quantum gravity. Thus we can look at how these puzzles are addressed in any given theory and decide if the relevant features should be part of our final understanding of quantum theory. 

In this article we will list these black hole puzzles and explain how each of the puzzles is resolved in string theory. We will then extract the relevant  lessons in a form that does not involve the technical intricacies of string theory, with the hope that these lessons can be used in any final approach to quantum gravity. The `fuzzball paradigm' which has emerged in string theory gives a picture of the black hole interior that is radically different from the one we expect semiclassically. This unexpected change at classical length scales has deep implications for the nature of the quantum gravity vacuum, with a potential impact on issues in cosmology like the Hubble tension, the cosmological constant and inflation. 

Black holes present us with three inter-related puzzles:

(A) {\it The entropy puzzle:} Gedanken experiments suggest that a black hole should have an entropy given by its surface area \cite{Bekenstein:1973ur}
\begin{equation}
S_{bek}=\frac{c^3}{\hbar}\frac{A}{4G}
\label{one}
\end{equation}
But black holes `have no hair', so what is this entropy counting? On the other extreme, Wheeler's bags-of-gold construction suggests that we can put an arbitrarily large  entropy inside the horizon. So the puzzle is: does (\ref{one}) represent a count of microstates for the black hole, in the way we understand in usual statistical mechanics?

(B) {\it The information paradox:} The vacuum around the horizon is unstable to the creation of entangled pairs. One member of the pair (which we call $b$) escapes to infinity as Hawking radiation, while the other member (which we call $c$) has negative energy and falls into the hole to lower its mass. The two members of the pair are in an entangled state which we may schematically model as
\begin{equation}
|\psi\rangle_{pair}=\frac{1}{\sqrt{2}}\left ( |0\rangle_b|0\rangle_c+|1\rangle_b|1\rangle_c\right)
\label{two}
\end{equation}
Thus the entanglement of the radiation with the remaining hole keeps increasing monotonically, leading to a sharp puzzle at the endpoint of evaporation: if the hole disappears, we are left with radiation that is entangled, but there is nothing that it is entangled {\it with}. Such a situation cannot be described by {\it any} wavefunction, leading to a breakdown of quantum unitarity \cite{Hawking:1975vcx}.

One might try to evade this problem by postulating small (hitherto unknown) quantum gravity effects that induce subtle correlations between the large number of  emitted quanta, with the hope that these small corrections would somehow remove the   entanglement of the radiation with the hole. But the small corrections theorem shows that such a resolution is not possible \cite{Mathur:2009hf}. Suppose (i) the emission at each step is close to the semiclassical one
\begin{equation}
|\psi\rangle_{pair}=\frac{1}{\sqrt{2}}\left ( |0\rangle_{b_i}|0\rangle_{c_i}+|1\rangle_{b_i}|1\rangle_{c_i}\right)+|\delta\psi_i\rangle, ~~~~\Big | |\delta \psi_i\rangle \Big |<\epsilon
\label{three}
\end{equation}
and 
(ii) there is no significant change to the state of a quantum after it recedes sufficiently far from the hole (i.e., no large distance nonlocality). Then the entanglement $S_{ent}(N)$ at emission step $N$ must keep rising monotonically as 
\begin{equation}
S_{ent}(N+1)>S_{ent}(N)+\log 2 -2\epsilon
\label{four}
\end{equation}

One can evade the violation of quantum unitarity by postulating the quantum gravity effects stop the evaporation when the hole reaches planck size, trapping the initial matter and all the infalling quanta $\{ c_i\}$ in a planck sized `remnant'. (Such remnants have been modeled as baby universes joined to the rest of spacetime by a planck sized neck.) But in string theory we cannot have remnants. Since we can reach the remnant by starting with an arbitrarily large hole, the remnant should have an unbounded number of possible internal states for an energy $E\lesssim m_p$.  But in string theory we believe that we know all the states at planck energy: we just have a few possible states of strings and branes. Further, the conjecture of AdS/CFT duality says that the number of states of such a remnant will equal the number of states in the dual gauge theory on a sphere below the energy scale corresponding to $E=m_p$; the number of such states in the gauge theory is finite.

Thus in string theory the entanglement of the radiation with the hole {\it must} go down to zero as we approach the endpoint of evaporation. Since small corrections to horizon dynamics cannot change the monotonic rise (\ref{four}), we need order {\it unity} corrections; i.e., the horizon cannot be a normal semiclassical region of spacetime. How should we explain such a big change in a region where the classical curvature was low: ${\mathcal R}\ll l_p^{-2}$?

(C) {\it The infall problem:} What does an infalling observer feel as he falls through the horizon? Semiclassical dynamics suggests that he feels nothing special, and carries his information inside the horizon. But then how will this information emerge, given the `inward pointing' nature of light cones inside the horizon? Does his information get `bleached' or `duplicated' as he crosses into the horizon? Is the traditional Penrose diagram a reliable guide to the structure and dynamics of the quantum black hole? 

Let us now see how string theory resolves these puzzles.

\section{Resolving the puzzles in string theory} 

The first difficulty that one encounters in a canonical quantization of  gravity is that loop amplitudes diverge, and cannot be renormalized. String theory, on the other hand, gives finite loop amplitudes. {\it The essential feature allowing a taming of the divergences is the existence of an infinite number of particle species, which arise as different vibration modes of the string.}

A second difficulty in any theory of quantum gravity is the divergence of the vacuum energy. String theory adopts the idea of supersymmetry to cancel this divergence. String theory is a unique theory with no free parameters: the spacetime dimension is $9+1$, and all the fundamental objects (strings, branes etc.) are uniquely determined, as are their tensions and interactions. 

Remarkably, this elaborate and unique structure can be obtained very simply (in retrospect) from classical supergravity,  by adding a simple requirement. The classical theory has certain discrete symmetries, called S and T dualities. {\it If we require that these symmetries be preserved at the quantum level, we obtain string theory.}

Let us now turn to the black hole puzzles.

(A') {\it Resolving the entropy puzzle:} Black holes are expected to involve strong effects of quantum gravity, so it appears difficult to understand and count black hole microstates. It turns out however that in string theory, for certain black holes, we can count the microstates without knowing their actual structure. The coupling in the theory is given by a scalar field (the dilaton), so we can study the microstates at weak coupling or at strong coupling.  We consider extremal black holes which have $M=Q$. Such holes are supersymmetric states of the theory, and an index argument says that the number of supersymmetric bound states with a given charge $Q$ cannot change as we vary the coupling. At weak coupling it is comparatively simple to count the degeneracy ${\mathcal N}$ of such a bound state, so we obtain a microscopic entropy $S_{micro}=\log {\mathcal N}$. Suppose we assume that at strong coupling this string state makes a classical extremal black hole with charge $Q$. Then we can compute the Bekenstein entropy $S_{bek}$  from the horizon area of this classical geometry. In all the cases which have been examined, one finds that \cite{Strominger:1996sh}
\begin{equation}
S_{micro}=S_{bek}
\end{equation}
This agreement is found to extend to the near-extremal holes which have been studied \cite{Callan:1996dv}. Based on these agreements, it appears that in string theory $S_{bek}$ indeed corresponds to a count of microstates of the black hole; in particular, we should not be able to construct bags-of-gold in the theory.

The above agreement is remarkable because it reproduces black hole entropy in terms of the count of states of well-defined objects (strings and branes). If we consider a theory with a given number of particle species, then the entropy at high energies in $d+1$ dimensions scales as $S\sim E^{d/(d+1)}$; in particular, we can never get $S\sim E^\mu$ with $\mu>1$. The black hole in 3+1 dimensions, on the other hand, has $S_{bek}\sim E^2$, so we have to ask how the value $\mu=2$ is obtained in the computation of $S_{micro}$ in string theory. This power $\mu>1$ was possible because when we have extended objects, we do not have a fixed number of particle species. Exciting a string or brane to higher and higher energies generates more and more new species of particles, and this leads to the rapid growth of the entropy with energy $E$.

{\it Thus the same feature of string theory which removed loop divergences at the planck scale (an infinite number of particle species) comes back in the domain of black holes $E\gg m_p$ to reproduce exactly the entropy of black holes.}

(B') {\it Resolving the information paradox:} To actually understand the structure of the  hole, we must consider the microstates that we counted above at {\it strong} coupling, where we expect the strings and branes to generate a black hole. Over the years it has indeed become possible to construct many families of brane bound states at strong coupling, and to examine their gravitational structure.  In particular all the microstates of the simplest  hole -- the 2-charge extremal hole in 4+1 dimensions -- have been constructed explicitly. In all these cases, we find that the microstate never develops a horizon; instead we find a horizon sized quantum object -- called a fuzzball -- which radiates from its surface like a normal body \cite{Lunin:2001jy, Lunin:2002iz, Kanitscheider:2007wq, Mathur:2005zp, Bena:2007kg, Bena:2017xbt}. Thus we do not get the rising entanglement (\ref{four}), and there is no information paradox.

This structure of black hole microstates is remarkable because in classical gravity we expect all the matter to rush to a central singularity, leaving the rest of the black hole interior in a vacuum state. If we make the black hole out of strings and branes, a naive expectation would be that the size of the brane bound state is $\sim l_p$, and this smears the central singularity from $r=0$ to $r\lesssim l_p$. The source of this naive expectation is that $l_p$ is the only length scale that we can make from the fundamental constants $c, \hbar,G$. But a  black hole is made from a large number $N$ of elementary objects; the larger the hole, the larger the value of $N$. Thus we should ask if the relevant length scale of quantum gravity effects in our theory is $\sim l_p$ or $\sim N^\alpha l_p$ for some $\alpha>0$.  What we find is that in string theory we have the latter possibility, with $\alpha$ being such the the relevant length scales are of order the horizon radius.  This  circumstance prevents existence of horizons in the theory, removing the information paradox. In \cite{Gibbons:2013tqa, Mathur:2016ffb} it is shown how the various features of string theory bypass the various no-hair arguments and expectations like the Buchdahl bound.

{\it Thus in theories where the effective number of particle species increases with the energy, the size of brane  bound states can grow with the number $N$ of quanta in the state, so that the states never collapse inside their own horizon. This resolves the information paradox.}

(C') {\it Resolving the infall problem:} Since curvatures are low at the horizon, it appears that a collapsing shell will keep shrinking towards $r=0$, leaving a vacuum region outside the horizon. But in string theory we find that there is a nonzero amplitude for the shell to tunnel into one of the fuzzball microstates. The tunneling probability is small, since we are tunneling between two macroscopic configurations. For an estimate, we set all length and time scales to be $\sim GM$, getting a classical action 
\begin{equation}
S_{cl}\sim \frac{1}{G}\int {\mathcal R} d^4 x\sim \frac{1}{G} \frac{1}{(GM)^2} (GM)^4\sim GM^2\sim \left ( \frac{M}{m_p}\right )^2
\end{equation}
This implies a small tunneling probability $P\sim Exp[-2S_{cl}]$ to any fuzzball state. But there are a large number of fuzzball states to tunnel to 
\begin{equation}
{\mathcal N}\sim e^{S_{bek}}\sim e^{GM^2}
\end{equation}
We see that due to the large entropy of the hole, it is possible to have $P{\mathcal N}\sim 1$, making the tunneling effect a significant correction to the semiclassical dynamics \cite{Mathur:2008kg, Kraus:2015zda, Bena:2015dpt}. The classical hole does not remain an approximate eigenstate of the full quantum gravity theory; instead it becomes a broad resonance. A collapsing shell tunnels into a linear combination of fuzzball microstates in a time which is much shorter than the Hawking evaporation timescale. The fuzzballs radiate from their surface like normal bodies, so there is no information paradox. 

We see that the collapsing shell state $|\psi_{shell}\rangle$ and the fuzzball state it tunnels to $|\psi_{fuzz}\rangle$ are two different branches of the quantum gravity wavefunctional, with probabilities that add up to unity: $P_{shell}(t)+P_{fuzz}(t)=1$. Thus the situation is like that of a Schrodinger cat wavefunction, where the radioactive decay of an atom triggers a lethal cyanide dose. The probability of the cat being alive drops from unity towards zero, while the probability of the cat being dead rises from zero towards unity. It is clear that we should not combine both branches of the cat wavefunction into one semiclassical description. With the black hole, the part of the wavefunction $|\psi_{shell}\rangle$ described the shell going through its horizon and thus creating a black hole interior, while the part $|\psi_{fuzz}\rangle$ described a fuzzball which was like a star having no horizon or region interior to this horizon. Combining both parts into a common Penrose diagram would be an error; thus the traditional Penrose diagram does not give the correct description of the quantum hole. The reason why a macroscopic object (the black hole) ended up in a Schrodinger cat type of wavefunction can again be traced back to the large degeneracy of microstates, which enhanced the tunneling rate to a value that competes with the rate of  classical evolution. 

Suppose one were to ask: what does an observer falling in with the shell feel? The answer can be found by looking at the corresponding state which describes the alive cat.  The alive cat does not feel that it is partly dead; its just the probability that the cat is in this alive state that decreases smoothly towards zero. In a similar manner, the infalling observer does not feel anything unusual as he continues his journey into the black hole 
interior, but the probability that he exists in that state goes smoothly to zero \cite{Mathur:2024mtf}.

{\it Thus the infall problem is resolved by understanding the role of the large number of microstates implies by the Bekenstein entropy. The full quantum gravity wavefunctional of  the hole spreads over all these states, which have the same quantum numbers as the classical hole, invalidating the traditional black hole geometry as an adequate semiclassical description.} 

\section{Summary}

String theory provides an elegant resolution to the puzzles that arise in the semiclassical description of black holes. We have described these puzzles and outlined their resolution. We have phrased these resolutions in abstract terms, so that the relevant ideas can be extended to other theories of quantum gravity. The large value of the Bekenstein entropy indicates that the number of effective particle species should grow quickly with energy. In string theory this growth is obtained because the elementary excitations are extended objects like strings and branes. This large degeneracy of states softens away the loop divergences at the planck scale, and also reproduces the Bekenstein entropy as the degeneracy of states at $E\gg m_p$. Bound states in such a theory have a size that grows with the number $N$ of elementary excitations in the state, so that the bound state never becomes smaller than its horizon radius.  Thus we get `fuzzballs' rather than black holes with horizons, evading the information paradox. The spread of the quantum gravity wavefunctional over this large set of microstates invalidates  the semiclassical approximation at the horizon scale which is a region of low curvature in the classical geometry: the smallness of the tunneling amplitude is offset by the largeness of the degeneracy of states that we can tunnel to.

With this understanding of black holes, we note that similar scales arise in cosmology. The mass inside the cosmological horizon $H^{-1}$ is of the order that would  make a black hole with radius $H^{-1}$. Given the order unity corrections to semiclassical dynamics that we find in the black hole interior, it is plausible that we should have corresponding departures from semiclassical dynamics at the scale of the cosmological horizon. Such corrections can impact issues like the Hubble tension, which suggests that there is Early Dark Energy at the radiation-dust transition, and the problem of the cosmological constant. For a discussion of how the lessons from black holes might resolve these puzzles, see \cite{Mathur:2024mtf}. 

\bibliographystyle{utphys}
\bibliography{bhio_mathur}

\title{Quantum Black Holes: From Regularization to Information Paradoxes}

\author{Niayesh Afshordi\inst{1,2,3} \and Stefano Liberati\inst{4,5}}

\institute{\textit{Department of Physics and Astronomy, University of Waterloo, Waterloo, Ontario, N2L 3G1, Canada} \and \textit{Waterloo Centre for Astrophysics, University of Waterloo, Waterloo, ON, N2L 3G1, Canada}\and 
\textit{Perimeter Institute for Theoretical Physics, 31 Caroline St N, Waterloo, Ontario, N2L 2Y5, Canada}
\and
\textit{SISSA, Via Bonomea 265, 34136 Trieste, Italy and INFN Sezione di Trieste} \and \textit{IFPU- Institute for Fundamental Physics of the Universe, Via Beirut 2, 34014 Trieste, Italy}}

\maketitle

\begin{abstract}
Quantum black holes, a broad class of objects that refine the solutions of general relativity by incorporating semiclassical and/or quantum gravitational effects, have recently attracted renewed attention within the scientific community. This resurgence of interest is largely driven by advances in gravitational wave astronomy, which have opened the possibility of testing some of these models in the near future. In this essay, we provide a concise overview of the key discussions that emerged during the ``Black Hole Inside and Out" meeting, held in August 2024 in Copenhagen. We report these ideas, their connections to the information paradox, and the potential use of analogue gravity as a test bed for these concepts.
\end{abstract}

\section{Introduction}

{\em “We all agree that your theory is crazy. The question which divides us is whether it is crazy enough to have a chance of being correct. My own feeling is that it is not crazy enough.”} --- Niels Bohr

Quantum black holes are fascinating objects that bridge the domains of general relativity (GR) and quantum gravity (QG). While classical GR describes black holes as regions of spacetime with gravitational fields so strong that nothing can escape, this theory fails near their centers, as it leads to formation of singularities. Quantum gravity aims to resolve this issue by providing a consistent framework that avoids the infinities associated with singularities or even modify the structure of a black hole at horizon scales, sometimes opening the possibility to generate a horizonless ultra-compact object.

In recent years, regular (singularity-free) black holes and, more generally, black hole mimickers (with or without horizons) have garnered renewed attention in the scientific community. This resurgence of interest can be largely attributed to advances in gravitational wave astronomy and imaging campaigns, spearheaded by experiments such as LIGO, Virgo, and the Event Horizon Telescope, as well as the promising future capabilities of observatories like LISA, Cosmic Explorer, and the Einstein Telescope. These experimental developments have opened new avenues for probing the nature of black holes, providing empirical data that could potentially reveal deviations from the classical solutions predicted by General Relativity (GR).

The study of quantum black holes offers a unique opportunity to explore the interplay between GR and quantum mechanics. By investigating these objects, we can address some of the most profound questions about the universe, such as the nature of spacetime, the fate of information, and the role of quantum effects in extreme gravitational fields. They also force us to switch the attention from stationary geometries, characterized by teleological objects such as event and Cauchy horizons, to local dynamical notions of trapped regions characterized by inner and outer dynamical horizons, with several implications not only for their classical behavior but also for long-standing puzzles like that of the information loss.

This essay explores briefly the various aspects of quantum black holes, horizonless ultra-compact objects, quantum hair, and the information paradox, drawing from recent theoretical developments and experimental prospects.

\section{Regular Black Holes}
One of the key challenges of quantum gravity is the regularization of singularities that appear in classical black hole solutions. Regular Black Holes (RBH), as reviewed by Alessia Palatania in this meeting, are models that eliminate these singularities by introducing a modified interior structure, often motivated by quantum corrections. These solutions have become increasingly popular in the literature, as they suggest that black holes might not be the catastrophic objects we once thought, but rather stable entities with finite energy densities at their cores.

A significant question arising from RBH models is whether this regularization of singularities is a generic feature of quantum gravity or if it is specific to certain models. For instance, some RBH proposals, like the Bardeen black hole, are well-known examples where singularities are replaced by a de Sitter core, and similar solutions might arise from quantum gravitational models, see e.g.~\cite{Alesci:2020zfi}.

Note, however, that renormalizable generalized gravity actions --- such as quadratic or projectable Horava gravities --- might still entail singular black hole solutions (e.g.~\cite{Podolsky:2019gro,Lara:2021jul}). This leaves open the question of whether singularity regularization is generically achievable only in non-perturbative discrete quantum gravitational theories and is precluded in models based on a continuous spacetime even if UV complete. Alternatively, it may be that the proper ``observables'' of such theories remain well behaved, even when GR observables like curvature and geodesics do not. 

Another important question concerns certain singularities, like the one associated with a negative mass Schwarzschild solution. These are commonly excluded and are considered nonphysical, as their regularized version cannot have a stable ground state~\cite{Horowitz:1995ta}. However, this might also signify that quantum gravity should do more than regularize such spacetimes, it should also avoid any breakdown of global hyperbolicity, which is generically violated in these cases. In addition, achieving global hyperbolicity in QG-regularized spacetimes appears beneficial, as it inherently upholds strong cosmic censorship and Hawking's chronology protection conjecture. This would, for instance, avert causal paradoxes linked with spacetimes that contain closed timelike curves.

The investigations into regular black holes have yielded several remarkable outcomes, some of which are quite far-reaching. One key finding is that regular black hole spacetimes can be categorized into a limited number of geometric families~\cite{Carballo-Rubio:2019nel,Carballo-Rubio:2019fnb}. Specifically, these geometries either feature pairs of outer and inner horizons or consist of an outer horizon with a spacelike transition bounce throat morphing a trapped region into an anti-trapped one (solutions of this kind are often called black bounces or hidden wormhole black holes). 

Allowing for integrable singularities~\cite{Lukash:2013ts,Casadio:2024lgw} (curvature singularities not implying infinite tidal forces when crossed) or for hybrid solutions mixing the aforementioned two kinds of regular black hole, one can basically cover most, if not all, the examples in the extant literature. Nevertheless, this restricted classification suggests that the space of possible regular black hole solutions is more constrained than previously thought.

Another significant lesson from these studies is that, despite being generated by quantum gravity effects that regularize the singularities inside GR black holes, all of these geometries exhibit deviations from classical black hole solutions even outside the outer horizon. Such deviations can be tiny if the regularization scale is fixed to be Planckian, but can be detectable if it is not fixed to be so (and, as we shall see later, this is a concrete possibility). This challenges the conventional wisdom that any phenomena occurring inside the horizon is hidden from external observers and, therefore, undetectable. The fact that these geometries are globally different from the GR one, and not just locally modified, might have profound implications for our understanding of black holes and their observational signatures.




\section{Horizonless Ultra Compact Objects}

Horizonless Ultra-Compact Objects (UCOs) represent another intriguing possibility in the context of quantum gravity. UCOs are compact objects that, unlike black holes, do not possess event horizons. Instead, they may sport light rings, regions in which light can orbit the object indefinitely. These objects can be found both as the horizonless limit of some of the well-known one parameter families of regular black holes (see e.g.~\cite{Carballo-Rubio:2022nuj}), or can be introduced by looking at self-consistent stars solution in semiclassical gravity~\cite{Arrechea:2023oax}. However, they are most naturally motivated by the possibility of large quantum gravity effects that may destroy causal event horizons, e.g., to avoid the black hole information paradox (see below) \cite{Abedi:2020ujo}.

One of the key challenges in studying UCOs is understanding how they can dynamically form. In classical gravity, trapping horizons are a natural consequence of the collapse of sufficiently massive objects. However, in quantum gravity, the existence of horizons is less certain. Some models suggest that quantum effects could prevent the formation of horizons, leading to the creation of horizonless UCOs instead (for example, in the form of a fuzzball~\cite{Mathur:2005zp}), or instabilities associated with regular black holes (see discussion below) might produce them as the end point of a transient trapped region. However, we still lack a concrete dynamical scenario for their formation.


The search for UCOs and their potential observational signatures, such as GW echoes~\cite{Abedi:2016hgu,Cardoso:2016rao}, is an exciting frontier in astrophysics. Observations of these echoes could provide strong evidence for the existence of horizonless objects and, by extension, for the breakdown of classical general relativity in the strong-field regime. However, it must be stressed that the detectability and characteristics of such echoes can be strongly model-dependent~\cite{Oshita:2019sat,Wang:2019rcf,Oshita:2023tlm,Vellucci:2022hpl,Arrechea:2024nlp} and that, in this sense, a more accurate modeling will be needed in the future.

Future experiments using Very Long Baseline Interferometry (VLBI) may also play a crucial role in testing these ideas and refining our understanding of compact objects in the universe. For example, by testing the possible re-emission of radiation by a UCO~\cite{EventHorizonTelescope:2022xqj,Carballo-Rubio:2022imz,Carballo-Rubio:2023fjj} or the characteristics of their shadows that are expected to sport subtle differences with the one of objects endowed with trapping horizons~\cite{Carballo-Rubio:2023ekp}.

\section{Instabilities, evolving horizons and compact objects}

A very important realization in recent years is that most, if not all, of the aforementioned geometries (with or without horizons) cannot be eternal. In general, they are characterized by a combination of instabilities that will render their geometry dynamical and possibly evolving on time scales much shorter than the one normally associated with evaporation by Hawking radiation($\sim M^3$ where $M$ is the black hole mass\footnote{Note that this is also the order of magnitude of the Page time, which appears in several proposals for the resolution of the information loss problem.}).

A notable category of these regular geometries is characterized by inner trapping horizons, which have recently been found to be subject to exponentially rapid accumulation of energy in their vicinity~\cite{Carballo-Rubio:2024dca}. This can be seen as a transient version of the usual mass-inflation phenomenon characterizing inner horizons in stationary geometries, where inner horizons are also Cauchy horizons. Furthermore, in stationary geometries, inner/Cauchy horizons were also found to suffer a semiclassical instability associated with the Unruh state imposed by the regularity of the outer trapping horizon~\cite{Balbinot:2023vcm,McMaken:2024tpc,McMaken:2024fvq}. Let us stress that all of these instabilities are quite general and insensitive to the presence or absence of a singular core, so much so that they also apply to Kerr black holes.

The most obvious conclusion is that these instabilities will lead to a non-negligible backreaction on the black hole geometry. However, it is not yet clear whether, as a consequence of these combined instabilities, the inner horizon will move inward or outward. 

It might end up moving inward and (assuming again singularity avoidance via quantum gravity effects), possibly producing a bounce (see, e.g.~\cite{Barcelo:2016hgb,Malafarina:2017csn,Bianchi:2018mml}) or moving outward and making the trapped region close~\cite{Barcelo:2022gii}. In this case, one could get a long-lived quasi-extremal black hole -- as both mass inflation and the semi-classical instability switch off as the surface gravities of both the inner and outer horizons tend to zero -- or an exactly extremal one (which could be still unstable, e.g., due to the Aretakis instability~\cite{Aretakis:2012ei}). The trapped region might even end up closing in a finite time, leaving a corresponding UCO. 

However, even such UCOs may be prone to instabilities! For instance, in the absence of dissipative properties, they could experience ergoregion instabilities, where the object’s rotation amplifies perturbations. Another potential issue arises from the generic presence of a stable light ring in addition to the usual unstable one, which could also lead to instability. These processes might eventually cause the UCO to collapse, forming a trapped region, or expand into a less compact configuration where these instabilities are either avoided or mitigated, allowing the UCO to remain at least metastable (long-lived).

Such signals are the time domain counterpart of the frequency domain feature of long living quasi-normal modes (QNMs) characterizing all of these objects as a consequence of the fact that if they sport the usual unstable light ring producing the shadows observed via very long baseline experiments, such as the Event Horizon Telescope (EHT), then they must also possess an inner stable light ring~\cite{DiFilippo:2024mnc}. 

Very long living QNM are usually considered harbingers of nonlinear instabilities which might imply a large backreaction on the object and a dynamical evolution, possibly toward a black hole configuration or alternatively towards a less compact one without light rings (which would then be a ``bad" black hole mimickers as it would not be able to produce a black-hole-like shadow). However, very few computations have been carried out in this sense (see, e.g.~\cite{Cunha:2022gde}) and any conclusion would be still premature.

We should note that the stability of UCOs under arbitrary accretion flows would generically require some non-local physics, given that for local interactions a sufficiently high accretion rate could require the UCO to expand superluminally in order to keep its own compactness from overriding the black hole one~\cite{Carballo-Rubio:2018vin}. Alternatively, one would have to postulate that these objects (if long-lived) can momentarily form transient, but unstable, trapped regions when brought out of equilibrium, which, however, revert to the UCO state at late times in the absence of accretion (possibly due to the aforementioned inner horizon instability).

Finally, let us comment on the fact that most of these UCO features/instabilities might act on timescales that are highly model-dependent. For example, in the realistic case of an imperfect reflective surface, the core of such objects can play a crucial role. While in the limiting horizonless configuration of the regular black hole spacetimes, such a core is low curvature for typical astrophysical black hole candidate masses, in other models it might be characterized by a very high curvature, which would make the crossing time for radiation or gravitational wave very long, effectively drastically tame the aforementioned instabilities and greatly reduce the possibility of echo detections (see, e.g.~\cite{Arrechea:2024nlp} but also \cite{Oshita:2023tlm} for an alternate perspective).

In summary, the aforementioned instabilities suggest a highly dynamic nature for regular black holes and UCOs (as well as for singular black holes in general relativity!). This dynamism may indicate a connection between these objects, representing distinct phases of the final stages of a non-singular gravitational collapse.

\section{Quantum Hair and Information Paradoxes}
One of the most debated topics in black hole physics is the information paradox. In classical GR, black holes are characterized by their simplicity: according to the no-hair theorem, a black hole can be fully described by only three parameters: mass, charge, and spin. However, quantum gravity may allow for additional features, often referred to as ``quantum hair'', which could encode information about the matter that formed the black hole.

The presence of quantum hair could potentially resolve the information paradox, which arises because, in classical GR, all information about the matter that falls into a black hole is lost to the outside universe. However, in quantum mechanics, information must be preserved, leading to a conflict between the two theories. One proposed solution is that quantum hair allows black holes to retain subtle information about their past states, which could be released in the form of radiation over time.

Several competing ideas have been proposed to resolve the information paradox:
\begin{itemize}
    \item \textbf{Regular Black Holes:} A point of view, advocated by Abhay Ashtekar at the conference, was that black holes may retain the relevant information until the end of evaporation, thanks to the extremely long  leaves of constant radius within trapping regions. These long leaves are indeed crucial to evade the holographic entropy bound.
    
    \item \textbf{Fuzzballs:} Another proposal, championed by Samir Mathur at this conference, suggests that classical stars would quantum tunnel into a large ensemble of horizonless quantum microstates prior to forming event horizon. The thermal radiation from this ensemble of ``fuzzball geometries'' would replicate Hawking radiation and carry away the information. 

    \item{\bf Islands:} The point of view advocated by Netta Englehardt was that in the semi-classical approximation, the Page time is associated with a phase transition where the degrees of freedom within an ``island'' inside the horizon become identical to those of Hawking radiation. The data that escapes from the horizon in this scenario is so thoroughly scrambled that deciphering it necessitates an exponentially powerful computer. 
   \end{itemize}

These ideas are not just theoretical curiosities; they could have observable consequences. For example, the presence of quantum hair might lead to deviations from the classical Kerr solution, which describes rotating black holes in GR. These deviations could potentially be detected in gravitational wave echoes, modified images, or love numbers, using current or future experiments, such as  the Laser Interferometer Space Antenna (LISA), the Einstein Telescope (ET), or even Event Horizon Telescope (EHT), or observations of black hole geodesy.

\section{Black Holes in Analogue Gravity}
In addition to studying black holes in astrophysical settings, researchers have explored the possibility of simulating black holes in laboratory environments. These analogue systems~\cite{Barcelo:2005fc}, such as acoustic black holes created in fluids, cannot reproduce a background independent geometrodynamics like that of General Relativity. However, they may serve as a test bed for classical and quantum field theory on curved spacetime predictions. Additionally, they provide useful toy models of emergent gravity scenarios in which we can learn some lessons about the interplay between gravitational physics and its UV completion.

One of the most significant achievements of analogue gravity experiments is the reproduction of Hawking radiation, the process by which black holes emit radiation due to quantum effects near the event horizon. By creating ``dumb holes" (acoustic analogues of black holes), researchers have been able to observe similar radiation in the laboratory. These results, together with their theoretical understanding, lent support to the robustness of Hawking's prediction against the detailed features of spacetime beyond the Planck scale.

With respect of the aforementioned black hole mimickers, analogue gravity could perhaps play a similar role. For example, an inner-horizon instability for an analogue black hole could be studied and albeit its back-reaction would be in general ruled by a different (simpler) geometrodynamics still it might provide useful insights for the gravitational problem. A similar study can be performed for extremal horizon instabilities that are often claimed to have a universal character (see e.g.~\cite{Gralla:2018xzo}).

Similarly, one could raise the question of whether UCOs could be simulated in these systems and if such analogue UCOs would show analogues of the gravitational wave echoes. If so, the corresponding experiments might provide valuable insights into the nature of these phenomena.


In summary, it seems that the potential for analogue systems to teach us more about black hole mimickers is far from being fully exploited. Although these systems do not involve actual gravitational equations, they can still mimic several interesting phenomena and provide a testing ground for theoretical ideas that would otherwise be inaccessible through direct observation.

\section{Conclusion}

The study of quantum black holes and related phenomena continues to push the boundaries of our understanding of the universe. From the regularization of singularities in black holes to the potential for horizonless ultra-compact objects, quantum gravity offers a wealth of new ideas that challenge the classical picture of black holes as simple, isolated objects.

At the heart of these debates lies the question of testability. Are stories about what happens inside black holes scientific if they cannot be tested? The same question could be asked of other speculative ideas, such as the inflationary landscape in cosmology or proposed resolutions to the information paradox. Although these ideas may seem abstract, we hope we have presented enough evidence here that they can be tested in the near future and that the quantum nature of black holes might not be out of reach forever. 

We are confident that, as observational techniques improve, particularly in the realm of gravitational wave astronomy and black hole imaging, we may finally be able to test some of these bold predictions. Until then, it is up to us to devise new ideas and methods to test them. All in all, the quest for understanding quantum black holes remains one of the most exciting challenges in modern physics.

\bibliographystyle{utphys}
\bibliography{refs}

\chapter*{Affiliation list}
\begingroup
\let\clearpage\relax
\centering
\begin{flushleft}

\noindent
$^{1}$ High Energy Physics Department, Institute for Mathematics, Astrophysics,and Particle Physics, Radboud University, Nijmegen, The Netherlands\\
$^{2}$ CP3-Origins, University of Southern Denmark, Campusvej 55, DK-5230 Odense M, Denmark\\
$^{3}$ Niels Bohr International Academy, Niels Bohr Institute, Blegdamsvej 17, 2100 Copenhagen, Denmark\\
$^{4}$ CENTRA, Departamento de F\'isica, Instituto Superior T\'ecnico – IST,
Universidade de Lisboa – UL, Avenida Rovisco Pais 1, 1049-001 Lisboa, Portugal\\
$^{5}$ Institute of Theoretical Physics, Faculty of Mathematics and Physics, Charles University, V.~Holešovičkách 2, 180 00 Prague 8, Czech Republic\\
$^{6}$ Institute for Theoretical Physics, University of Heidelberg, Philosophenweg 16, 69120 Heidelberg, Germany\\
$^{7}$Department of Physics and Astronomy, University of Waterloo, Waterloo, Ontario, N2L 3G1, Canada\\
$^{8}$Waterloo Centre for Astrophysics, University of Waterloo, Waterloo, ON, N2L 3G1, Canada\\
$^{9}$Perimeter Institute for Theoretical Physics, 31 Caroline St N, Waterloo, Ontario, N2L 2Y5, Canada\\
$^{10}$ Physics Department, Penn State, University Park, PA 16802, USA\\
$^{11}$ SISSA, Via Bonomea 265, 34136 Trieste, Italy and INFN Sezione di Trieste\\
$^{12}$ IFPU - Institute for Fundamental Physics of the Universe, Via Beirut 2, 34014 Trieste, Italy\\
$^{13}$ William H. Miller III Department of Physics and Astronomy, Johns Hopkins University, Baltimore, Maryland 21218, USA\\
$^{14}$ School of Physics and Astronomy and Institute for Gravitational Wave Astronomy, University of Birmingham, Edgbaston, Birmingham, B15 2TT, United Kingdom\\
$^{15}$ Department of Mathematics, Princeton University, Fine Hall, Washington Road, Princeton~NJ 08544, United States of America\\
$^{16}$ Department of Pure Mathematics and Mathematical
Statistics, University of Cambridge, Wilberforce Road, Cambridge CB3 0WA, United Kingdom\\
$^{17}$ Dipartimento di Fisica, Universitá di Napoli and INFN Sezione di Napoli, 
Compl. Univ. di Monte S. Angelo, Edificio G, Via Cinthia,
I-80126, Napoli, Italy\\
$^{18}$ Departamento de Matem\'aticas, Universidad Carlos III de Madrid. Avda. de la Universidad 30, 28911 Leganes, Spain\\
$^{19}$ Instituci\'o Catalana de Recerca i Estudis Avançats (ICREA), Passeig Lluis Companys, 23, 08010 Barcelona, Spain\\ 
$^{20}$ Departament de F\'isica Qu\`antica i Astrof\`isica and Institut de Ci\`encies del Cosmos, Universitat de Barcelona, 08028 Barcelona, Spain\\
$^{21}$ Department of Physics, King’s College London, University of London, Strand, London, WC2R 2LS, UK\\
$^{22}$ Departamento de Matem\'atica da Universidade de Aveiro and
Centre for Research and
Development in Mathematics and Applications (CIDMA), Campus de Santiago, 3810-193
Aveiro, Portugal\\
$^{23}$ Institute of Physics, University of Oldenburg, Oldenburg, Germany\\
$^{24}$ Department of Physics, The Ohio State University, Columbus, OH 43210, USA\\
$^{25}$ GRAPPA, Anton Pannekoek Institute for Astronomy and Institute of High-Energy Physics, University of Amsterdam, Science Park 904, 1098 XH Amsterdam, The Netherlands\\
$^{26}$ Nikhef, Science Park 105, 1098 XG Amsterdam, The Netherlands\\
$^{27}$ Dipartimento di Fisica, Sapienza Università di Roma and INFN, Sezione di Roma, Piazzale Aldo Moro 5, 00185, Roma, Italy\\
$^{28}$ Department of Physics and Princeton Gravity Initiative, Princeton University, Princeton, New Jersey 08544, USA\\
$^{29}$ Kavli Institute for the Physics and Mathematics of the Universe (WPI), The University of Tokyo, Kashiwa, Chiba 277-8583, Japan\\
$^{30}$ Center for Gravitational Physics and Quantum Information, Yukawa Institute for Theoretical Physics, Kyoto University, Kyoto 606-8502, Japan\\
$^{31}$ Leung Center for Cosmology and Particle Astrophysics, National Taiwan University, Taipei 10617, Taiwan\\
$^{32}$ Black Hole Initiative at Harvard University, 20 Garden Street, Cambridge, MA 02138, USA\\
$^{33}$ Center for Astrophysics | Harvard \& Smithsonian, 60 Garden Street, Cambridge, MA 02138, USA\\
$^{34}$ Department of Physics and Astronomy, University of British Columbia, Vancouver, BC, Canada V6T 1Z1\\
$^{35}$ Victoria University of Wellington, Wellington 6140, New Zealand\\
$^{36}$ Enrico Fermi Institute and Department of Physics, University of Chicago 933 E. 56th St. Chicago, IL 60637

\end{flushleft}
\endgroup


\end{document}